\documentclass[traditabstract]{aa} 

\usepackage{graphicx}
\usepackage{txfonts}
\usepackage{subfigure}
\usepackage{subfig}
\usepackage{amssymb}
\usepackage{natbib}
\bibpunct{(}{)}{;}{a}{}{,} 
\usepackage{url}

\begin{document}
\titlerunning {Nitrogen hydrides   in interstellar gas II}
\title{Nitrogen hydrides   in interstellar gas}
\subtitle{II. Analysis of {\emph{Herschel}}\thanks{\emph{Herschel} is an ESA space observatory with science instruments provided by European-led  Principal Investigator consortia and with important participation from NASA.}/HIFI  observations  towards W49N and G10.6$-$0.4 (W31C)}  
\authorrunning{C.M.~Persson et~al.}   
            \author{  
 \author{C.M.~Persson   
          \inst{1},
          M.~De~Luca\inst{2},
          B.~Mookerjea\inst{3}, 
          A.O.H.~Olofsson\inst{1},   
          J.H.~Black\inst{1}, 
          M.~Gerin\inst{2}, 
 E.~Herbst\inst{4}, 
   T.A.~Bell\inst{5},
A.~Coutens\inst{6,7}  
      B.~Godard\inst{5}, 
J.R.~Goicoechea\inst{5}, 
   G.E.~Hassel\inst{8}, 
           P.~Hily-Blant\inst{9},   
K.M.~Menten\inst{10},    
H.S.P.~M\"uller\inst{11}, 
J.C.~Pearson\inst{12},  
S.~Yu\inst{12}  
          }   
          }
   \offprints{carina.persson@chalmers.se}
   \institute{Chalmers University of Technology, Department of Earth and Space Sciences, Onsala Space Observatory,  SE-439 92 Onsala, Sweden.    \email{\url{carina.persson@chalmers.se}} 
\and LERMA-LRA, UMR 8112 du CNRS, Observatoire de Paris, \'Ecole Normale
Sup\'erieure, UPMC \& UCP, 24 rue Lhomond, 75231 Paris Cedex 05, France 
\and   Tata Institute of Fundamental Research, Homi Bhabha Road, Mumbai 400005, India   
\and   Department of Chemistry, University of Virginia, McCormick Road, Charlottesville, VA 22904, USA 
\and Centro de Astrobiolog\`{i}a, CSIC-INTA, 28850, Madrid, Spain 
\and Universit\'{e} de Toulouse, UPS-OMP, Toulouse, France 
\and CNRS, IRAP, 9 Av. Colonel Roche, BP 44346, F-31028 Toulouse Cedex 4, France
\and Department of Physics \& Astronomy, Siena College, Loudonville, NY  12211,   USA  
\and Laboratoire d’Astrophysique de Grenoble, UMR 5571-CNRS,  Universit´e Joseph Fourier, Grenoble, France  
\and  Max-Planck-Institut f\"ur Radioastronomie, Auf dem H\"ugel 69, D-53121 Bonn, Germany  
\and I. Physikalisches Institut, Universit\"at zu K\"oln, Z\"ulpicher Str. 77, 50937 K\"oln, Germany
\and  Jet Propulsion Laboratory, California Institute of Technology, 4800 Oak Grove Dr., Pasadena CA 91109, USA 
}

   \date{Received Dec 20, 2011 /  Accepted May 30, 2012}

  \abstract
{As a part of the \emph{Herschel} key
programme PRISMAS, we have used
the \emph{Herschel}-HIFI instrument    to observe interstellar nitrogen hydrides along the sight-lines  towards eight  high-mass 
star-forming regions   in order to elucidate the production pathways leading to nitrogen-bearing species in diffuse gas. 
Here, we report  observations towards W49N of the NH \mbox{$N$\,=\,1\,--\,0},  \mbox{$J$\,=\,2\,--\,1}, and  \mbox{$J$\,=\,1\,--\,0},   
\mbox{ortho-NH$_2$} \mbox{$N_{K_a, K_c}J$\,=\,1$_{1,1}3/2$\,--\,0$_{0,0}$1/2}, \mbox{ortho-NH$_3$} \mbox{$J_K$\,=\,1$_0$\,--\,0$_0$} and 
\mbox{2$_0$\,--\,1$_0$},  \mbox{para-NH$_3$} \mbox{$J_{K}$\,=\,2$_{1}$\,--\,1$_{1}$} transitions, and unsuccessful searches for NH$^+$. 
All detections show absorption   by foreground material 
over a wide range of velocities, as well as absorption associated directly with the hot-core source itself.
As in the previously 
published observations towards G10.6$-$0.4, 
the NH,  NH$_2$ and NH$_3$  spectra towards W49N show strikingly similar and non-saturated absorption features.
We decompose the absorption of the foreground material towards W49N  into different velocity components  
in order to investigate whether the relative abundances vary 
among the velocity components, and, in addition,  we   re-analyse the   absorption lines towards G10.6$-$0.4~in the same manner. 
Abundances, with respect to molecular hydrogen,  in each velocity component 
are estimated using CH,  
which is found to correlate with H$_2$ in the solar neighbourhood diffuse gas.  	The analysis   points to a co-existence of the nitrogen hydrides in   diffuse or translucent  interstellar gas
with a high molecular fraction. 
Towards both sources, we 
find   that NH is  always at least as abundant as both \mbox{o-NH$_2$} and  \mbox{o-NH$_3$}, in  sharp  contrast 
to previous results for dark clouds. 
We find relatively constant \mbox{$N$(NH)/$N$(o-NH$_3$)} and \mbox{$N$(o-NH$_2$)/$N$(o-NH$_3$)} ratios 
with mean values   of 3.2 and 1.9 towards W49N, and 5.4 and 2.2 towards G10.6$-$0.4, respectively. 
The    mean abundance of o-NH$_3$ is  $\sim$2$\times$10$^{-9}$  towards both sources. 
The nitrogen hydrides also show  linear correlations with    CN and HNC towards both sources, and 
looser correlations
with CH. 
The upper limits on the  NH$^+$ abundance  indicate  column densities   \mbox{$\lesssim$2\,--\,14\,\%} 
of $N$(NH), which is in contrast to 
the behaviour of the abundances of CH$^+$ and 
OH$^+$ relative to the values determined for  the corresponding neutrals CH and OH.     
Surprisingly low values of the ammonia  ortho-to-para  ratio 
are found in both sources,  \mbox{$\approx$\,0.5\,--\,0.7$\pm$0.1}, in the strongest absorption components.  This result cannot be explained 
by current models     
as we had expected to find a value of unity or higher.  
}


   \maketitle
%

\section{Introduction}
 
Nitrogen is among the six most
abundant elements in the universe and, despite its fundamental role 
in the   chemistry of  molecules connected with life, the chemical network of
nitrogen in the interstellar medium is still poorly
understood due to severe difficulties to observe key
molecules from the ground.
Today  about 55 molecules containing nitrogen have been discovered in interstellar space 
and a few more in circumstellar envelopes\footnote{www.astrochymist.org, www.astro.uni-koeln.de/cdms/molecules}. 
The major reservoir of nitrogen is believed to be in the form of atomic N and molecular N$_2$. The latter is extremely 
difficult to observe since it has no permanent dipole moment \citep{2001ApJ...548..836S}. 
In dense molecular clouds
N$_2$H$^+$ is instead often used as a tracer of  N$_2$, which 
remained undetected until   \citet{2004Natur.429..636K} reported  far-ultraviolet 
observations towards HD\,124314. 
The total abundance of nitrogen therefore still relies, to a high degree,   on chemical modelling and observations of   
nitrogen-bearing species other than  N$_2$.

In order to
constrain the nitrogen formation pathways, observations of 
\emph{nitrogen hydrides} are crucial since they are at the root of the nitrogen chemical network, appearing in 
its first steps in chains of reactions that lead to other more complex species. 
The  abundances of these species are thus key diagnostics for the nitrogen chemistry.
The nitrogen hydrides are, however, also problematic to observe since their   ground state  rotational transitions   lie at sub-mm 
wavelengths and are thus very difficult, or impossible,  to reach from the ground.
Key species, such as  imidogen (NH)  and amidogen (NH$_2$), have therefore previously not been widely observed,   and there 
is still no detection of the   NH$^+$ radical.

Although very few observations exist of 
NH and  NH$_2$~in interstellar space, they 
are well known  
in comets 
\citep[e.g.][]{1941ApJ....94..320S, 1998Icar..136..268M, 1993ApJ...404..348F}, and have been observed  in stellar photospheres 
\citep[e.g.][]{1969PASP...81..657S, 1989hra1.book.....F}
via their electronic, vibration-rotation, and high rotational transitions. 
The first detection of 
interstellar NH was made by \citet{1991ApJ...376L..49M}
 by optical absorption spectroscopy.  
Subsequent 
observations  by \citet{1997MNRAS.291L..53C} and \citet{2009MNRAS.400..392W}
have yielded 
several lines of sight in  diffuse and translucent gas where column densities of NH, CH, CN, and H$_2$  
have been directly measured. 
The average value of the column density ratio in these diffuse and translucent sight-lines
is $N$(NH)/$N$(H$_2$)\,=\,3$\times$10$^{-9}$.

Interstellar  NH$_2$ was first detected by \citet{1993ApJ...416L..83V}  in 
absorption towards Sgr\,B2 in three fine-structure components of the 
para-symmetry form of NH$_2$, the  \mbox{$N_{K_a, K_c}$\,=\,$1_{1,0}-1_{0,1}$} 
transition, with partially resolved hyperfine 
structure at frequencies 461 to 469~GHz. 
The Infrared Space Observatory (ISO) was later used to observe   unresolved 
absorption lines of both 
ortho and para symmetry forms of  NH$_2$, as well as NH, towards this source  through the use of 
the long-wavelength spectrometer  
\citep{2000ApJ...534L.199C, 2004ApJ...600..214G, 2007MNRAS.377.1122P}.

In contrast to NH and  NH$_2$, ammonia (NH$_3$) has been extensively observed for more than 40 years and
was in fact the first polyatomic molecule to be identified in interstellar space 
\citep{1968PhRvL..21.1701C} 	 by means of its $J\!=\!K$ inversion transitions at cm wavelengths 
($K$ is the quantum number of the projection of total
angular momentum $J$ on the molecule's symmetry axis). 
The ammonia molecule has, however, two symmetries, like  NH$_2$,  which arise due to the 
possible orientations of the hydrogen spins and behave like two
distinct species: \mbox{ortho-NH$_3$} (all H spins parallel, $K$\,=\,3$n$ where $n$ is an integer 
$\geq 0$) and 
\mbox{para-NH$_3$} (not all H spins parallel, $K\neq3n$). 
Important information about the ammonia formation pathways could   be inferred 
from observations of both symmetries. 
Unfortunately,  only the para form has relatively low-excitation transitions accessible from ground since 
the  $K$\,=\,0 ladder of energy levels has no inversion splitting and the $J_K\!=\!3_3$ inversion 
transitions' lower energy level is 122\,K above ground. 
Ortho inversion lines ($K$\,=\,3$n$, $n \ge 1$) can thus only be observed in relatively warm molecular gas.  
Only a few ammonia observations of  the cold diffuse interstellar gas exist, using  
para inversion lines   
\citep{1990ApJS...72..303N,1994A&A...289..579T, 2006A&A...448..253L}   which leaves the ammonia formation mechanism poorly constrained in diffuse gas.  
The
(0,0) ammonia ground state, with ortho symmetry,   
can  only be studied by \mbox{sub-mm} rotational transitions.
Observations of the fundamental rotational transition of 
\mbox{ortho-NH$_3$} \mbox{$J_K$\,=\,1$_0$\,--\,0$_0$} at 572~GHz, which 
has similar upper state energy as the inversion lines, but
several orders of magnitudes higher critical densities,  thus  requires space telescopes. The Kuiper Airborne Observatory 
\citep{1983ApJ...271L..27K} performed the first observations of the \mbox{$J_K$\,=\,1$_0$\,--\,0$_0$} transition using heterodyne receivers, 
and later on the Odin satellite continued such observations 
in a number of environments,  for instance in 
photo-dissociation  and star-forming regions \citep[e.g.][]{2003A&A...402L..69L,  2009A&A...494..637P}, circumstellar 
envelopes \citep{2006ApJ...637..791H},  diffuse clouds in absorption towards Sgr\,B2  \citep{2010A&A...522A..19W},  and    in comets     \citep{2007P&SS...55.1058B}.

 \begin{table*}[\!htb] 
\centering
\caption{Observed transitions towards W49N and G10.6$-$0.4.
}
\begin{tabular} {lrccrrrllcc} 
 \hline\hline
     \noalign{\smallskip}
Species 	& Frequency\tablefootmark{a}	& Transition &	Band\tablefootmark{b} & $T_\mathrm{sys}$\tablefootmark{c}  &
  \multicolumn{2}{c}{$t{_\mathrm{int}}$\tablefootmark{d}}   	& \multicolumn{2}{c} {$T_\mathrm{C}$\tablefootmark{e}}   
 &\multicolumn{2}{c}{$1\sigma$/$T_\mathrm{C}$\tablefootmark{f}}    \\    \noalign{\smallskip}
&&& && G10.6 & W49N & G10.6 & W49N  &  G10.6 & W49N   \\
& (GHz) &&  & (K) & (s)&  (s)& (K) & (K) &     \\
     \noalign{\smallskip}
     \hline
\noalign{\smallskip}  

 NH\tablefootmark{g}   	 
   &	946.476	& $N_J\!=\!1_0\leftarrow0_1$ &	  3b	&  416&  234 &  116& 2.5 &3.4     &  0.021 &  0.027\\
   &974.478	&$N_J\!=\!1_2\leftarrow0_1$      &	4a & 338 	 &186   & 103	&2.6& 3.9	 	&  0.018  &  0.018   \\
o-NH$_2$\tablefootmark{h}	&	952.578	&  $N_{K_a,K_c} J = 1_{1,1} 3/2 \gets 0_{0,0} 1/2$           &	3b& 230  &92 &68& 2.6  &  3.6	 	&0.018  &  0.017  \\

o-NH$_3$\tablefootmark{i}	&	572.498&  $J_K$\,=\,1$_0 \leftarrow 0_0$		& 1b  & 87 & 1\,024  &94& 0.61  &  0.93  		 	&  0.013  &   0.025 \\

&	1\,214.859    & $J_{K}$\,=\,2$_{0} \leftarrow 1_{0}$   & 5a  & 1\,024   & 293 & 196 & 3.4 &  5.2  	  & 0.032  &  0.025 \\
p-NH$_3$	 	&	1\,215.246   &$J_{K}$\,=\, 2$_{1}  \leftarrow 1_{1}$     &5a & 1\,024     &293&  196&  3.4  & 5.2    	 &0.032  &  0.025 \\

 NH$^{+}$ & 	1\,012.540&   $^2\Pi_{1/2}$ $N$\,=\,1$\leftarrow$1  $J$\,=\,3/2\,$\leftarrow$\,1/2 	& 4a &  327	 &171& 91& 3.0 & 4.4    &0.013& 0.013  \\   

 
    \noalign{\smallskip}
\hline 
\label{Table: transitions}
\end{tabular}
\tablefoot{
\tablefoottext{a}{The frequencies refer  to the strongest hyperfine structure components.} \tablefoottext{b}{HIFI consists of 7 different 
mixer bands and two  double sideband spectrometers. All transitions were observed in the upper sideband except NH$^+$.} 
\tablefoottext{c}{System 
temperature.} 
\tablefoottext{d}{The averaged on-source integration  time for each transition.} 
\tablefoottext{e}{The single sideband (SSB) continuum intensity (the observed double sideband (DSB) continuum 
divided by two).}  
\tablefoottext{f}{The rms noise, at a resolution of 1.1\,MHz, divided by $T_\mathrm{C}$.} 
\tablefoottext{g}{The quantum numbers for the rotational 
transition $N_J$ are $F_1$\,=\,$I_H$\,+\,$J$ and $F$\,=\,$I_N$\,+\,$F_N$ \citep{1997A&A...322L...1K}.}  
\tablefoottext{h}{The quantum numbers 
for the rotational transition $N_{K_a, K_c}J$ are $F_1$\,=\,$I_N$\,+\,$J$ and $F$\,=\,$I_H$\,+\,$F_1$ 
\citep{1999JMoSp.195..177M}.} 
\tablefoottext{i}{The rotational energy is given by the two principal quantum numbers  ($J, K$), 
corresponding to the total angular momentum and its projection along the molecular axis \citep[e.g.][]{1983ARA&A..21..239H, 1993JChPh..98.4662R}.  
} 
}
\end{table*}

 
With the launch  of the \emph{Herschel}
Space Observatory   \citep{Pilbratt2010} in May 2009,    unique opportunities to perform observations of  
transitions between 157 and 625\,$\mu$m (0.48\,--\,1.9\,THz)  became feasible  with the Heterodyne Instrument for the Far-Infrared 
\citep[HIFI;][]{Graauw2010} 
owing to its very high sensitivity and spectral resolution. This allowed, for the first time, searches for spectrally resolved, 
rotational  transitions involving the ground states of  NH$^+$, NH,  NH$_2$, 
and NH$_3$~with the same instrument.  
Several  observations   have already  been reported. For instance, using  \emph{Herschel}-HIFI, 
\citet{2010A&A...521L..42B} found very high column densities of NH and ND in the cold envelope of the class\,0  protostar 
IRAS16293-2422  
(2$\times$10$^{14}$\,cm$^{-2}$~and $\sim$1.3$\times$10$^{14}$\,cm$^{-2}$, respectively).
\citet{2010A&A...521L..52H} found  NH:NH$_2$:NH$_3$ abundance ratios  of $\sim$5:1:300 towards the same source.

The PRISMAS\footnote{http://astro.ens.fr/?PRISMAS}  key programme (PRobing InterStellar Molecules with Absorption line Studies) is targeting absorption lines in the 
line-of-sight towards eight bright sub-millimetre-wave continuum sources using \emph{Herschel}-HIFI: G10.6$-$0.4~(W31C), W49N, W51, G34.3+0.1, DR21(OH), SgrA (+ 50~km~s$^{-1}$ cloud), G005.9-0.4 (W28A) and W33A.   High-resolution absorption 
line spectroscopy is generally a very sensitive  and model-independent method for measuring column densities of interstellar molecules,   
and a powerful tool to probe the diffuse interstellar gas clouds with no or little excitation.

The first results and analysis of absorption lines of nitrogen hydrides along the sight-line towards the massive star-forming 
region G10.6$-$0.4~(W31C) 
have already been presented in 
\citet[][hereafter paper~I]{2010A&A...521L..45P}. 
Similar abundances with respect to the total amount of hydrogen $N_\mathrm{H}$\,=\,2\,$N(\mathrm{H_2}$)+$N(\mathrm{H}$),   were found for all three species:   approximately  
$6\times 10^{-9}$,  $3\times 10^{-9}$, and $3\times 10^{-9}$ for NH, NH$_2$, and NH$_3$, respectively. They were estimated across the whole line-of-sight and using the high temperature ortho-to-para limits of 
three and one for NH$_2$ and NH$_3$, respectively.    NH$^+$ was  
not detected at a 1$\sigma$ rms level of 74\,mK with a resolution of 1.1~MHz. 
The abundance patterns that we see in diffuse molecular gas are thus clearly very different 
from those in IRAS16293-2422 and Sgr~B2, where NH:NH$_2$:NH$_3\sim$1:10:100 and the fractional abundance
of NH is a few times 10$^{-9}$. The Sgr~B2 results may, however, 
not be representative of 
cold dark clouds since  this source is very   complex and  atypical.

The 
unexpectedly high NH abundance 
has been difficult to explain with chemical models and 
both the NH and  NH$_2$ production 
by purely gas-phase
processes  is inhibited by a lack of a sufficient source of N$^+$. 
The models  
fail to   simultaneously  predict the absolute and   relative abundances of the nitrogen hydrides. 
Typical steady state  \emph{dark cloud} models 
($n$\,=\,1$\times$10$^{3}$\,--\,5$\times$10$^{4}$~cm$^{-3}$, 
$T$\,=\,10\,--\,40~K, $A_\mathrm{V}\gtrsim10$),   predict an NH$_2$ abundance of 
\mbox{(1-10)$\times$10$^{-8}$}, an	 NH abundance 10 times lower, 
and an NH$_2$/NH$_3$~ratio of  \mbox{0.3\,--\,1.5} for a wide range of assumptions  
\citep[e.g.][]{1989ApJS...69..241L,1991A&AS...87..585M}, 
but these are not directly 
applicable
to diffuse molecular gas. Examples of chemical models for diffuse cloud conditions are found in 
Fig.~A.1 and A.2 in paper~I. These models were also unable to explain the observed 
abundances and ratios.   
Processes on dust  grains 
have   previously been proposed as a way to increase the NH production  
\citep{1991ApJ...376L..49M, 1993MNRAS.260..420W}. 
Such models, however, often predict up to 1\,000 times more NH$_3$ than NH$_2$ \citep[][paper~I]{1993MNRAS.263..589H}.   
The importance of grain surface chemistry in diffuse clouds is also not clear 
since 
water ice mantles have not been
detected in diffuse gas, and  strong (UV) radiation fields counteract molecular 
formation on grains. 
Grain surface production of NH is therefore still debated, and would, if   true,   change our understanding
of surface  chemistry in diffuse gas. 
On the other hand, 
if  grains indeed were unimportant in diffuse gas nitrogen chemistry it would imply 
 that  either 
key gas-phase reactions must  have been overlooked,   or that the uncertainty of some 
reaction rates could make a difference.
Both   
additional high-quality observations and chemical modelling are needed to solve this problem.
 
In this paper, we present new observations and  analyses of absorption lines in the line-of-sight 
towards the high-mass star-forming region  W49N, and, we also  
re-analyse the absorption towards G10.6$-$0.4~in more detail.  
W49N is one of the  most  luminous high-mass star-forming regions in the Galaxy ($\sim$10$^7$\,L$_\odot$) with a core that 
contains more than a dozen ultra-compact \ion{H}{II} regions 
\citep{1984ApJ...283..632D, 1990ApJ...351..189D, 2000ApJ...540..308D}. 
It is located on the far side of the Galaxy at a distance of 11.4\,kpc with Galactic coordinates $l$\,=\,43.17$^\circ$ and 
$b$\,=\,0.012$^\circ$,
in one of the most
massive giant molecular clouds ($\sim$10$^6$\,M$_\odot$) in the Milky Way. The source velocity is about +8\,km\,s$^{-1}$~and the 
foreground gas  along the
line-of-sight is detected at  \mbox{$v_\mathrm{LSR}\approx 30-75$\,km\,s$^{-1}$}, revealing gas at two locations 
in the near and far side of the Sagittarius spiral arm 
around 40 and 60\,km\,s$^{-1}$~\citep{1985ApJ...297..751D}.
The  ultra-compact \ion{H}{II} region
G10.6$-$0.4~in the star-forming W31 complex  is an extremely luminous sub-millimetre and infrared continuum source.
The source is located within the so-called 30\,km\,s$^{-1}$~arm at a kinematic distance of 4.8\,kpc  
\citep{2003ApJ...587..701F}. 
The gas associated directly with G10.6$-$0.4~is detected at a systemic source velocity of  
\mbox{$v_\mathrm{LSR}\approx -1$\,km\,s$^{-1}$}, determined from OH maser emission observations, while the foreground gas is 
detected at \mbox{$v_\mathrm{LSR}\approx 10-55$\,km\,s$^{-1}$}.

Section~\ref{observations} summarises the observations and data reduction,
and
the results from our \emph{Herschel} observations are found in Sect.~\ref{section: results}.
The hyperfine structure (hfs) components of the nitrogen hydrides are discussed in Sect.~4. 
In Sect.~\ref{Section: Analysis of abundances in different velocity components} we use three different methods to decompose the 
absorption lines   along the sight-lines towards both sources    in different velocity components, 
and  estimate  column densities, $N$,  and  relative abundances, $X$,   in each   component. 
We also compare the column densities of the nitrogen hydrides, both with each other to investigate possible correlations, and with other species tracing regions with  both low and high molecular fractions. 
Section~\ref{OPR ammonia} presents our estimates of the ortho-to-para ratio (OPR) of NH$_3$. 
We end this paper with a summary and outlook in Sect.~\ref{section summary}.
Note that the analysis of the \emph{background} source emissions and absorptions    is left for a future paper.

 \section{Observations and data reduction} \label{observations}
All the reported observations are a part of the more extended PRISMAS programme towards G10.6$-$0.4~and  W49N. 
The nitrogen observations, which took place between March~2 and April~18, 2010, are summarised in Table~\ref{Table: transitions}.  
All \emph{Herschel} identification numbers (OBSID's) are found in 
Table~\ref{Table: obsid} (on-line material).   
Note that  NH \mbox{$N_J$\,=\,1$_0$\,--\,0$_1$} was observed together with CH$_2$ at 946~GHz, and 
ortho-NH$_3$ \mbox{$J_K$\,=\,2$_0$\,--\,1$_0$} was observed  together with   
para-NH$_3$ \mbox{$J_K$\,=\,2$_1$\,--\,1$_1$} (the absorptions from the source are 
97~km~s$^{-1}$ apart).     
Before the launch of \emph{Herschel}, no observations of these NH$_3$ transitions,  
 nor the  NH$^+$ \mbox{$^2\Pi_{1/2}$  $J$\,=\,3/2\,$\leftarrow$\,1/2}  transition, had been performed. 
 We do not, however, analyse the \mbox{ortho-NH$_3$}  \mbox{$J_K$\,=\,2$_0$\,--\,1$_0$} transition in this paper since
this line   shows absorption only at the background source velocities.

We used the
dual beam switch mode and the wide band spectrometer (WBS) with a bandwidth of 4$\times$1~GHz and an effective spectral  
resolution of 1.1~MHz. 
The corresponding  velocity resolution 
is about 0.3~km\,s$^{-1}$~at the highest frequencies and 0.6~km~s$^{-1}$~at 572~GHz. In addition, simultaneous observations were 
performed using the high 
resolution spectrometer (HRS) with an effective spectral resolution of 0.25\,MHz \mbox{($\Delta v\!\sim\!0.1$~km~s$^{-1}$)}  
and a bandwidth of 240\,--\,340~MHz except for 
the 1\,214 and 1\,215~GHz lines which had   an effective resolution of 0.5~MHz 
\mbox{($\Delta v\!\sim\!0.1$\,km\,s$^{-1}$)}   
and a bandwidth of 780~MHz in order to cover both lines in the same band.   
Note that the channel separation of the observations is 0.5~MHz in the WBS and 
0.12~MHz in the HRS   (0.24~MHz for 
the  1\,215~GHz line).
Each line was observed with three different frequency settings of the local oscillator (LO) 
corresponding to a change of approximately 15~km~s$^{-1}$~to
determine the sideband origin
of the lines. During all observations two
orthogonal polarisations  were used.

The pointings were centred at   $\alpha$\,=\,19$^\mathrm{h}$10$^\mathrm{m}$13\,$\fs$2, 
$\delta$\,=\,09$^\circ$\,06$\arcmin$\,12$\arcsec$ ($J$2000) for  W49N, and 
 $\alpha$\,=\,18$^\mathrm{h}$10$^\mathrm{m}$28\,$\fs$7, 
$\delta$\,=\,$-$19$^\circ$\,55$\arcmin$\,5$\arcsec$ ($J$2000) 
for G10.6$-$0.4. 
The recommended values for half-power beam width of the telescope are  
$44\farcs 2$, $22\farcs 1$, and $18\farcs 9$~at 480, 960 and 1\,120~GHz, respectively.  
The 
reference beams were located within 3\arcmin~on either side of the source. 
The total calibration uncertainties are $\lesssim$9\% for band~1 and $\lesssim$13\% for 
band~5, including the sideband gain 
ratio uncertainty which is 3\,--\,4\% for band~1b (ortho-NH$_3$), and 4\,--\,6\% for band~5a 
(para-NH$_3$). All errors are added in quadrature.   
Detailed information about 
the HIFI calibration including  beam efficiency, mixer sideband ratio, pointing,  
etc., can be  found  
on the Herschel internet
 site\footnote{http://herschel.esac.esa.int/Docs/HIFI/html/ch5.html}.
The in-flight performance is  
described by \citet{2012A&A...537A..17R}. Since we are interested in absorption 
lines and their relative  
strength compared  to the continuum in this paper, the forward efficiency of 96\% 
and the 
main beam-efficiency of 64\,--\,76\% (between 1\,120 and 480~GHz)  do
 not affect our results and we 
have therefore used a   $T_\mathrm{A}^*$  intensity scale throughout this paper.

The data were processed using the standard \emph{Herschel} Interactive Processing Environment  
  \citep[HIPE,\footnote{HIPE is a joint development by the \emph{Herschel} Science Ground
Segment Consortium, consisting of ESA, the NASA \emph{Herschel} Science Centre, and the HIFI, PACS and
SPIRE consortia.}][]{2010ASPC..434..139O},
version 5.1,   up to  level~2 which provides fully calibrated spectra.  
The data quality is excellent with very low intensity ripples in most cases, typically below a few percent
of the double sideband continuum. For NH$^+$ in both sources and both 
polarisations, and \mbox{o-NH$_3$} at 572~GHz in the vertical polarisation towards G10.6$-$0.4, 
we corrected the data  from ripples using the \mbox{\emph{fitHifiFringe}} tool in HIPE  
(in paper~I we only used one polarisation for 
these two 
transitions). 
The FITS  files were then exported to  the spectral line   software package  {\tt xs}\footnote{Developed by 
Per Bergman at Onsala Space Observatory, Sweden; {\tt http://www.chalmers.se/rss/oso-en/observations/data-\\reduction-software}} 
which was used in the subsequent data reduction. 

The two polarisations  are generally in agreement to better than 10\%.
The three LO-tunings  
are also  in  very good agreements 
without contamination from the image sideband with one exception  
in  W49N around the   \mbox{o-NH$_3$} at 572~GHz line 
(see Sect.~\ref{subsubsection: SO2 in NH3}).  
In all other spectra we 
averaged the three LO-tunings and both polarisations.

For the above line identification work we  used the Cologne Database for Molecular Spectroscopy\footnote{\tt{http://www.cdms.de}} 
\citep[CDMS,][]{2005JMoSt.742..215M}, the  Jet Propulsion Laboratory catalogue  \footnote{\tt{http://spec.jpl.nasa.gov}} 
\citep[JPL,][]{1998JQSRT..60..883P},  and  Frank  
Lovas'  Spectral Line Atlas of Interstellar Molecules\footnote{\tt{http://physics.nist.gov/}} (SLAIM).  
Laboratory measurements of NH transitions have been performed by  
e.g. 
\citet{1997A&A...322L...1K} and  \citet{2004JMoSp.225..189F}, and measurements of 
o-NH$_2$~by e.g. \citet{1999JMoSp.195..177M}. 
A review on ammonia is found in \citet{1983ARA&A..21..239H}, and an investigation 
of the hyperfine structure components of ortho-NH$_3$~at 
572~GHz was recently performed by 
 \citet{2009A&A...507.1707C}.
NH$^+$ measurements have been made by \citet{1986CPL...132..213V} and 
\citet{2009JChPh.131c4311H}.
 
 \begin{figure} 
\includegraphics[scale=0.61]{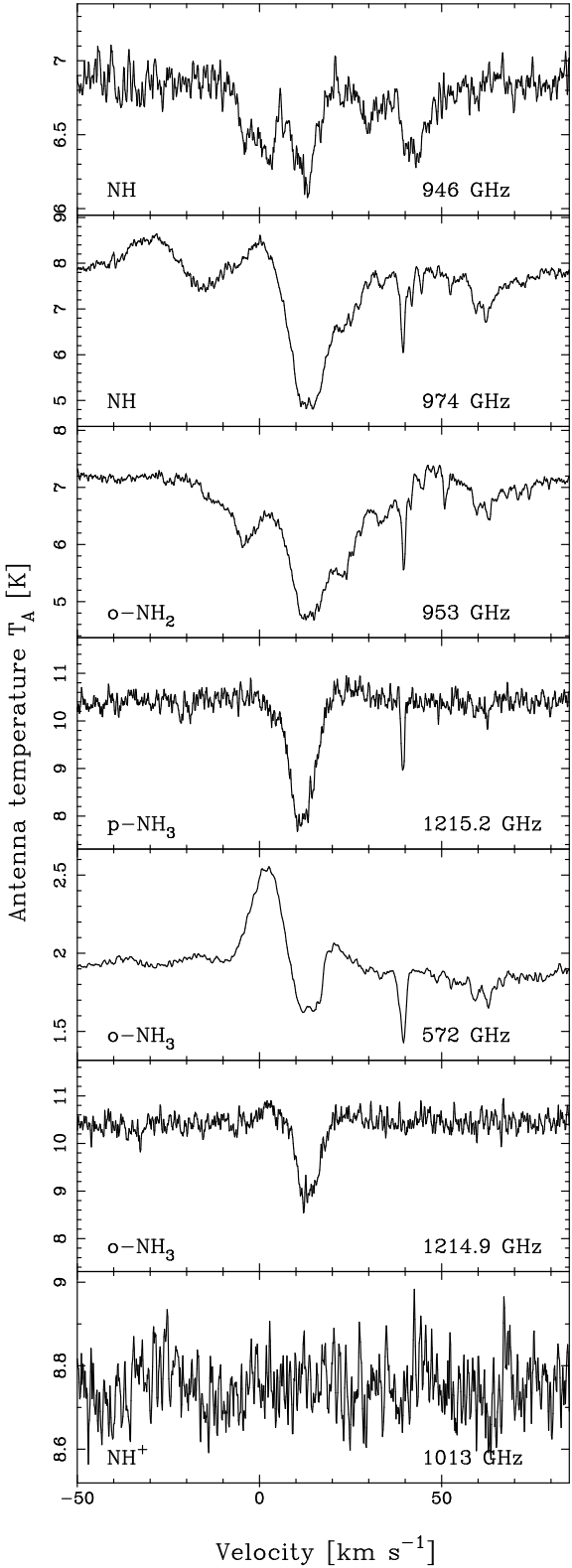} 
\caption{\emph{Nitrogen hydrides towards W49N}: Double sideband  WBS spectra  of  NH, ortho-NH$_2$,  ortho- and para-NH$_3$,   
and NH$^+$  
over the LSR velocity range -50 to 85~km~s$^{-1}$. Quantum numbers are found in Table~\ref{Table: transitions}. 
Note that we only analyse the absorption in the velocity range 
30\,--\,75~km~s$^{-1}$ in this paper and leave the source absorption and emission for a future paper.} 
 \label{Fig: DSB W49N hydrides}
\end{figure}

 \section{Observational results} \label{section: results}

Figures~\ref{Fig: DSB W49N hydrides} and \ref{Fig: W31C All original hydrides} (on-line material) 
show the  averaged double sideband
WBS spectra of the transitions listed in Table~\ref{Table: transitions} 
towards    
W49N and  G10.6$-$0.4, respectively (special treatment of o-NH$_3$ 
1$_0$\,--\,0$_0$ in W49N is found in Sect.~\ref{subsubsection: SO2 in NH3}).    
All detected species show very similar   absorption patterns over a wide range 
of velocities,  
although   W49N has fewer   absorbing velocity components in the line-of-sight. 
No detections of NH$^+$ have been found in any of the two sources.

The resulting noise and  single sideband (SSB) continuum levels, for each transition, 
are   found in Table~\ref{Table: transitions}.
Note that since HIFI uses double side band (DSB) receivers, the observed continuum has to be divided by two to 
obtain the SSB continuum.
The sideband gain ratio is   assumed to be unity throughout this paper. 
This has proven to be a good assumption based on PRISMAS observations of saturated absorption lines like HF that establish the true zero levels 
\citep{2010A&A...518L.108N}.

\subsection{Background source emission} 

Even though we only analyse the absorption lines in the foreground clouds in this paper 
we still have to take the  source emission into account in the analysis. 
Several spectra show broad emission features from the background sources 
which  
extend far into the velocity range
pertaining to the line-of-sight clouds.

Ammonia in particular   shows strong emission at the background 
source velocities with self-absorption 
around +13 and 
0\,km\,s$^{-1}$~for W49N and G10.6$-$0.4, respectively.  
A large part of the NH  974~GHz emission at the source velocities in both sources is most likely  
not  from NH, but   
from HCN \mbox{$J$\,=\,11\,--\,10} at 974.487~GHz,  only 
9\,MHz above the NH line  
corresponding to 2.5~km~s$^{-1}$. The PRISMAS observations of these sources have 
detected three additional HCN emission lines:  
\mbox{$J$\,=\,6\,--\,5} 
 at 532\,GHz,  
\mbox{$J$\,=\,7\,--\,6} at 620~GHz, and  \mbox{$J$\,=\,10\,--\,9} at 886~GHz which 
strengthens  our identification. 
In addition, we detect no NH emission
in the 946~GHz transition which is a further confirmation that
the emission is due to another species. 
HCN in foreground gas is unlikely to produce measurable absorption in the 
$J=11 - 10$ transition because the excitation energy of the lower state, 
$E_{10}/k$\,=\,234~K, is much too high to be populated at the density and 
temperature of diffuse or translucent gas.

In the analysis  performed in
Sect.~\ref{Section: Analysis of abundances in different velocity components} 
using three different methods, 
we  
have removed  
the broad emission lines that overlap with the absorption in the line-of-sight 
by means of Gaussian fits.

\subsubsection{Emission line contamination}\label{subsubsection: SO2 in NH3}  
 
In  W49N, close to the   o-NH$_3$ line at 572~GHz,    
an emission line  from the lower sideband, 
identified as 
the \mbox{$J_{K_a, K_c}$\,=\,16$_{6,10}$\,--\,16$_{5,11}$} SO$_2$ line, 
moves across the ammonia absorption lines in the different LO-tunings.    
Thus for the   o-NH$_3$ 572~GHz transition in W49N we were only able to use the A and 
C LO-settings in Table~\ref{Table: obsid}, which also had to be cut to 
remove the parts contaminated by the SO$_2$ line.

The o-NH$_2$~absorption  towards  W49N is contaminated by a an 
emission line in the same sideband from the source.
We identify the emission line as a blend of three unresolved 
hfs components of 
NO $^2\Pi_{1/2}\;J=9.5^f\to 8.5^f$, $F=10.5-9.5$, $9.5-8.5$, and $8.5-9.5$, 
at 952.464201~GHz.  
The NO emission line is removed  from the o-NH$_2$~absorption 
(described in Sect.~\ref{NO removal of NH2}, on-line material), 
and in all following figures in  this paper  we 
show o-NH$_2$ in  W49N  with removed   NO emission. 
 The NO identification is strengthened by the PRISMAS observations of 
three additional NO lines towards W49N  shown in Fig.~\ref{Fig: W49N NO} (on-line material). 
No emission of NO is found towards G10.6$-$0.4.

     \begin{figure}[\!ht] 
\centering
\includegraphics[scale = 0.5]{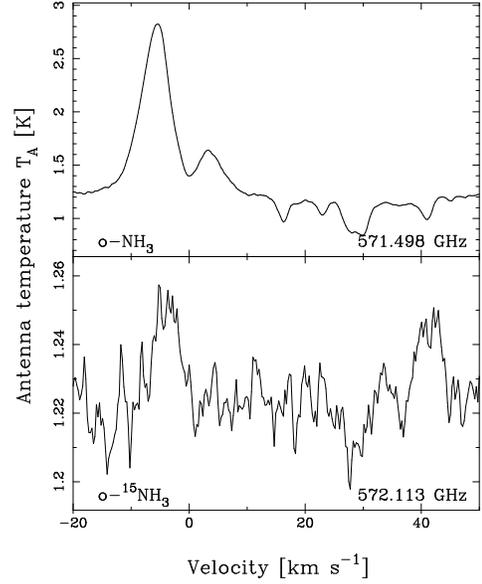}
 \caption{\emph{W31C: \mbox{o-$^{14}$NH$_3$} and  \mbox{o-$^{15}$NH$_3$}.} Both isotopologues 
show emission from the source. Absorption  
in the line-of-sight
is  only seen in the o-$^{14}$NH$_3$~spectra.
The emission line at \mbox{$v_\mathrm{LSR}\!\approx\!41$\,km\,s$^{-1}$}~in 
the o-$^{15}$NH$_3$~spectra,
is an SO$_2$ line from the upper sideband.}
 \label{Fig: W31C NH3 and 15NH3}
\end{figure}

\subsubsection{The $^{15}$NH$_3$ isotopologue: only  in G10.6$-$0.4}

Since our observations also covered the frequency of  \mbox{o-$^{15}$NH$_3$}  \mbox{$J_K$\,=\,1$_0$\,--\,0$_0$} at 572.113~GHz, we have
checked the averaged spectra for emission and/or absorption features of this isotopologue.   
In Figs.~\ref{Fig: W31C NH3 and 15NH3} and   \ref{Fig: W31C NH3 and 15NH3 normalised}  
(on-line material) we  show 
a  5$\sigma$  emission  feature in G10.6$-$0.4 with  
an amplitude of 40~mK and a line-width of about 4~km~s$^{-1}$, which we identify  
as \mbox{o-$^{15}$NH$_3$}  \mbox{1$_0$\,--\,0$_0$,}  
giving a source velocity of $\sim -$4~km~s$^{-1}$. 
Checking the possibility of absorption lines of  o-$^{15}$NH$_3$~in the line-of-sight towards  
G10.6$-$0.4, we note that there seems to be an absorption feature  
corresponding to
the  
strongest velocity components of  \mbox{o-$^{14}$NH$_3$} at $\sim$27\,--\,32~km~s$^{-1}$.
This feature 
is, however,  
 probably not a real detection     
since it would otherwise imply that \mbox{$^{14}$N/$^{15}$N\,$\sim$\,20.} 
This  is about 20 times lower than 
the   $^{14}$N/$^{15}$N ratio in the local ISM which is about $\sim$450 
\citep{1994ARA&A..32..191W}, or 
441$\pm$5 in the Solar nebula \citep{2011Sci...332.1533M}.  
Odin observations of absorption towards Sgr~B2 show 
$^{14}$NH$_3$/$^{15}$NH$_3 > 600$ \citep{2010A&A...522A..19W}.  
Lower ratios
have been found, for instance, in 
  low mass cores, 334$\pm$50 \citep{2010ApJ...710L..49L}, and on Earth, $\sim$270.  
\citet{2011IAUS..280P..76A} find values of \mbox{$\sim$100\,--\,350} using CN and HNC in ten  
dense molecular clouds located at various galactic distances from the Galactic Centre,  
 but this is still
several times higher than inferred from a possible  o-$^{15}$NH$_3$~absorption. 
The $^{14}$N/$^{15}$N ratio is also difficult to determine and the results using cyanides  may differ from the  
ratio inferred from other species such as N$_2$H$^+$, and may not reflect the true $^{14}$N/$^{15}$N 
abundance ratio.   
More observations are needed to lower the noise in order to 
obtain a real detection of \mbox{$^{15}$NH$_3$} and to   
determine 
the true $^{14}$N/$^{15}$N ratio towards G10.6$-$0.4.

In W49N, no o-$^{15}$NH$_3$~detection is made due to blending SO$_2$ lines from the upper sideband which are both stronger and broader 
than in G10.6$-$0.4. In addition, the  o-$^{14}$NH$_3$~emission is more than two times weaker than in G10.6$-$0.4, and the spectrum 
has   a three times higher noise level which does not allow a detection of   o-$^{15}$NH$_3$ 
emission from the source.

\section{Hyperfine structure components}\label{section: hfs}

\begin{figure} 
   \centering
\includegraphics[scale=0.43]{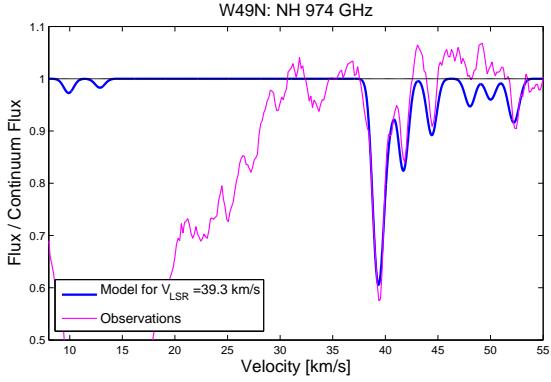} 
\caption{\emph{NH hfs components}: Normalised   WBS spectrum towards W49N. Also shown are Gaussian fits with a line width of 1~km~s$^{-1}$~of 
the NH 974~GHz hfs components in the +39~km~s$^{-1}$~velocity component.} 
 \label{Fig: NH hfs lines in 39 kms in W49N}
\end{figure}

\begin{figure}
    \centering
\includegraphics[scale=0.43]{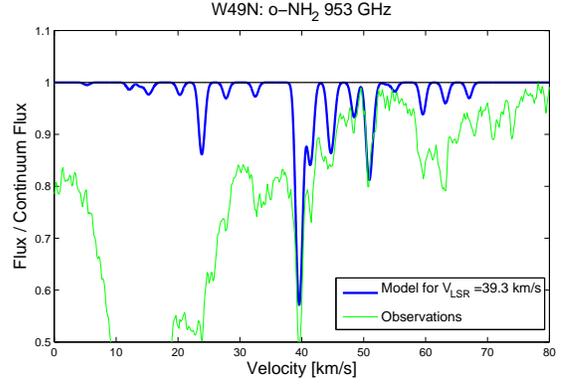} 
\caption{\emph{Ortho-NH$_2$~hfs components}: Normalised  WBS spectrum towards W49N. Also shown are Gaussian fits with a line width of 1~km~s$^{-1}$~of 
the  o-NH$_2$~953~GHz hfs components in the +39~km~s$^{-1}$~velocity component.} 
 \label{Fig: NH2 hfs lines in 39 kms in W49N}
\end{figure}

Both NH and NH$_2$ have numerous hyperfine structure (hfs) components which require  a high resolution spectrometer, 
such as HIFI, 
to be spectrally resolved. 
The on-line Tables~\ref{Table: NH hfs transitions}\,--\,\ref{Table: NH2 hfs transitions} list 
the   hfs components of the NH 974 and 946~GHz, and the \mbox{o-NH$_2$} 953~GHz  transitions which have 21, 9, and 30
hfs components, respectively. 
The observed NH 974~GHz and o-NH$_2$ transitions  in  the +39~km~s$^{-1}$~velocity 
component  towards  W49N
are   shown in 
Figs.~\ref{Fig: NH hfs lines in 39 kms in W49N} and \ref{Fig: NH2 hfs lines in 39 kms in W49N}
together with   models of respective transition, including hfs components, using  Gaussian opacity profiles and a     
line width of 1~km~s$^{-1}$.   
Here, the intensities have   been normalised to the continuum in  
single sideband  as $T_\mathrm{A}$/$T_\mathrm{C}$-1 
assuming a sideband gain ratio of unity where $T_\mathrm{A}$ is the 
observed intensity  and 
$T_\mathrm{C}$ is the SSB  continuum as measured in line-free regions in the spectra. 
The large number of hfs components, in addition to many different, overlapping, 
velocity components in the line-of-sight,   complicates 
the analysis substantially.  In particular, the o-NH$_2$~hfs components in  the 
line-of-sight absorption  overlap  
with the source emission/absorption, thus requiring a   
source model in the analysis.  
 

\begin{figure}
   \centering
\includegraphics[scale = 0.43]{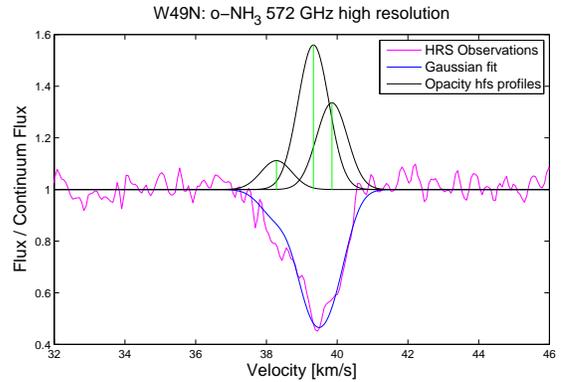} 
\caption{\emph{Ortho-NH$_3$~hfs components}: Normalised high resolution spectra  ($\Delta v$\,=\,0.13~km~s$^{-1}$)
 towards W49N 
in the +39~km~s$^{-1}$~velocity component.
Three hfs components are 
seen  at 38.4, 39.4 and 40.0~km~s$^{-1}$. Also shown are the positions and relative intensities
of the hfs components and their fitted opacity profiles with a line width of 1.0~km~s$^{-1}$.}
 \label{Fig: NH3 hfs lines}
\end{figure}

The much   simpler   hyperfine structure 
of  
the ortho- and  para-NH$_3$ ground state rotational transitions
has  never   been  resolved  in space before \emph{Herschel}. 
The Odin satellite was able to observe the 572~GHz ammonia transition, but the velocity resolution of about 0.5~km~s$^{-1}$~was 
not enough to spectrally resolve the hfs components   even though the asymmetric line shapes hinted the hfs components 
\citep[e.g.][]{2003A&A...402L..73L}. 
Figure~\ref{Fig: NH3 hfs lines} shows an example of our observations of the 572~GHz  
o-NH$_3$ transition in 
the +39~km~s$^{-1}$~absorption velocity component towards  W49N with 
the HRS and a velocity resolution 
of 0.13~km~s$^{-1}$. 
Three hfs components are seen  at 38.4, 39.4 and 40.0~km~s$^{-1}$~although the line width 
of 1.0~km~s$^{-1}$~prevents
a detailed check of the relative intensities. 
In the Gaussian fit we therefore use 0.2, 1 and 0.6 as 
relative intensities,   with   corresponding relative 
offsets at -1.05, 0.0, and 0.52~km~s$^{-1}$~found by \citet{2009A&A...507.1707C}  
(Table~\ref{Table: NH3 hfs components}, on-line
material).  
The positions and relative intensities
of the hfs components and the fitted opacity profiles are also shown in 
Fig.~\ref{Fig: NH3 hfs lines}. 
The six hfs components in the para-NH$_3$ line at 1\,215~GHz are even more closely spaced 
and  not resolved (Table~\ref{Table: NH3 1215 hfs components}).

The WBS ammonia observations     do 
not spectrally resolve the  NH$_3$~hfs components ($\Delta v$\,=\,0.6~km~s$^{-1}$~at 572~GHz 
and $\Delta v$\,=\,0.3~km~s$^{-1}$~at 1\,215~GHz),   
but we still take  the hfs components into account in the following NH$_3$~modelling  
since they produce  a slightly asymmetric line shape and  a systematic line broadening.
\\
\\

\section{Analysis: Velocity decomposition, column densities, and abundances}
\label{Section: Analysis of abundances in different velocity components}

In paper~I we estimated the abundances towards  G10.6$-$0.4~simply by evaluating the integrated opacity of each line over 
the  velocity  range \mbox{11\,--\,54\,km\,s$^{-1}$}.
This approach   gave a first   estimate of the average abundance in the full line-of-sight, but 
did not take into account possible  abundance variations  in the different velocity components. 
In addition, no source model  was included,  and thus the blend of the 
NH and o-NH$_2$ hfs components from the foreground absorptions  
 with the source absorption was neglected.

In this paper we use 
three different approaches  to decompose the absorption into separate velocity components, and to obtain
column densities and abundances in each component. 
Each method has its own strengths and weaknesses, and 
differences in the results can be considered an estimate of the errors of the methods. 
If   all three methods point  to   the same result, it will be considered as robust. 
We note, however, that the uncertainties of the data are higher towards  W49N  than for G10.6$-$0.4, since 
we have removed a SO$_2$ line from the upper sideband in the ortho-NH$_3$~absorption,
and an  
 NO line in the o-NH$_2$~absorption.

In this work we also only estimate 
$N$  and $X$  for the ortho-symmetries of   NH$_2$ and  NH$_3$   since our own measurements 
of the ammonia OPR in Sect.~\ref{OPR ammonia} do not point to the high temperature limit,  
but instead suggest  an OPR \emph{lower} than unity, which cannot be straightforwardly explained. If this unexpected  
low ammonia OPR is true, then the  
processes that affect the ammonia OPR perhaps also could affect the OPR of NH$_2$   in diffuse clouds   about which we have no 
information. And since we want to compare relative abundances, we choose to focus on the 
ortho-symmetries since both are observed in    NH$_2$ and  NH$_3$, and also since the  \mbox{o-NH$_3$}  
observations have higher S/N than the 
\mbox{p-NH$_3$}.

To obtain integrated opacities of each transition, in each velocity
component, the first method uses Gaussian fitting  (Sect.~\ref{henriks method}), while the
second one uses the observed spectra of ortho-NH$_3$ and CH as templates for
other species (Sect.~\ref{massimos method}). Both methods calculate the opacity 
as \mbox{$\tau\!=\!-\ln{(T_\mathrm{A}/T_\mathrm{C} -1)}$}.   
The derived integrated opacities are  then used to estimate  column densities by means of the non-LTE 
(Local Thermodynamic Equilibrium) 
{\tt RADEX} code (Sect.~\ref{Section: RADEX}). 
As a third method, we have  used XCLASS (Sect.~\ref{Bhaswatis method}) which, in 
contrast to {\tt RADEX}, assumes fixed excitation temperatures, to obtain column densities in each velocity component \mbox{directly}.

In order to be able to use our results as comparisons  with chemical models, we need to
derive the abundance of the species with respect to the total amount of   gas. 
The difficulty is to obtain reliable estimates of $N$(H) and/or $N$(H$_2$). 
In paper~I, we compared the nitrogen hydrides column densities
to the total amount of hydrogen in the line-of-sight, 
 $N_\mathrm{H}$\,=\,$N$(H)+2$N$(H$_2$). 
This can not be done in the present work since  the neutral hydrogen absorption shows 
too  broad and overlapping
absorption components 
over much larger
velocity ranges  as compared to the nitrogen hydrides.  
In order to estimate the amount of molecular hydrogen in each (narrow) velocity component,  
 we
therefore 
consider the comparison with CH \citep{2010A&A...521L..16G}  as an abundance determination. This can be made 
 using the CH/H$_2$ correlation 
(CH/H$_2$\,=\,3.5$\times$10$^{-8}$)  
observed by \citet{2008ApJ...687.1075S}
 in the solar neighbourhood
diffuse medium.  
This correlation is assumed to be valid in regions which are dominated by 
ultraviolet radiation where $N$(CH)\,$\lesssim$\,2$\times$10$^{14}$~cm$^{-2}$.
In this way we can use our own PRISMAS observations of CH and 
measure abundance ratios obtained with the same instrument.

The physical properties of the bulk of the absorbing gas in our sight-lines are \citep{2010A&A...521L..16G}, according
to the definitions of \citet{2006ARA&A..44..367S},  a mixture of diffuse ($N_\mathrm{H}\!\!\lesssim\!\!500$~cm$^{-3}$~and 
\mbox{$A_\mathrm{v}\!\lesssim\!1$)}  
and translucent gas
(\mbox{$500\!\lesssim N_\mathrm{H}\!\lesssim\!5\,000$~cm$^{-3}$}~and a shielding \mbox{$1\lesssim A_\mathrm{v}\! \lesssim \!5$}). 
And since  the comparison later in this section of the nitrogen hydrides with other species tracing both high and low density gas, 
points to an existence in the denser parts of the gas, 
molecular hydrogen is most likely
the dominant form of hydrogen in these components.

Assuming that the nitrogen hydrides co-exist with HF and H$_2$O in the line-of-sight material we  use 
the 
fact that the saturated absorptions of  
the latter two species reach the zero level, for a sideband ratio of unity, to support our assumption that 
the foreground absorbing material completely covers  the continuum  within the beam. 
Comparison spectra of o-NH$_3$  and o-H$_2$O are shown in 
Figs.~\ref{Fig: W49N comparison species}\,--\,\ref{Fig: W31C comparison species}, and   with 
HF, ortho- and para-H$_2$O in  
Figs.~\ref{fig: comparison of W31C-W49N NH, NH2 vs 572, 572 and 1215, CH}\,--\,\ref{fig: comparison of W31C-W49N 572 vs CH, water, h2o+} (on-line material).

\begin{table}[\!ht] 
\centering
\caption{\emph{Conversion factors}\tablefootmark{a}.
}
\begin{tabular} {lrccl  } 
 \hline\hline
     \noalign{\smallskip}
 Species $x$ &  Transition &  $T_\mathrm{ex}$  &
$N(x)$($\int  \tau d v$\,=\,1.0~km~s$^{-1}$)   \\    \noalign{\smallskip}
   &  (GHz)& (K) & (cm$^{-2}$)   \\
     \noalign{\smallskip}
     \hline
     \noalign{\smallskip}

NH   & 946.476& 3.9 &  3.65$\times$10$^{13}$ & $a$\\
& 974.478&4.0 &  7.47$\times$10$^{12}$ &$a$\\
o-NH$_2$   & 952.578&  3.9&   3.62$\times$10$^{12}$  &$b$ \\
o-NH$_3$   & 572.498 & 2.8 &  3.92$\times$10$^{12}$  & ${b}$ \\
 p-NH$_3$     &  1\,215.245 &4.9  &  1.65$\times$10$^{13}$  & ${b}$\\
NH$^+$ &  1\,012.524& 4.2  & 6.48$\times$10$^{12}$ & \\
CH             &  532.724&2.9  & 2.28$\times$10$^{13}$ &$b$ \\
HNC     &  90.664 &2.8 & 1.80$\times$10$^{12}$  &${b}$\\
CN     &  113.169 &2.8 & 1.90$\times$10$^{13}$  &${b}$\\
\noalign{\smallskip}
\hline 
\label{Table: columns opacity one}
\end{tabular}
\tablefoot{
\tablefoottext{a}{The excitation temperatures and conversion factors from integrated opacity, 
$\int  \tau d v$\,=\,1.0~km~s$^{-1}$, 
to column densities, $N(x)$, 
are calculated using {\tt RADEX} with $T_\mathrm{k}$\,=\,30~K 
and $n$(H$_2$)\,=\,500~cm$^{-3}$ which corresponds to 
\mbox{$n_H\ga 10^3$~cm$^{-3}$}.} 
\tablefoottext{b}{Integrated over all hfs components.}   
}
\end{table} 

   \subsection{{\tt RADEX}}  \label{Section: RADEX}

Observations of absorption lines can provide accurate determinations of molecular column densities if the observed 
transitions trace all of the populated states. 
In the case of NH$_3$ we observe two transitions of the ortho (A-symmetry) form, 
arising in the $J_K=0_0$ and $1_0$ states, and one transition of the para 
(E-symmetry) form arising in the lower $1_1$ rotation-inversion level. Although 
the upper $1_1$ inversion level is less metastable than the lower one, it may 
still be significantly populated. The metastable $2_2$ para levels and $3_3$ 
ortho levels may also have significant populations in diffuse molecular clouds 
 at low 
density, \mbox{$n_H\lesssim10^4$~cm$^{-3}$},  owing to 
infrared pumping and formation.
Because we are directly sensitive to three lower 
states and lack direct information about the excitation temperatures of the 
observed transitions, we use the 
non-equilibrium {{\tt RADEX}} code  \citep{2007A&A...468..627V}
to relate the total 
column density of ammonia to the observed integrated optical depths.  
The models provide integrated opacities of both observed and unobserved excited 
levels and quantify possible effects of stimulated emission, chemistry, and electron collisions.  
Results for one reference model are summarised in Table~\ref{Table: columns opacity one}, where conversion factors 
for all nitrogen hydrides and comparison species   
are listed in terms of the total column density $N(x)$ of molecule $x$ that is needed to achieve an integrated optical 
depth $\int \tau dv = 1.0$~km~s$^{-1}$ in the specified transition. In this model we 
treat  ortho- and para-symmetries separately.  
The integrated optical depth is a sum over all 
hyperfine structure in the transition.

We have assumed  diffuse molecular cloud conditions with  a kinetic temperature of  30~K 
and  a  density of molecular hydrogen   \mbox{$n$(H$_2$)\,=\,500~cm$^{-3}$},  which corresponds to 
\mbox{$n_H\ga 10^3$~cm$^{-3}$}.   
The resulting conversion factors for the rotational ground state transitions are  not very sensitive  to density, temperature 
or   electron collisions. 
The critical densities of  the   
fundamental rotational  transitions of NH, NH$_2$, and
NH$_3$
are   high, $n_\mathrm{crit} \! \sim$10$^8$\,cm$^{-3}$.
This means the upper energy levels of these transitions must be
excited almost entirely radiatively at the low  densities in diffuse gas. 
We find, however, that electron collisions can be responsible for 
excitation temperatures of $\sim\!4$~K in the cm-wave inversion transitions rather than 
$\lesssim\!3$~K 
that would be found in conventional excitation analyses at low densities, 
$n({\rm H}_2)\lesssim 10^3$~cm$^{-3}$.  
Where collision rates are unknown, we have made guesses scaled in proportion to radiative line 
strengths.
The background radiation field 
includes both the 2.725~K cosmic microwave background  and a model of the Galactic infrared 
radiation in the  solar neighborhood.    
The resulting excitation temperatures  of the observed sub-millimetre 
transitions are typically \mbox{3\,--\,4~K}, which
are small enough compared to $h\nu/k$ that no correction for emission is required. 
Note that the lower state  of the 572~GHz transition contains $95\%$ of the ortho molecules while the lower state  
of the 1215.2~GHz transition occupies only 
$43\%$ of the para molecule.

 \subsection{Method~I: Multi-Gaussian fit of the nitrogen hydrides simultaneously}  \label{henriks method}

We have     modelled the observed spectra of NH, \mbox{o-NH$_2$}, and \mbox{o-NH$_3$} 
using Gaussian
optical depth profiles. These were generated for each hfs component of each  
species   in separate velocity intervals,
 and were made to fit the observations under
the condition that the LSR velocity and width  in each velocity component  must be the same for all  
molecules. 
This can be made  assuming that all species co-exist  in the same velocity 
ranges and therefore show absorption at the same 
velocities.  
This assumption is supported by the 
striking similarities of the strongest NH, \mbox{o-NH$_2$} and \mbox{o-NH$_3$} 
absorptions, 
despite the complicated hyperfine structure patterns  of 
NH and \mbox{o-NH$_2$}, shown in normalised comparison spectra   in 
Fig.~\ref{fig: comparison of W31C-W49N NH, NH2 vs 572, 572 and 1215, CH} 
(on-line material). 
We note that   a good 
fit therefore is a good indicator if this assumption is valid.

The observed line profiles are thus  modelled by Gaussian components of both the hfs components 
and all velocity components according to 
\begin{equation}
\frac{T_\mathrm{A}}{T_\mathrm{C}} = \exp\,\biggl[-\sum_{i=1}^{N_{v}}   \sum_{j=1}^{N_\mathrm{hfs}}  
\tau_\mathrm{hfs}(j) \, \tau_0(i) \,  \exp[-4 \ln(2)\,\biggl[\frac{v-v_\mathrm{0}(i)-v_\mathrm{hfs}(j)}
{\Delta v}
\biggr]^2 \biggr]\ ,
\end{equation}
where $N_v$ is the number of the modelled velocity components (same  for all
species),  $N_\mathrm{hfs}$ is the number of hfs components (which differs between species), 
$\tau_\mathrm{hfs}$ is the theoretical relative line strength of each hfs component, $\tau_0$ is 
the opacity of the strongest hfs component in 
each velocity component, 
$v_\mathrm{0}$ is the LSR velocity of each velocity component, $v_\mathrm{hfs}$ is the relative velocity 
offset of the hfs components with respect to the strongest, and $\Delta v$ is the FWHM (full width of half maximum). 
We assume that  \mbox{$\tau_\mathrm{hfs}$\,=\,$A_{ul}\times g_u / (A_{ul}\mathrm{(main)}\times g_u \mathrm{(main)})$} for 
the hfs components of NH and o-NH$_2$, since the  
many overlapping velocity components prevent  a check of the relative intensities of these components.

We have used both 
ortho- and para-NH$_{3}$, which are not significantly complicated by hfs splitting,    
to determine the minimum number of velocity components needed, 
and also to set-up reasonable initial guesses for the line properties
necessary for the fitting procedure to converge at all.  
The fitting was done for each velocity component at a time in a loop
until all fits had converged. 
The fits are most likely not unique. Provided that the fits are good and that we use the same velocity 
parameters for all species, 
this is, however, not considered important as we do not ascribe any physical meaning to the Gaussian components.

The results from Method~I are peak opacities, line widths and centre velocities for each velocity component and species.
The integrated opacity in each velocity component was assumed to be well represented by   
$\int\tau \,\mathrm{d}  v  = 1.06 \,\Delta v \, \tau_\mathrm{peak}$.
The non-detections of NH$^+$ are used to put upper limits on its integrated opacities (3$\sigma$). 
We then used the 
conversion factors in 
Table~\ref{Table: columns opacity one} to calculate the column densities   
tabulated in 
Table~\ref{Table: Method I resulting ratios and columns}.  
Typical errors in the G10.6$-$0.4 resulting column densities for NH, o-NH$_2$ and o-NH$_3$ are between 7 and 15\%. In the  
$v_\mathrm{LSR}$\,=\,45~km~s$^{-1}$ component, the errors are 22\% for all three species, and 23\% in the  
$v_\mathrm{LSR}$\,=\,39~km~s$^{-1}$ component for NH. 
The errors in W49N varies between 8 to 11\% for all three species, except in the 33~km~s$^{-1}$ component where the
uncertainties are 37, 26 and 31\% for NH, o-NH$_2$ and o-NH$_ 3$, respectively. 
The errors include calibration uncertainties (see Sect.~\ref{OPR ammonia}) 
as well as uncertainties from the fitting procedure.

In addition to the  nitrogen hydrides, we also modelled the  PRISMAS  
CH 532~GHz transition,
and  the CN 113~GHz and HNC 91~GHz transitions observed with the IRAM 30~m antenna \citep{2010Godard}.
These species were not included in the fitting procedure described above.
Instead we
have used the resulting velocity parameters  from the fits of 
the nitrogen hydrides ($v_\mathrm{LSR}$ and line widths)
since we
want to investigate possible correlations and make abundance determinations only in the 
same parts of velocity space as the nitrogen hydrides.    
The resulting CH column densities are found in 
Table~\ref{Table: Method I resulting ratios and columns},
and the CN and HNC column densities  in the on-line 
Table~\ref{Table: Method I, III CN and HNC columns}.

\begin{figure}
   \centering
\includegraphics[scale=0.82]{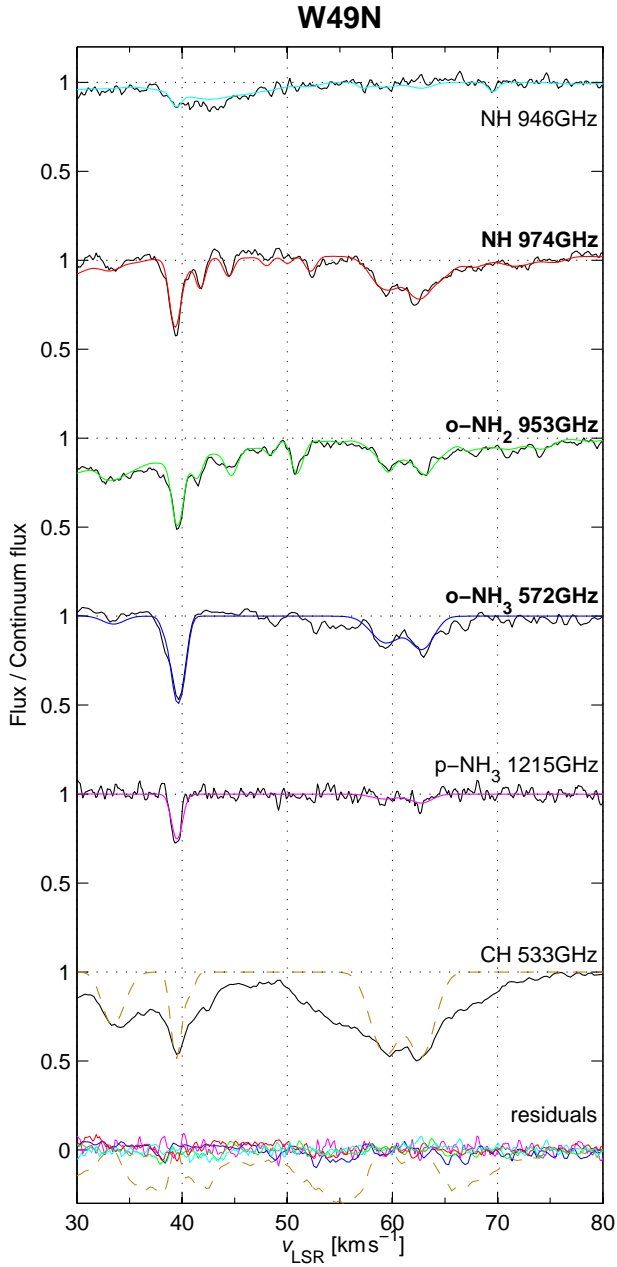} 
\caption{\emph{W49N: Fits of Method~I} and  the SSB normalised spectra of the nitrogen hydrides and CH. 
The black lines are the observations, and the coloured lines are the model fits. In the bottom the residuals are plotted
on top of each other with respective colour. The three transitions used in Method~I to determine the $v_\mathrm{LSR}$ and line widths of the velocity components are marked in bold.}
 \label{Fig: W49N N-hydrides normalised} 
\end{figure}

\begin{figure}
   \centering
\includegraphics[scale=0.8]{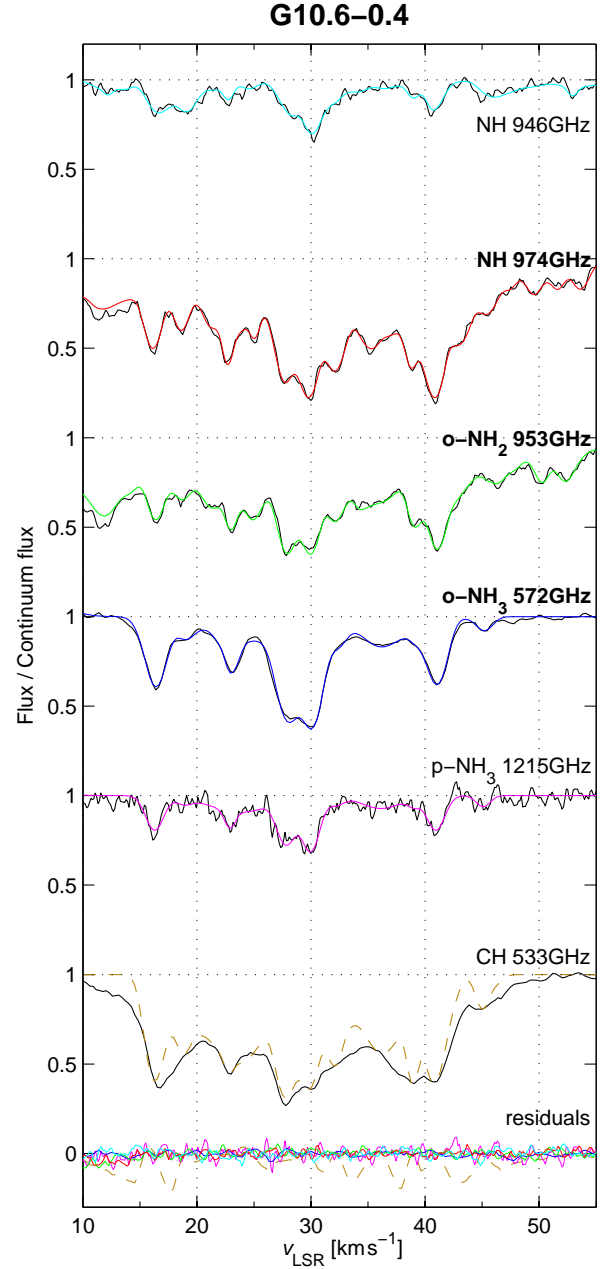}  
\caption{\emph{G10.6$-$0.4: Fits of Method~I} and  the SSB normalised spectra of the nitrogen hydrides and CH.
Same notation as in Fig.~\ref{Fig: W49N N-hydrides normalised}.}
 \label{Fig: W31C N-hydrides normalised}
\end{figure}

Figures~\ref{Fig: W49N N-hydrides normalised} and \ref{Fig: W31C N-hydrides normalised}
show the fits and residuals of the nitrogen hydrides together with CH.   
The resulting fits are well reproducing the
observations for all species except for CH 
suggesting that CH   exist in a more widely spread gas than the  nitrogen hydrides.  
 
The CH column densities are then used together with the relation
 [CH]/[H$_2$]\,=\,3.5$\times$10$^{-8}$ \citep{2008ApJ...687.1075S} to estimate the abundances  listed in 
Table~\ref{Table: Method I resulting ratios and columns}.

   \begin{table*}[\!ht] 
\centering
\caption{\emph{Method~I results\tablefootmark{a}.}} 
\begin{tabular} {lc ccc ccc c llll } 
 \hline\hline
\noalign{\smallskip}
      \multicolumn{13}{c} {W49N} \\     \noalign{\smallskip}
$v_\mathrm{LSR}$ & $\Delta v$ & $N$(NH)    & $N$(oNH$_2$)   & 
$N$(oNH$_3$) &$N$(NH$^+$) &$N$(CH)  &  ${{\rm NH}\over{{\rm oNH}_3}}$   &${{\rm oNH}_2\over{{\rm oNH}_3}}$   &                    
 $X$(NH)\tablefootmark{b}    &   $X$(oNH$_2$)\tablefootmark{b}    &    
$X$(oNH$_3$)\tablefootmark{b} &    $X$(NH$^+$)\tablefootmark{b} 	   \\
(km\,s$^{-1}$) &(km\,s$^{-1}$)  & (cm$^{-2}$)  &(cm$^{-2}$) &  (cm$^{-2}$) &  (cm$^{-2}$)  &  (cm$^{-2}$)   \\     
     \noalign{\smallskip}
     \hline
     \noalign{\smallskip}
 
33.3  &   2.1  &   1.5e12  &   2.1e12   &   4.6e11   &$\lesssim$3.3e11 &1.6e13&  3.3  &    4.6  & 3.3e-9&  4.6e-9  &  1.0e-9   & $\lesssim$7.2e-10\\ 
39.5  &   1.1  &   1.0e13  &   7.3e12   &   4.3e12   &$\lesssim$2.4e11 &2.3e13&  2.3  &    1.7  & 1.5e-8&  1.1e-8  &  6.5e-9   & $\lesssim$3.7e-10\\ 
59.2  &   2.9 &    9.2e12  &   4.1e12   &   2.1e12   &$\lesssim$7.5e11 &6.0e13&  4.4  &    2.0  & 5.4e-9&  2.4e-9  &  1.2e-9   & $\lesssim$4.4e-10\\ 
62.7   &  2.3 &    8.0e12  &  3.9e12   &  2.2e12   &$\lesssim$6.3e11 &5.1e13&  3.6   &   1.8  & 5.5e-9&  2.7e-9 &  1.5e-9   & $\lesssim$4.3e-10\\

\noalign{\smallskip} \noalign{\smallskip}\noalign{\smallskip}
Total:  &\ldots &2.9e13  & 1.7e13  &   9.1e12  &$\lesssim$2.0e12& 1.5e14&  3.2 &  1.9 &  6.7e-9  &  4.1e-9 &   2.1e-9 &  $\lesssim$4.6e-10 \\
Mean: &\ldots&7.2e12& 4.4e12 & 2.3e12 &$\lesssim$4.9e11& 3.8e13& 3.2 &  1.9  &  6.7e-9  & 4.1e-9  & 2.1e-9  &  $\lesssim$4.6e-10 \\
Median: &\ldots&8.6e12& 4.0e12 & 2.2e12 &$\lesssim$4.8e11& 3.7e13& 4.0 &  1.9  &  8.1e-9  &  3.8e-9 & 2.0e-9 &  $\lesssim$4.5e-10  \\

     \noalign{\smallskip}
       \hline  \hline
     \noalign{\smallskip}     \noalign{\smallskip}
      \multicolumn{13}{c} {G10.6$-$0.4} \\     \noalign{\smallskip}
$v_\mathrm{LSR}$ & $\Delta v$ & $N$(NH)    & $N$(oNH$_2$)   & 
$N$(oNH$_3$) &$N$(NH$^+$) &$N$(CH)  &  ${{\rm NH}\over{{\rm oNH}_3}}$   &${{\rm oNH}_2\over{{\rm oNH}_3}}$   &                    
 $X$(NH)\tablefootmark{b}    &   $X$(oNH$_2$)\tablefootmark{b}    &    
$X$(oNH$_3$)\tablefootmark{b} &    $X$(NH$^+$)\tablefootmark{b} 	   \\
(km\,s$^{-1}$) &(km\,s$^{-1}$)  & (cm$^{-2}$)  &(cm$^{-2}$) &  (cm$^{-2}$) &  (cm$^{-2}$)  &  (cm$^{-2}$)   \\       
     \noalign{\smallskip}
     \hline
     \noalign{\smallskip}
 16.2   &    1.7        &   2.0e13   &   9.6e12   &     4.2e12    &$\lesssim$4.7e11 &  5.8e13    &   4.8      &      2.3   & 1.2e-8  &   5.8e-9  &   2.5e-9  & $\lesssim$2.8e-10\\  
 18.8    &   1.5         &   6.1e12  &   3.1e12    &  8.7e11    &$\lesssim$4.1e11   &2.6e13     & 7.0      &        3.6  & 8.2e-9 &  4.2e-9  & 1.2e-9 & $\lesssim$5.5e-10 \\  
 22.1    &   4.4            & 1.6e13    &  6.9e12     &  1.8e12   & $\lesssim$1.2e12 &7.7e13    &  8.9       &    3.8  & 7.3e-9 & 3.1e-9  &  8.2e-10   & $\lesssim$5.5e-10\\ 
 22.8    &   1.0              &  1.0e13   &    3.6e12    &   1.5e12 & $\lesssim$2.7e11 & 7.1e12        &  6.7      &    2.4  & 4.9e-8 & 1.8e-8  &  7.4e-9   & $\lesssim$1.3e-9\\
 24.8    &   3.1              &   6.2e12  &   4.3e12    &   1.4e12 &$\lesssim$8.4e11  & 3.7e13       &  4.4       &      3.1  & 5.9e-9 & 4.1e-9  &  1.3e-9  & $\lesssim$8.0e-10 \\
 27.8    &   1.9                &  2.7e13   &   1.2e13      &  7.3e12    &$\lesssim$5.0e11 & 7.4e13        &   3.7       &    1.6   & 1.3e-8&  5.7e-9 & 3.5e-9    & $\lesssim$2.4e-10\\
 29.9    &   1.7                 &2.9e13   &    1.3e13   &   7.9e12      & $\lesssim$4.6e11  &5.6e13          &  3.7      &    1.6  & 1.8e-8  & 8.1e-9   & 4.9e-9  & $\lesssim$2.9e-10\\
 32.1    &   1.5                  &9.0e12  &  3.5e12     &   1.3e12 & $\lesssim$4.1e11    &3.5e13         &    6.9    &         2.7 & 9.0e-9 &  3.5e-9 &  1.3e-9   & $\lesssim$4.1e-10 \\
 36.1   &    4.0           &   2.0e13   &   6.5e12    &     3.2e12   & $\lesssim$1.1e12   &7.8e13          &   6.3    &   2.0  & 9.0e-9 &  2.9e-9 &  1.4e-9   & $\lesssim$4.9e-10\\
 39.0    &   1.2               & 1.1e13  &    3.0e12      &   4.7e11    & $\lesssim$3.1e11  & 3.3e13         &    23    &      6.4  & 1.2e-8 & 3.2e-9  &  5.0e-10  & $\lesssim$3.3e-10\\
 40.9     &  1.7             &3.0e13   &  1.1e13    &  4.1e12   & $\lesssim$4.7e11     &5.4e13     &   7.3      &  2.7  & 1.9e-8  &  7.1e-9  &  2.7e-9   & $\lesssim$3.0e-10 \\
 45.0     &  1.4           & 3.0e12  &   1.1e12  &    5.9e11    &$\lesssim$3.6e11  & 1.1e13         &  5.1      & 1.9  & 9.6e-9 &  3.5e-9  &  1.9e-9   & $\lesssim$1.1e-9\\

\noalign{\smallskip} \noalign{\smallskip}\noalign{\smallskip}
Total:  & \ldots&1.9e14  & 7.8e13     &   3.5e13  &$\lesssim$6.8e12  &  5.5e14  & 5.4 &  2.2  &  1.2e-8  &  5.0e-9  &  2.2e-9 &  $\lesssim$4.4e-10     \\
Mean: &\ldots&1.6e13& 6.5e12          & 2.9e12    &$\lesssim$5.7e11  &4.6e13   & 5.4 &  2.2  &   1.2e-8  &  5.0e-9  &  2.2e-9 &  $\lesssim$4.4e-10 \\
Median: &\ldots&1.4e13& 5.4e12        & 1.7e12    &$\lesssim$4.7e11  &4.6e13   & 8.2 &  3.3 &   1.0e-8  &  4.2e-9  &  1.3e-9 &  $\lesssim$3.6e-10  \\

    \noalign{\smallskip}
\hline 
\label{Table: Method I resulting ratios and columns}
\end{tabular}
\tablefoot{ 
\tablefoottext{a}{LSR velocity, $v_\mathrm{LSR}$, line width, $\Delta v$,  
column densities, $N(x)$, and abundances, $X(x)$\,=\,$N$($x$)/$N$(H$_2$),  of each velocity component, 
including 3$\sigma$ upper limits of $N$(NH$^+$).  Note that we  estimate $N$
and $X$ only  for the ortho-symmetry forms of NH$_2$ and  NH$_3$.}
\tablefoottext{b}{Using [CH]/[H$_2$]\,=\,3.5$\times$10$^{-8}$ \citep{2008ApJ...687.1075S}.} 
}
\end{table*}

\subsection{Method~II: ortho-NH$_3$~and CH as templates for other species} \label{massimos method}

This method directly compares the opacity line profile of 
the non-saturated o-NH$_3$~line at 572~GHz with the 
line profiles of  NH and o-NH$_2$, convolved with  respective hyperfine structure.
This assumes  that all species co-exist in the same velocity ranges, and that 
the  opacity ratios  depend  only on their column density ratios,  which may vary from one velocity 
component to another. 
No opacity profile is assumed for a given absorbing cloud so, unlike what is done by methods I  and III,
we do not attempt to decompose the velocity structure of the
template line into a number of Gaussian curves. We perform,
instead, simple cuts of the template profile into large velocity
bins.

The hfs  of the  o-NH$_3$ line is  
not resolved in the WBS data and  introduces only a small 
broadening of the lines that is dealt with by 
smoothing the NH and o-NH$_2$ spectra. 
Before
comparing the opacity profiles, we mask out those portions
of the spectra where a simple scaling between template and
target is not expected, such as the velocity range associated
to the background sources,  and remove the emission line wing by Gaussian fitting.
We note that
the removal of the line wing introduces very large uncertainties
towards  W49N  around 33~km~s$^{-1}$~where a weak absorption component is visible.
This method therefore disregards the results in this velocity component.

The velocity bins into which the o-NH$_3$ opacity profile
is split are chosen to be significantly larger than the original 
resolution, but narrow enough to separate the most prominent velocity features (typically a few km~s$^{-1}$, see 
Table~\ref{Table: Model II ratios results}).
Each velocity bin of the o-NH$_3$ template is then convolved, channel by channel,  with the NH and o-NH$_2$ hfs components.  
We thus obtain an intermediate   opacity model of how the NH and o-NH$_2$ spectra would appear if absorption were limited to the 
selected velocity bin. 
The observed  o-NH$_3$ opacity is then  fitted by the IDL Least Squares Fitting routine MPFIT \citep{2009ASPC..411..251M} 
with a linear combination of the intermediate models. The output of the  fits consists of the opacity ratios, between the 
o-NH$_3$ and the modelled NH and o-NH$_2$ spectra, in the various velocity bins.

To obtain column densities in each velocity bin, we sum the  opacities of 
all channels of the template transition,     
and then apply the {\tt RADEX} conversion factors  from 
Table~\ref{Table: columns opacity one}. 
The resulting abundance ratios are finally used to obtain column 
densities of the other species. Typical errors (including both calibration and fitting uncertainties) are \mbox{5\,--\,15\%} 
except for the \mbox{43.5\,--\,50~km~s$^{-1}$} velocity bin in G10.6$-$0.4 
and  the \mbox{50\,--\,57~km~s$^{-1}$} velocity bin in W49N which both have an uncertainty of 
approximately 25\%.

The main advantage of this method is that a  small number of free parameters is needed 
to obtain a reasonable fit since 
it uses, as much as possible, the information carried by the template spectrum.  The model 
also avoids the problem of non-uniqueness of Gaussian decompositions of the absorption profiles and allows a straightforward comparison 
of column densities  evaluated for exactly the same velocity intervals.
On the other hand, such rigidity in the definition of the velocity bins does not allow us to reliably retrieve the column 
density ratios if the absorption comes from very different gas volumes resulting in significant velocity shifts of the 
absorption features between the two species. We note that the fit quality, estimated with $\int |\mathrm{residuals}|\times \mathrm{d} v \, / \int \tau\times \mathrm{d} v$,   
is a good indicator of such cases.    
Note also that residual baseline structures,  like those produced by standing 
waves or by an imperfect removal of ammonia emission line wings, 
may introduce artificial discrepancies between the template spectra 
and the 
modelled species. 
  
In addition to the above modelling,   we   use  the o-NH$_3$ as a template for CN and HNC \citep{2010Godard},
and also use 
the deconvolved CH  spectrum at 532.724~GHz \citep{2010A&A...521L..16G} as a 
template to model all three nitrogen hydrides,
in order to obtain an estimate of the abundance with respect to molecular hydrogen.  

The resulting column densities  and relative abundances  in different velocity bins are given in 
Table~\ref{Table: Model II ratios results}, except for the CN and HNC results which are 
found in on-line Table~\ref{Table: Method II CN and HNC columns}.

 The model fits compared to the normalised spectra in both sources are shown in 
Figs.~\ref{Fig: W49N method I NH and NH2 from NH3}\,--\,\ref{Fig: W31C method I NH, NH2, NH3 from CH}
 (on-line material).  
The fit qualities of  the comparisons between 
the o-NH$_3$, NH and o-NH$_2$ are excellent, except in the narrow and deep 
+39~km~s$^{-1}$~component towards  W49N  which was very difficult to model.  
The modelling of 
the nitrogen hydrides   using CH as a template show on the other hand rather low 
fit
qualities, 
suggesting that 
the assumption that  the
nitrogen hydrides and CH only come from the same gas is less justified, also supported by Method~I.

\subsection{Method~III: XCLASS} \label{Bhaswatis method}

Finally, we have generated synthetic spectra separately for   NH,
o-NH$_2$, o-NH$_3$, CH, CN, and HNC using the software 
XCLASS\footnote{\tt{https://www.astro.uni-koeln.de/projects/schilke/\\XCLASS}} 
\citep[created by P. Schilke, and described by][and references therein]{2005ApJS..156..127C} 
by considering for each species all
hfs components and velocity components and the 
assumption of a fixed excitation 
temperature.  The synthetic spectra generated with XCLASS were fitted to
the observed spectra using MAGIX, an iterating engine that allows
automatic minimization to constrain model parameters.  XCLASS accesses
the CDMS and JPL molecular databases, and models each molecule with the
following free parameters: source size, temperature, column density,
line width and velocity offset relative to the systemic velocity of the
source, and derives the column density corresponding to the different
velocity components detected in absorption and emission in the observed
spectra.  The source size refers to the relative size of the 
absorbing cloud vs. the continuum source, which is assumed to be fully covered within the beam.    
We have assumed a Gaussian profile for each hfs component, and an
excitation temperature of 4~K.  
Since Method~III explicitly assumes a fixed excitation temperature 
for the estimate of column
densities, it is significantly different from methods~I and II.

Figures~\ref{Fig: w49n guass fits n-hydrides method III} and \ref{Fig: w31c guass fits n-hydrides method III},  
in the on-line material, show the model fits  together with the observed SSB normalised spectra.
Resulting column densities and abundance ratios of the nitrogen hydrides and CH  
are found in the on-line Table~\ref{Table: Method III resulting ratios and columns}, and
the CN and HNC column densities    in on-line Table~\ref{Table: Method I, III CN and HNC columns}. 

\subsection{Comparison of results} \label{Subsection: comparison of results}

The results from all three methods agree very well, especially considering
that the  velocity bins used  are not the same.

The total $N$(CH) from all  
methods   
are lower than 
found by \citet{2010A&A...521L..16G}, which is expected since
we only use parts of the CH absorptions. Comparisons of CH in different velocity
bins give much more similar results in both sources. 
Since we use CH as a tracer of H$_2$, we also compare our estimate of 
$N$(H$_2$) with \citet[][]{1989A&A...222..231N} towards   W49N: 
for the +39~km~s$^{-1}$~component they suggest  
\mbox{$N$(H$_2$)\,=\,3$\times$10$^{21}$~cm$^{-2}$} and they also quote 
\mbox{$N$($^{13}$CO)\,=\,7\,--\,30$\times$10$^{15}$~cm$^{-2}$} which 
gives \mbox{$N$(H$_2$)\,=\,7\,--\,30$\times$10$^{21}$~cm$^{-2}$}  using 
a standard conversion factor of 10$^{-6}$.
They also suggest  column densities towards W49N between 
\mbox{50\,--\,70~km~s$^{-1}$:} \mbox{$N$(H$_2$)\,=\,5$\times$10$^{21}$~cm$^{-2}$} and 
\mbox{$N$(H$_2$)\,=\,5.5\,--\,23$\times$10$^{21}$~cm$^{-2}$} from $^{13}$CO observations.
All these suggestions for $N$(H$_2$) are several times higher than inferred from 
our CH observations. 
This is, however, not surprising because the $^{13}$CO conversion factor is 
rather uncertain  at 
such low column densities. \citet{2004ApJ...605..247P} also obtain higher $N$(H$_2$) than we do 
towards W49N.
The total amount of $N$(H$_2$) in the sight-lines towards our sources,  can also 
be compared with estimates from the K-band extinction \citep{2006A&A...453..635M}: 
\mbox{$N$(H$_2$)\,=\,1.3$\times$10$^{22}$} and 7.5$\times$10$^{21}$~cm$^{-2}$ 
towards  W49N and G10.6$-$0.4, respectively.

The resulting NH  mean  abundances of all three methods  in all velocity components  are    
5.5$\times$10$^{-9}$ and  1.1$\times$10$^{-8}$ in W49N and G10.6$-$0.4, 
respectively. This can be compared to the average 
 $N$(NH)/$N$(H$_2$) value of 3$\times$10$^{-9}$ in diffuse and translucent 
sight-lines found by \citet{2009MNRAS.400..392W}.
The  mean  values of o-NH$_2$ are  3.1$\times$10$^{-9}$ and 4.5$\times$10$^{-9}$ 
in W49N and G10.6$-$0.4, respectively,
and  1.5$\times$10$^{-9}$ and 1.9$\times$10$^{-9}$ for o-NH$_3$.

Our upper limits on the NH$^+$ abundances, relative to 
molecular hydrogen, are  similar   in both sources with   mean value 
$N$(NH$^+$)/$N$(H$_2$) $\lesssim$4$\times$10$^{-10}$. 
This is orders of magnitudes lower than previous findings from ultraviolet observations towards $\rho$\,Oph  
 \citep{1979Ap&SS..66..453S, 1982A&A...113..199D},  but still much higher than the predictions from  
the chemical models presented in Paper~I: approximately 10$^{-13}$\,--\,10$^{-14}$. 
The upper limits of the NH$^+$ abundance compared to NH are $\lesssim$2\,--\,14\,\%  
in the different velocity components 
(except in the uncertain velocity
feature around +33~km~s$^{-1}$ in W49N), with a  mean ratio of~$\lesssim$6\%. 
This is in contrast to the  behaviour of CH$^+$ 
and
marginally with that of  OH$^+$\citep{2010A&A...518A..26W, 2010A&A...518L.110G} 
with respect to the corresponding neutrals.   
The CH$^+$ radical reaches   comparable 
column densities to those of CH \citep[e.g.][]{1995ApJS...99..107C}, 
and $N$(OH$^+$) is a factor 30 below $N$(OH) in the visible data \citep{2010ApJ...719L..20K}.

Since our  different approaches to deconvolve the velocity components for the observed species agree   well,  
we estimate that the uncertainties in our derived absolute column densities  
are 
$\lesssim$\,20\,--\,50\%.  
The column densities for CH    have larger uncertainties, but we  estimate that
the absolute $N$(CH) results are correct within a factor of approximately two.   
Furthermore, the  scatter in the CH to H$_2$  relationship  is estimated by 
\citet{2008ApJ...687.1075S}  to a factor of 1.6.   In summary, 
we believe that the abundance determinations from CH are correct within a factor of a few. 
The abundance ratios of the nitrogen hydrides, relative to each other, are on the other hand 
more accurately determined.

 The results of all three methods confirm  our conclusion of the analysis 
of G10.6$-$0.4 in paper~I  that NH is  
more abundant than  NH$_2$ and  NH$_3$  by a factor of  approximately 
\mbox{1.5\,--\,3.5}, assuming 
the high temperature ortho-to-para limits of three and one, respectively. 
Note that the mean abundances of \mbox{o-NH$_3$} are similar towards both sources, $\sim$2$\times$10$^{-9}$, 
in contrast to 
\mbox{o-NH$_2$}, and, in particular  NH, which have higher mean abundances towards G10.6$-$0.4 than towards W49N.

\subsection{Comparison and correlations of species} \label{Comparison and correlations of species}

\begin{figure}
   \centering
\resizebox{\hsize}{!}{ 
\includegraphics{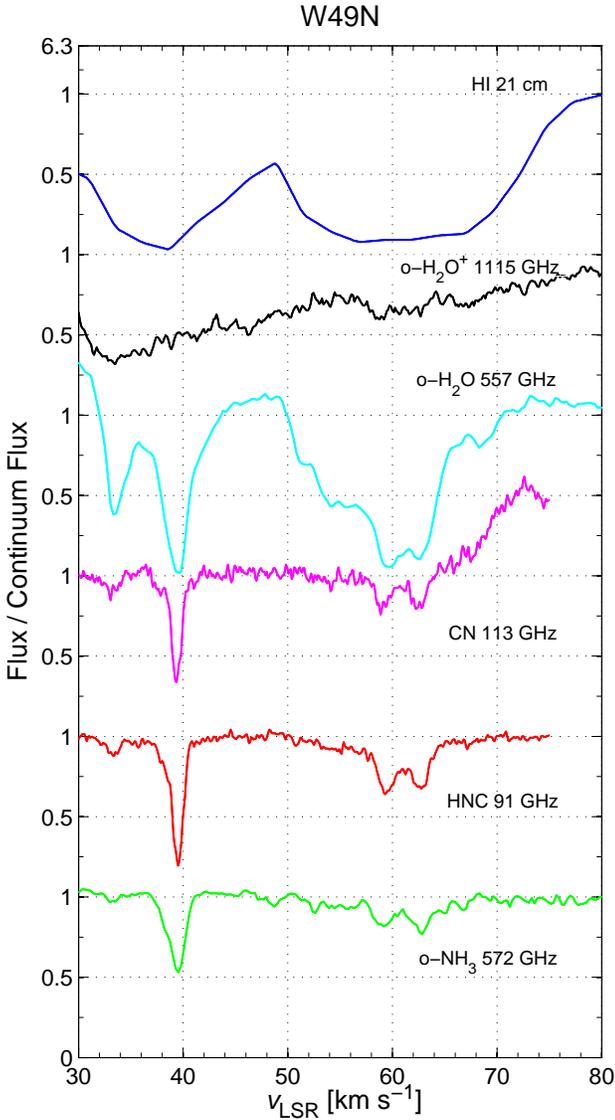}}
\caption{\emph{Line-of-sight absorptions towards W49N:} Normalised (SSB) spectra of ortho-ammonia and   species 
tracing both high and low molecular 
fractions. Strong similarities are seen between ammonia, CN, HNC and water, but not with H$_2$O$^+$ and HI tracing the atomic
diffuse gas.   
} 
\label{Fig: W49N comparison species}
\end{figure}

\begin{figure}
   \centering
\resizebox{\hsize}{!}{\includegraphics{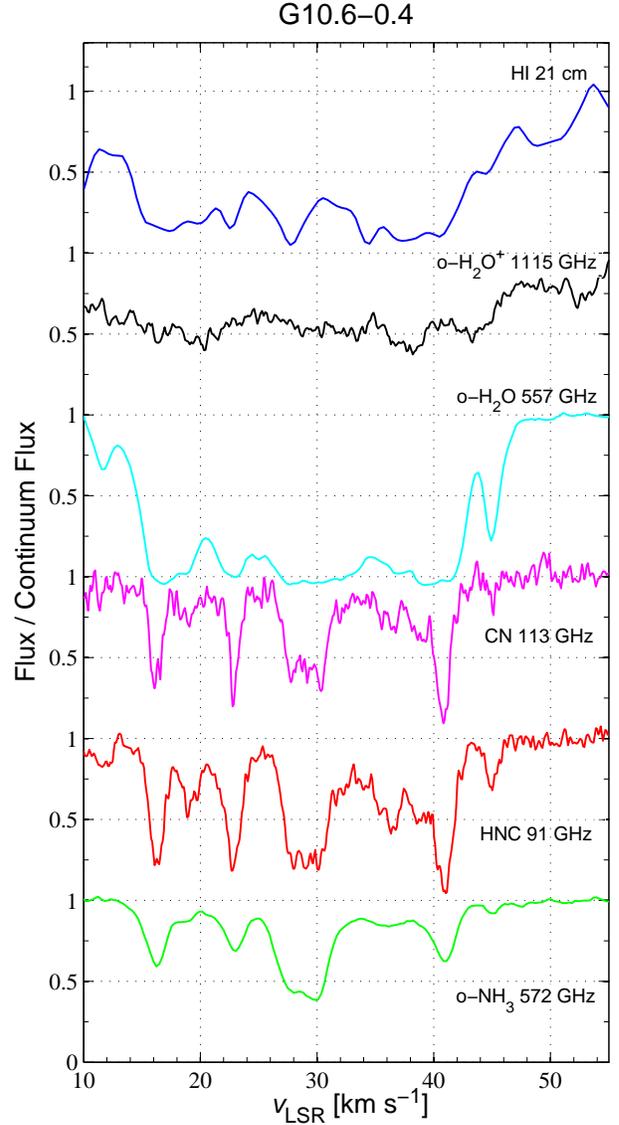}} 
\caption{\emph{Line-of-sight absorptions towards G10.6$-$0.4:} Notation as in Fig.~\ref{Fig: W49N comparison species}.}
 \label{Fig: W31C comparison species}
\end{figure}

Figures   \ref{Fig: W49N comparison species}  and \ref{Fig: W31C comparison species} 
 show   \mbox{o-NH$_3$} and five   SSB normalised comparison spectra, 
 and  in the on-line material  Figs.~\ref{fig: comparison of W31C-W49N NH, NH2 vs 572, 572 and 1215, CH}\,--\,
\ref{fig: comparison of W31C-W49N 572 vs CH, water, h2o+} we show   
additional comparisons of the nitrogen hydrides
with the deconvolved CH absorption, the H\ion{I} 21 cm line, H$_2$O$^+$, OH$^+$, HNC, CN, 
ortho- and para-H$_2$O, HF, and HCO$^+$.

The CH spectra  show  absorption over a much wider range of velocities  in both sources than  
the nitrogen hydrides.
Ammonia largely follows  the CH absorptions, but there are also large differences in some parts of the 
spectra where there is no or very little absorption 
of NH$_3$, while CH shows a much broader and stronger absorption, for instance at 
\mbox{35\,--\,38}, \mbox{41\,--\,55} and \mbox{65\,--\,70~km~s$^{-1}$} towards  W49N.  Ammonia seems to 
trace  CH slightly better towards G10.6$-$0.4 than  W49N. 
This comparison suggests   that CH exists in both relatively low and high density gas, while  
the nitrogen hydrides only exist in  the parts of the interstellar gas with a relatively high  density.

These differences  are even more pronounced when comparing with 
neutral hydrogen    
as observed by the VLA $\lambda$21~cm absorption by  
\citet{2003ApJ...587..701F}, which shows  absorption over a very wide range of velocities 
with a resolution of 2.5~km~s$^{-1}$.  
The much more extended velocity space coverage of the HI absorption is expected
since not all foreground clouds have molecular gas.

The ammonia absorption  also follow  similar trends as  
HCO$^+$ $J$\,=\,1$\leftarrow$0 \citep[][]{2010Godard}, and  H$_ 2$O, which is known to trace clouds of high molecular fraction 
\citep{2010A&A...518L.108N, 2010A&A...521L..12S}. 
HCO$^+$ and  H$_ 2$O seem to have a rather constant abundance ratio in these sight-lines, in contrast
to their abundances with respect to ammonia. 
The comparison of ammonia to  
 H$_ 2$O$^+$ and OH$^+$, which mostly reside  in lower density  gas containing considerable amounts
of atomic hydrogen 
\citep{2010A&A...518L.110G, 2010A&A...521L..10N}, shows no similarities.

When we compare ammonia with CN and HNC we find 
very similar absorption patterns.  
The CN \mbox{$J$\,=\,1$\leftarrow$0} \mbox{$F$\,=\,1/2\,--\,3/2, 1/2\,--\,1/2} single hyperfine component 
and  the HNC $J$\,=\,1$\leftarrow$0 line were observed  with the IRAM 30~m antenna  
\citep{2010Godard}. Note, that HNC has three hfs components that, however, lie very close 
in velocities (0.2 and
0.5~km~s$^{-1}$), which mostly leads to a broadening  of the absorption features.
These species mainly reside in denser gas than CH, and both are   closely connected to the NH and NH$_2$ chemistry    
\citep[e.g.][]{1990MNRAS.246..183N, 2001A&A...370..576L}.


\subsubsection{Column density correlations}
 
\begin{figure*}[\!ht]
\centering
\subfigure[]{
\includegraphics[width=.4\textwidth]{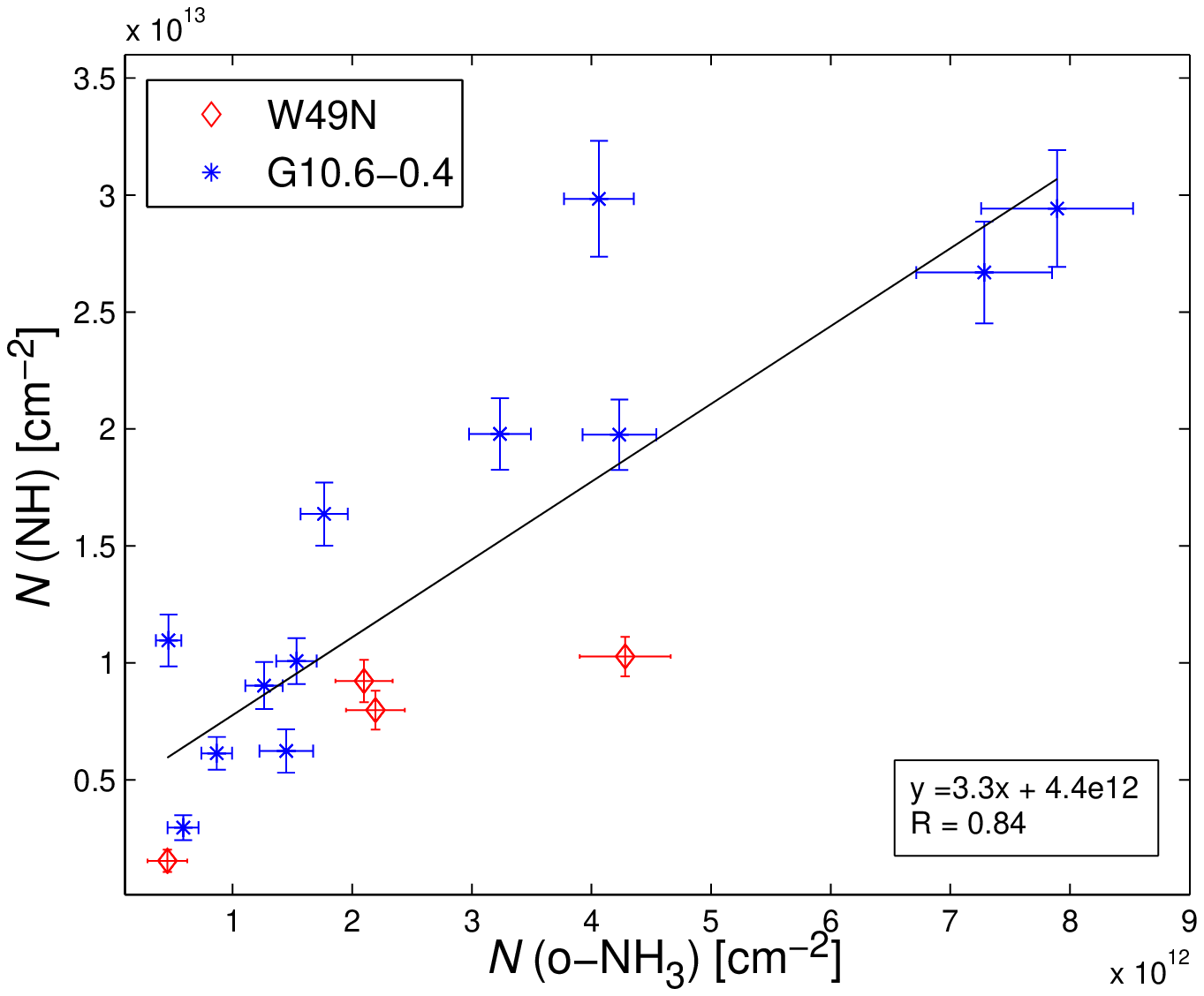}}   
\hspace{.3in}
\subfigure[]{
\includegraphics[width=.4\textwidth]{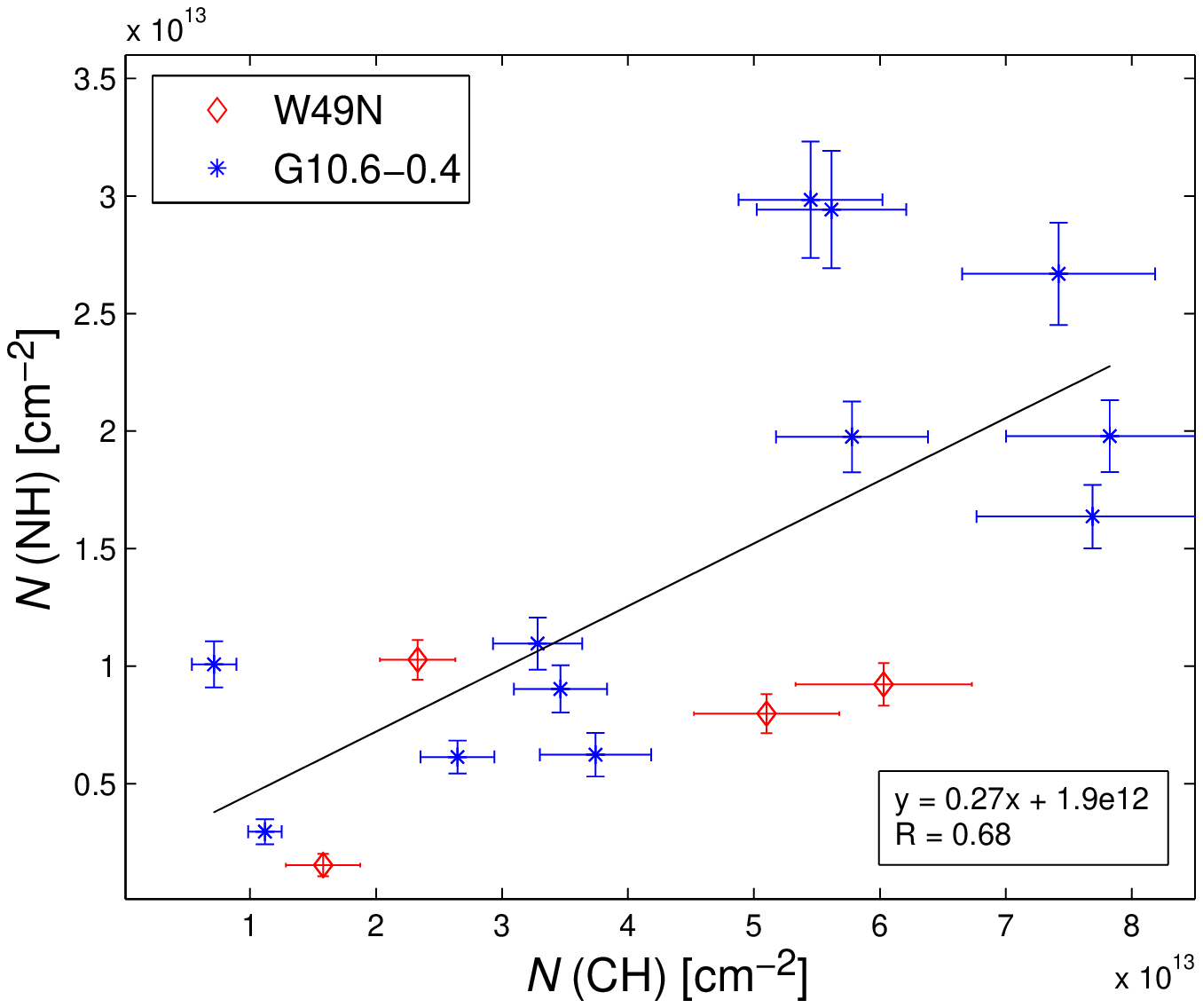}}  
\vspace{.3in}
\subfigure[]{ 
\includegraphics[width=.4\textwidth]{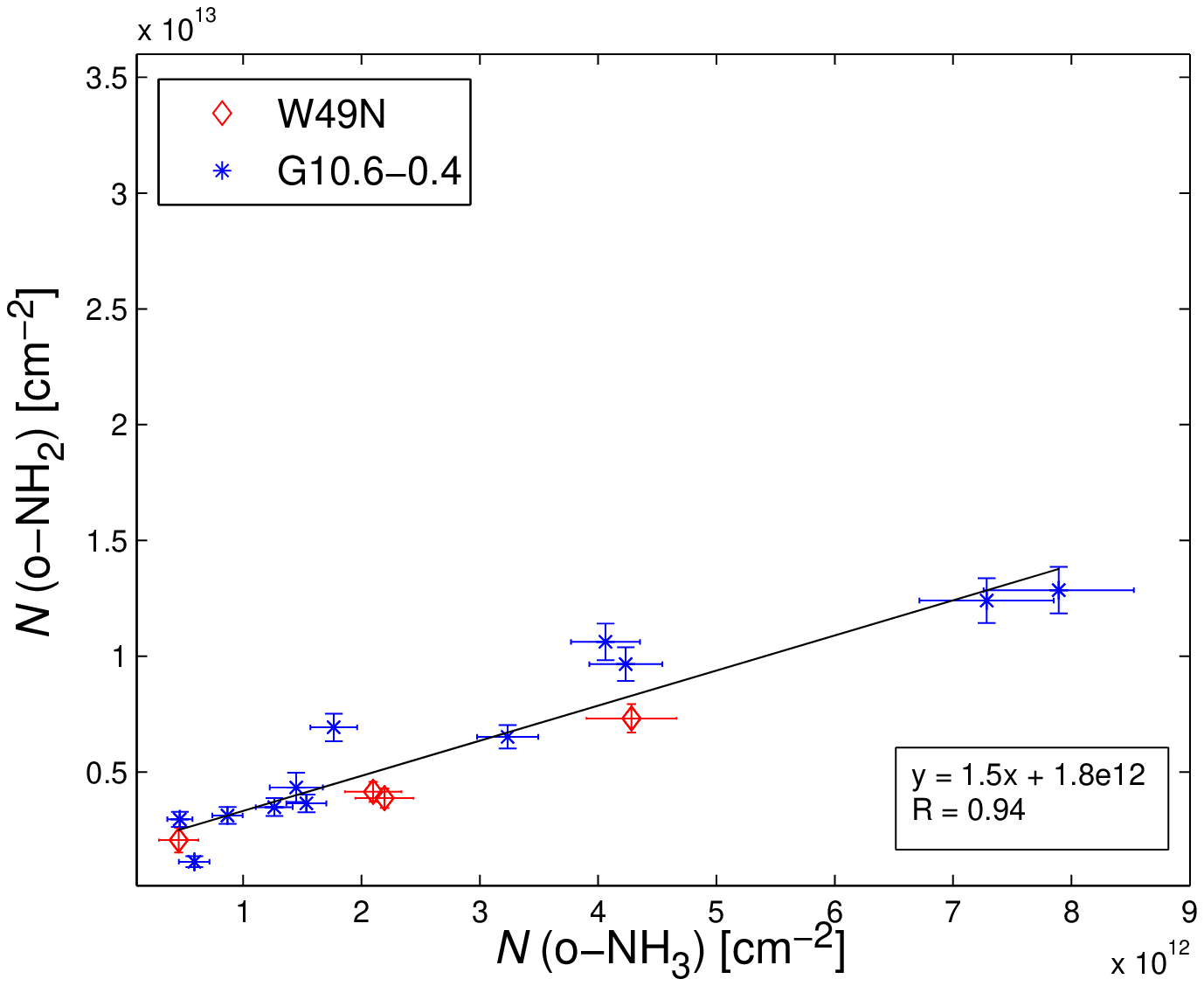}}     
\hspace{.3in}
\subfigure[]{
\includegraphics[width=.4\textwidth]{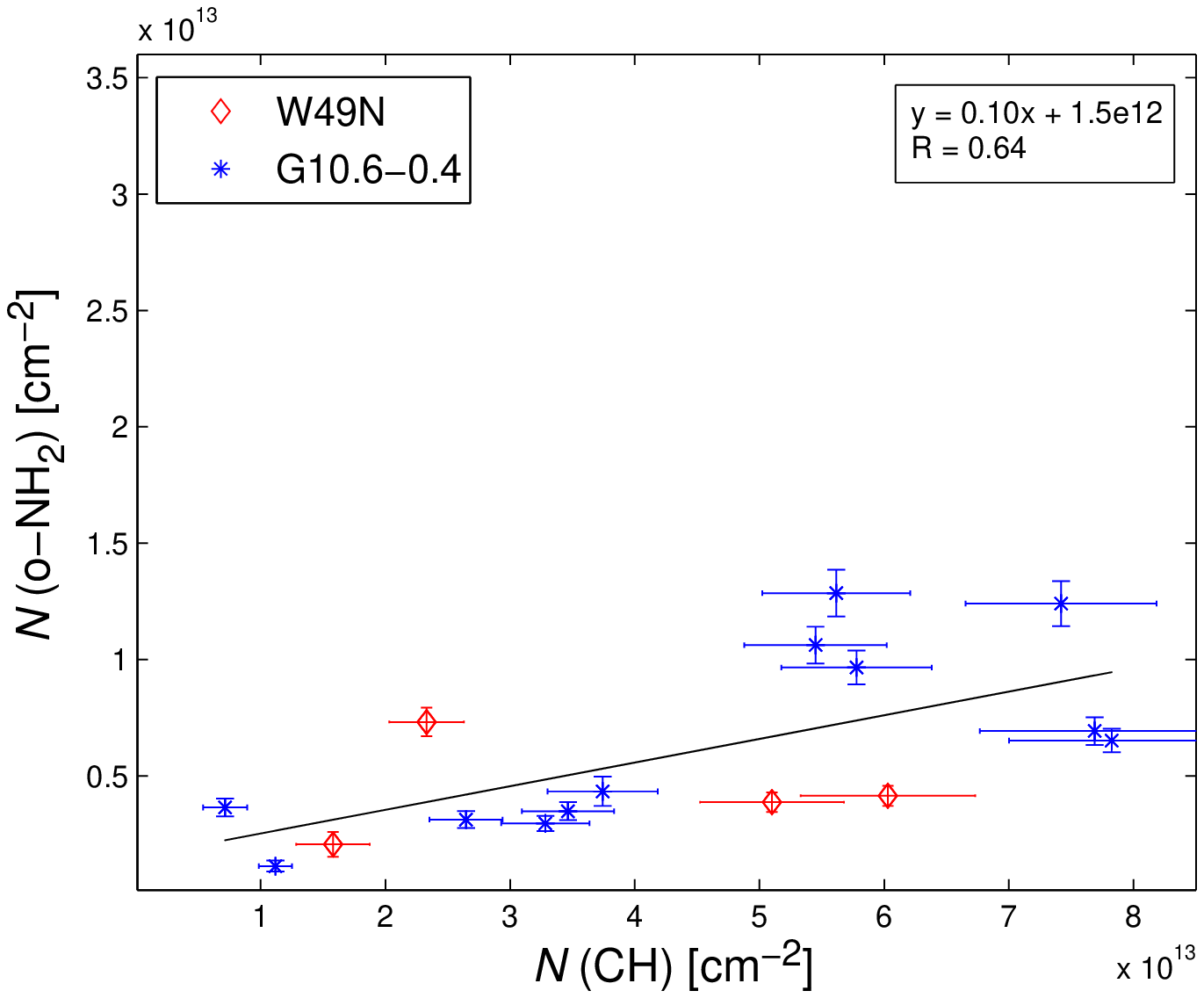}}   
\vspace{.3in}
\subfigure[]{ 
\includegraphics[width=.4\textwidth]{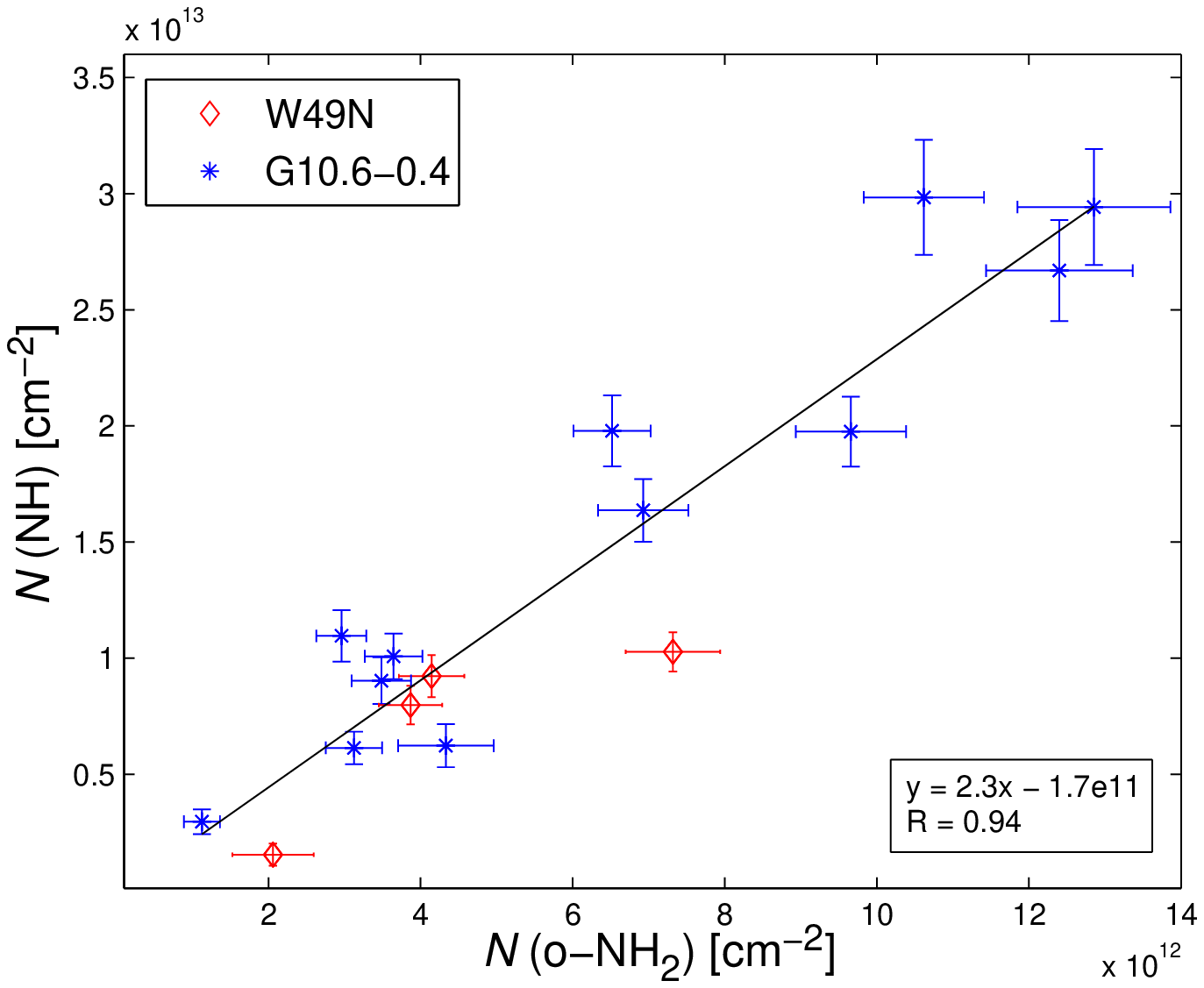}} 
\hspace{.3in}
\subfigure[]{
\includegraphics[width=.4\textwidth]{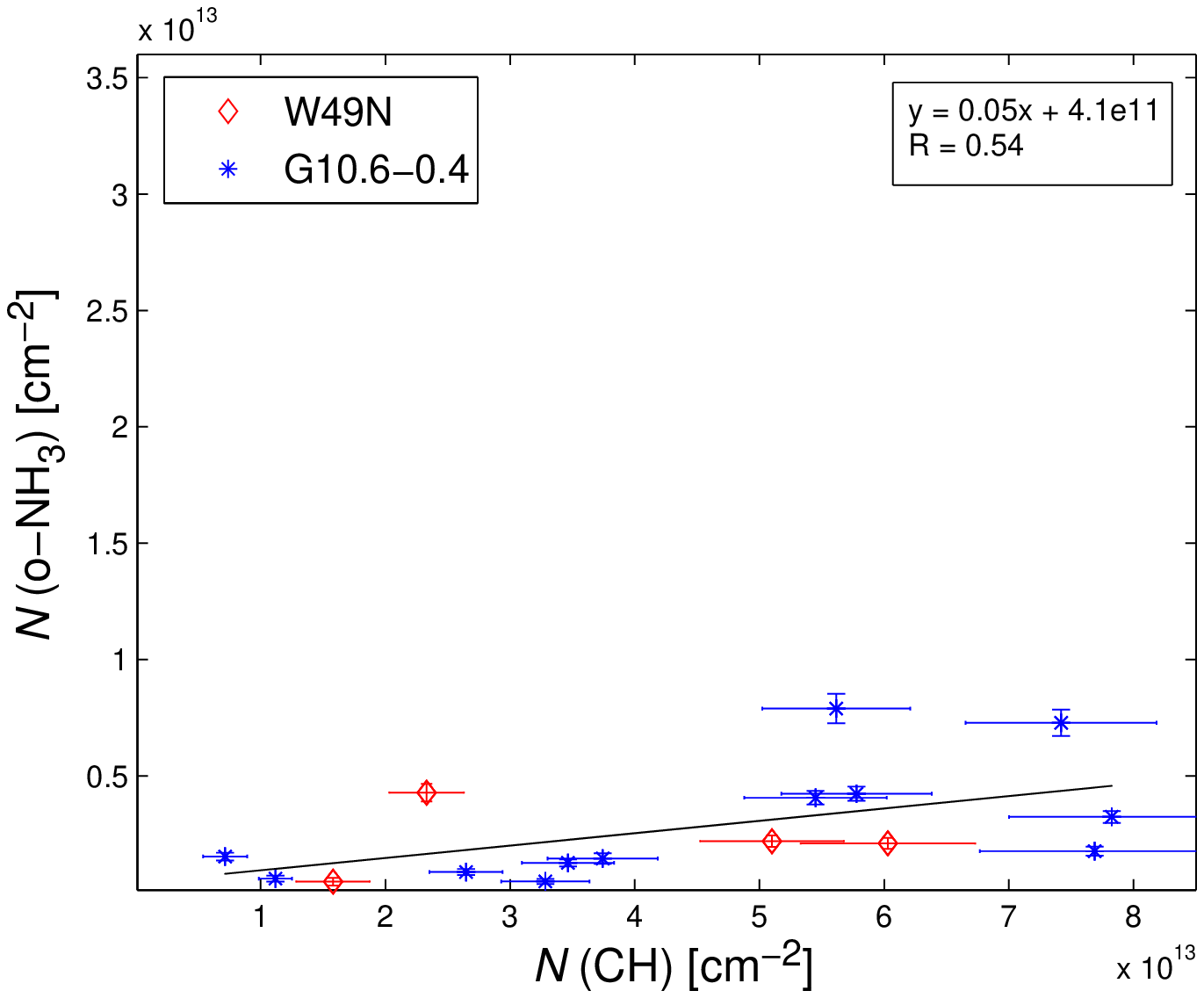}}  
\caption{\emph{Column density comparison plots.} On the left hand side: NH, o-NH$_2$ and o-NH$_3$.     
On the right hand side:  nitrogen hydrides vs. CH. The blue lines show linear least square fits to all data points 
towards both sources. The correlation coefficients and the linear fit equations   are given in respective figure.}
\label{fig: column density plots 1}
\end{figure*}

\begin{figure*}[\!ht]
\centering
\subfigure[]{
\includegraphics[width=.4\textwidth]{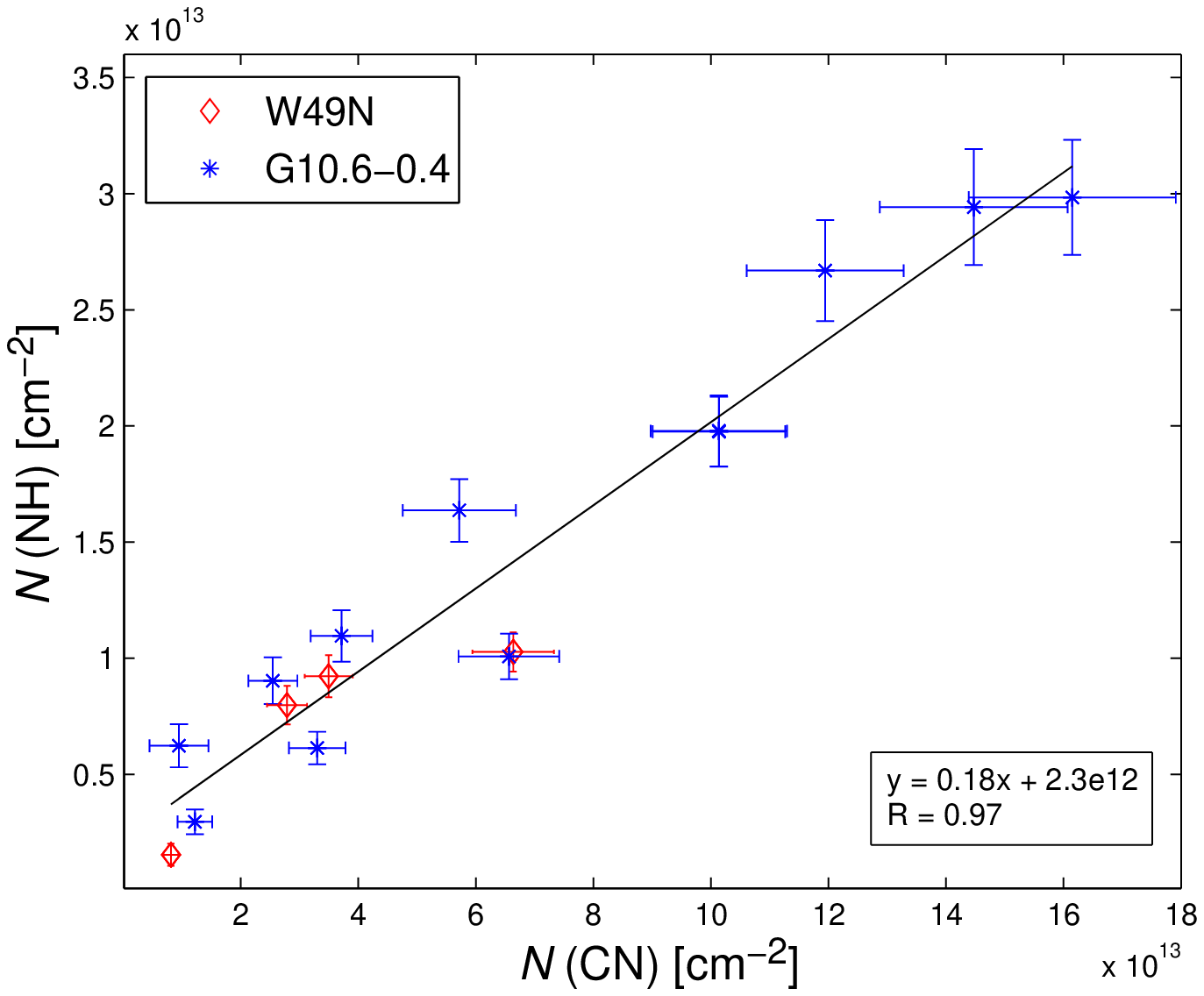}}    
\hspace{.3in}
\subfigure[]{
\includegraphics[width=.4\textwidth]{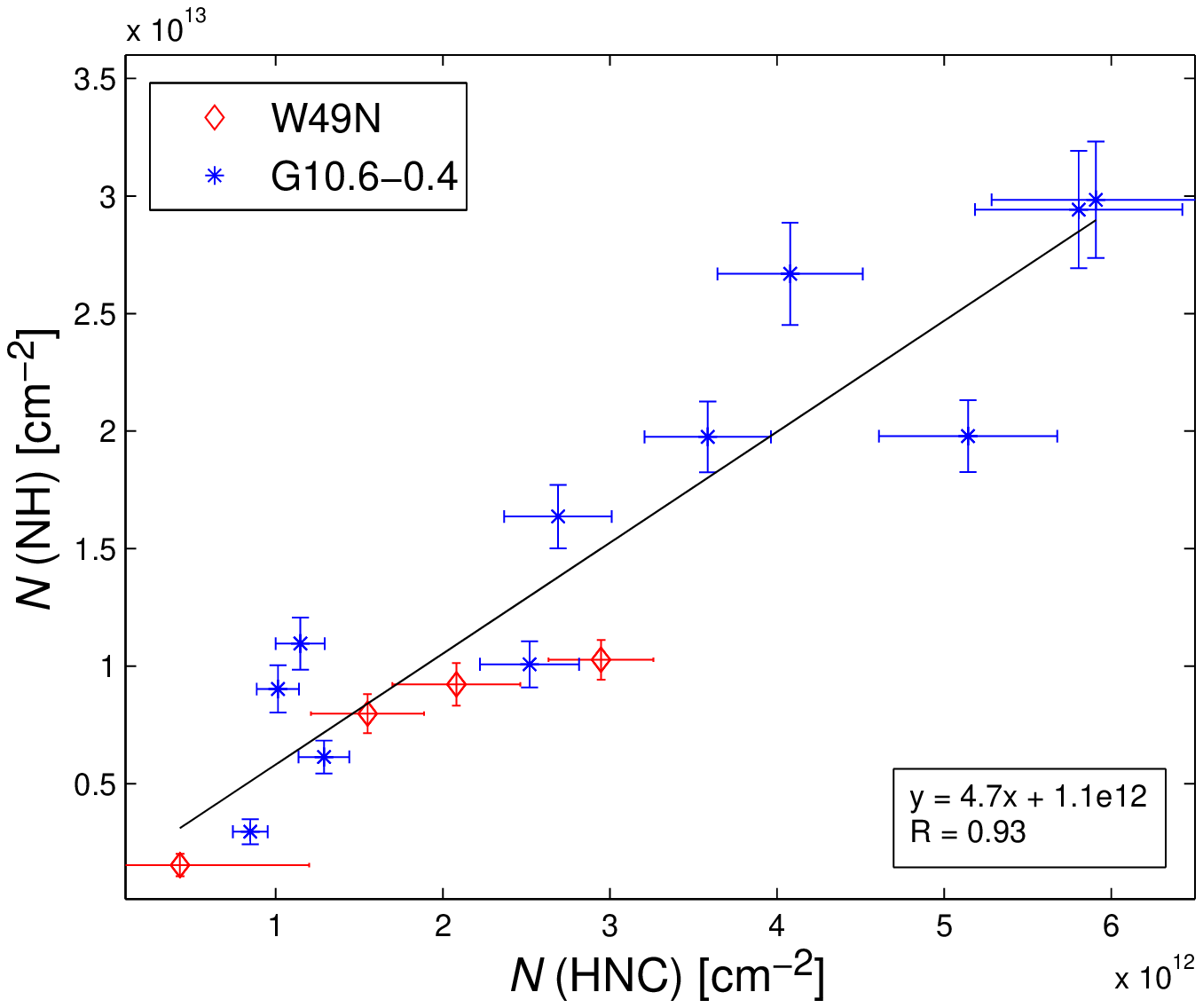}}   
\vspace{.3in}
\subfigure[]{
\includegraphics[width=.4\textwidth]{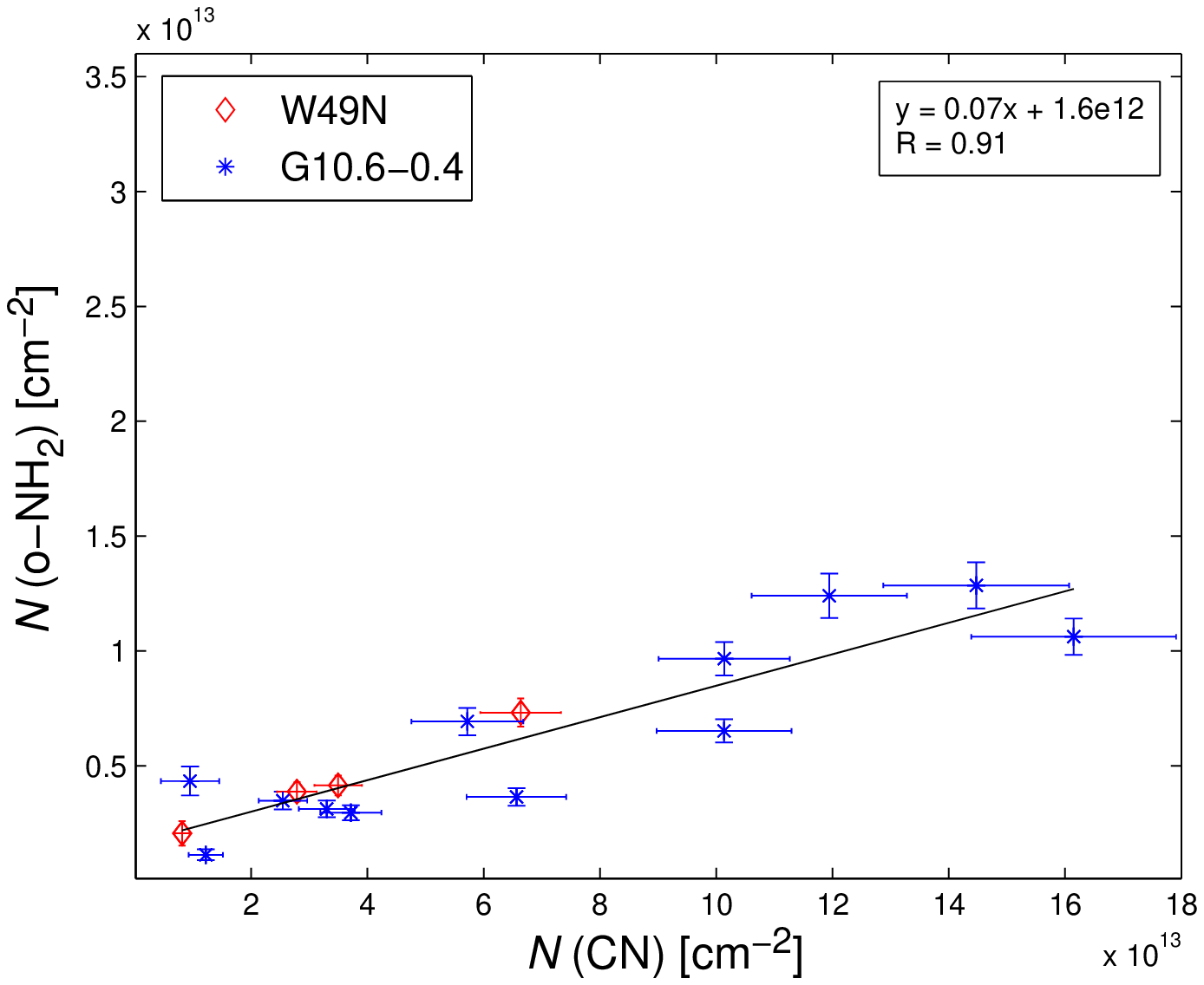}}    
\hspace{.3in}
\subfigure[]{
\includegraphics[width=.4\textwidth]{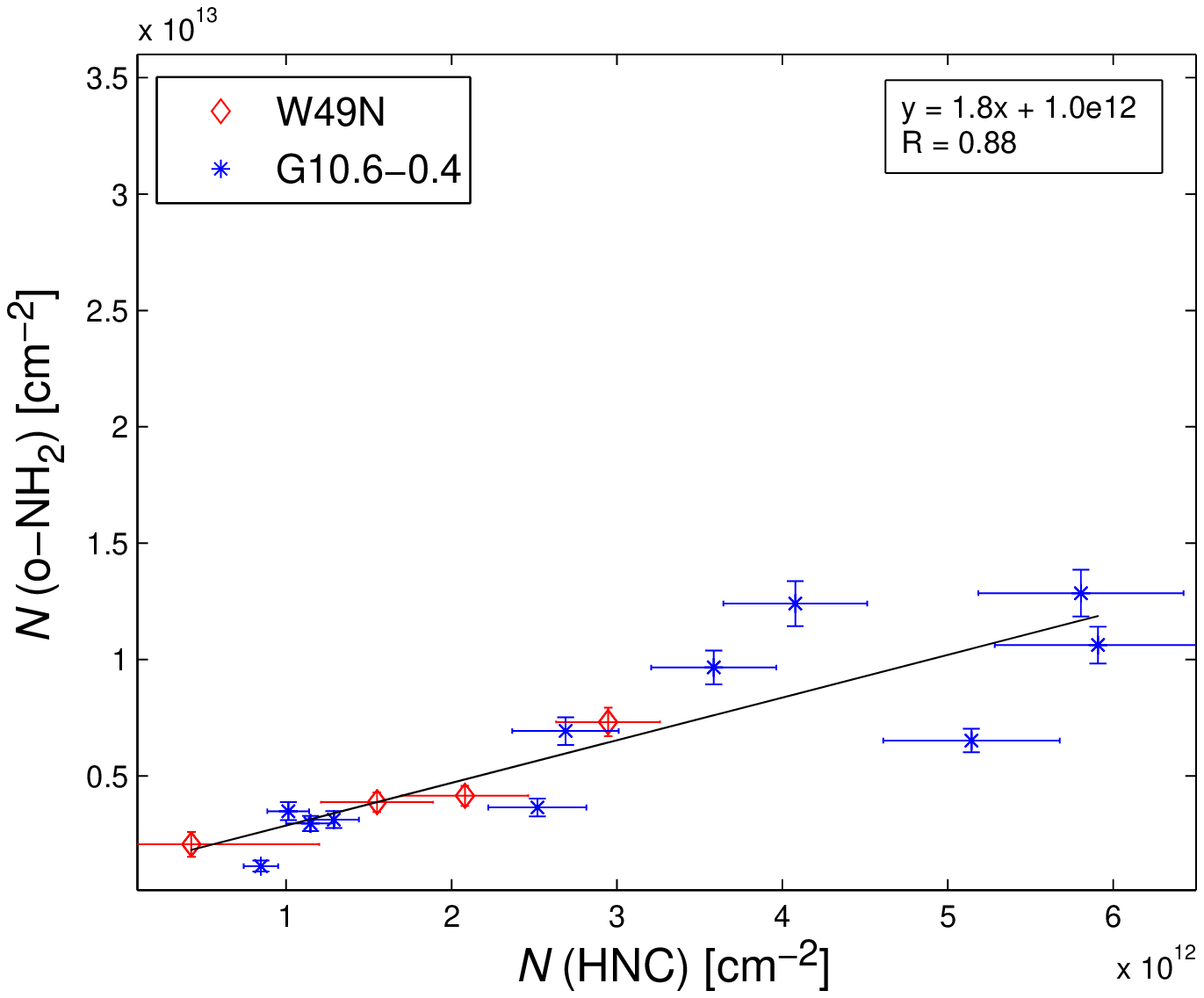}}   
\vspace{.3in}
\subfigure[]{ 
\includegraphics[width=.4\textwidth]{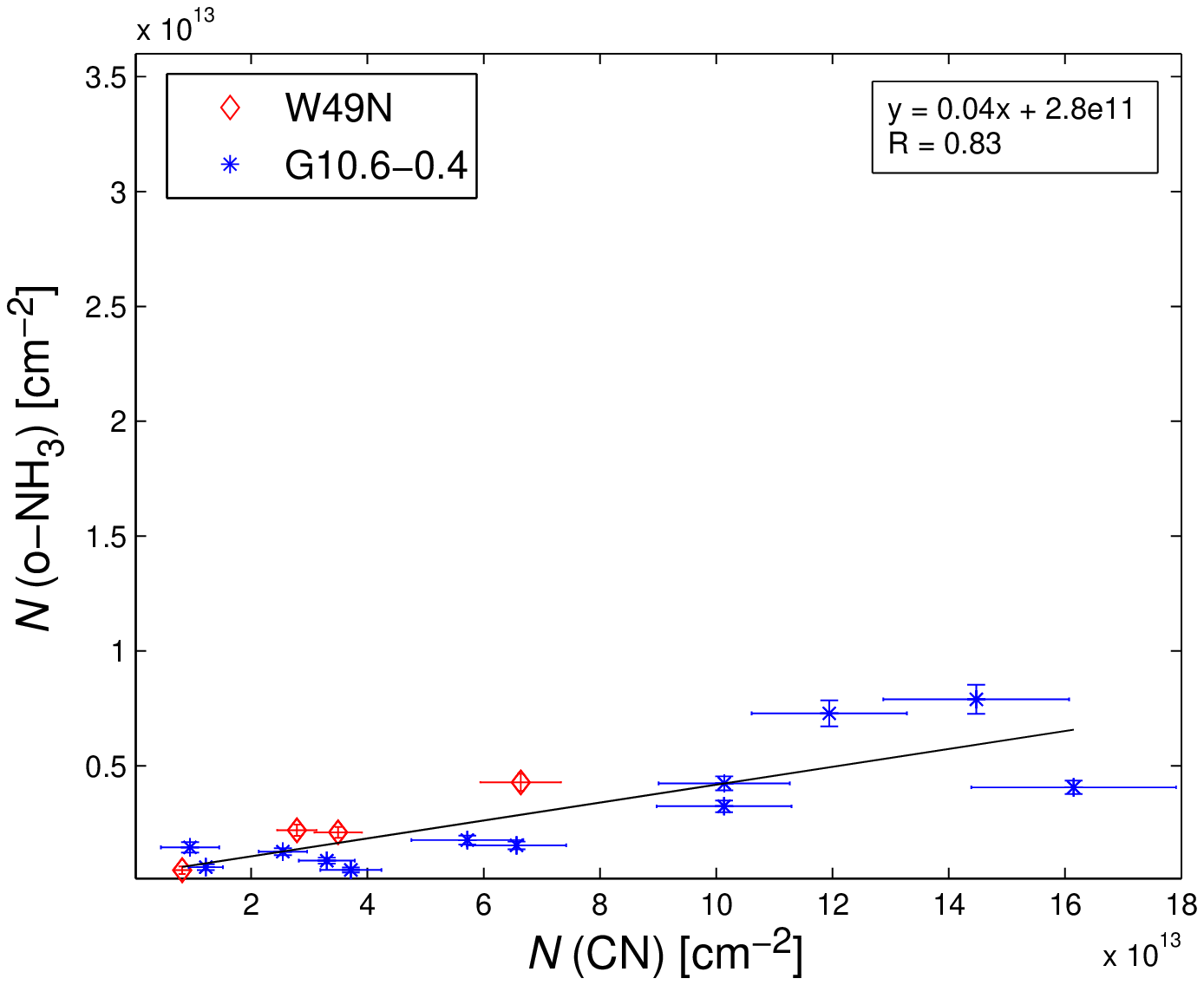}}   
\hspace{.3in}
\subfigure[]{
\includegraphics[width=.4\textwidth]{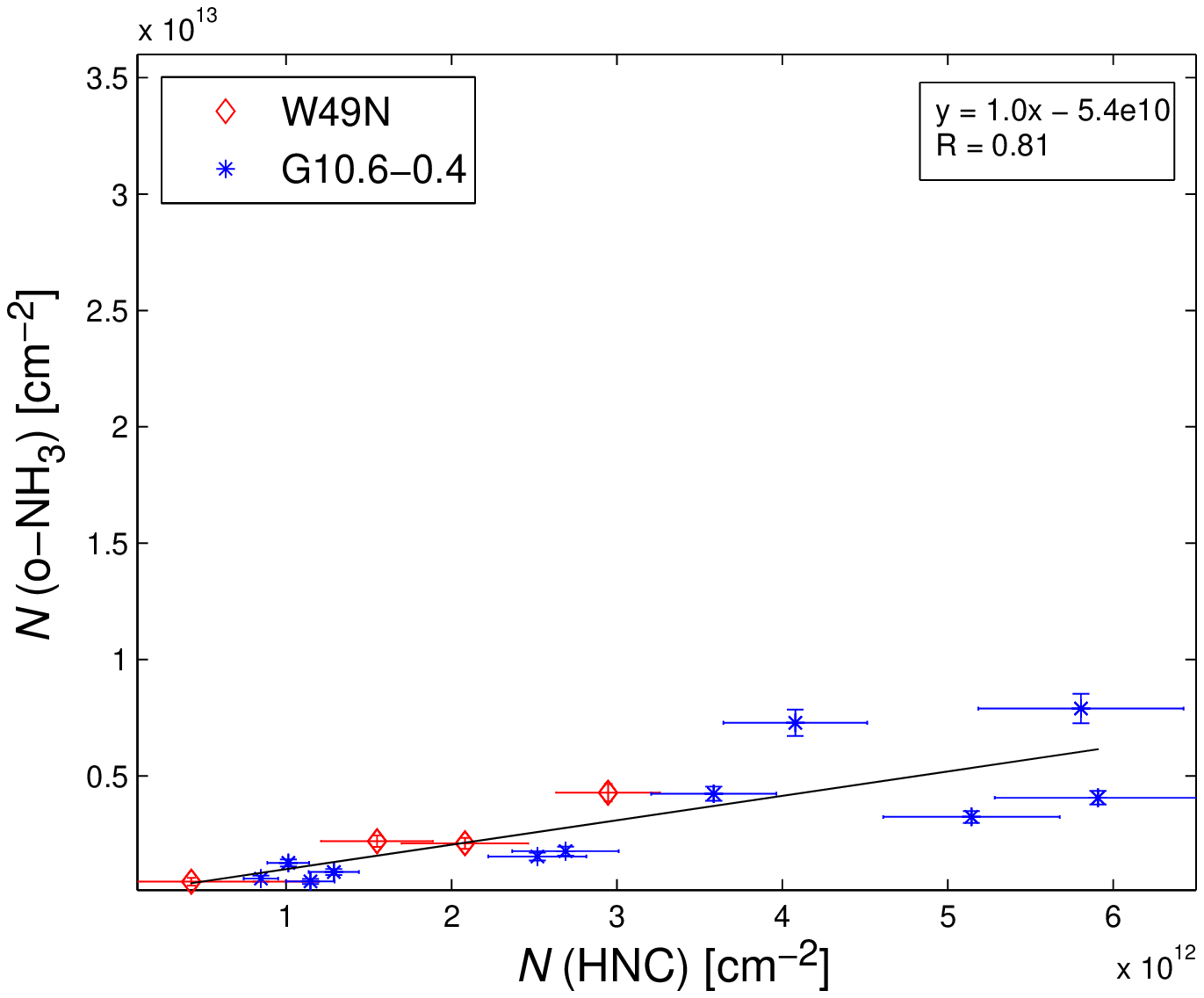}}   
\caption{\emph{Column density comparison plots.} On the left hand side we show comparisons of the nitrogen hydrides  with CN, and
on the right hand side comparisons   with HNC. Notation as in Fig.~\ref{fig: column density plots 1}.}
\label{fig: column density plots 2}
\end{figure*}

In order to quantitatively examine abundance correlations of the nitrogen hydrides, CH, HNC and CN, we 
show  column density plots   
in Figs.~\ref{fig: column density plots 1}\,--\,\ref{fig: column density plots 2}. 
We have here plotted the results from Method~I. The parameters of the linear least square fits to the 
data are found in the figures in 
addition to the correlation coefficient $R$. 
In the on-line material we also show column density plots with all three methods 
(Figs.~\ref{Fig: column density plots 1}\,--\,\ref{Fig: column density plots 4}). 

Column density plots  
of the nitrogen hydrides are shown 
on the left hand side in  Fig.~\ref{fig: column density plots 1}.  
We note that the scatter  is larger for  $N$(NH) vs. \mbox{$N$(o-NH$_3$)}	 than compared 
to \mbox{$N$(o-NH$_2$)} vs. \mbox{$N$(o-NH$_3$)}   which show a rather tight correlation. This may indicate 
that NH does not entirely exist in the same gas as \mbox{o-NH$_2$} and    \mbox{o-NH$_3$}. 
There is also a possibility that the NH and o-NH$_3$ 
correlation may   only  be valid for      
$N$(\mbox{o-NH$_3$})\,$\lesssim$\,5$\times$10$^{12}$~cm$^{-2}$.   
The  
$N$(NH) towards G10.6$-$0.4  appear to increase up to a maximum of $\sim$3$\times$10$^{13}$~cm$^{-2}$. 
Also  $N$(\mbox{o-NH$_2$}) show a slight 
tendency of this behaviour towards G10.6$-$0.4 with a maximum of $\sim$1.2$\times$10$^{13}$~cm$^{-2}$.  
This possible "chemical saturation" corresponds to     a few $\mathrm{A}_{V}$ (estimated from our CH measurements and 
\mbox{$N_\mathrm{H} \sim 1.7 \times 10^{21}$~cm$^{-2}$ mag$^{-1} \times \mathrm{A}_{V}$)}	.
This 
may be explained by a more efficient ammonia production at higher column densities, or that NH, and \mbox{o-NH$_2$}, are somehow consumed in the ammonia formation.

On the right hand side in Fig.~\ref{fig: column density plots 1}  we show column density plots of the 
nitrogen hydrides vs. CH. 
The spread of the data points 
reflects the difficulty  of finding a velocity
structure common to hydrides and CH. 
 
In Fig.~\ref{fig: column density plots 2}  we 
show column density plots of the nitrogen
hydrides vs. CN on the left  and vs. HNC on the right. 
All three nitrogen hydrides show   linear correlations with both  CN and HNC.  
In dark cloud chemistry, CN and HNC   are closely related to NH and  NH$_2$~through the 
reactions of \mbox{C + NH$_2$\,$\rightarrow$\,HNC + H}, \mbox{N + CH\,$\rightarrow$\,CN + H}, and 
\mbox{C + NH\,$\rightarrow$\,CN + H} \citep[][]{1990MNRAS.246..183N}.
Also \citet{2009MNRAS.400..392W} found a slightly better correlation between 
NH and CN than compared to
CH,  and no correlations with 
species such as CH$^+$.

\section{Ortho-to-para ratio  of NH$_3$}  \label{OPR ammonia}

There are two ways to produce
molecules in interstellar space: in the gas-phase or on grain surfaces.
If ammonia is formed in the gas-phase,  by highly exoergic processes,  
the   ammonia 
ortho-to-para   ratio has been expected to be very close to  the  statistical equilibrium value of 
1.0 
(the spin statistical value 
is 4\,--\,2 for ortho-para, but the number of para states is on the other hand almost a 
factor of two larger).
If ammonia is formed at  temperatures lower than 40~K on cold dust grains, and 
then desorbed when the grains are heated above 100~K, 
 the OPR may differ from unity since the lowest 
ortho level is 22~K  below the lowest para level. 
If  no conversion processes between the two symmetries exist, 
the OPR of ammonia is expected to increase \emph{above} unity at low formation temperatures. 
 
Prior to Herschel, ortho-to-para ratios have been derived from measurements 
of inversion transitions, the $(1,1)$ and $(2,2)$ transitions involving
para states, and the $(3,3)$ and $(6,6)$ transitions probing highly
excited, metastable ortho states. For example,  \citet{1999ApJ...525L.105U}  
have found OPR\,=\,1.3\,--\,1.7 in the L~1157 outflow from the observation of six 
inversion lines $(J,K)=(1,1)$ to $(6,6)$. Similarly,
\citet{2009PASJ...61.1023N}  have inferred an even higher value of 
OPR\,=\,1.5\,--\,3.5 in the central molecular zone of the Galaxy.

Using \emph{Herschel}-HIFI observations of the  fundamental  rotational  
transitions  of both   
ortho- and para-ammonia, 
it is for the first time possible to estimate the ammonia OPR in cold and 
diffuse interstellar gas of low excitation.   
Our results do, however, point to the surprising result of an OPR \emph{lower} 
than unity.

\begin{figure}
   \centering
\includegraphics[scale=0.6]{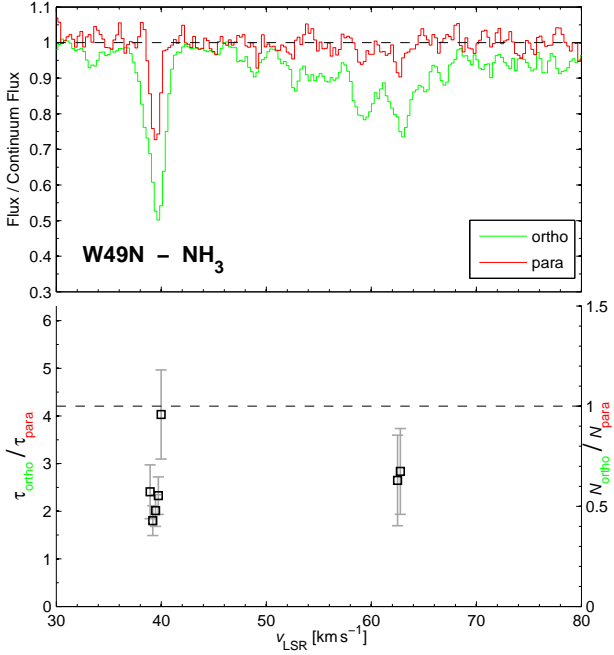}  
\caption{{\emph W49N.}  \emph{(Upper)} WBS   spectra of \mbox{1$_0$\,--\,0$_0$} ortho-NH$_3$ and \mbox{2$_1$\,--\,1$_1$} 
para-NH$_3$ normalised to SSB 
continuum. \emph{(Lower)} The optical depth ratios are shown    for absorptions   
larger than 3$\sigma$ as a function of 
LSR velocity. The 
column density ratios (OPR), estimated with {\tt RADEX}, are given by the right hand  y-axis. 
The horisontal dashed line marks  
an ortho-to-para optical depth ratio of 
4.2  corresponding to a column density
ratio  of unity. 
} 
\label{Fig: W49N OPR NH3}
\end{figure}

\begin{figure}
   \centering
\includegraphics[scale=0.6]{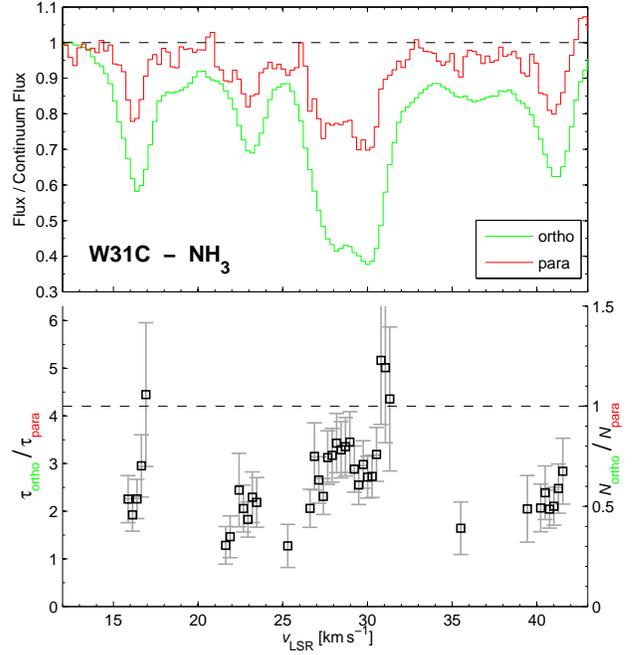}
\caption{{\emph G10.6$-$0.4 (W31C).} Notation as in Fig.~\ref{Fig: W49N OPR NH3}.
}
 \label{Fig: W31C OPR NH3}
\end{figure}

This is shown in Figs.~\ref{Fig: W49N OPR NH3} and \ref{Fig: W31C OPR NH3}
where the upper panels show the normalised  1$_0$\,--\,0$_0$ ortho-NH$_3$ 
and 2$_1$\,--\,1$_1$  para-NH$_3$ spectra, and 
the lower panels show the corresponding optical depth ratios for absorptions larger than 3$\sigma$   
  as a function of LSR velocity towards both sources 
(channel widths are 0.26~km~s$^{-1}$). An ortho-to-para optical depth ratio of 
4.2 corresponds to a column density
ratio  of unity using the {\tt RADEX} conversion factors in 
Table~\ref{Table: columns opacity one}. The resulting column density ratios are 
given by the right hand y-axis in both figures. The para line has lower S/N  
than the ortho line in 
both sources   and several
velocity components are  weak. Still, most column density ratios are 
found to be below
unity  within  the uncertainties.

We have also used Method~I and II to estimate the OPR.  
Method~I gives \mbox{OPR(NH$_3$)\,=\,0.7$\pm$0.1} and  \mbox{0.6\,--\,0.7$\pm$0.1}  
in 
the strongest velocity
components towards  W49N  (+39~km~s$^{-1}$ component) and  G10.6$-$0.4  
(+16, 28, 30 and 41~km~s$^{-1}$ components), respectively. 
The results from Method~II  gives  similar results  
towards  G10.6$-$0.4:  \mbox{0.5\,--\,0.7$\pm$0.1} in the velocity bins 
\mbox{$v_\mathrm{LSR}$\,=\,12.5\,--\,20}, \mbox{25\,--\,29}, \mbox{29\,--\,31} and \mbox{39.5\,--\,43.5~km~s$^{-1}$}; but lower results     towards 
  W49N in $v_\mathrm{LSR}$\,=\,36\,--\,42~km~s$^{-1}$,  0.4$\pm$0.2.

The errors are dominated by the noise and the uncertainty in the sideband gain ratio. 
We have used  the errors in sideband gain ratios stated on the \emph{Herschel} 
internet site (4\% for ortho and 6\% for para).  
There is also 
an  additional error arising from the rather large sideband separation of 12~GHz  
which means 
that the assumption of equal continuum temperature is not   fully valid. 
This effect  mimics a sideband gain ratio different from unity and is 
also taken into account  
(4 and 2\% for ortho and para, respectively, estimated by   
$T_\mathrm{C,L}/T_\mathrm{C,U} = (\nu_\mathrm{L}/\nu_\mathrm{U})^\beta$ 
with $\beta = 2$).  
The   errors   from the noise 
 is estimated by 
$\delta \tau$\,=\,$\exp(\tau) \times \delta I /I_0$,   and 
the errors from the sideband gain ratio 
and differences in the continuum in the sidebands    are 
estimated by
$\delta \tau = \epsilon - ln(1 + \epsilon \exp(\tau))$ where
$\epsilon$ is both errors added in quadrature. 
The derivation of this formula assumes that the depth of the absorption line
is correct,  which means that 
$T_\mathrm{C}$(SSB)\,--\,$T_\mathrm{A}$ is conserved  but the continuum level is not.  
The noise and calibration errors are finally   added in quadrature.  
The error estimates 
for Method~I and II 
add respective uncertainties from the methods to    
the  calibration errors.

Some previous measurements of ammonia have resulted in an OPR lower than unity, although
not in diffuse gas. In Orion~KL      
\citet{1988AA...201..285H} obtained   
OPR(NH$_3$)\,=\,0.5  from inversion 
emission lines, and  
\citet{1985AA...146..134H}  found OPR($^{15}$NH$_3$)\,=\,0.7. The latter low value was 
suggested to be caused by an 
excitation effect: a possible over-abundance of the unobserved $K$\,=\,0 state. 
This explanation cannot be applied in our case, 
since we actually observe the 
$K$\,=\,0 ortho state.    
\citet{1995A&A...294..815F} modelled several NH$_3$ inversion  
transitions in the frame of C-shock  models towards the warm and dense Sgr\,B2 envelope 
and their best OPR of ammonia  was $\sim$0.5, also lower than 
the statistical value of unity. 
\citet{2002A&A...383..603C}, however,  observed 21   high excitation ortho- and para-ammonia lines, 
both metastable and non-metastable levels,
towards Sgr\,B2 using ISO, and derived an ammonia OPR of unity.

There are to our knowledge no previous estimate  of the OPR(NH$_3$) in \emph{diffuse} gas, 
and it 
may very well be very different from the OPR in dense gas. 
We can, however, compare our  
strong absorption of the para-NH$_3$~line   to 
the observations by
 \citet{1994A&A...289..579T} who observed    
the two lowest inversion lines of para-NH$_3$~towards  W49N. 
Their estimate of \mbox{$N$(p-NH$_3$)} is 2.5$\times 10^{12}$~cm$^{-2}$ which is less than half our value,  
$\sim$6$\times 10^{12}$~cm$^{-2}$. 
If this column density  
 is correct,  it would imply an OPR higher than unity in this velocity 
component in W49N. 
In our excitation model, however, the integrated optical depth observed by 
\citet{1994A&A...289..579T} in the $1_1$ inversion line at $+39$~km~s$^{-1}$ implies  a  
\mbox{$N$(p-NH$_3)$} in good agreement with our value, since we obtain a higher excitation temperature   than the 2.7~K 
 used by \citet{1994A&A...289..579T}
for the $1_1$ inversion transition.

We have no clear explanation yet of our surprising result and we can only speculate about its origin. 
Either there are to us unknown instrumental effects, or that our assumption that the line-of-sight gas completely
covers the background continuum within the beam is not correct, or 
there must exist some physical or chemical processes that affect the OPR   in diffuse gas.   
In general, the fact that interstellar clouds are weak plasmas allows for  
conversion processes at 
low temperatures that would not be present in a purely neutral gas.  
What OPR these processes lead 
to must be determined by   careful modelling.
In diffuse molecular clouds of relatively low density, 
$n({\rm H}_2) \lesssim 10^3$, but relatively high ionisation fraction, 
$n(e)/n({\rm H}_2)\ga10^{-4}$, the rate of destruction of NH$_3$ by 
reactions with C$^+$, H$^+$, and H$_3^+$, can be as high as $10^{-9}$~s$^{-1}$ 
at kinetic temperatures $T\!\sim\!30$~K. The latter two ions
can also interchange ortho and para states directly by proton substitution, perhaps
at a comparable rate. These rates of destruction and interchange can 
approach the rates of radiative and collisional excitation out of the
metastable levels, which means that the excitation and chemistry of 
NH$_3$  should  be treated together in a self-consistent fashion. Under these
conditions, NH$_3$ is a non-equilibrium system and a ``spin temperature''
derived from an ortho-to-para ratio is not expected to be meaningful.
  
The NH$_4^+$ molecule could also affect the ammonia OPR in a similar manner.  
The fastest gas-phase reaction sequence of ammonia is 
\begin{equation}
\mathrm{N^+}   \rightarrow^\mathrm{H_2}  \mathrm{NH^+}  \rightarrow^\mathrm{H_2} \mathrm{NH_2^+}  \rightarrow^\mathrm{H_2} \mathrm{NH_3^+} \rightarrow^\mathrm{H_2} \mathrm{NH_4^+}  \rightarrow^\mathrm{e^-}  \mathrm{NH_3}, \, \mathrm{NH_2}, 
\end{equation}
where N$^+$ is formed by cosmic ray ionisation, or by reactions of He$^+$ with 
N$_2$ or CN which are formed by neutral-neutral
reactions.  
Whether or not ammonia is produced on grains or in the gas, its destruction 
in the gas by 
protonating ions such as H$_3^+$ also leads to NH$_4^+$, which can have a variety of nuclear spin states  
depending on the overall nuclear spin of H$_3^+$.   
The ammonia OPR could then simply reflect the spin states of NH$_4^+$.

In addition, some laboratory evidence exists that dissociative
recombination reactions have different rate coefficients depending upon
the nuclear spin configuration of the molecular ion.
 
Observations of the ortho-to-para ratio of NH$_3$ could thus provide  a valuable  
insight into the competing processes of formation and destruction, 
radiative and collisional excitation, and reactive interchange processes.

\section{Summary} \label{section summary}

Our spectrally resolved  rotational transitions of NH, o-NH$_2$, ortho- and para-NH$_3$ 
along the   sight-lines towards the high-mass star-forming 
regions  W49N  and G10.6$-$0.4 show remarkable similarities
of line profiles and abundances.
We find  similar abundances of all three species and 
a co-existence in   diffuse or translucent  interstellar gas 
with a high molecular fraction. 
The mean abundance of \mbox{ortho-NH$_ 3$} abundance is \mbox{$\sim$2$\times$10$^{-9}$} towards both sources.  
The  mean ratios of all three methods   of the nitrogen hydrides 
in all velocity components,      
are  \mbox{$N$(NH)/$N$(o-NH$_3)$\,=\,5.9 and 3.5},  and  \mbox{$N$(o-NH$_2$)/$N$(o-NH$_3$)\,=\,2.4} and 2.0, 
towards G10.6$-$0.4 and W49N, respectively.   
This is in sharp contrast to previous observations of the nitrogen hydrides in dark clouds 
where the 
ammonia abundances are found to be $\sim$100 times higher than $X$(NH), 
and $\sim$10\,--\,300 	
times higher than $X$(NH$_2$).  
NH and \mbox{o-NH$_2$} are  found to be  linearly correlated with  \mbox{o-NH$_3$} at least for       
$N$(\mbox{o-NH$_3$})\,$\lesssim$\,5$\times$10$^{12}$~cm$^{-2}$ which corresponds to a few $A_\mathrm{V}$.  
Upper limits of $N$(NH$^+$) in both sources indicate a $N$(NH$^+$)/$N$(NH) ratio of $\lesssim$2\,--\,14\,\%, with a mean 
of $\lesssim$6\,\%.

Linear correlations are also found for all three nitrogen hydrides with  respect to
CH, CN and HNC, although CH displays a  more loose correlation  than the latter two species. 
The nitrogen hydrides also largely follow the absorption pattern in Doppler velocity space of   
HCO$^+$ and
water, a species also known to trace regions of a high molecular fraction. 

We have obtained a surprisingly low  ortho-to-para ratio of ammonia, 
\mbox{$\approx$\,0.5\,--\,0.7$\pm$0.1}, in the strongest velocity components, which is below the 
high-temperature limit of unity. 
No clear explanation  has been found. 
More observations are needed of both the    rotational  transitions  and the 
inversion
lines with ground-based facilities, to be able to make firm conclusions about the  ammonia OPR in diffuse gas. 
 
We will continue to investigate the absorption lines in the sight-lines towards the other six PRISMAS sources. 
This will allow an analysis of the nitrogen chemistry 
at
various galactic distances from the Galactic Centre.   
We will also use new  Open~Time~1 (OT1) \emph{Herschel}-HIFI data of higher 
excitation lines to analyse the hot core sources  which will be compared and contrasted  with the
diffuse interstellar gas. 
The ortho-to-para ratio of NH$_3$ will also be further 
investigated both in  the sources and in the diffuse gas, 
in addition to the  OPR  of  NH$_2$, for which  new  OT1 
data in four of the PRISMAS sources will be analysed and compared to the ammonia OPR. 
 

 

 

\begin{acknowledgements}
The \emph{Herschel} spacecraft was designed, built, tested, and launched under a contract to ESA managed by 
the Herschel/Planck Project team by an industrial consortium under the overall responsibility of the prime 
contractor Thales Alenia Space (Cannes), and including Astrium (Friedrichshafen) responsible for the payload 
module and for system testing at spacecraft level, Thales Alenia Space (Turin) responsible for the service module, 
and Astrium (Toulouse) responsible for the telescope, with in excess of a hundred subcontractors.
HIFI has been designed and built by a consortium of institutes and university departments from across
Europe, Canada and the United States under the leadership of SRON Netherlands Institute for Space
Research, Groningen, The Netherlands and with major contributions from Germany, France and the US.
Consortium members are: Canada: CSA, U.Waterloo; France: CESR, LAB, LERMA, IRAM; Germany:
KOSMA, MPIfR, MPS; Ireland, NUI Maynooth; Italy: ASI, IFSI-INAF, Osservatorio Astrofisico di Arcetri-
INAF; Netherlands: SRON, TUD; Poland: CAMK, CBK; Spain: Observatorio Astronómico Nacional (IGN),
Centro de Astrobiología (CSIC-INTA). Sweden: Chalmers University of Technology - MC2, RSS \& GARD;
Onsala Space Observatory; Swedish National Space Board, Stockholm University - Stockholm Observatory;
Switzerland: ETH Zurich, FHNW; USA: Caltech, JPL, NHSC.
CP and JHB acknowledge generous support from the Swedish National Space Board. 
MdL and MG acknowledge funding by CNES and by the ANR SCHISM project
(ANR-09-BLAN-0231-01). 
T.A.B, B.G. and J.R.G.  thank the Spanish MICINN for funding support
through grants
AYA2009-07304 and CSD2009-00038. 
H.S.P.M. is very grateful to the Bundesministerium f\"ur Bildung und
Forschung (BMBF) for financial support aimed at maintaining the
Cologne Database for Molecular Spectroscopy, CDMS. This support has been
administered by the Deutsches Zentrum f\"ur Luft- und Raumfahrt (DLR).
We also thank the  referee Harvey Liszt whose constructive comments led to a significant improvement of
the paper. 

\end{acknowledgements}

\bibliographystyle{aa-package/bibtex/aa}
\bibliography{references}

\begin{thebibliography}{85}
\expandafter\ifx\csname natexlab\endcsname\relax\def\natexlab#1{#1}\fi

\bibitem[{{Adande} \& {Ziurys}(2011)}]{2011IAUS..280P..76A}
{Adande}, G. \& {Ziurys}, L. 2011, in IAU Symposium, Vol. 280, IAU Symposium,
  76P

\bibitem[{{Bacmann} {et~al.}(2010){Bacmann}, {Caux}, {Hily-Blant}, {Parise},
  {Pagani}, {Bottinelli}, {Maret}, {Vastel}, {Ceccarelli}, {Cernicharo},
  {Henning}, {Castets}, {Coutens}, {Bergin}, {Blake}, {Crimier}, {Demyk},
  {Dominik}, {Gerin}, {Hennebelle}, {Kahane}, {Klotz}, {Melnick}, {Schilke},
  {Wakelam}, {Walters}, {Baudry}, {Bell}, {Benedettini}, {Boogert}, {Cabrit},
  {Caselli}, {Codella}, {Comito}, {Encrenaz}, {Falgarone}, {Fuente},
  {Goldsmith}, {Helmich}, {Herbst}, {Jacq}, {Kama}, {Langer}, {Lefloch}, {Lis},
  {Lord}, {Lorenzani}, {Neufeld}, {Nisini}, {Pacheco}, {Pearson}, {Phillips},
  {Salez}, {Saraceno}, {Schuster}, {Tielens}, {van der Tak}, {van der Wiel},
  {Viti}, {Wyrowski}, {Yorke}, {Faure}, {Benz}, {Coeur-Joly}, {Cros},
  {G{\"u}sten}, \& {Ravera}}]{2010A&A...521L..42B}
{Bacmann}, A., {Caux}, E., {Hily-Blant}, P., {et~al.} 2010, \aap, 521, L42

\bibitem[{{Biver} {et~al.}(2007){Biver}, {Bockel{\'e}e-Morvan}, {Crovisier},
  {Lecacheux}, {Frisk}, {Hjalmarson}, {Olberg}, {Flor{\'e}n}, {Sandqvist}, \&
  {Kwok}}]{2007P&SS...55.1058B}
{Biver}, N., {Bockel{\'e}e-Morvan}, D., {Crovisier}, J., {et~al.} 2007,
  \planss, 55, 1058

\bibitem[{{Cazzoli} {et~al.}(2009){Cazzoli}, {Dore}, \&
  {Puzzarini}}]{2009A&A...507.1707C}
{Cazzoli}, G., {Dore}, L., \& {Puzzarini}, C. 2009, \aap, 507, 1707

\bibitem[{{Ceccarelli} {et~al.}(2002){Ceccarelli}, {Baluteau}, {Walmsley},
  {Swinyard}, {Caux}, {Sidher}, {Cox}, {Gry}, {Kessler}, \&
  {Prusti}}]{2002A&A...383..603C}
{Ceccarelli}, C., {Baluteau}, J.-P., {Walmsley}, M., {et~al.} 2002, \aap, 383,
  603

\bibitem[{{Cernicharo} {et~al.}(2000){Cernicharo}, {Goicoechea}, \&
  {Caux}}]{2000ApJ...534L.199C}
{Cernicharo}, J., {Goicoechea}, J.~R., \& {Caux}, E. 2000, \apjl, 534, L199

\bibitem[{{Cheung} {et~al.}(1968){Cheung}, {Rank}, {Townes}, {Thornton}, \&
  {Welch}}]{1968PhRvL..21.1701C}
{Cheung}, A.~C., {Rank}, D.~M., {Townes}, C.~H., {Thornton}, D.~D., \& {Welch},
  W.~J. 1968, Physical Review Letters, 21, 1701

\bibitem[{{Comito} {et~al.}(2005){Comito}, {Schilke}, {Phillips}, {Lis},
  {Motte}, \& {Mehringer}}]{2005ApJS..156..127C}
{Comito}, C., {Schilke}, P., {Phillips}, T.~G., {et~al.} 2005, \apjs, 156, 127

\bibitem[{{Coudert} \& {Roueff}(2006)}]{2006A&A...449..855C}
{Coudert}, L.~H. \& {Roueff}, E. 2006, \aap, 449, 855

\bibitem[{{Coudert} \& {Roueff}(2009)}]{2009A&A...499..347C}
{Coudert}, L.~H. \& {Roueff}, E. 2009, \aap, 499, 347

\bibitem[{{Crane} {et~al.}(1995){Crane}, {Lambert}, \&
  {Sheffer}}]{1995ApJS...99..107C}
{Crane}, P., {Lambert}, D.~L., \& {Sheffer}, Y. 1995, \apjs, 99, 107

\bibitem[{{Crawford} \& {Williams}(1997)}]{1997MNRAS.291L..53C}
{Crawford}, I.~A. \& {Williams}, D.~A. 1997, \mnras, 291, L53

\bibitem[{{Dame} \& {Thaddeus}(1985)}]{1985ApJ...297..751D}
{Dame}, T.~M. \& {Thaddeus}, P. 1985, \apj, 297, 751

\bibitem[{{de Almeida} \& {Singh}(1982)}]{1982A&A...113..199D}
{de Almeida}, A.~A. \& {Singh}, P.~D. 1982, \aap, 113, 199

\bibitem[{{de Graauw} {et~al.}(2010){de Graauw}, {Helmich}, {Phillips},
  {Stutzki}, {Caux}, \& {Whyborn}}]{Graauw2010}
{de Graauw}, T., {Helmich}, F.~P., {Phillips}, T.~G., {et~al.} 2010, \aap, 518,
  L4

\bibitem[{{De Pree} {et~al.}(2000){De Pree}, {Wilner}, {Goss}, {Welch}, \&
  {McGrath}}]{2000ApJ...540..308D}
{De Pree}, C.~G., {Wilner}, D.~J., {Goss}, W.~M., {Welch}, W.~J., \& {McGrath},
  E. 2000, \apj, 540, 308

\bibitem[{{Dickel} \& {Goss}(1990)}]{1990ApJ...351..189D}
{Dickel}, H.~R. \& {Goss}, W.~M. 1990, \apj, 351, 189

\bibitem[{{Dreher} {et~al.}(1984){Dreher}, {Johnston}, {Welch}, \&
  {Walker}}]{1984ApJ...283..632D}
{Dreher}, J.~W., {Johnston}, K.~J., {Welch}, W.~J., \& {Walker}, R.~C. 1984,
  \apj, 283, 632

\bibitem[{{Farmer} \& {Norton}(1989)}]{1989hra1.book.....F}
{Farmer}, C.~B. \& {Norton}, R.~H. 1989, {A high-resolution atlas of the
  infrared spectrum of the sun and the earth atmosphere from space. A
  compilation of ATMOS spectra of the region from 650 to 4800 cm$^{-1}$ (2.3 to
  16 {$\mu$}m). Vol. I. The sun.}, ed. {Farmer, C.~B.~\& Norton, R.~H.}

\bibitem[{{Feldman} {et~al.}(1993){Feldman}, {Fournier}, {Grinin}, \&
  {Zvereva}}]{1993ApJ...404..348F}
{Feldman}, P.~D., {Fournier}, K.~B., {Grinin}, V.~P., \& {Zvereva}, A.~M. 1993,
  \apj, 404, 348

\bibitem[{{Fish} {et~al.}(2003){Fish}, {Reid}, {Wilner}, \&
  {Churchwell}}]{2003ApJ...587..701F}
{Fish}, V.~L., {Reid}, M.~J., {Wilner}, D.~J., \& {Churchwell}, E. 2003, \apj,
  587, 701

\bibitem[{{Flores-Mijangos} {et~al.}(2004){Flores-Mijangos}, {Brown},
  {Matsushima}, {Odashima}, {Takagi}, {Zink}, \&
  {Evenson}}]{2004JMoSp.225..189F}
{Flores-Mijangos}, J., {Brown}, J.~M., {Matsushima}, F., {et~al.} 2004, Journal
  of Molecular Spectroscopy, 225, 189

\bibitem[{{Flower} {et~al.}(1995){Flower}, {Pineau des Forets}, \&
  {Walmsley}}]{1995A&A...294..815F}
{Flower}, D.~R., {Pineau des Forets}, G., \& {Walmsley}, C.~M. 1995, \aap, 294,
  815

\bibitem[{{Gerin} {et~al.}(2010{\natexlab{a}}){Gerin}, {de Luca}, {Black},
  {Goicoechea}, {Herbst}, {Neufeld}, {Falgarone}, {Godard}, {Pearson}, {Lis},
  {Phillips}, {Bell}, {Sonnentrucker}, {Boulanger}, {Cernicharo}, {Coutens},
  {Dartois}, {Encrenaz}, {Giesen}, {Goldsmith}, {Gupta}, {Gry}, {Hennebelle},
  {Hily-Blant}, {Joblin}, {Kazmierczak}, {Kolos}, {Krelowski},
  {Martin-Pintado}, {Monje}, {Mookerjea}, {Perault}, {Persson}, {Plume},
  {Rimmer}, {Salez}, {Schmidt}, {Stutzki}, {Teyssier}, {Vastel}, {Yu},
  {Contursi}, {Menten}, {Geballe}, {Schlemmer}, {Shipman}, {Tielens},
  {Philipp-May}, {Cros}, {Zmuidzinas}, {Samoska}, {Klein}, \&
  {Lorenzani}}]{2010A&A...518L.110G}
{Gerin}, M., {de Luca}, M., {Black}, J., {et~al.} 2010{\natexlab{a}}, \aap,
  518, L110

\bibitem[{{Gerin} {et~al.}(2010{\natexlab{b}}){Gerin}, {de Luca}, {Goicoechea},
  {Herbst}, {Falgarone}, {Godard}, {Bell}, {Coutens}, {Ka{\'z}mierczak},
  {Sonnentrucker}, {Black}, {Neufeld}, {Phillips}, {Pearson}, {Rimmer},
  {Hassel}, {Lis}, {Vastel}, {Boulanger}, {Cernicharo}, {Dartois}, {Encrenaz},
  {Giesen}, {Goldsmith}, {Gupta}, {Gry}, {Hennebelle}, {Hily-Blant}, {Joblin},
  {Ko{\l}os}, {Kre{\l}owski}, {Mart{\'{\i}}n-Pintado}, {Monje}, {Mookerjea},
  {Perault}, {Persson}, {Plume}, {Salez}, {Schmidt}, {Stutzki}, {Teyssier},
  {Yu}, {Contursi}, {Menten}, {Geballe}, {Schlemmer}, {Morris}, {Hatch},
  {Imram}, {Ward}, {Caux}, {G{\"u}sten}, {Klein}, {Roelfsema}, {Dieleman},
  {Schieder}, {Honingh}, \& {Zmuidzinas}}]{2010A&A...521L..16G}
{Gerin}, M., {de Luca}, M., {Goicoechea}, J.~R., {et~al.} 2010{\natexlab{b}},
  \aap, 521, L16

\bibitem[{{Godard} {et~al.}(2010){Godard}, {Falgarone}, {Gerin}, {Hily-Blant},
  \& {de Luca}}]{2010Godard}
{Godard}, B., {Falgarone}, E., {Gerin}, M., {Hily-Blant}, P., \& {de Luca}, M.
  2010, \aap, 520, A20+

\bibitem[{{Goicoechea} {et~al.}(2004){Goicoechea},
  {Rodr{\'{\i}}guez-Fern{\'a}ndez}, \& {Cernicharo}}]{2004ApJ...600..214G}
{Goicoechea}, J.~R., {Rodr{\'{\i}}guez-Fern{\'a}ndez}, N.~J., \& {Cernicharo},
  J. 2004, \apj, 600, 214

\bibitem[{{Hasegawa} \& {Herbst}(1993)}]{1993MNRAS.263..589H}
{Hasegawa}, T.~I. \& {Herbst}, E. 1993, \mnras, 263, 589

\bibitem[{{Hasegawa} {et~al.}(2006){Hasegawa}, {Kwok}, {Koning}, {Volk},
  {Justtanont}, {Olofsson}, {Sch{\"o}ier}, {Sandqvist}, {Hjalmarson}, {Olberg},
  {Winnberg}, {Nyman}, \& {Frisk}}]{2006ApJ...637..791H}
{Hasegawa}, T.~I., {Kwok}, S., {Koning}, N., {et~al.} 2006, \apj, 637, 791

\bibitem[{{Hermsen} {et~al.}(1985){Hermsen}, {Wilson}, {Walmsley}, \&
  {Batrla}}]{1985AA...146..134H}
{Hermsen}, W., {Wilson}, T.~L., {Walmsley}, C.~M., \& {Batrla}, W. 1985, \aap,
  146, 134

\bibitem[{{Hermsen} {et~al.}(1988){Hermsen}, {Wilson}, {Walmsley}, \&
  {Henkel}}]{1988AA...201..285H}
{Hermsen}, W., {Wilson}, T.~L., {Walmsley}, C.~M., \& {Henkel}, C. 1988, \aap,
  201, 285

\bibitem[{{Hily-Blant} {et~al.}(2010){Hily-Blant}, {Maret}, {Bacmann},
  {Bottinelli}, {Parise}, {Caux}, {Faure}, {Bergin}, {Blake}, {Castets},
  {Ceccarelli}, {Cernicharo}, {Coutens}, {Crimier}, {Demyk}, {Dominik},
  {Gerin}, {Hennebelle}, {Henning}, {Kahane}, {Klotz}, {Melnick}, {Pagani},
  {Schilke}, {Vastel}, {Wakelam}, {Walters}, {Baudry}, {Bell}, {Benedettini},
  {Boogert}, {Cabrit}, {Caselli}, {Codella}, {Comito}, {Encrenaz}, {Falgarone},
  {Fuente}, {Goldsmith}, {Helmich}, {Herbst}, {Jacq}, {Kama}, {Langer},
  {Lefloch}, {Lis}, {Lord}, {Lorenzani}, {Neufeld}, {Nisini}, {Pacheco},
  {Phillips}, {Salez}, {Saraceno}, {Schuster}, {Tielens}, {van der Tak}, {van
  der Wiel}, {Viti}, {Wyrowski}, \& {Yorke}}]{2010A&A...521L..52H}
{Hily-Blant}, P., {Maret}, S., {Bacmann}, A., {et~al.} 2010, \aap, 521, L52

\bibitem[{{Ho} \& {Townes}(1983)}]{1983ARA&A..21..239H}
{Ho}, P.~T.~P. \& {Townes}, C.~H. 1983, \araa, 21, 239

\bibitem[{{H{\"u}bers} {et~al.}(2009){H{\"u}bers}, {Evenson}, {Hill}, \&
  {Brown}}]{2009JChPh.131c4311H}
{H{\"u}bers}, H., {Evenson}, K.~M., {Hill}, C., \& {Brown}, J.~M. 2009, \jcp,
  131, 034311

\bibitem[{{Keene} {et~al.}(1983){Keene}, {Blake}, \&
  {Phillips}}]{1983ApJ...271L..27K}
{Keene}, J., {Blake}, G.~A., \& {Phillips}, T.~G. 1983, \apjl, 271, L27

\bibitem[{{Klaus} {et~al.}(1997){Klaus}, {Takano}, \&
  {Winnewisser}}]{1997A&A...322L...1K}
{Klaus}, T., {Takano}, S., \& {Winnewisser}, G. 1997, \aap, 322, L1

\bibitem[{{Knauth} {et~al.}(2004){Knauth}, {Andersson}, {McCandliss}, \&
  {Warren Moos}}]{2004Natur.429..636K}
{Knauth}, D.~C., {Andersson}, B., {McCandliss}, S.~R., \& {Warren Moos}, H.
  2004, \nat, 429, 636

\bibitem[{{Kre{\l}owski} {et~al.}(2010){Kre{\l}owski}, {Beletsky}, \&
  {Galazutdinov}}]{2010ApJ...719L..20K}
{Kre{\l}owski}, J., {Beletsky}, Y., \& {Galazutdinov}, G.~A. 2010, \apjl, 719,
  L20

\bibitem[{{Langer} \& {Graedel}(1989)}]{1989ApJS...69..241L}
{Langer}, W.~D. \& {Graedel}, T.~E. 1989, \apjs, 69, 241

\bibitem[{{Larsson} {et~al.}(2003){Larsson}, {Liseau}, {Bergman}, {Bernath},
  {Black}, {Booth}, {Buat}, {Curry}, {Encrenaz}, {Falgarone}, {Feldman},
  {Fich}, {Flor{\'e}n}, {Frisk}, {Gerin}, {Gregersen}, {Harju}, {Hasegawa},
  {Johansson}, {Kwok}, {Lecacheux}, {Liljestr{\"o}m}, {Mattila}, {Mitchell},
  {Nordh}, {Olberg}, {Olofsson}, {Pagani}, {Plume}, {Ristorcelli}, {Sandqvist},
  {Sch{\'e}ele}, {Tothill}, {Volk}, {Wilson}, \&
  {Hjalmarson}}]{2003A&A...402L..69L}
{Larsson}, B., {Liseau}, R., {Bergman}, P., {et~al.} 2003, \aap, 402, L69

\bibitem[{{Lis} {et~al.}(2010){Lis}, {Wootten}, {Gerin}, \&
  {Roueff}}]{2010ApJ...710L..49L}
{Lis}, D.~C., {Wootten}, A., {Gerin}, M., \& {Roueff}, E. 2010, \apjl, 710, L49

\bibitem[{{Liseau} {et~al.}(2003){Liseau}, {Larsson}, {Brandeker}, {Bergman},
  {Bernath}, {Black}, {Booth}, {Buat}, {Curry}, {Encrenaz}, {Falgarone},
  {Feldman}, {Fich}, {Flor{\'e}n}, {Frisk}, {Gerin}, {Gregersen}, {Harju},
  {Hasegawa}, {Hjalmarson}, {Johansson}, {Kwok}, {Lecacheux}, {Liljestr{\"o}m},
  {Mattila}, {Mitchell}, {Nordh}, {Olberg}, {Olofsson}, {Pagani}, {Plume},
  {Ristorcelli}, {Sandqvist}, {Sch{\'e}ele}, {Serra}, {Tothill}, {Volk}, \&
  {Wilson}}]{2003A&A...402L..73L}
{Liseau}, R., {Larsson}, B., {Brandeker}, A., {et~al.} 2003, \aap, 402, L73

\bibitem[{{Liszt} \& {Lucas}(2001)}]{2001A&A...370..576L}
{Liszt}, H. \& {Lucas}, R. 2001, \aap, 370, 576

\bibitem[{{Liszt} {et~al.}(2006){Liszt}, {Lucas}, \&
  {Pety}}]{2006A&A...448..253L}
{Liszt}, H.~S., {Lucas}, R., \& {Pety}, J. 2006, \aap, 448, 253

\bibitem[{{Markwardt}(2009)}]{2009ASPC..411..251M}
{Markwardt}, C.~B. 2009, in Astronomical Society of the Pacific Conference
  Series, Vol. 411, Astronomical Data Analysis Software and Systems XVIII, ed.
  {D.~A.~Bohlender, D.~Durand, \& P.~Dowler}, 251--+

\bibitem[{{Marshall} {et~al.}(2006){Marshall}, {Robin}, {Reyl{\'e}},
  {Schultheis}, \& {Picaud}}]{2006A&A...453..635M}
{Marshall}, D.~J., {Robin}, A.~C., {Reyl{\'e}}, C., {Schultheis}, M., \&
  {Picaud}, S. 2006, \aap, 453, 635

\bibitem[{{Marty} {et~al.}(2011){Marty}, {Chaussidon}, {Wiens}, {Jurewicz}, \&
  {Burnett}}]{2011Sci...332.1533M}
{Marty}, B., {Chaussidon}, M., {Wiens}, R.~C., {Jurewicz}, A.~J.~G., \&
  {Burnett}, D.~S. 2011, Science, 332, 1533

\bibitem[{{Meier} {et~al.}(1998){Meier}, {Wellnitz}, {Kim}, \&
  {A'Hearn}}]{1998Icar..136..268M}
{Meier}, R., {Wellnitz}, D., {Kim}, S.~J., \& {A'Hearn}, M.~F. 1998, Icarus,
  136, 268

\bibitem[{{Meyer} \& {Roth}(1991)}]{1991ApJ...376L..49M}
{Meyer}, D.~M. \& {Roth}, K.~C. 1991, \apjl, 376, L49

\bibitem[{{Millar} {et~al.}(1991){Millar}, {Bennett}, {Rawlings}, {Brown}, \&
  {Charnley}}]{1991A&AS...87..585M}
{Millar}, T.~J., {Bennett}, A., {Rawlings}, J.~M.~C., {Brown}, P.~D., \&
  {Charnley}, S.~B. 1991, \aaps, 87, 585

\bibitem[{{M{\"u}ller} {et~al.}(1999){M{\"u}ller}, {Klein}, {Belov},
  {Winnewisser}, {Morino}, {Yamada}, \& {Saito}}]{1999JMoSp.195..177M}
{M{\"u}ller}, H.~S.~P., {Klein}, H., {Belov}, S.~P., {et~al.} 1999, Journal of
  Molecular Spectroscopy, 195, 177

\bibitem[{{M{\"u}ller} {et~al.}(2005){M{\"u}ller}, {Schl{\"o}der}, {Stutzki},
  \& {Winnewisser}}]{2005JMoSt.742..215M}
{M{\"u}ller}, H.~S.~P., {Schl{\"o}der}, F., {Stutzki}, J., \& {Winnewisser}, G.
  2005, Journal of Molecular Structure, 742, 215

\bibitem[{{Nagayama} {et~al.}(2009){Nagayama}, {Omodaka}, {Handa}, {Toujima},
  {Sofue}, {Sawada}, {Kobayashi}, \& {Koyama}}]{2009PASJ...61.1023N}
{Nagayama}, T., {Omodaka}, T., {Handa}, T., {et~al.} 2009, \pasj, 61, 1023

\bibitem[{{Nash}(1990)}]{1990ApJS...72..303N}
{Nash}, A.~G. 1990, \apjs, 72, 303

\bibitem[{{Nejad} {et~al.}(1990){Nejad}, {Williams}, \&
  {Charnley}}]{1990MNRAS.246..183N}
{Nejad}, L.~A.~M., {Williams}, D.~A., \& {Charnley}, S.~B. 1990, \mnras, 246,
  183

\bibitem[{{Neufeld} {et~al.}(2010{\natexlab{a}}){Neufeld}, {Goicoechea},
  {Sonnentrucker}, {Black}, {Pearson}, {Yu}, {Phillips}, {Lis}, {de Luca},
  {Herbst}, {Rimmer}, {Gerin}, {Bell}, {Boulanger}, {Cernicharo}, {Coutens},
  {Dartois}, {Kazmierczak}, {Encrenaz}, {Falgarone}, {Geballe}, {Giesen},
  {Godard}, {Goldsmith}, {Gry}, {Gupta}, {Hennebelle}, {Hily-Blant}, {Joblin},
  {Ko{\l}os}, {Kre{\l}owski}, {Mart{\'{\i}}n-Pintado}, {Menten}, {Monje},
  {Mookerjea}, {Perault}, {Persson}, {Plume}, {Salez}, {Schlemmer}, {Schmidt},
  {Stutzki}, {Teyssier}, {Vastel}, {Cros}, {Klein}, {Lorenzani}, {Philipp},
  {Samoska}, {Shipman}, {Tielens}, {Szczerba}, \&
  {Zmuidzinas}}]{2010A&A...521L..10N}
{Neufeld}, D.~A., {Goicoechea}, J.~R., {Sonnentrucker}, P., {et~al.}
  2010{\natexlab{a}}, \aap, 521, L10

\bibitem[{{Neufeld} {et~al.}(2010{\natexlab{b}}){Neufeld}, {Sonnentrucker},
  {Phillips}, {Lis}, {de Luca}, {Goicoechea}, {Black}, {Gerin}, {Bell},
  {Boulanger}, {Cernicharo}, {Coutens}, {Dartois}, {Kazmierczak}, {Encrenaz},
  {Falgarone}, {Geballe}, {Giesen}, {Godard}, {Goldsmith}, {Gry}, {Gupta},
  {Hennebelle}, {Herbst}, {Hily-Blant}, {Joblin}, {Ko{\l}os}, {Kre{\l}owski},
  {Mart{\'{\i}}n-Pintado}, {Menten}, {Monje}, {Mookerjea}, {Pearson},
  {Perault}, {Persson}, {Plume}, {Salez}, {Schlemmer}, {Schmidt}, {Stutzki},
  {Teyssier}, {Vastel}, {Yu}, {Cais}, {Caux}, {Liseau}, {Morris}, \&
  {Planesas}}]{2010A&A...518L.108N}
{Neufeld}, D.~A., {Sonnentrucker}, P., {Phillips}, T.~G., {et~al.}
  2010{\natexlab{b}}, \aap, 518, L108

\bibitem[{{Nyman} \& {Millar}(1989)}]{1989A&A...222..231N}
{Nyman}, L.-A. \& {Millar}, T.~J. 1989, \aap, 222, 231

\bibitem[{{Ott}(2010)}]{2010ASPC..434..139O}
{Ott}, S. 2010, in Astronomical Society of the Pacific Conference Series, Vol.
  434, Astronomical Data Analysis Software and Systems XIX, ed. Y.~{Mizumoto},
  K.-I. {Morita}, \& M.~{Ohishi}, 139

\bibitem[{{Persson} {et~al.}(2010){Persson}, {Black}, {Cernicharo},
  {Goicoechea}, {Hassel}, {Herbst}, {Gerin}, {de Luca}, {Bell}, {Coutens},
  {Falgarone}, {Goldsmith}, {Gupta}, {Ka{\'z}mierczak}, {Lis}, {Mookerjea},
  {Neufeld}, {Pearson}, {Phillips}, {Sonnentrucker}, {Stutzki}, {Vastel}, {Yu},
  {Boulanger}, {Dartois}, {Encrenaz}, {Geballe}, {Giesen}, {Godard}, {Gry},
  {Hennebelle}, {Hily-Blant}, {Joblin}, {Ko{\l}os}, {Kre{\l}owski},
  {Mart{\'{\i}}n-Pintado}, {Menten}, {Monje}, {Perault}, {Plume}, {Salez},
  {Schlemmer}, {Schmidt}, {Teyssier}, {P{\'e}ron}, {Cais}, {Gaufre}, {Cros},
  {Ravera}, {Morris}, {Lord}, \& {Planesas}}]{2010A&A...521L..45P}
{Persson}, C.~M., {Black}, J.~H., {Cernicharo}, J., {et~al.} 2010, \aap, 521,
  L45 (Paper~I)

\bibitem[{{Persson} {et~al.}(2009){Persson}, {Olberg}, {.~Hjalmarson},
  {Spaans}, {Black}, {Frisk}, {Liljestr{\"o}m}, {Olofsson}, {Poelman}, \&
  {Sandqvist}}]{2009A&A...494..637P}
{Persson}, C.~M., {Olberg}, M., {.~Hjalmarson}, {\AA}., {et~al.} 2009, \aap,
  494, 637

\bibitem[{{Pickett} {et~al.}(1998){Pickett}, {Poynter}, {Cohen}, {Delitsky},
  {Pearson}, \& {Muller}}]{1998JQSRT..60..883P}
{Pickett}, H.~M., {Poynter}, I.~R.~L., {Cohen}, E.~A., {et~al.} 1998, Journal
  of Quantitative Spectroscopy and Radiative Transfer, 60, 883

\bibitem[{{Pilbratt} {et~al.}(2010){Pilbratt}, {Riedinger}, {Passvogel},
  {Crone}, {Doyle}, \& {Gageur}}]{Pilbratt2010}
{Pilbratt}, G., {Riedinger}, J.~R., {Passvogel}, T., {et~al.} 2010, \aap, 518,
  L1

\bibitem[{{Plume} {et~al.}(2004){Plume}, {Kaufman}, {Neufeld}, {Snell},
  {Hollenbach}, {Goldsmith}, {Howe}, {Bergin}, {Melnick}, \&
  {Bensch}}]{2004ApJ...605..247P}
{Plume}, R., {Kaufman}, M.~J., {Neufeld}, D.~A., {et~al.} 2004, \apj, 605, 247

\bibitem[{{Polehampton} {et~al.}(2007){Polehampton}, {Baluteau}, {Swinyard},
  {Goicoechea}, {Brown}, {White}, {Cernicharo}, \&
  {Grundy}}]{2007MNRAS.377.1122P}
{Polehampton}, E.~T., {Baluteau}, J., {Swinyard}, B.~M., {et~al.} 2007, \mnras,
  377, 1122

\bibitem[{{Rist} {et~al.}(1993){Rist}, {Alexander}, \&
  {Valiron}}]{1993JChPh..98.4662R}
{Rist}, C., {Alexander}, M.~H., \& {Valiron}, P. 1993, \jcp, 98, 4662

\bibitem[{{Roelfsema} {et~al.}(2012){Roelfsema}, {Helmich}, {Teyssier},
  {Ossenkopf}, {Morris}, {Olberg}, {Shipman}, {Risacher}, {Akyilmaz},
  {Assendorp}, {Avruch}, {Beintema}, {Biver}, {Boogert}, {Borys}, {Braine},
  {Caris}, {Caux}, {Cernicharo}, {Coeur-Joly}, {Comito}, {de Lange},
  {Delforge}, {Dieleman}, {Dubbeldam}, {de Graauw}, {Edwards}, {Fich},
  {Flederus}, {Gal}, {di Giorgio}, {Herpin}, {Higgins}, {Hoac}, {Huisman},
  {Jarchow}, {Jellema}, {de Jonge}, {Kester}, {Klein}, {Kooi}, {Kramer},
  {Laauwen}, {Larsson}, {Leinz}, {Lord}, {Lorenzani}, {Luinge}, {Marston},
  {Mart{\'{\i}}n-Pintado}, {McCoey}, {Melchior}, {Michalska}, {Moreno},
  {M{\"u}ller}, {Nowosielski}, {Okada}, {Orlea{\'n}ski}, {Phillips}, {Pearson},
  {Rabois}, {Ravera}, {Rector}, {Rengel}, {Sagawa}, {Salomons},
  {S{\'a}nchez-Su{\'a}rez}, {Schieder}, {Schl{\"o}der}, {Schm{\"u}lling},
  {Soldati}, {Stutzki}, {Thomas}, {Tielens}, {Vastel}, {Wildeman}, {Xie},
  {Xilouris}, {Wafelbakker}, {Whyborn}, {Zaal}, {Bell}, {Bjerkeli}, {De Beck},
  {Cavali{\'e}}, {Crockett}, {Hily-Blant}, {Kama}, {Kaminski}, {Lefl{\'o}ch},
  {Lombaert}, {de Luca}, {Makai}, {Marseille}, {Nagy}, {Pacheco}, {van der
  Wiel}, {Wang}, \& {Y{\i}ld{\i}z}}]{2012A&A...537A..17R}
{Roelfsema}, P.~R., {Helmich}, F.~P., {Teyssier}, D., {et~al.} 2012, \aap, 537,
  A17

\bibitem[{{Sandford} {et~al.}(2001){Sandford}, {Bernstein}, {Allamandola},
  {Goorvitch}, \& {Teixeira}}]{2001ApJ...548..836S}
{Sandford}, S.~A., {Bernstein}, M.~P., {Allamandola}, L.~J., {Goorvitch}, D.,
  \& {Teixeira}, T.~C.~V.~S. 2001, \apj, 548, 836

\bibitem[{{Schmitt}(1969)}]{1969PASP...81..657S}
{Schmitt}, J.~L. 1969, \pasp, 81, 657

\bibitem[{{Sheffer} {et~al.}(2008){Sheffer}, {Rogers}, {Federman}, {Abel},
  {Gredel}, {Lambert}, \& {Shaw}}]{2008ApJ...687.1075S}
{Sheffer}, Y., {Rogers}, M., {Federman}, S.~R., {et~al.} 2008, \apj, 687, 1075

\bibitem[{{Snow} \& {McCall}(2006)}]{2006ARA&A..44..367S}
{Snow}, T.~P. \& {McCall}, B.~J. 2006, \araa, 44, 367

\bibitem[{{Snow}(1979)}]{1979Ap&SS..66..453S}
{Snow}, Jr., T.~P. 1979, \apss, 66, 453

\bibitem[{{Sonnentrucker} {et~al.}(2010){Sonnentrucker}, {Neufeld}, {Phillips},
  {Gerin}, {Lis}, {de Luca}, {Goicoechea}, {Black}, {Bell}, {Boulanger},
  {Cernicharo}, {Coutens}, {Dartois}, {Ka{\'z}mierczak}, {Encrenaz},
  {Falgarone}, {Geballe}, {Giesen}, {Godard}, {Goldsmith}, {Gry}, {Gupta},
  {Hennebelle}, {Herbst}, {Hily-Blant}, {Joblin}, {Ko{\l}os}, {Kre{\l}owski},
  {Mart{\'{\i}}n-Pintado}, {Menten}, {Monje}, {Mookerjea}, {Pearson},
  {Perault}, {Persson}, {Plume}, {Salez}, {Schlemmer}, {Schmidt}, {Stutzki},
  {Teyssier}, {Vastel}, {Yu}, {Caux}, {G{\"u}sten}, {Hatch}, {Klein}, {Mehdi},
  {Morris}, \& {Ward}}]{2010A&A...521L..12S}
{Sonnentrucker}, P., {Neufeld}, D.~A., {Phillips}, T.~G., {et~al.} 2010, \aap,
  521, L12

\bibitem[{{Swings} {et~al.}(1941){Swings}, {Elvey}, \&
  {Babcock}}]{1941ApJ....94..320S}
{Swings}, P., {Elvey}, C.~T., \& {Babcock}, H.~W. 1941, \apj, 94, 320

\bibitem[{{Tieftrunk} {et~al.}(1994){Tieftrunk}, {Pineau des Forets},
  {Schilke}, \& {Walmsley}}]{1994A&A...289..579T}
{Tieftrunk}, A., {Pineau des Forets}, G., {Schilke}, P., \& {Walmsley}, C.~M.
  1994, \aap, 289, 579

\bibitem[{{Umemoto} {et~al.}(1999){Umemoto}, {Mikami}, {Yamamoto}, \&
  {Hirano}}]{1999ApJ...525L.105U}
{Umemoto}, T., {Mikami}, H., {Yamamoto}, S., \& {Hirano}, N. 1999, \apjl, 525,
  L105

\bibitem[{{van der Tak} {et~al.}(2007){van der Tak}, {Black}, {Sch{\"o}ier},
  {Jansen}, \& {van Dishoeck}}]{2007A&A...468..627V}
{van der Tak}, F.~F.~S., {Black}, J.~H., {Sch{\"o}ier}, F.~L., {Jansen}, D.~J.,
  \& {van Dishoeck}, E.~F. 2007, \aap, 468, 627

\bibitem[{{van Dishoeck} {et~al.}(1993){van Dishoeck}, {Jansen}, {Schilke}, \&
  {Phillips}}]{1993ApJ...416L..83V}
{van Dishoeck}, E.~F., {Jansen}, D.~J., {Schilke}, P., \& {Phillips}, T.~G.
  1993, \apjl, 416, L83

\bibitem[{{Varberg} {et~al.}(1999){Varberg}, {Stroh}, \&
  {Evenson}}]{1999JMoSp.196....5V}
{Varberg}, T.~D., {Stroh}, F., \& {Evenson}, K.~M. 1999, Journal of Molecular
  Spectroscopy, 196, 5

\bibitem[{{Verhoeve} {et~al.}(1986){Verhoeve}, {Ter Meulen}, {Meerts}, \&
  {Dyamanus}}]{1986CPL...132..213V}
{Verhoeve}, P., {Ter Meulen}, J.~J., {Meerts}, W.~L., \& {Dyamanus}, A. 1986,
  Chemical Physics Letters, 132, 213

\bibitem[{{Wagenblast} {et~al.}(1993){Wagenblast}, {Williams}, {Millar}, \&
  {Nejad}}]{1993MNRAS.260..420W}
{Wagenblast}, R., {Williams}, D.~A., {Millar}, T.~J., \& {Nejad}, L.~A.~M.
  1993, \mnras, 260, 420

\bibitem[{{Weselak} {et~al.}(2009){Weselak}, {Galazutdinov}, {Beletsky}, \&
  {Kre{\l}owski}}]{2009MNRAS.400..392W}
{Weselak}, T., {Galazutdinov}, G.~A., {Beletsky}, Y., \& {Kre{\l}owski}, J.
  2009, \mnras, 400, 392

\bibitem[{{Wilson} \& {Rood}(1994)}]{1994ARA&A..32..191W}
{Wilson}, T.~L. \& {Rood}, R. 1994, \araa, 32, 191

\bibitem[{{Wirstr{\"o}m} {et~al.}(2010){Wirstr{\"o}m}, {Bergman}, {Black},
  {Hjalmarson}, {Larsson}, {Olofsson}, {Encrenaz}, {Falgarone}, {Frisk},
  {Olberg}, \& {Sandqvist}}]{2010A&A...522A..19W}
{Wirstr{\"o}m}, E.~S., {Bergman}, P., {Black}, J.~H., {et~al.} 2010, \aap, 522,
  A19

\bibitem[{{Wyrowski} {et~al.}(2010){Wyrowski}, {Menten}, {G{\"u}sten}, \&
  {Belloche}}]{2010A&A...518A..26W}
{Wyrowski}, F., {Menten}, K.~M., {G{\"u}sten}, R., \& {Belloche}, A. 2010,
  \aap, 518, A26

\end{thebibliography}

\Online
\appendix

\section{Herschel observations}

 \begin{table*}[\!htb] 
\centering
\caption{\emph{Herschel} OBSID's of the observed transitions analysed in this paper.
}
\begin{tabular} {llrccccc } 
 \hline\hline
     \noalign{\smallskip}
Source & Species 	& Frequency & Band & LO-setting$^a$ & Date &	OBSID    \\    \noalign{\smallskip}

&  & (GHz)    \\
     \noalign{\smallskip}
     \hline
\noalign{\smallskip}

W49N & NH$^b$ & 946.476 & 3b  & A &2010-04-13   & 1342194700 \\ 
     &        &         &     & B &             & 1342194701 \\
     &        &         &     & C &             & 1342194702 \\
\noalign{\smallskip} 

     & NH       & 974.478 & 4a  & A &2010-04-18  & 1342195004 \\ 
     &          &         &     & B &            & 1342195005 \\
     &          &         &     & C &            & 1342195006 \\
\noalign{\smallskip}

     & o-NH$_2$ & 952.578 & 3b  & A &2010-04-13  & 1342194706 \\ 
     &          &         &     & B &            & 1342194707 \\
     &          &         &     & C &            & 1342194708 \\
\noalign{\smallskip}

     & o-NH$_3$ & 572.498 & 1b & A & 2010-04-11   & 1342194517 \\  
     &          &         &    & B &              & 1342194518 \\
     &          &         &    & C &              & 1342194519 \\
\noalign{\smallskip}

     & p-NH$_3^c$ & 1\,215.246 & 5a & A &2010-04-18  & 1342195067 \\
     &            &            &    & B &            & 1342195068 \\
     &            &            &    & C &            & 1342195069 \\
\noalign{\smallskip}

     & NH$^+$     & 1\,012.540 & 4a & A & 2010-04-18  & 1342194998 \\ 
     &            &            &    & B &             & 1342194999 \\
     &            &            &    & C &             & 1342195000 \\

\noalign{\smallskip} \noalign{\smallskip} \noalign{\smallskip} 
G10.6-0.4 & NH$^b$ & 946.476 & 3b  & A &2010-03-18   & 1342192316 \\ 
     &        &         &     & B &             & 1342192317 \\
     &        &         &     & C &             & 1342192318 \\
\noalign{\smallskip} 

     & NH       & 974.478 & 4a  & A &2010-03-03  & 1342191620 \\ 
     &          &         &     & B &            & 1342191621 \\
     &          &         &     & C &            & 1342191622 \\
\noalign{\smallskip}

     & o-NH$_2$ & 952.578 & 3b  & A &2010-03-18  & 1342192319 \\ 
     &          &         &     & B &            & 1342192320 \\
     &          &         &     & C &            & 1342192321 \\
\noalign{\smallskip} 

     & o-NH$_3$ & 572.498 & 1b & A &2010-03-02  & 1342191578 \\
     &          &         &    & B &            & 1342191579 \\
     &          &         &    & C &            & 1342191580 \\
\noalign{\smallskip} 

     & p-NH$_3^c$ & 1\,215.246 & 5a & A &2010-03-05  & 1342191697 \\
     &            &            &    & B &            & 1342191698 \\
     &            &            &    & C &            & 1342191699 \\

\noalign{\smallskip}

     & NH$^+$     & 1\,012.540 & 4a & A & 2010-03-03  & 1342191623 \\ 
     &            &            &    & B &             & 1342191624 \\
     &            &            &    & C &             & 1342191625 \\

    \noalign{\smallskip}
\hline 
\label{Table: obsid}
\end{tabular}
\tablefoot{
\tablefoottext{a}{Three different frequency settings of the LO were performed,  
with approximately 15~km~s$^{-1}$ 
between each setting in order to determine the  sideband origin of the signals.}
\tablefoottext{b}{Observed in the same setting as CH$_2$ at 946~GHz.}
\tablefoottext{c}{The ortho-NH$_3$ 2$_0$\,--\,1$_0$ transition at 1214.859~GHz is observed
in the same setting as para-NH$_3$ 2$_1$\,--\,1$_1$ at 1215.246~GHz}   
}
\end{table*}

 \clearpage

\section{Emission line contamination of o-NH$_2$ towards W49N} \label{NO removal of NH2} 
\begin{figure}[\!ht] 
\centering
\includegraphics[scale=0.5]{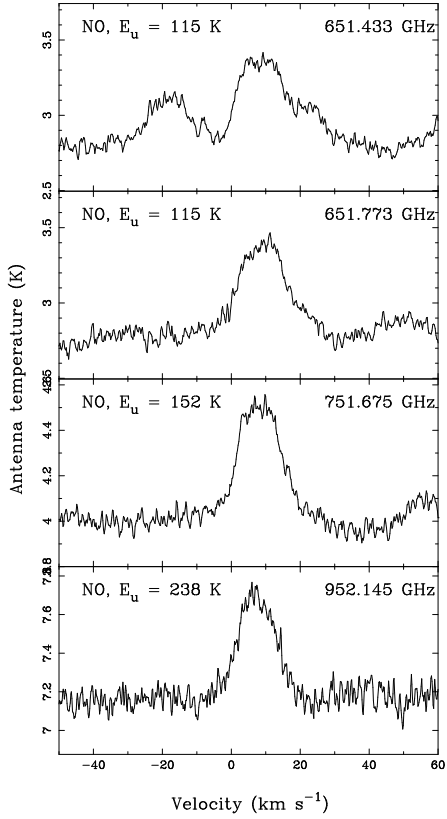}
\caption{\emph{W49N}: Double sideband WBS spectra      of four NO emission lines from 
the source itself. 
Each  NO line consists of 3 hfs components, with   
quatum numbers   $^2\Pi_{1/2}\;J=6.5^e\to 5.5^e$ (651~GHz), $J=6.5^f\to 5.5^f$ (652~GHz),  $J=7.5^e\to 6.5^e$ (752~GHz), 
and $J=9.5^e\to 8.5^e$ (952~GHz).    
The
emission line at $-$18\,km\,s$^{-1}$~blended with  NO 651.433~GHz is SO 
11$_{11}$\,--\,11$_{10}$.
}
 \label{Fig: W49N NO}
\end{figure}

\begin{figure}[\!ht] 
   \centering
\includegraphics[scale = 0.5]{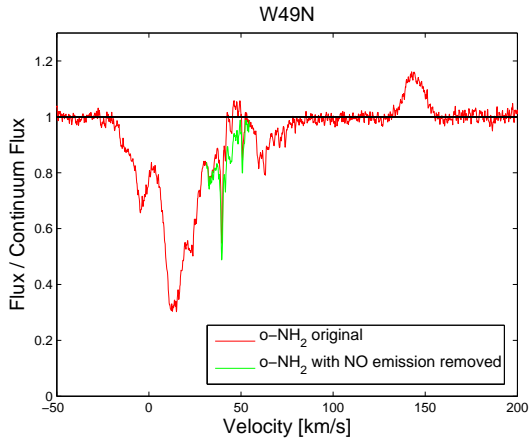}
\caption{\emph{NO emission from  W49N in the line-of-sight o-NH$_2$~absorption.}: 
Normalised SSB spectrum of o-NH$_2$. The emission line at 
142\,km\,s$^{-1}$~is from 
NO at $\nu_0$\,=\,952.145\,GHz. The NO $\nu_0$\,=\,952.464\,GHz emission feature is seen 
at 47\,km\,s$^{-1}$~in the o-NH$_2$ absorption. 
The two NO 
lines have similar line strengths.
The red line shows the SSB normalised  o-NH$_2$~line 
without removal of the NO-line, and the   green shows the spectra with removed NO-line.
}
 \label{Fig:W49N NH2 and NO}
\end{figure}

The o-NH$_2$~absorption  towards  W49N is contaminated by an
emission line in the same sideband from the source, a more complicated situation 
than an emission line from the other sideband.   
In Fig.~\ref{Fig:W49N NH2 and NO}  an emission line is clearly visible    
around 
47\,km\,s$^{-1}$ in the o-NH$_2$ spectrum.  
Here, the intensities have   been normalised to the continuum in  
single sideband  as $T_\mathrm{A}$/$T_\mathrm{C}$-1 
assuming a sideband gain ratio of unity where $T_\mathrm{A}$ is the 
observed intensity  and 
$T_\mathrm{C}$ is the SSB  continuum as measured in line-free regions in the spectra. 
We identify the emission line as a blend of three unresolved 
hfs components of 
NO $^2\Pi_{1/2}\;J=9.5^f\to 8.5^f$, $F=10.5-9.5$, $9.5-8.5$, and $8.5-9.5$, 
at 952.464201~GHz   \citep[weighted mean frequency, cf.][]{1999JMoSp.196....5V}. 
The emission line appearing near
142\,km\,s$^{-1}$ in this figure is consequently identified as the hfs blend in the 
lower half of the same spin-rotation
doublet of NO $^2\Pi_{1/2}\;J=9.5^e\to 8.5^e$ at an average rest
frequency of 952.145441~GHz.
The PRISMAS observations have also found three additional NO lines  
in  W49N  shown in Fig.~\ref{Fig: W49N NO}, each one 
consisting of  
unresolved  hfs components, while no emission of NO is found 
in   G10.6$-$0.4.

\begin{figure}[\!ht] 
\centering
\resizebox{\hsize}{!}{\includegraphics{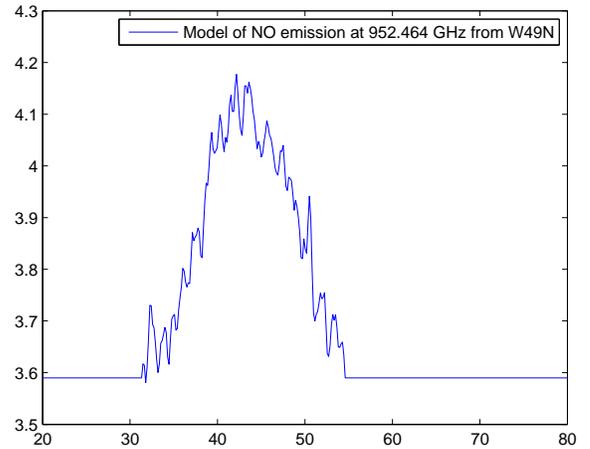}} 
\caption{\emph{W49N}: Model of the 952.464\,GHz NO line    to be removed from the o-NH$_2$ absorption lines. 
The model is using the NO  952.145\,GHz   emission line observed at  $v_\mathrm{LSR}$\,=\,+7.7\,km\,s$^{-1}$~and only shifted in velocity to a  
$v_\mathrm{LSR}$ of +43.5\,km\,s$^{-1}$~corresponding to a $v_\mathrm{LSR}$ of +7.7\,km\,s$^{-1}$~for that line. }
 \label{Fig: W49N: NO model}
\end{figure}

 Since the two NO lines seen in Fig.~\ref{Fig:W49N NH2 and NO}  have almost equal line strength, and are 
also   observed with the same instrument in the same band, we have used the observed 952.145~GHz transition as a template 
to remove the interfering NO line at 952.464~GHz from the o-NH$_2$~absorption.
  In order to do this we use
 \begin{equation}\label{NO removal}
 T_\mathrm{norm} = \frac{T_\mathrm{A} - T_\mathrm{C}} {T_\mathrm{C} + T_\mathrm{NO} },
 \end{equation}
to calculate the normalised SSB intensity $T_\mathrm{norm}$  in K, where  $T_\mathrm{C}$ is the 
SSB continuum, and  
$T_\mathrm{NO}$ is 
the intensity  
of the NO line \mbox{($T_\mathrm{A} - T_\mathrm{C, DSB}$)}. The model NO line, shown in Fig.~\ref{Fig: W49N: NO model}, 
is then moved to the velocity of the 
emission line. 
The resulting absorption line spectrum of o-NH$_2$~towards W49N  with removed NO emission is shown in green in Fig.~\ref{Fig:W49N NH2 and NO}.

\clearpage

\section{Hyperfine structure components}

\begin{table}[\!htb] 
\centering
\caption{Hyperfine structure components of NH  $^3\Sigma^-$ \mbox{$N\!=\!1\leftarrow0$},  \mbox{$J\!=\!2\leftarrow1$}. 
}
\begin{tabular} {cccc  } 
 \hline\hline
     \noalign{\smallskip}

 Frequency	   & $A_{ul}$& $\Delta v $\tablefootmark{a} 	  & Rel. Intensity  \\    \noalign{\smallskip}  
(MHz) &(s$^{-1}$)& (km\,s$^{-1}$) & $\frac {A_{ul}\times g_u} {A_{ul}\mathrm{(main)}\times g_u \mathrm{(main)}}$\tablefootmark{b} \\
     \noalign{\smallskip}
     \hline
     \noalign{\smallskip}

   974315.58  &   1.78e-6   & 50.08&  0.0001 \\
   974342.57 &   6.02e-6   & 41.78 &  0.0007 \\
  974354.64  &  4.54e-6   & 38.07&  0.0003\\
   974410.56 &   8.91e-5   & 20.87&  0.006 \\
   974411.39 &   5.13e-4   & 20.61 &  0.018 \\
   974436.35 &   2.28e-3   & 12.93 &  0.164 \\
   974437.54  &   1.27e-3   & 12.56 &  0.137 \\
   974444.04  &  1.81e-3  & 10.56 &  0.130 \\
   974450.44 &   4.80e-3   &  8.60 &  0.173\\
   974462.22 &   5.04e-4   &  4.97 &  0.363\\
   974471.00 &   5.67e-3   &  2.27 &  0.612\\
   974475.41  &  3.38e-3   &  0.91 &  0.243 \\
   974478.38  &   6.94e-3   & 0      &  1.0 \\   
   974479.34 &   6.01e-3   & -0.30 &  0.649 \\
   974531.32  &   2.59e-4   &-16.29 &  0.019\\
   974539.82 &   6.46e-4  &-18.90 &  0.023 \\
   974558.07 &  9.82e-4   &-24.52&  0.035 \\
   974564.78  &   7.67e-4   &-26.58 &  0.055 \\
   974574.43 &   8.15e-4   &-29.55&  0.088\\
   974583.03  &   2.59e-4  &-32.20&  0.019\\
   974607.78 &  1.21e-4  &-39.81 &  0.013 \\  

     \noalign{\smallskip}
\hline 
\label{Table: NH hfs transitions}
\end{tabular}
\tablefoot{ 
\tablefoottext{a}{The velocity offset from the strongest hfs component at 974478.38\,MHz.} 
\tablefoottext{b}{The sum of the relative intensities  
of the 21 hfs components is 3.75.}  
}
\end{table}

\begin{table}[\!htb] 
\centering
\caption{Hyperfine structure components of NH $^3\Sigma^-$ \mbox{$N\!=\!1\leftarrow0$},  \mbox{$J\!=\!0\leftarrow1$}. 
}
\begin{tabular} {cccc  } 
 \hline\hline
     \noalign{\smallskip}

 Frequency	   & $A_{ul}$& $\Delta v $\tablefootmark{a} 	  & Rel. Intensity  \\    \noalign{\smallskip}  
(MHz) &(s$^{-1}$)& (km\,s$^{-1}$) & $\frac {A_{ul}\times g_u} {A_{ul}\mathrm{(main)}\times g_u \mathrm{(main)}}$\tablefootmark{b} \\
     \noalign{\smallskip}
     \hline
     \noalign{\smallskip}

   946380.79  &   2.40e-3 & 30.100375183108216  &   0.369  \\
   946380.79   &  9.58e-4 & 30.100375183108216&  0.295\\
   946419.79   &  3.66e-4 & 17.746741564471662  & 0.056   \\
   946419.98 &   8.99e-4 & 17.688682017264846 &  0.276  \\
   946475.82  &   3.25e-3 &                  0  & 1.0  \\
   946509.24   &   1.93e-3& -10.586919289910648  &0.297 \\
   946509.24  &   1.21e-3 &-10.586919289910648 &  0.372  \\
   946527.48  &   1.81e-3& -16.363416629849031 &  0.278 \\
   946527.56  &   1.84e-4& -16.389231431253044 &  0.057 \\
 
     \noalign{\smallskip}
\hline 
\label{Table: NH 946 hfs transitions}
\end{tabular}
\tablefoot{ 
\tablefoottext{a}{The velocity offset from the strongest hfs component at 946475.82\,MHz.} 
\tablefoottext{b}{The sum of the relative intensities  
of the 9 hfs components is 3.66.}     
}
\end{table} 
 
\begin{table}[\!htb] 
\centering
\caption{Hyperfine structure components of ortho-NH$_2$	  $1_{1,1}$\,--\,$0_{0,0}$. 
}
\begin{tabular} {cccc } 
 \hline\hline
     \noalign{\smallskip}

 Frequency	   & $A_{ul}$& $\Delta v $\tablefootmark{a} 	  & Rel. Intensity  \\    \noalign{\smallskip}  
(MHz) &(s$^{-1}$)& (km\,s$^{-1}$) & $\frac {A_{ul}\times g_u} {A_{ul}\mathrm{(main)}\times g_u \mathrm{(main)}}$\tablefootmark{b} \\
     \noalign{\smallskip}
     \hline
     \noalign{\smallskip}

   952435.66 &   4.86e-6&  44.91  & 0.0002     \\ 
   952446.99  & 1.33e-5  &41.34  &   0.0006  \\ 
   952463.69  &   3.07e-5 & 36.09  &     0.002   \\ 
   952490.73  &   3.5e-3  &  27.58   &    0.079   \\ 
   952502.06  &   1.28e-3  & 24.01  &     0.029  \\ 
   952503.09  &  1.79e-3  &  23.69   &      0.081  \\ 
   952514.42  &  3.65e-3  &  20.12   &     0.164   \\ 
   952528.90  &  2.03e-3  & 15.56   &      0.046  \\ 
   952533.03  &  2.23e-4  & 14.27   &      0.010   \\ 
   952540.23  &  6.53e-3  & 12.00   &     0.147  \\ 
   952542.21  &  7.02e-3  & 11.37   &      0.474   \\ 
   952549.73  & 2.63e-3   & 9.01    &     0.178  \\ 
   952560.41  &   3.32e-3  & 5.65   &     0.149   \\ 
   952562.12  &  6.42e-3  & 5.11   &     0.289  \\  
   952571.74  &  7.55e-3  & 2.08   &     0.340  \\ 
   952573.46  &  3.88e-3  & 1.54   &     0.174\\  
   952577.11  &  8.47e-3  & 0.39   &    0.570   \\ 
   952578.35  &  1.11e-2     &              0  & 1.0 \\ 
   952600.46  & 1.59e-3  &    -6.96   &   0.072\\ 
   952615.49  &   3.58e-3  &   -11.69   &     0.081\\ 
   952626.82  &  2.73e-3   &   -15.25   &      0.061 \\ 
   952627.84  &   2.67e-3   &   -15.57   &     0.120 \\ 
   952628.25  &   3.35e-3   &   -15.70   &      0.226 \\ 
   952639.17  &  1.40e-3  &   -19.14   &     0.063 \\ 
   952653.66  &  9.66e-3   &   -23.70   &     0.022\\ 
   952655.64  &  7.43e-4  &    -24.32   &     0.050 \\ 
   952659.49  &   4.40e-4  &   -25.54   &     0.020 \\
   952664.99  & 1.58e-3   &   -27.27   &   0.036 \\
   952686.88  &   2.97e-4  &    -34.16   &     0.013  \\ 
   952698.21  &   7.06e-5   &   -37.72  &    0.003 \\

     \noalign{\smallskip}
\hline 
\label{Table: NH2 hfs transitions}
\end{tabular}
\tablefoot{ 
\tablefoottext{a}{The velocity offset from the strongest hfs component at 952578.35\,MHz.} 
\tablefoottext{b}{The sum of the relative intensities  
of the 30 hfs components is 4.50.}     
}
\end{table}


\begin{table}[\!htb] 
\centering
\caption{Hyperfine structure components of ortho-NH$_3$~1$_0$\,--\,0$_0$. 
 }
\begin{tabular} {ccc   } 
 \hline\hline
     \noalign{\smallskip}

 Frequency\tablefootmark{a} 	   & $\Delta v$\tablefootmark{b} 	  & Rel. Intensity\tablefootmark{a,c}  \\    \noalign{\smallskip}  
(MHz) & (km\,s$^{-1}$) &   \\
     \noalign{\smallskip}
     \hline
     \noalign{\smallskip}

572.4971   & -1.0473 & 0.1999  \\
572.4984   &  0.0    &     1.0 \\
572.5002   &  0.5237 &  0.6 \\

     \noalign{\smallskip}
\hline 
\label{Table: NH3 hfs components}
\end{tabular}
\tablefoot{ 
\tablefoottext{a}{From \citet{2009A&A...507.1707C}.}
\tablefoottext{b}{The velocity offset from the strongest hfs component at 572.4984\,MHz.} 
\tablefoottext{c}{The sum of the relative intensities    of the three hfs components is 1.8.}     
}
\end{table}


\begin{table}[\!htb] 
\centering
\caption{Hyperfine structure components of para-NH$_3$~2$_1$\,--\,1$_1$. 
}

\begin{tabular} {cccc   } 
 \hline\hline
     \noalign{\smallskip}

 Frequency\tablefootmark{a}	   & $A_{ul}$& $\Delta v $\tablefootmark{b} 	  & Rel. Intensity\tablefootmark{a}  \\    \noalign{\smallskip}  
(MHz) &(s$^{-1}$)& (km\,s$^{-1}$) & $\frac {A_{ul}\times g_u} {A_{ul}\mathrm{(main)}\times g_u \mathrm{(main)}}$\tablefootmark{c} \\
     \noalign{\smallskip}
     \hline
     \noalign{\smallskip}

  1215.2443   &1.02e-2   &  0.31    &  0.536 \\
  1215.2449   &3.39e-3   &  0.16    &  0.179 \\
  1215.2453   &5.65e-3   &  0.06    &  0.179 \\
  1215.2456   &1.36e-2   &  0       &  1.0 \\
  1215.2459   &3.77e-4   & -0.09    &  0.012 \\
  1215.2468   &7.53e-3   & -0.32    &  0.238  \\

     \noalign{\smallskip}
\hline 
\label{Table: NH3 1215 hfs components}
\end{tabular}
\tablefoot{ 
\tablefoottext{a}{\citet{2006A&A...449..855C, 2009A&A...499..347C}.}
\tablefoottext{b}{The velocity offset from the strongest hfs component at 1215.2456~MHz.} 
\tablefoottext{c}{The sum of the relative intensities    of the six hfs components is 2.14.}     
}
\end{table}

 

\clearpage
\section{Method II results}

   \begin{table*}[\!ht] 
\centering
\caption{\emph{Results from Method~II}\tablefootmark{a}.
}
\begin{tabular} {ll c ccc c   ccc cc } 
 \hline\hline
     \noalign{\smallskip}

           \multicolumn{10}{c} {W49N} \\     \noalign{\smallskip}
$v_\mathrm{LSR}$ & $N$(NH)\tablefootmark{b}     & $N$(o-NH$_2$)\tablefootmark{b}     & $N$(o-NH$_3$)   &$N$(CH)   
& NH/o-NH$_3$      &  o-NH$_2$/o-NH$_3$    & $X$(NH)\tablefootmark{c}& 
$X$(o-NH$_2$)\tablefootmark{c} & $X$(o-NH$_3$)\tablefootmark{c} \\
(km\,s$^{-1}$) &  (cm$^{-2}$)  &(cm$^{-2}$) &  (cm$^{-2}$) &  (cm$^{-2}$)  &  \\     
     \noalign{\smallskip}
     \hline
     \noalign{\smallskip}
36.0 - 42.0     & 1.1e13 & 8.3e12 &4.1e12 & 6.3e13   &2.7    &   2.0   &   6.1e-9 & 4.6e-9  &  2.3e-9 \\ 
50.0 - 57.0     & 1.4e12&1.4e12&1.2e12  & 6.6e13  & 1.2         & 1.2 & 7.4e-10  & 7.4e-10 & 6.4e-10  \\
57.0 - 61.0     & 7.8e12&3.5e12  & 1.9e12  & 7.3e13  &4.1          & 1.8 &3.7e-9  & 1.7e-9   &   9.1e-10 \\
61.0 - 70.0     & 1.0e13&4.8e12 & 3.0e12 & 9.1e13  & 3.3        & 1.6 & 3.8e-9  & 1.8e-9   &  1.2e-9  \\

 \noalign{\smallskip}     \noalign{\smallskip}
Total &3.0e13 &1.8e13& 1.0e13 & 2.9e14  & 3.0  & 1.8  & 3.6e-9 & 2.2e-9 & 1.2e-9\\
Mean & 7.6e12 &4.5e12 &  2.6e12 & 7.3e13 & 3.0 & 1.8  & 3.6e-9  & 2.2e-9  & 1.2e-9\\
Median & 9.0e12&4.2e12 & 2.5e12 & 7.0e13& 3.6 & 1.7 & 4.5e-9	  & 2.1e-9 & 1.2e-9 \\
\noalign{\smallskip}
     \hline  \hline
     \noalign{\smallskip}     \noalign{\smallskip}

      \multicolumn{10}{c} {G10.6$-$0.4} \\     \noalign{\smallskip}
$v_\mathrm{LSR}$ & $N$(NH)\tablefootmark{b}     & $N$(o-NH$_2$)\tablefootmark{b}     & $N$(o-NH$_3$)   &$N$(CH)   
& NH/o-NH$_3$      &  o-NH$_2$/o-NH$_3$    & $X$(NH)\tablefootmark{c}& 
$X$(o-NH$_2$)\tablefootmark{c} & $X$(o-NH$_3$)\tablefootmark{c} \\
(km\,s$^{-1}$) &  (cm$^{-2}$)  &(cm$^{-2}$) &  (cm$^{-2}$) &  (cm$^{-2}$)  &  \\     
     \noalign{\smallskip}
     \hline
     \noalign{\smallskip}
12.5 -  20.0    & 2.9e13 & 1.4e13	&5.4e12	  & 1.2e14   &  5.4     &      2.6  & 
8.5e-9  &  4.1e-9   & 1.6e-9  \\ %
20.0    -  25.0 & 2.7e13  &1.2e13	& 3.8e12	  &9.3e13    &7.1     &        3.2 &  1.0e-8 &4.5e-9  & 1.4e-9  \\
25.0   -   29.0 & 3.0e13 &1.5e13	& 8.4e12	  &  1.2e14   &3.6      &      1.8 &   8.8e-9 &4.4e-9   & 2.5e-9  \\
29.0   -   31.0 & 2.8e13 &1.2e13	& 6.9e12	  &  5.4e13   &4.1    &       1.7 &     1.8e-8 &7.8e-9 & 4.5e-9 \\     
31.0  -    35.0 & 1.4e13 &5.0e12	& 2.8e12 	  & 7.1e13 & 5.0       &      1.8 & 
6.9e-9 &2.5e-9  & 1.4e-9  \\
35.0   -   39.5 & 2.6e13 & 8.2e12	& 3.0e12	  &  1.1e14  & 8.7      &       2.7 & 8.3e-9  &2.6e-9  & 9.6e-10  \\
39.5 -   43.5   & 3.2e13 &1.1e13	& 3.9e12 	& 6.3e13  &8.2          &       2.8  & 1.8e-8 &6.1e-9   &  2.2e-9 \\
43.5 -   50.0   & 1.8e12 &1.0e12	& 8.0e11	& 1.9e13  & 2.3     &      1.3 &   
3.3e-9    & 1.8e-9    & 1.5e-9 \\
 \noalign{\smallskip}     \noalign{\smallskip}
Total & 1.9e14 & 7.8e13 & 3.5e13 & 6.5e14 & 5.4  & 2.2& 1.0e-8  & 4.2e-9 & 1.9e-9\\
Mean & 2.3e13   &9.8e12 & 4.4e12& 8.1e13 & 5.4 & 2.2  & 1.0e-8 & 4.2e-9  & 1.9e-9 \\
Median & 2.7e13& 1.1e13 & 3.9e12& 8.2e13 & 7.1 & 3.0  &1.2e-8  & 4.9e-9  & 1.6e-9   \\

    \noalign{\smallskip}
\hline 
\label{Table: Model II ratios results}
\end{tabular}
\tablefoot{ 
\tablefoottext{a}{Column densities,  $N(x)$,  column  density ratios, and abundances, 
$X(x)$\,=\,$N$($x$)/$N$(H$_2$),   
in different 
velocity ranges.}
\tablefoottext{b}{Using  ortho-NH$_3$ \mbox{1$_0$\,--\,0$_0$}  as a template.} 
\tablefoottext{c}{Using [CH]/[H$_2$]\,=\,3.5$\times$10$^{-8}$ \citep{2008ApJ...687.1075S}.}     
}
\end{table*}

\clearpage

\section{Method III results}

  \begin{table*}[\!ht] 
\centering
\caption{\emph{Method~III results}\tablefootmark{a}. }
\begin{tabular} {lc ccc ccc ccc } 
 \hline\hline
\noalign{\smallskip}
      \multicolumn{11}{c} {W49N} \\     \noalign{\smallskip}
$v_\mathrm{LSR}$ & $\Delta v$ & $N$(NH)    & $N$(o-NH$_2$)   & $N$(o-NH$_3$) & $N$(CH) &  NH/o-NH$_3$  &  o-NH$_2$/o-NH$_3$    & 
 $X$(NH)\tablefootmark{b}& 
$X$(o-NH$_2$)\tablefootmark{b}& $X$(o-NH$_3$)\tablefootmark{b} \\
(km\,s$^{-1}$) &(km\,s$^{-1}$)  & (cm$^{-2}$)  &(cm$^{-2}$) &  (cm$^{-2}$)  &  (cm$^{-2}$) &   \\     
     \noalign{\smallskip}
     \hline
     \noalign{\smallskip}
  33.5 &  1.9 &  1.1e12  &  2.2e12  &  3.6e11  & 3.2e13 &  3.1   &  6.1   & 1.2e-9   &  2.4e-9     &  3.9e-10   \\ 
  39.6 &  1.1&   1.2e13 &   7.4e12  &  3.2e12   &3.3e13&  3.8   &  2.3   & 1.3e-8   &   7.8e-9   &  3.4e-9  \\ 
  59.4 &  3.0 &  1.1e13 &   3.5e12  &  1.7e12   & 7.0e13 &  6.5 &   2.1  &  5.5e-9  &  1.8e-9    &   8.5e-10 \\ 
  62.8 &  2.4&   9.7e12  &  3.2e12  &  1.9e12   & 5.8e13 &  5.1   &  1.7  & 5.9e-9  &   1.9e-9   &  1.1e-9  \\ 

\noalign{\smallskip} \noalign{\smallskip}\noalign{\smallskip}
Total:  &\ldots &3.4e13    &  1.6e13  &  7.2e12   & 1.9e14  &  4.7    &   2.3   & 6.1e-9  &  3.0e-9   &  1.3e-9 \\
Mean: &\ldots&  8.5e12  &  4.1e12   & 1.8e12  &  4.8e13    &  4.7    &   2.3    & 6.1e-9   &  3.0e-9   &  1.3e-9  \\
Median: &\ldots& 1.0e13 &  3.4e12   &  1.8e12    &  4.6e13  &  5.8    &   1.9    & 8.0e-9  &   2.6e-9   &  1.4e-9  \\

     \noalign{\smallskip}
       \hline  \hline
     \noalign{\smallskip}     \noalign{\smallskip}
      \multicolumn{11}{c} {G10.6$-$0.4} \\     \noalign{\smallskip}
$v_\mathrm{LSR}$ & $\Delta v$ & $N$(NH)    & $N$(o-NH$_2$)   & $N$(o-NH$_3$) & $N$(CH) &  NH/o-NH$_3$  &  o-NH$_2$/o-NH$_3$    & 
 $X$(NH)\tablefootmark{b}& 
$X$(o-NH$_2$)\tablefootmark{b}& $X$(o-NH$_3$)\tablefootmark{b} \\
(km\,s$^{-1}$) &(km\,s$^{-1}$)  & (cm$^{-2}$)  &(cm$^{-2}$) &  (cm$^{-2}$)  &  (cm$^{-2}$) &   \\     
     \noalign{\smallskip}
     \hline
     \noalign{\smallskip}tables
  16.3   &   1.9  &  2.4e13  &    9.6e12  &    3.4e12   &  7.7e13  &  7.1  &   2.8    & 1.1e-8   &  4.4e-9   &    1.5e-9   \\
  19.0   &   1.3  &  6.8e12  &    2.3e12  &   6.0e11   &  2.6e13   & 11  &   3.8    &  9.2e-9  &   3.1e-9   &    8.1e-10   \\
  22.2   &   4.2  &  1.9e13  &    7.4e12  &   1.7e12  &  9.0e13   &  11   &   4.4   & 7.4e-9   &  2.9e-9   &     6.6e-10  \\
  22.9   &   1.0  &  1.1e13  &   3.0e12  &   1.0e12  &  1.2e13   &  11  &   3.0   & 3.2e-8   &   8.8e-9   &    2.9e-9   \\
  25.0   &   3.3  &  8.3e12  &  4.8e12  &    1.1e12   & 4.0e13   & 7.5  &  4.4   & 7.3e-9   &   4.2e-9   &     9.6e-10  \\
  28.0   &   1.9  &  3.2e13 &   1.2e13   &   6.4e12   & 9.8e13   & 5.0  &   1.9   &  1.1e-8  &  4.3e-9    &    2.3e-9 \\
  30.0   &   1.5  &  3.1e13  &   1.4e13   &   6.8e12  &  6.1e13    & 4.6    &  2.1   & 1.8e-8   &  8.0e-9   &    3.9e-9 \\
  32.2   &   1.7  &  1.2e13  &    3.1e12  &   8.9e11   &  2.5e13   & 13     &  3.5   & 1.7e-8   &  4.3e-9    &   1.2e-9  \\
  36.2   &   3.8  &  2.2e13  &   7.3e12   &   2.5e12   & 9.4e13    & 8.8   &  2.9   &  8.2e-9   &   2.7e-9   &    9.3e-10  \\
  39.1   &   1.2   & 1.4e13  &    3.2e12  &   5.5e11   & 4.6e13   &  25   &   5.8   & 1.1e-8   &  2.4e-9    &   4.2e-10  \\
  41.0   &   1.7  &  3.5e13  &    1.1e13  &  3.1e12   &  8.5e13  &  11    &  3.5   & 1.4e-8   &  4.5e-9    &  1.3e-9  \\
  45.1   &   1.3  &  3.7e12  &    1.6e12  &    3.8e11   & 1.4e13   &  9.7  &   4.2   &  9.3e-9  &   4.0e.9   &   9.5e-10 \\

\noalign{\smallskip} \noalign{\smallskip}\noalign{\smallskip}
Total:  & \ldots &2.2e14  & 7.9e13 &   2.8e13  &   6.7e14&     7.7   &  2.8  & 1.1e-8  &  4.2e-9   &  1.5e-9 \\

Mean: &\ldots&1.8e13& 6.6e12 & 2.4e12 & 5.6e13& 7.7 &  2.8  &1.1e-8 &  4.2e-9     &   1.5e-9   \\

Median: &\ldots&1.7e13& 6.1e12 & 1.4e12 & 5.4e13& 12  &  4.3 &   1.1e-8 & 4.0e-9  &   9.2e-10  \\

    \noalign{\smallskip}
\hline 
\label{Table: Method III resulting ratios and columns}
\end{tabular} 
\tablefoot{ 
\tablefoottext{a}{LSR velocity, $v_\mathrm{LSR}$, line width, $\Delta v$,  
column densities, $N$($x$), column density ratios and abundances, $X(x)$\,=\,$N$($x$)/$N$(H$_2$),   of each velocity component.}
\tablefoottext{b}{$X(x)$\,=\,$N$($x$)/$N$(H$_2$) using [CH]/[H$_2$]\,=\,3.5$\times$10$^{-8}$ \citep{2008ApJ...687.1075S}.}      
}
\end{table*}

\clearpage

\section{Results for CN and HNC}

   \begin{table*}[\!ht] 
\centering
\caption{\emph{Method~I and III results for CN and HNC\tablefootmark{a}.} }
\begin{tabular} {l c | cc cc | c ccc } 
 \hline\hline
\noalign{\smallskip}
 W49N &&   \multicolumn{4}{c} {Method I}  &     \multicolumn{4}{c} {Method III} \\     \noalign{\smallskip}
$v_\mathrm{LSR}$ & $\Delta v$ & $N$(CN)    & $N$(HNC)    &  CN/o-NH$_3$  & HNC/o-NH$_3$   & $N$(CN)    & $N$(HNC)  &  CN/o-NH$_3$  & HNC/o-NH$_3$ \\
(km\,s$^{-1}$) &(km\,s$^{-1}$)  & (cm$^{-2}$)  &(cm$^{-2}$) & && (cm$^{-2}$)  &(cm$^{-2}$)\\
     \noalign{\smallskip}
     \hline
     \noalign{\smallskip}
  33.5 &  1.9 & 8.1e12	& 4.3e11	& 18	& 0.93	&7.5e12&  4.5e11  & 21  & 1.3	\\ 
  39.6 &  1.1 & 6.6e13	& 2.9e12	& 15	& 0.67	&6.3e13&  2.9e12  & 20  & 0.91	 \\ 
  59.4 &  3.0 & 3.5e13	& 2.1e12	& 17	& 1.0	&3.3e13&  2.2e12  & 19	& 1.3	\\ 
  62.8 &  2.4 & 2.8e13	& 1.5e12	& 13	& 0.68	&2.6e13&  1.7e12  & 14	& 0.89	 \\ 

\noalign{\smallskip} \noalign{\smallskip}\noalign{\smallskip}
Total:  &&   1.4e14 &  	6.9e12	& 15   & 0.76	&1.3e14	& 7.3e12 & 18  &  1.0	\\
Mean: &&   3.4e13 &   1.7e12	& 15   & 0.76   &3.2e13	& 1.8e12 & 18	&  1.0	\\
Median:   &&   3.2e13 &   1.8e12& 15   & 0.84   &3.0e13	& 2.0e12 & 16	&  1.1	\\

     \noalign{\smallskip}
       \hline  \hline
     \noalign{\smallskip}     \noalign{\smallskip}    
 G10.6$-$0.4 &&   \multicolumn{4}{c} {Method I}  &     \multicolumn{4}{c} {Method III} \\     \noalign{\smallskip}
$v_\mathrm{LSR}$ & $\Delta v$ & $N$(CN)    & $N$(HNC)    &  CN/o-NH$_3$  & HNC/o-NH$_3$   & $N$(CN)    & $N$(HNC)  &  CN/o-NH$_3$  & HNC/o-NH$_3$ \\
(km\,s$^{-1}$) &(km\,s$^{-1}$)  & (cm$^{-2}$)  &(cm$^{-2}$) & && (cm$^{-2}$)  &(cm$^{-2}$)\\     \noalign{\smallskip}
     \hline
     \noalign{\smallskip}

  16.3   &   1.9  & 1.0e14  & 3.6e12  & 24   & 0.86    & 9.0e13  &4.1e12  &26  & 1.2\\
  19.0   &   1.3  & 3.3e13  & 1.3e12  & 38   & 1.5    & 2.9e13  &1.5e12  &48  &2.5\\
  22.2   &   4.2  & 5.7e13  & 2.7e12  & 32   & 1.5    & 5.1e13  &3.1e12  &30  & 1.8\\
  22.9   &   1.0  & 6.6e13  & 2.5e12  & 44   & 1.7    & 5.8e13  &2.9e12  &58  & 2.9\\
  25.0   &   3.3  & 9.5e12  & \ldots  & 6.8  & \ldots  & 8.4e12  &1.7e12  &7.6   &1.5\\
  28.0   &   1.9  & 1.2e14  & 4.1e12  & 16  & 0.56    & 1.1e14  &4.7e12  &17   &0.73\\
  30.0   &   1.5  & 1.4e14  & 5.8e12  & 18  & 0.73    & 1.3e14  &6.7e12  &19   &1.0\\
  32.2   &   1.7  & 2.5e13  & 1.0e12  & 19   & 0.77    & 2.3e13  &1.2e12  &26  &1.3\\
  36.2   &   3.8  & 1.0e14  & 5.1e12  & 31   & 1.6    & 9.0e13  &5.9e12  &36  &2.4 \\
  39.1   &   1.2  & 3.7e13  & 1.1e12  & 79   & 2.3     & 3.3e13  &1.3e12  &60  &2.4\\
  41.0   &   1.7  & 1.6e14  & 5.9e12  & 39   & 1.4    & 1.4e14  &6.8e12  &45  &2.2\\
  45.1   &   1.3  & 1.2e13  & 8.5e11  & 20   & 1.4    & 1.1e14  &9.7e12  &289 &26\\

\noalign{\smallskip} \noalign{\smallskip}\noalign{\smallskip}
Total:  & & 8.6e14& 3.4e13 & 25 & 1.0 & 8.7e14 & 5.0e13 & 31  & 1.7  \\

Mean: &&   7.2e13 & 3.1e12 & 25 & 1.0 & 7.4e13  &3.6e12 & 53  &  2.6 \\

Median: && 6.2e13 & 2.7e12 & 37   & 1.5 & 7.3e13 & 4.1e12& 31  &  1.7\\
 
    \noalign{\smallskip}
\hline 
\label{Table: Method I, III CN and HNC columns}
\end{tabular}
\tablefoot{ 
\tablefoottext{a}{LSR velocity, $v_\mathrm{LSR}$, line width, $\Delta v$,  
and column densities, $N$,  of each velocity component, and column density ratios.}     
}
\end{table*}

   \begin{table*}[\!ht] 
\centering
\caption{\emph{CN and HNC results from Method~II\tablefootmark{a}.}
}
\begin{tabular} {l  c ccc} 
 \hline\hline
     \noalign{\smallskip}

           \multicolumn{5}{c} {W49N} \\     \noalign{\smallskip}
$v_\mathrm{LSR}$ & $N$(CN)\tablefootmark{b}      & $N$(HNC)\tablefootmark{b} & CN/o-NH$_3$  & HNC/o-NH$_3$ \\

(km\,s$^{-1}$) &  (cm$^{-2}$)  &(cm$^{-2}$) &   \\       
     \noalign{\smallskip}
     \hline
     \noalign{\smallskip}
36.0 - 42.0     &  5.3e13  & 2.9e12  & 13   & 0.71\\   
50.0 - 57.0     &  4.4e12  & 5.7e11  & 3.7  & 0.48\\
57.0 - 61.0     &  2.1e13  & 1.7e12  & 11   & 0.89\\
61.0 - 70.0     &  2.6e13  & 2.0e12  & 8.7  & 0.67\\

 \noalign{\smallskip}     \noalign{\smallskip}
Total  & 1.0e14  & 7.2e12 & 10 & 0.70\\
Mean   & 2.6e13  & 1.8e12 & 10 & 0.70\\
Median & 2.4e13  & 1.9e12 & 9.6 & 0.76\\
\noalign{\smallskip}
     \hline  \hline
     \noalign{\smallskip}     \noalign{\smallskip}

      \multicolumn{5}{c} {G10.6$-$0.4} \\     \noalign{\smallskip}
$v_\mathrm{LSR}$ & $N$(CN)\tablefootmark{b}      & $N$(HNC)\tablefootmark{b} & CN/o-NH$_3$  & HNC/o-NH$_3$ \\

(km\,s$^{-1}$) &  (cm$^{-2}$)  &(cm$^{-2}$) &   \\        
     \noalign{\smallskip}
     \hline
     \noalign{\smallskip}

12.5 -  20.0    &1.0e14  &5.8e12   & 19 & 1.1\\  
20.0    -  25.0 &9.1e13  &5.2e12   & 24  & 1.4 \\ 
25.0   -   29.0 &9.6e13  &5.6e12   & 11 & 0.67   \\ 
29.0   -   31.0 &8.8e13  &4.9e12   & 13 & 0.71   \\     
31.0  -    35.0 &4.2e13  &2.3e12   & 15 & 0.82   \\ 
35.0   -   39.5 &8.0e13  &5.3e12   & 27 & 1.8   \\ 
39.5 -   43.5   &1.4e14  &7.4e12   & 36  & 1.9   \\ 
43.5 -   50.0   &9.8e12  &1.1e12   & 12 & 1.4   \\ 
 \noalign{\smallskip}     \noalign{\smallskip}
Total & 6.5e14 & 3.8e13   & 18 & 1.1 \\ 
Mean &8.1e13 &  4.7e12  & 18 & 1.1\\ 
Median & 9.0e13 & 5.3e12   & 23 & 1.4\\ 
    \noalign{\smallskip}
\hline 
\label{Table: Method II CN and HNC columns}
\end{tabular}
\tablefoot{ 
\tablefoottext{a}{ Column densities and  ratios    
in different 
velocity ranges.} 
\tablefoottext{b}{Using  ortho-NH$_3$~1$_0$\,--\,0$_0$  as a template.}  
}
\end{table*}

\clearpage 
\section{Figures}

   \begin{figure}[\!ht] 
\centering
\includegraphics[scale = 0.6]{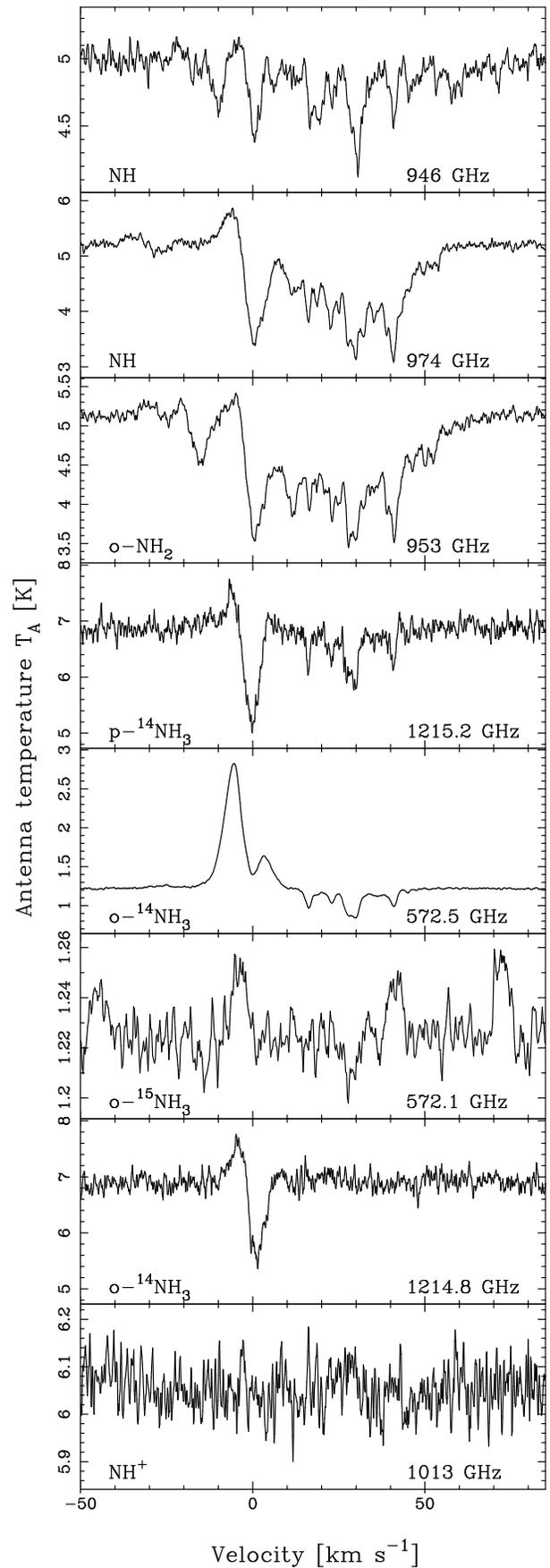}  
 \caption{\emph{G10.6$-$0.4 (W31C)}: Double sideband spectra of NH,  ortho-NH$_2$,  ortho- and para-$^{14}$NH$_3$, ortho-$^{15}$NH$_3$, and NH$^+$  
over the LSR velocity range -50 to 85\,km\,s$^{-1}$.}
 \label{Fig: W31C All original hydrides}
\end{figure}

   \begin{figure}[\!ht] 
\centering
\includegraphics[scale = 0.5]{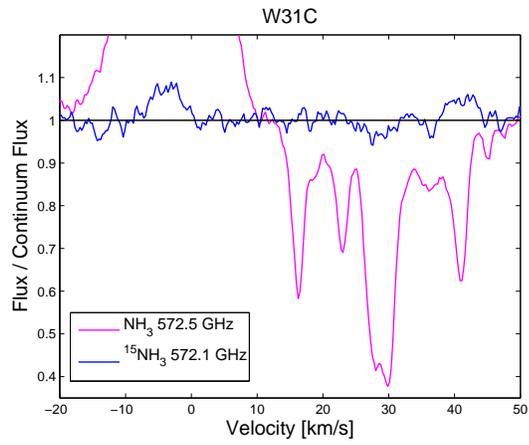}
 \caption{\emph{G10.6$-$0.4 (W31C):} Normalised ammonia and the ammonia isotopologue 
$^{15}$NH$_3$. (The emission line at $v_\mathrm{LSR}\!\approx\!41$ and 
72\,km\,s$^{-1}$~is an SO$_2$ line from the upper sideband.)}
 \label{Fig: W31C NH3 and 15NH3 normalised}
\end{figure}

\clearpage

   \begin{figure}[\!ht] 
\centering
\includegraphics[scale=0.5]{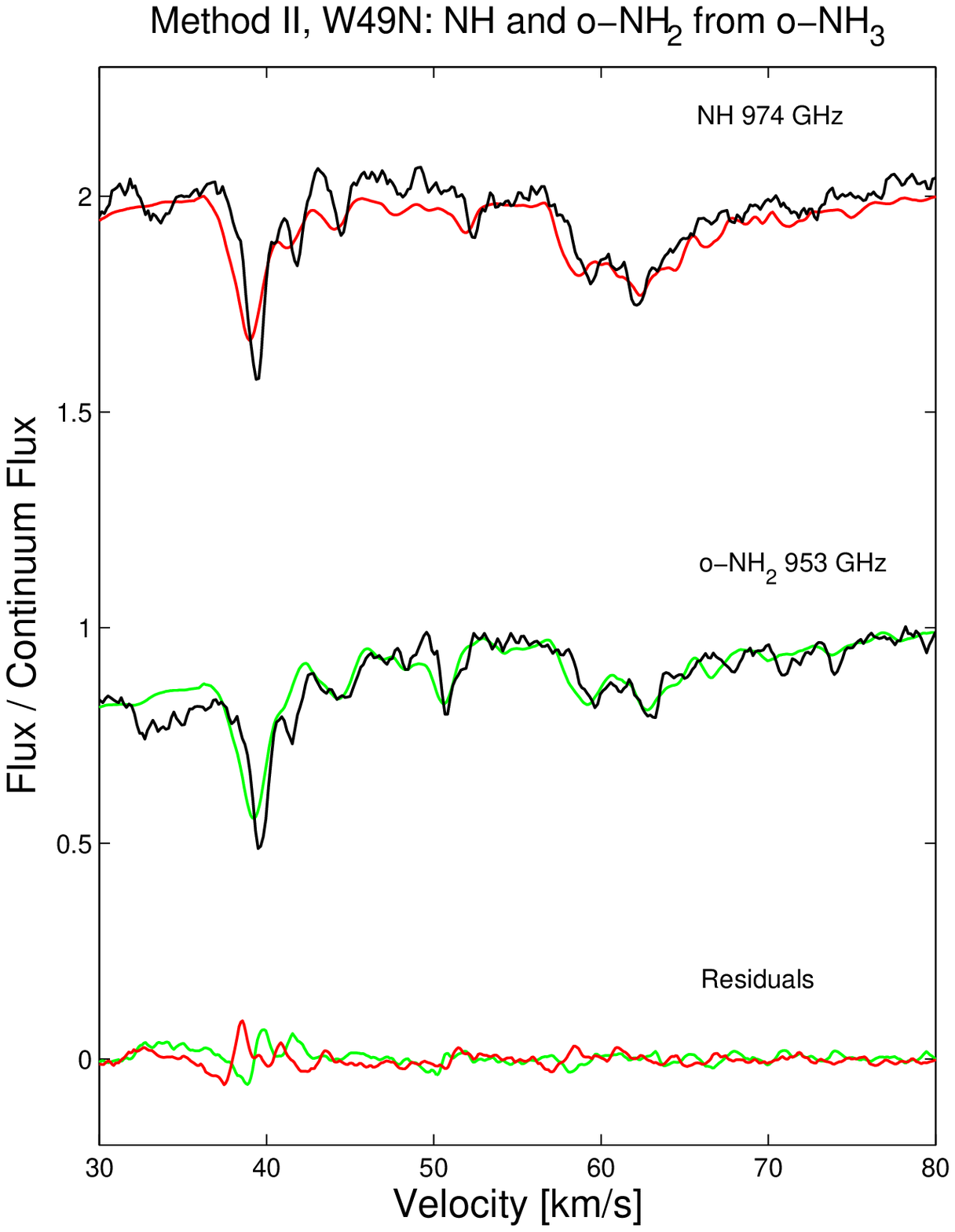}
 \caption{\emph{W49N:} Fits of method~II to NH and o-NH$_2$, and their observed spectra. Residuals are found at the bottom. 
Template spectrum is o-NH$_3$ at
572\,GHz.}
 \label{Fig: W49N method I NH and NH2 from NH3}
\end{figure}

    \begin{figure}[\!ht] 
\centering
\includegraphics[scale=0.5]{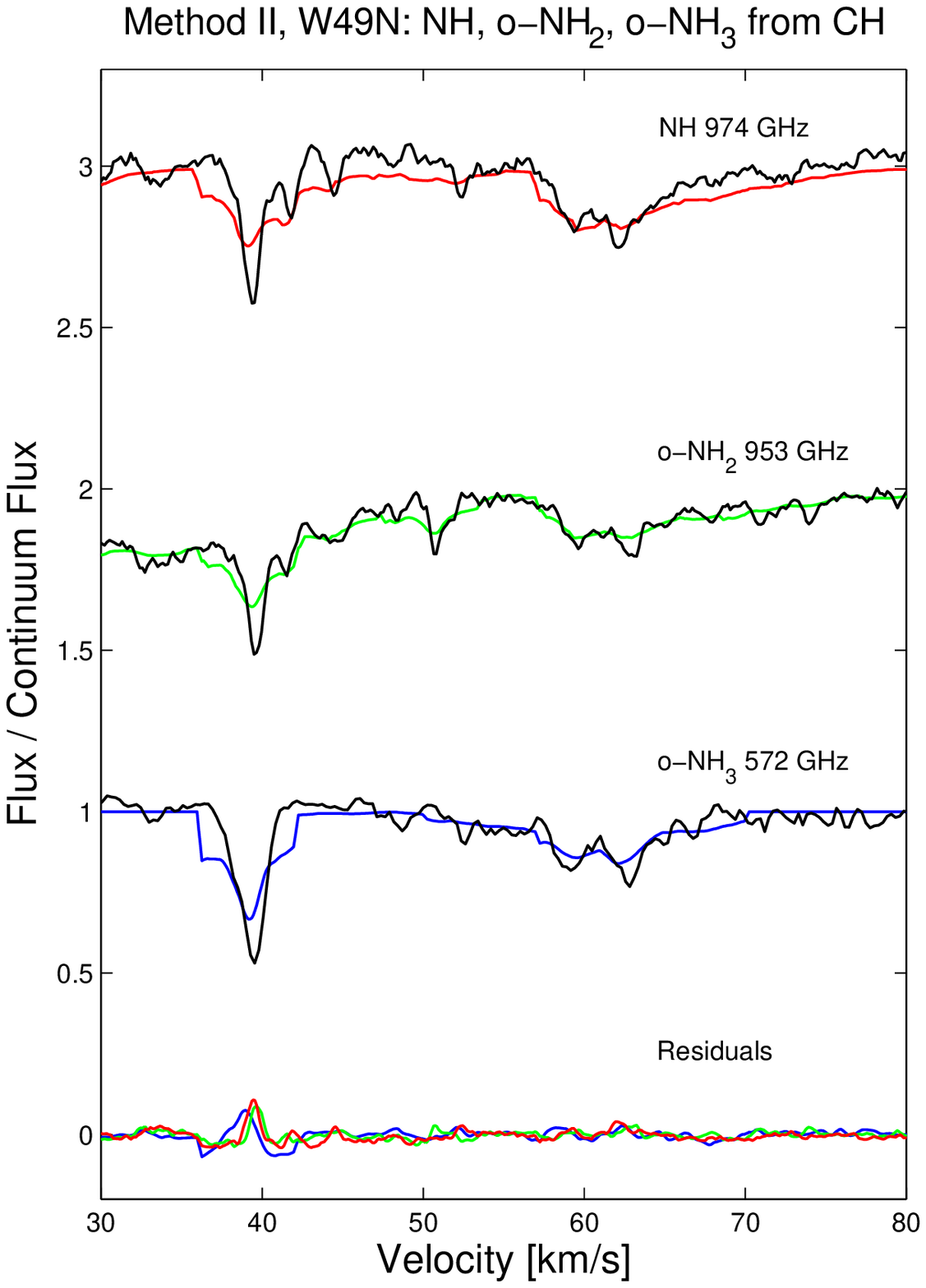} 
 \caption{\emph{W49N:} Fits of method~II to NH, o-NH$_2$ and o-NH$_3$, and their observed spectra. Residuals are found at the bottom. 
Template spectrum is CH at
532\,GHz.}
\label{Fig: W49N method I NH, NH2, NH3 from CH}
\end{figure}

    \begin{figure}[\!ht] 
\centering
\includegraphics[scale=0.5]{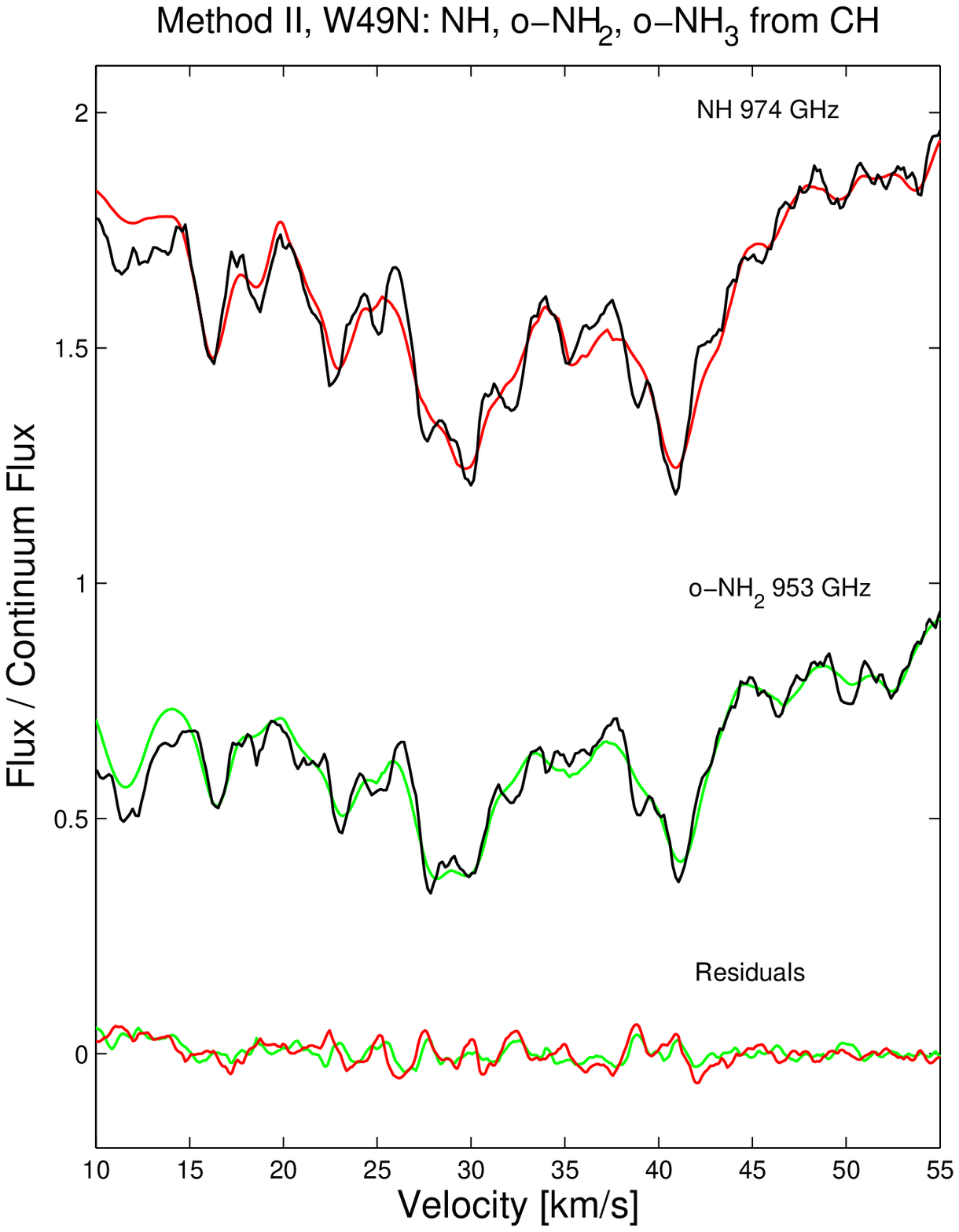} 
 \caption{\emph{G10.6$-$0.4 (W31C):} Fits of method~II to NH and \mbox{o-NH$_2$}, and their observed spectra. Residuals are found at the bottom.
Template spectrum is o-NH$_3$ at
572\,GHz.}
 \label{Fig: W31C method I NH and NH2 from NH3}
\end{figure}

    \begin{figure}[\!ht] 
\centering
\includegraphics[scale=0.5]{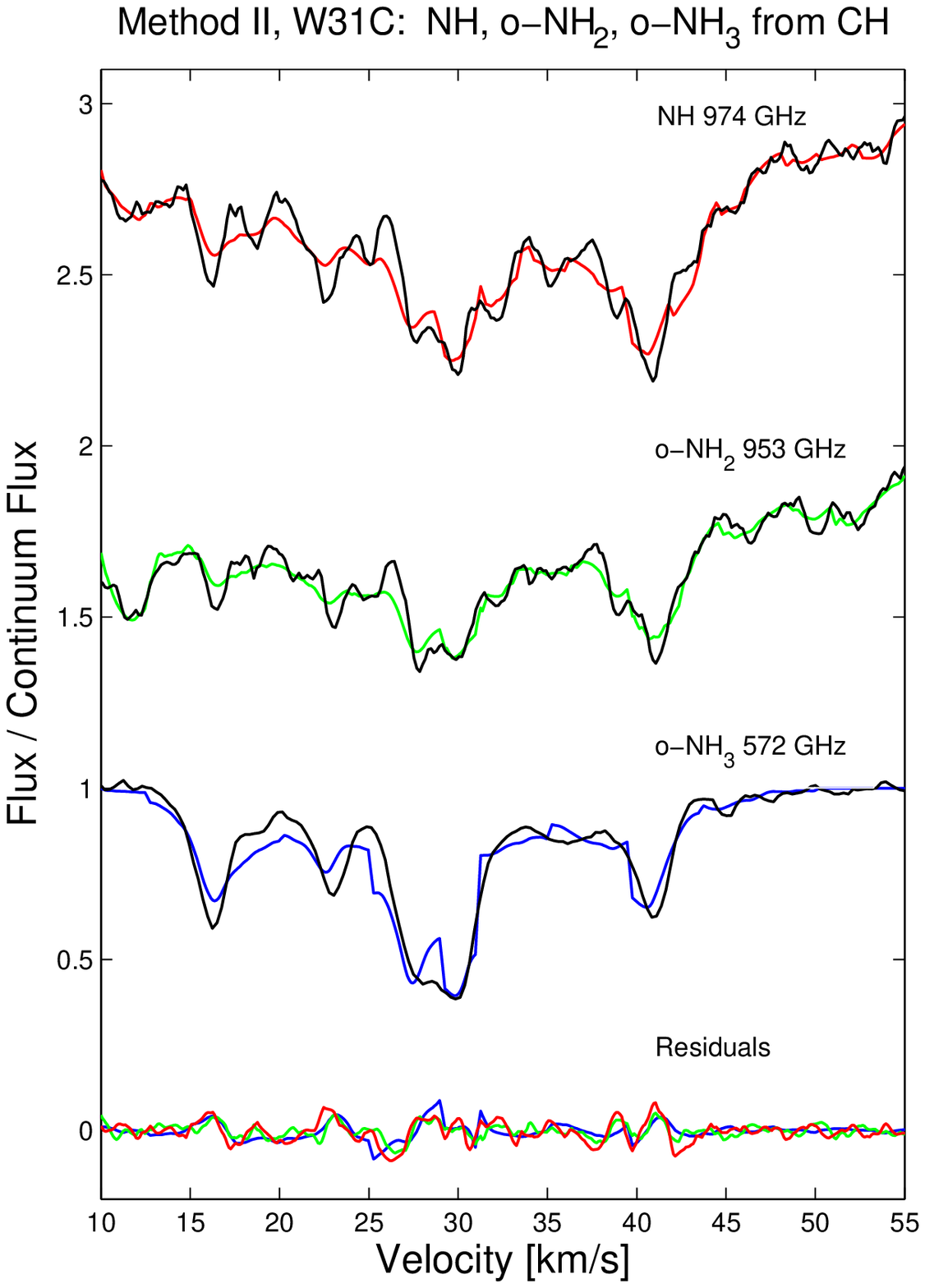} 
 \caption{\emph{G10.6$-$0.4 (W31C):} Fits of method~II to NH, \mbox{o-NH$_2$} and o-NH$_3$, and their observed spectra. Residuals are found at the bottom. 
Template spectrum is CH at
532\,GHz.}
\label{Fig: W31C method I NH, NH2, NH3 from CH}
\end{figure}
 
\clearpage
\begin{figure}[\!ht] 
\centering
\includegraphics[scale=0.75]{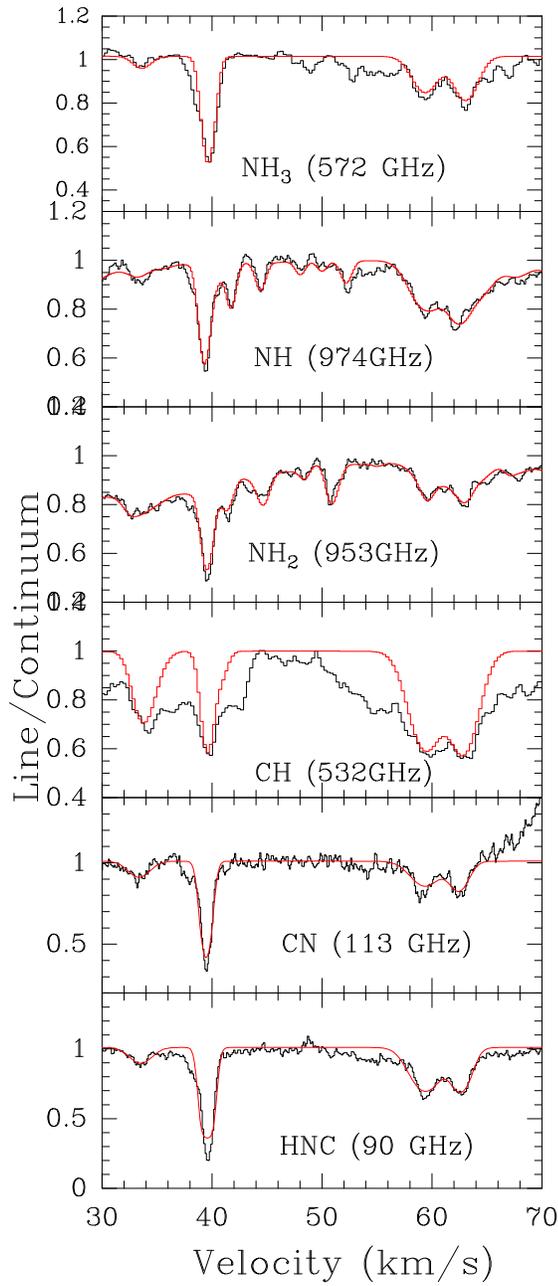}
\caption{\emph{W49N:} Method~III  (XCLASS) fits to the Nitrogen hydrides and CH.
}
 \label{Fig: w49n guass fits n-hydrides method III}
\end{figure}


\begin{figure}[\!ht] 
\centering
\includegraphics[scale=0.75]{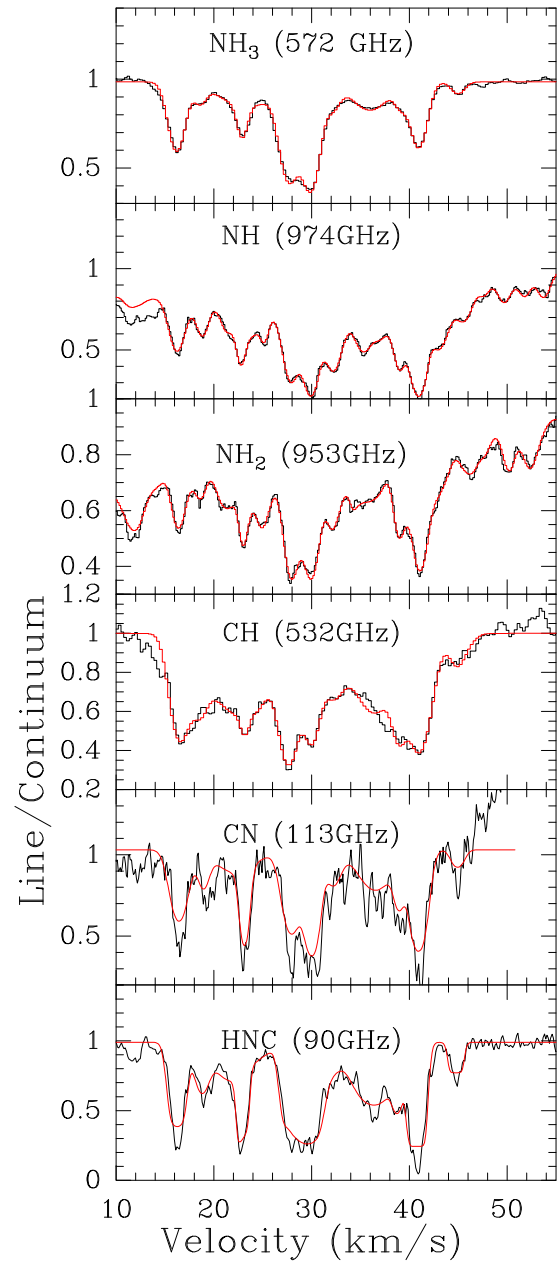}
\caption{\emph{G10.6$-$0.4 (W31C):} Method~III (XCLASS) fits to the Nitrogen hydrides and CH.
}
 \label{Fig: w31c guass fits n-hydrides method III}
\end{figure}

\clearpage

 \begin{figure*}[\!ht]
\centering
\subfigure[]{
\includegraphics[width=.35\textwidth]{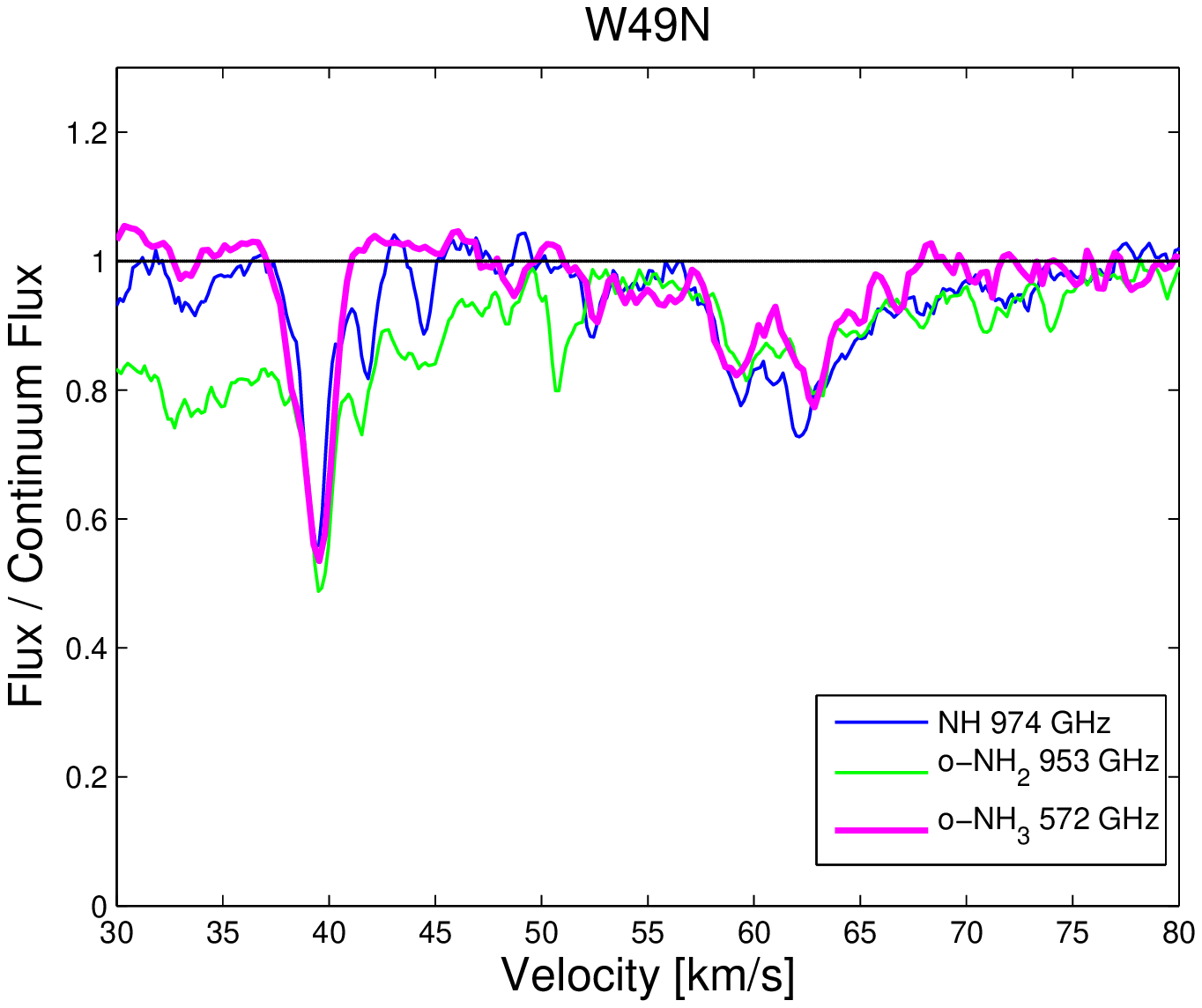}} 
\hspace{.3in}
\subfigure[]{
\includegraphics[width=.35\textwidth]{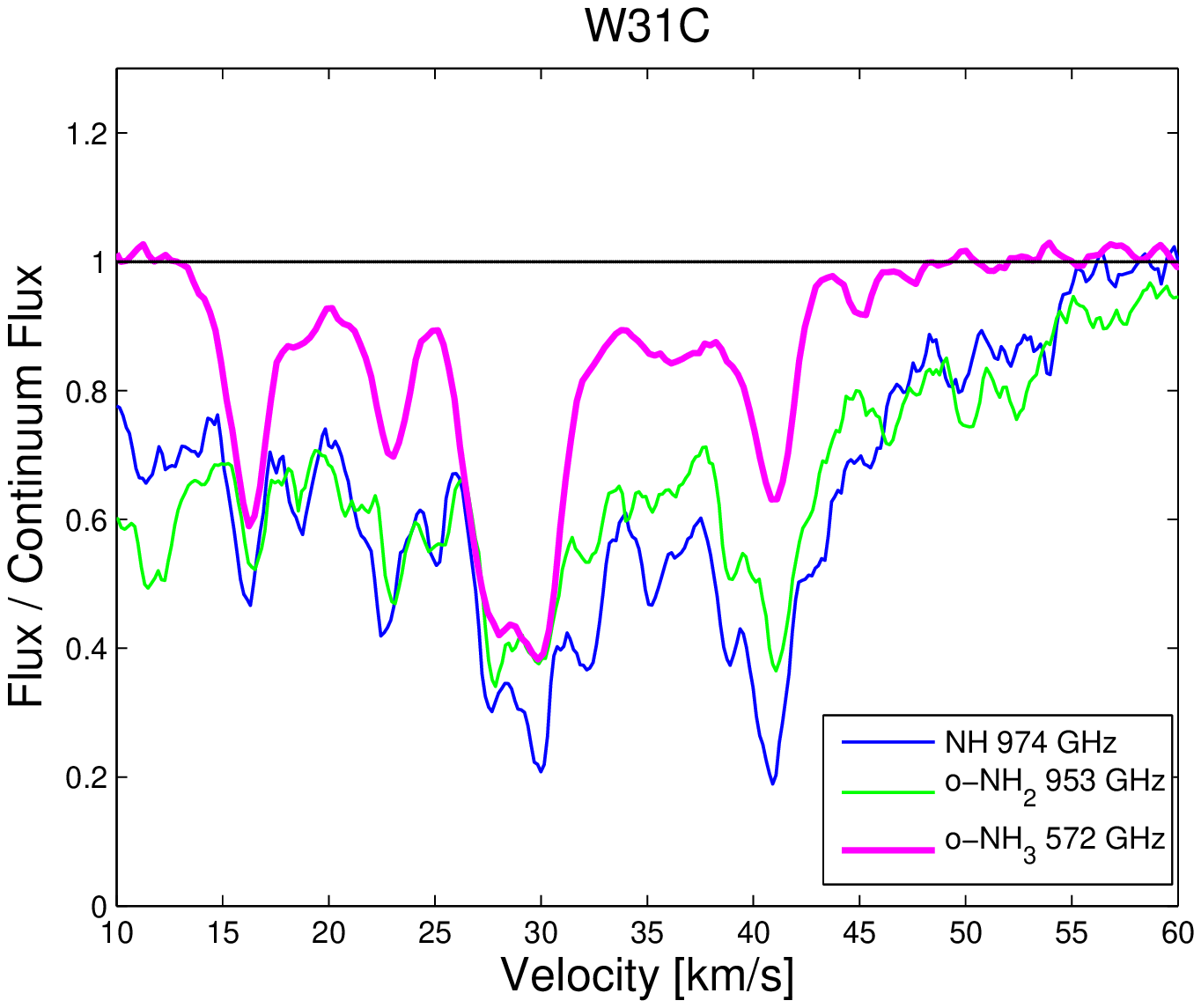}} 
\vspace{.3in}
\subfigure[]{ 
\includegraphics[width=.35\textwidth]{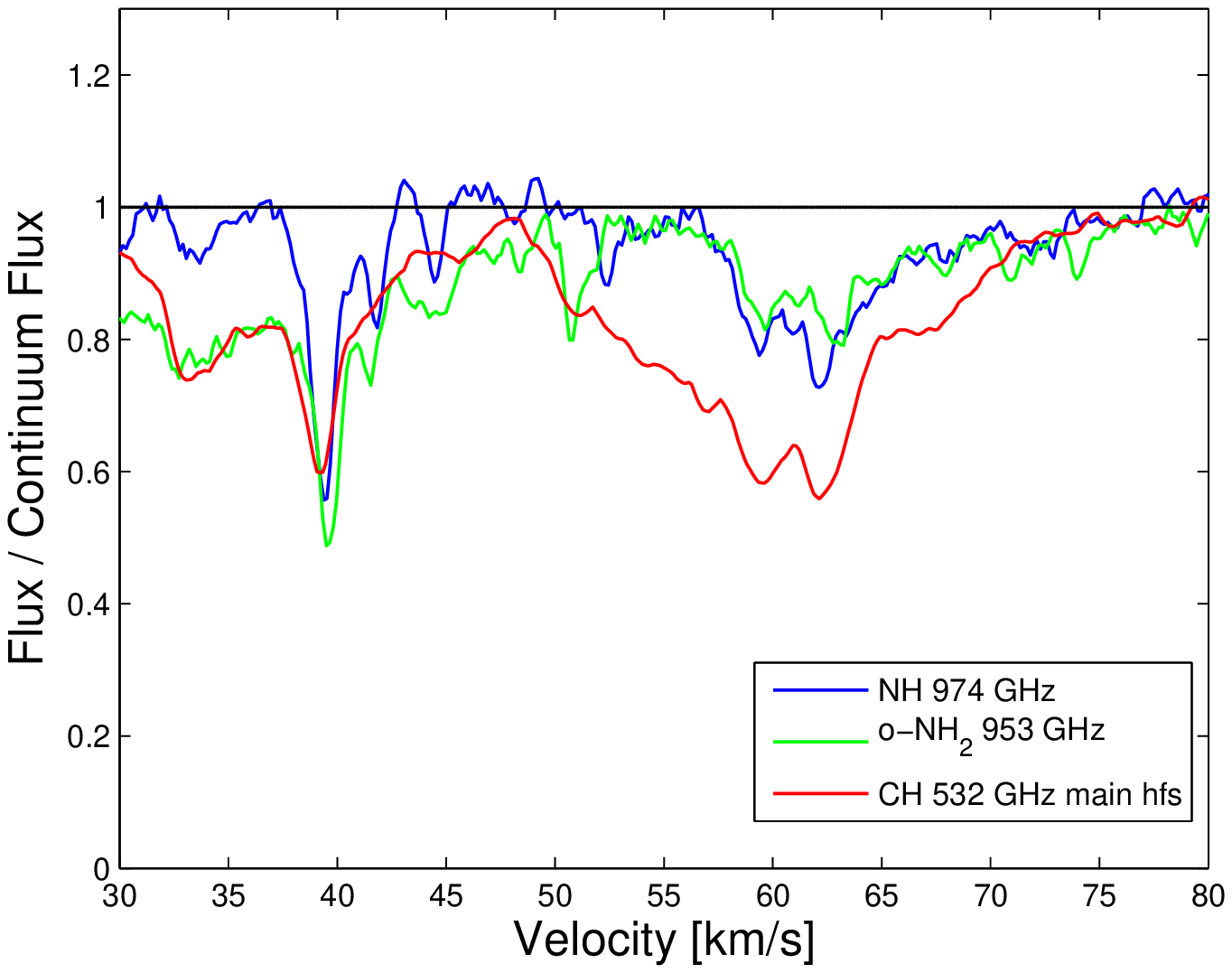}} 
\hspace{.3in}
\subfigure[]{
\includegraphics[width=.35\textwidth]{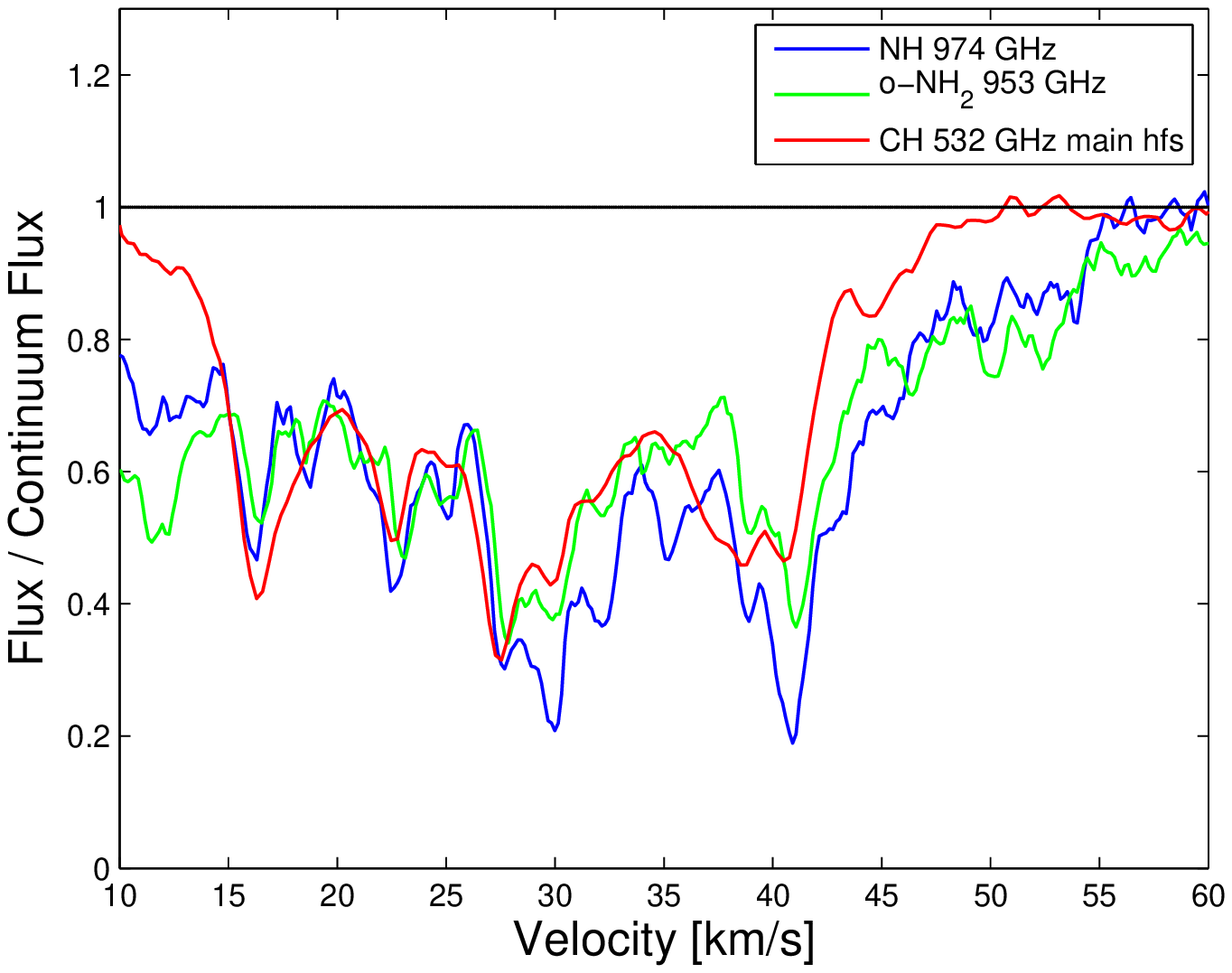}} 
\vspace{.3in}
\subfigure[]{ 
\includegraphics[width=.35\textwidth]{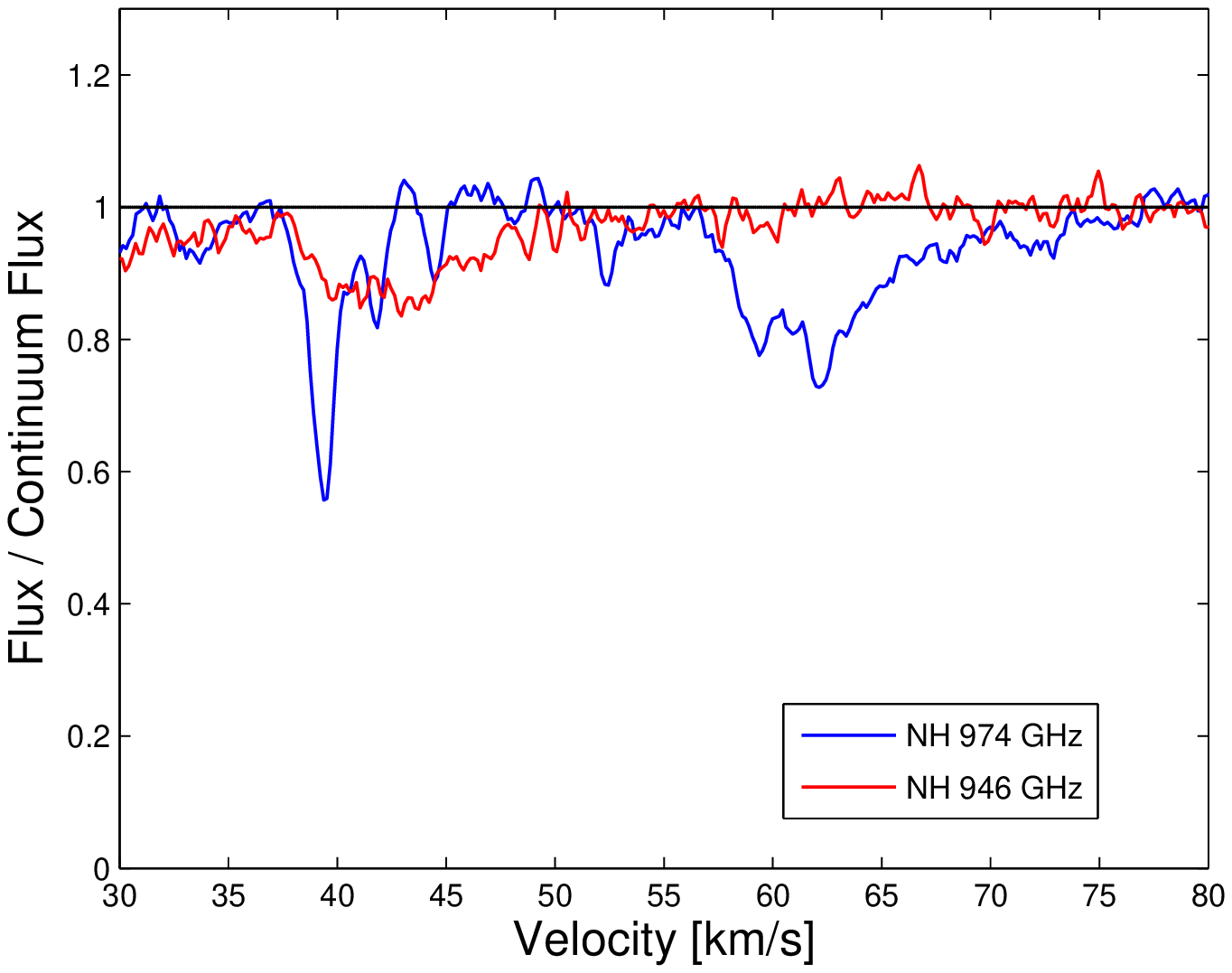}} 
\hspace{.3in}
\subfigure[]{
\includegraphics[width=.35\textwidth]{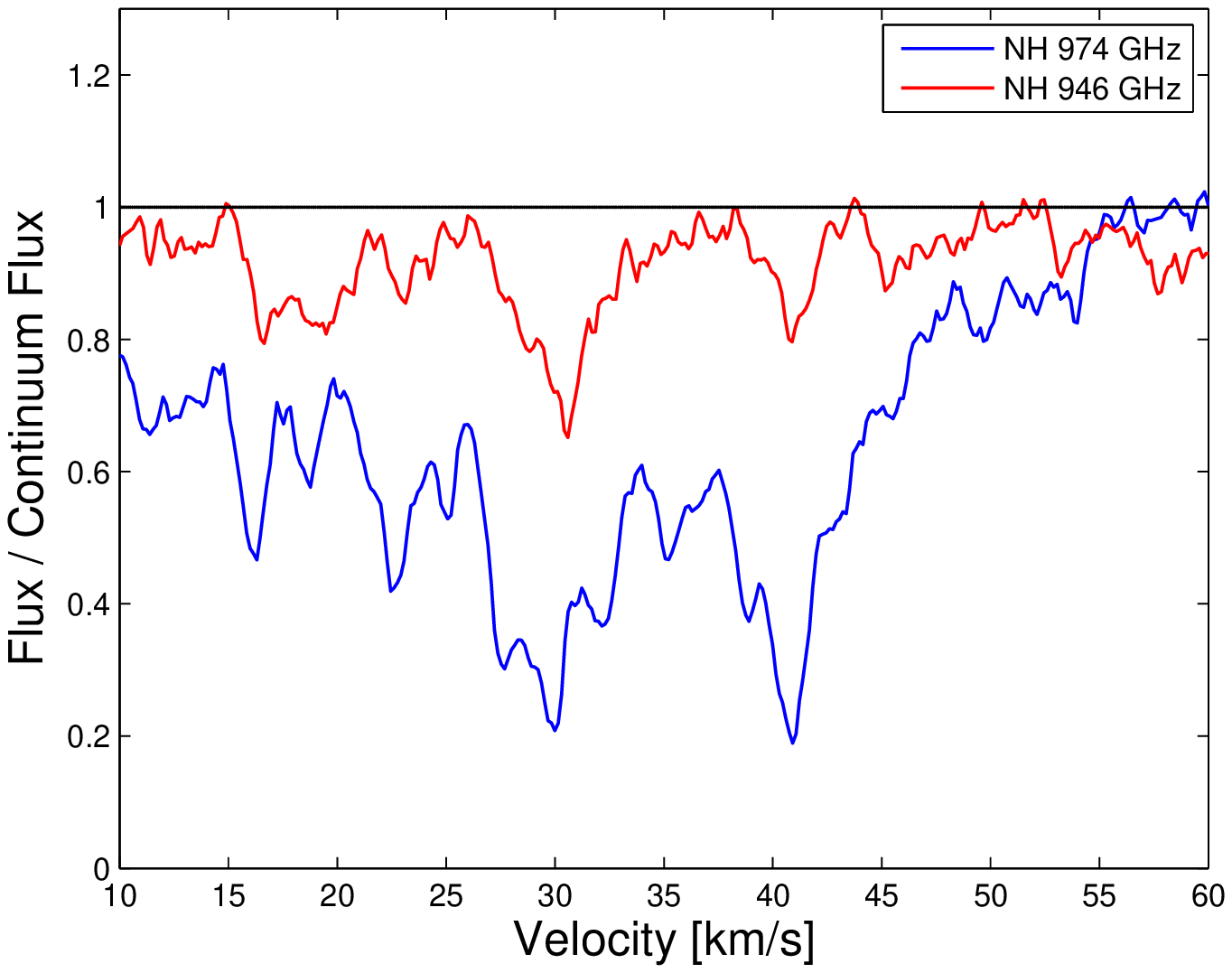}}
\vspace{.3in} 
\subfigure[]{ 
\includegraphics[width=.35\textwidth]{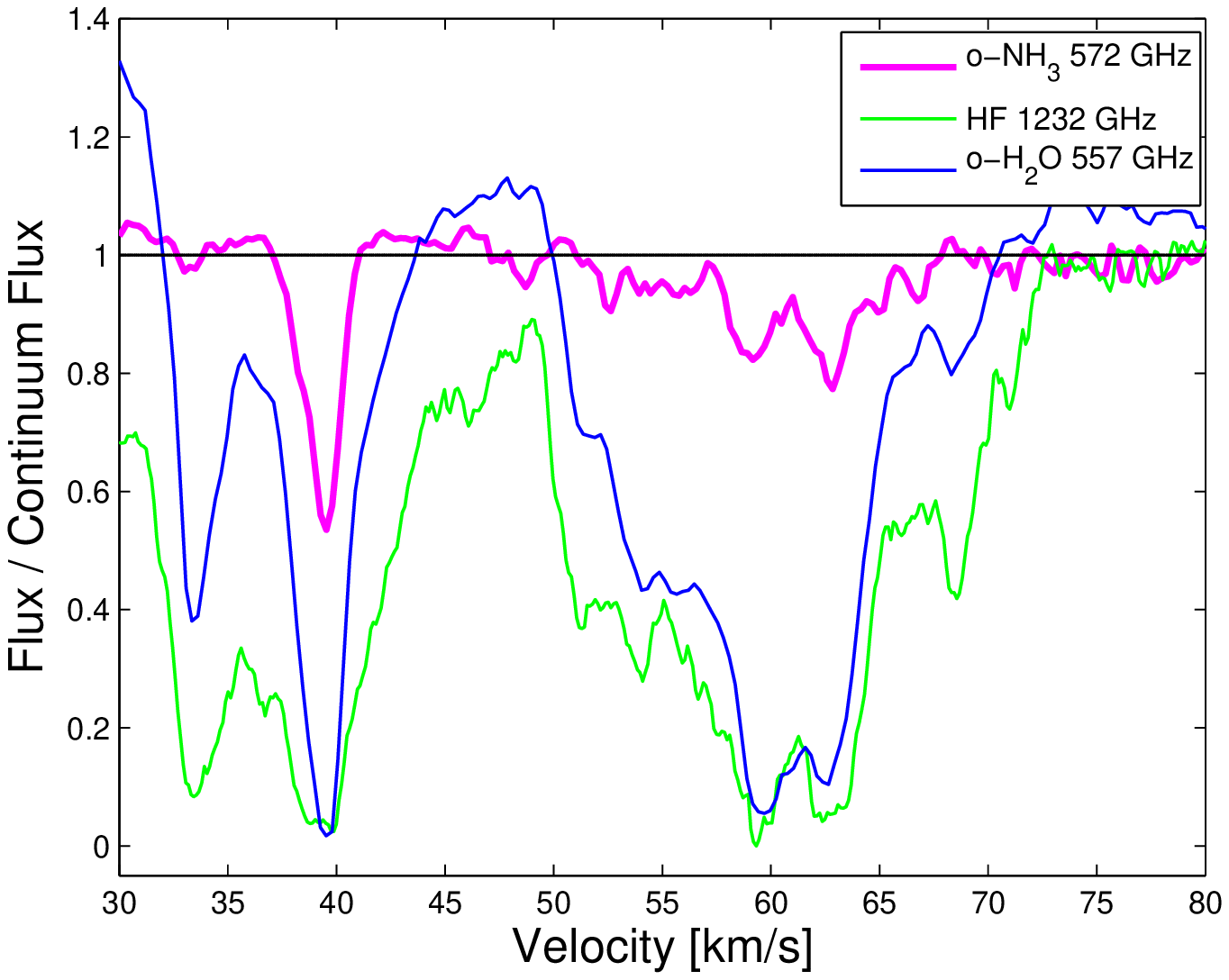}} 
\hspace{.3in}
\subfigure[]{
\includegraphics[width=.35\textwidth]{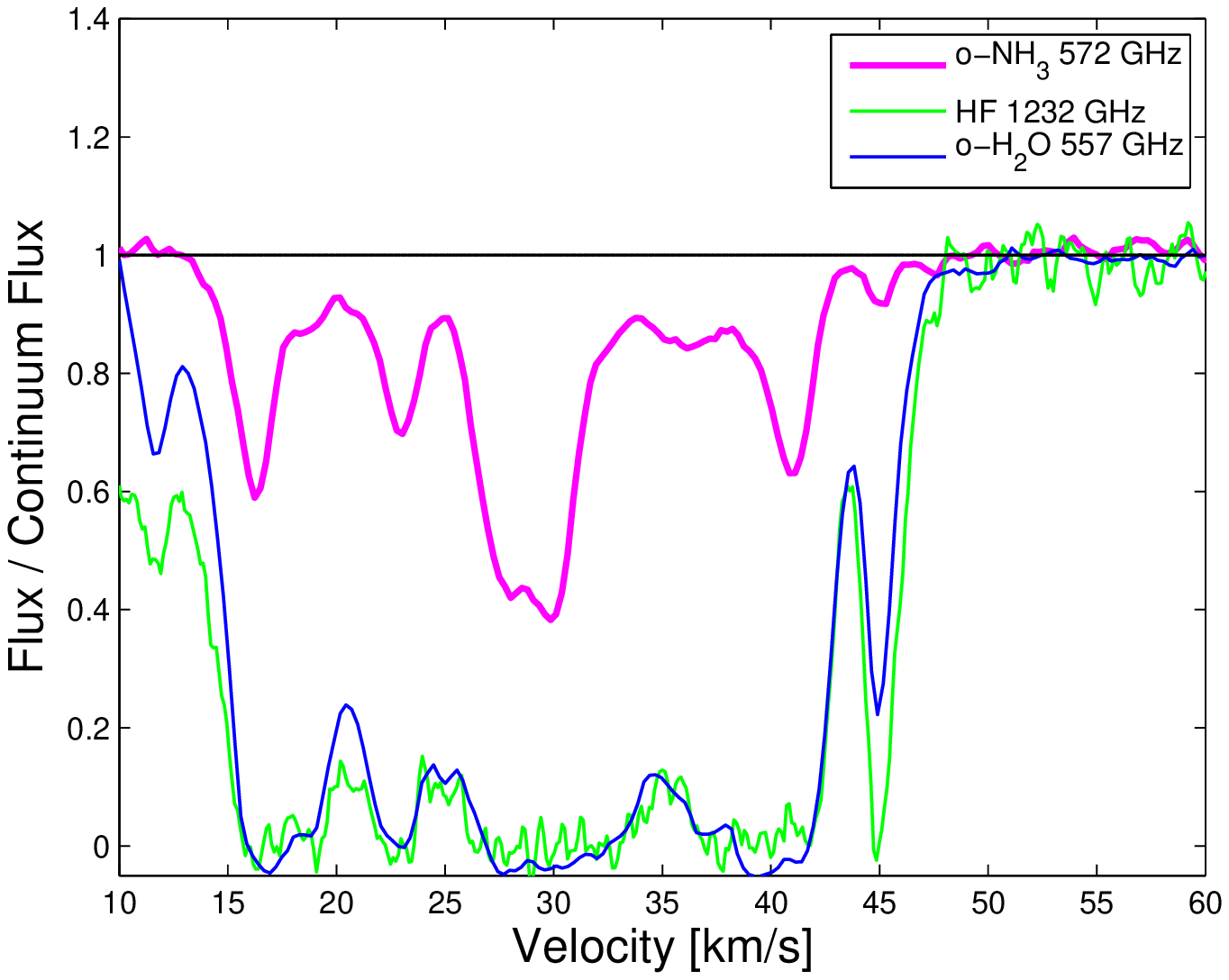}}
\caption{\emph{Comparison of absorption lines towards W49N and G10.6$-$0.4.} The intensities have been normalised to single sideband continuum. An NO emission line from the source (952.464\,GHz) has been removed from the o-NH$_2$~spectrum (see Sect.~\ref{NO removal of NH2}.).}
\label{fig: comparison of W31C-W49N NH, NH2 vs 572, 572 and 1215, CH}
\end{figure*}

\clearpage

\begin{figure*}[\!ht]
\centering
\subfigure[]{
\includegraphics[width=.35\textwidth]{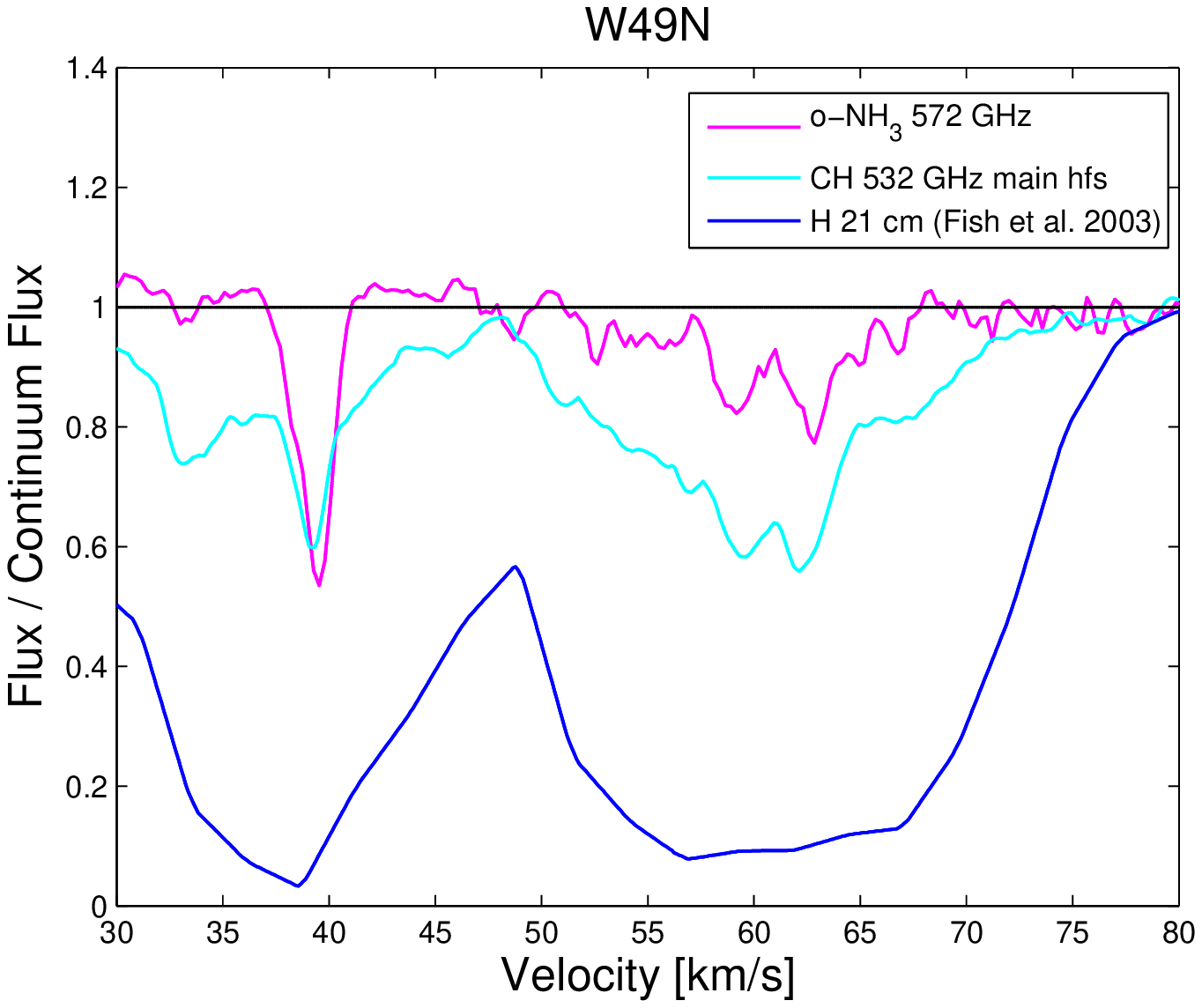}} 
\hspace{.3in}
\subfigure[]{
\includegraphics[width=.35\textwidth]{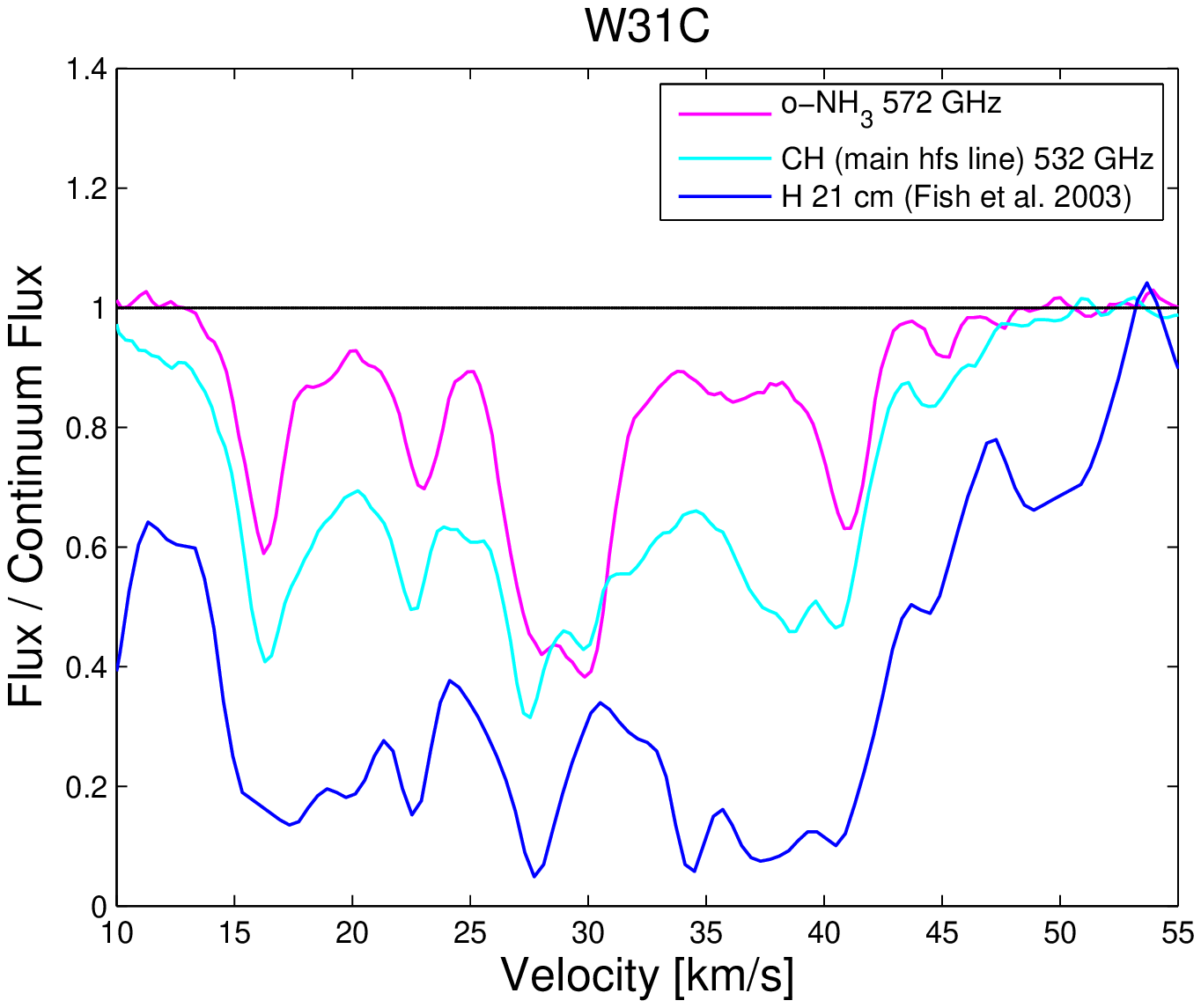}} 
\vspace{.3in}
\subfigure[]{ 
\includegraphics[width=.35\textwidth]{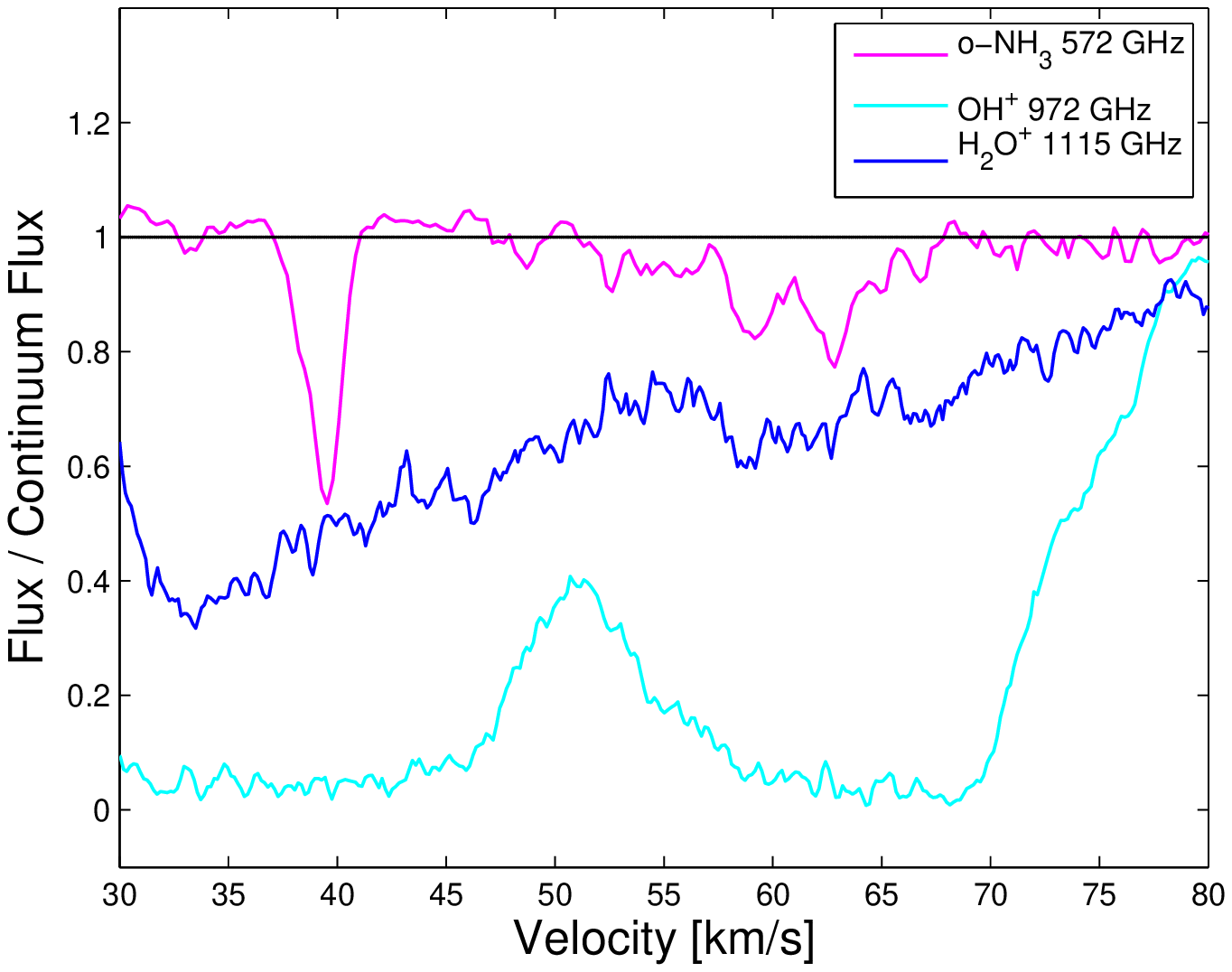}}
\hspace{.3in}
\subfigure[]{
\includegraphics[width=.35\textwidth]{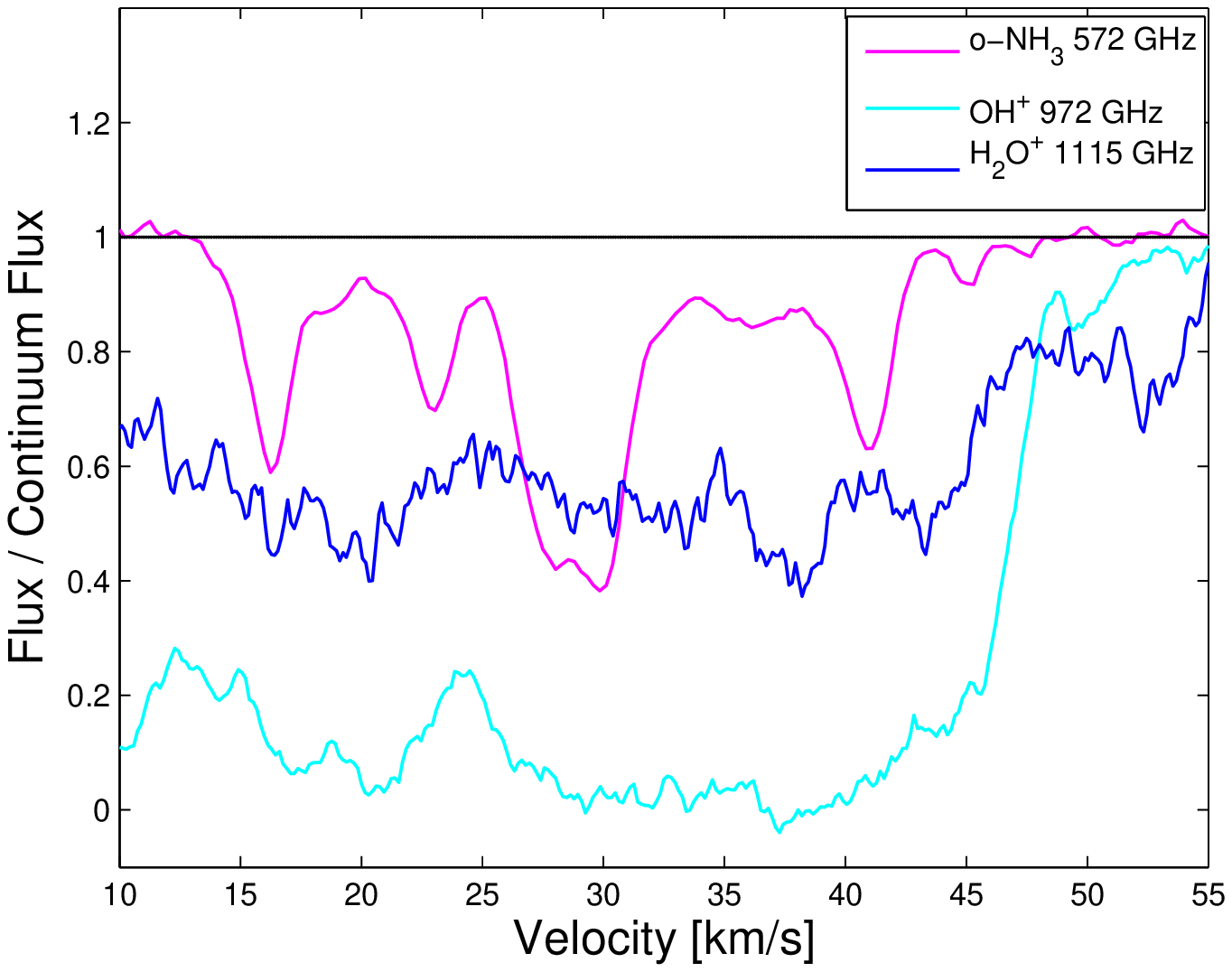}}
\vspace{.3in}
\subfigure[]{ 
\includegraphics[width=.35\textwidth]{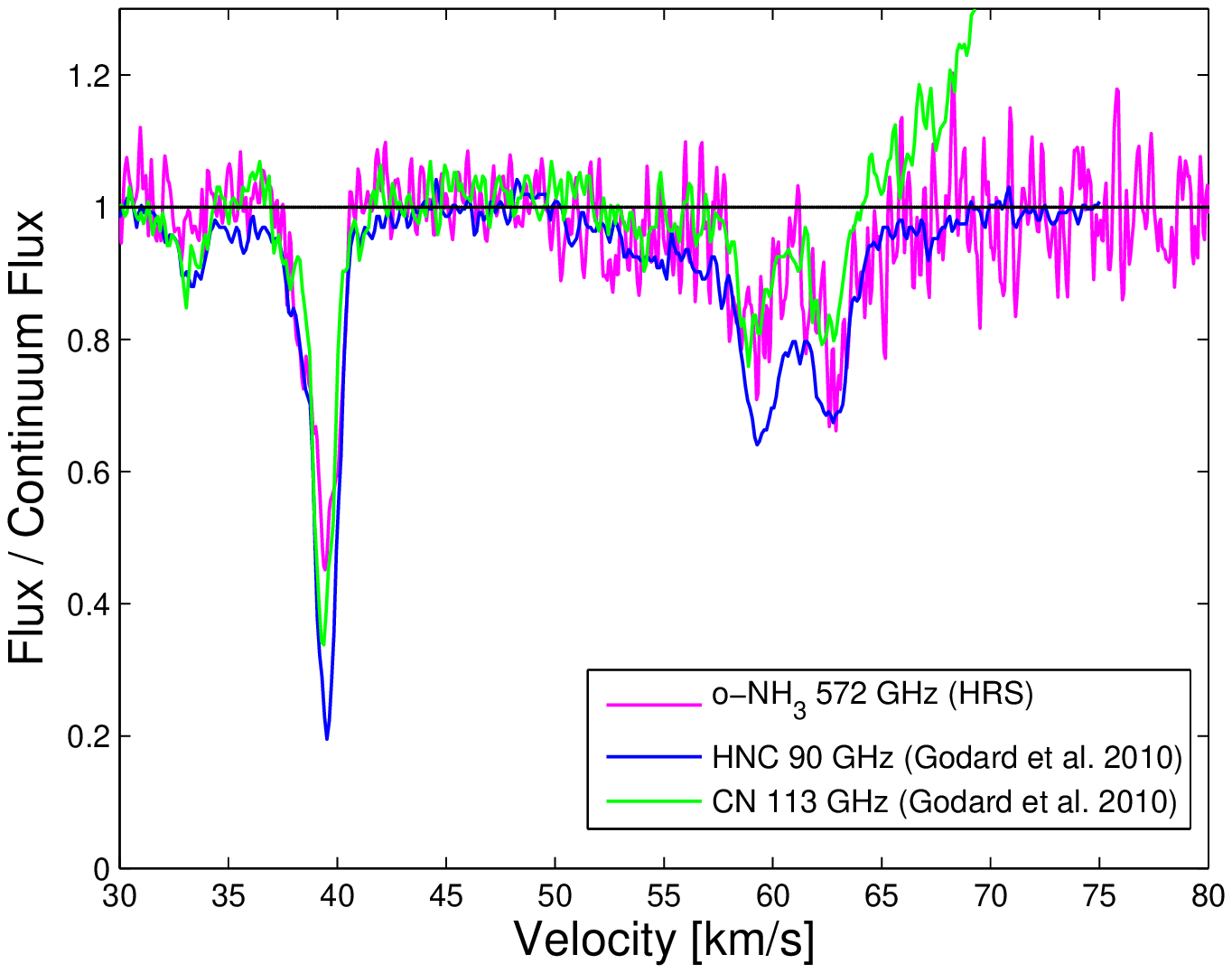}} 
\hspace{.3in}
\subfigure[]{
\includegraphics[width=.35\textwidth]{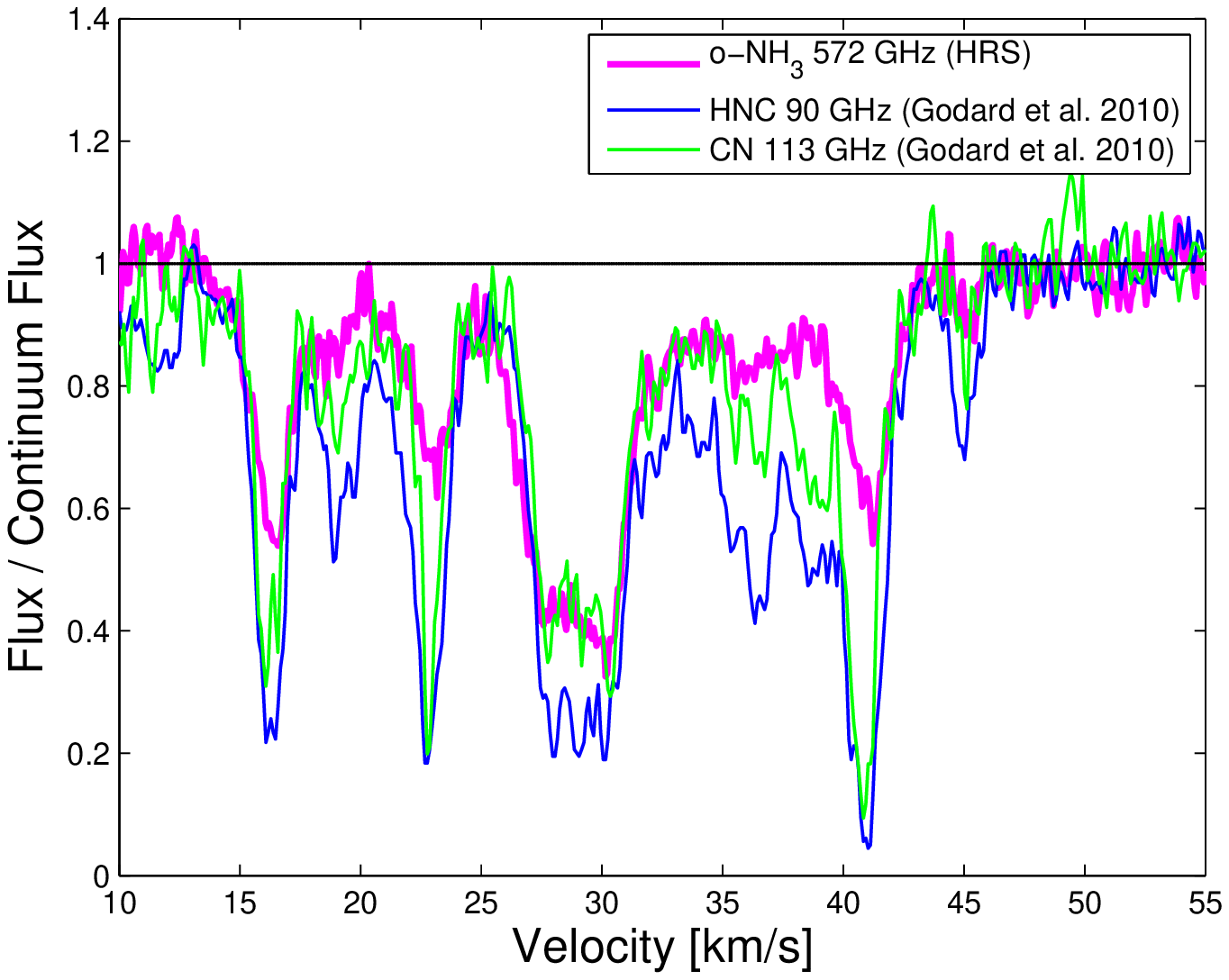}} 
\vspace{.3in}
\subfigure[]{ 
\includegraphics[width=.35\textwidth]{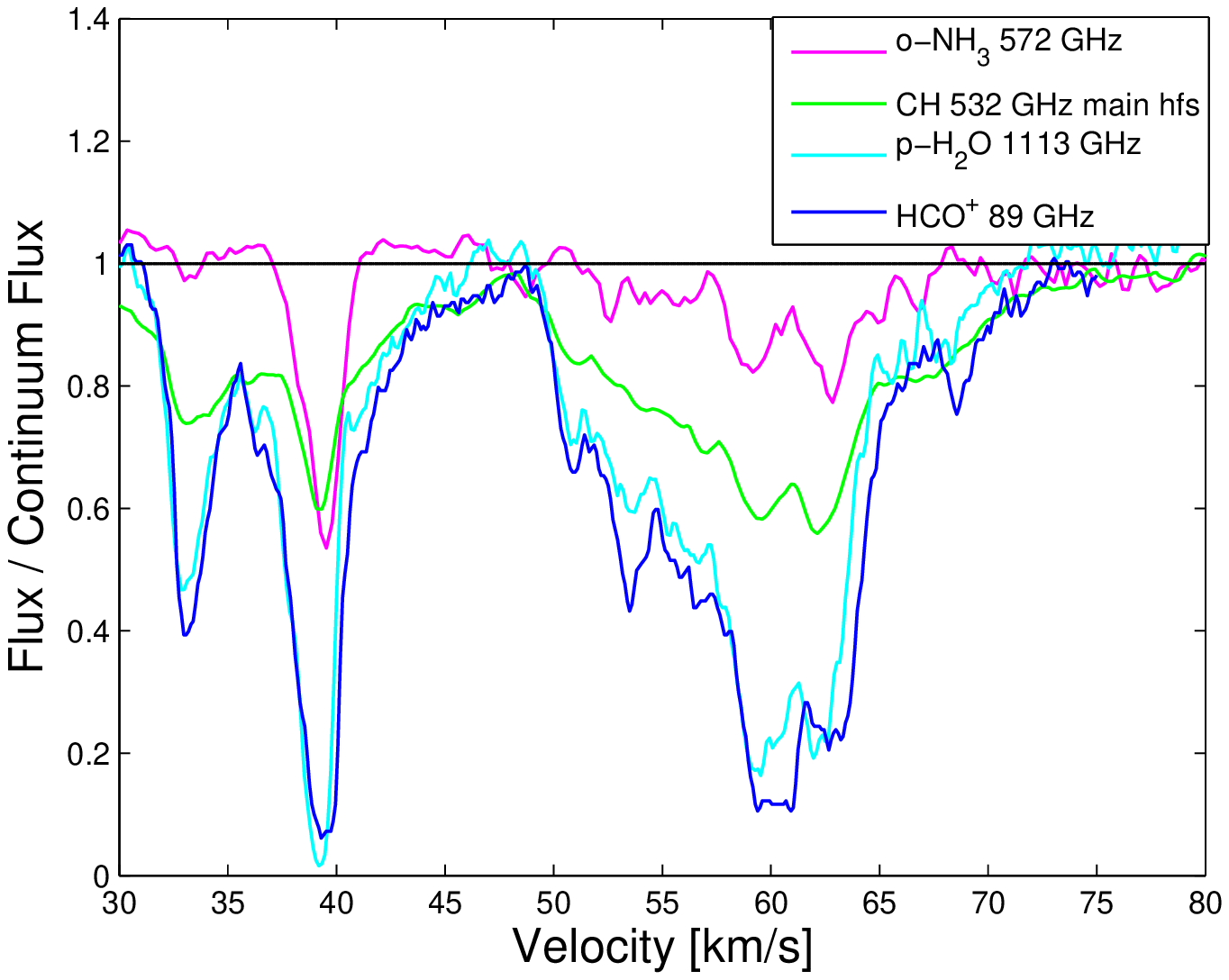}}
\hspace{.3in}
\subfigure[]{
\includegraphics[width=.35\textwidth]{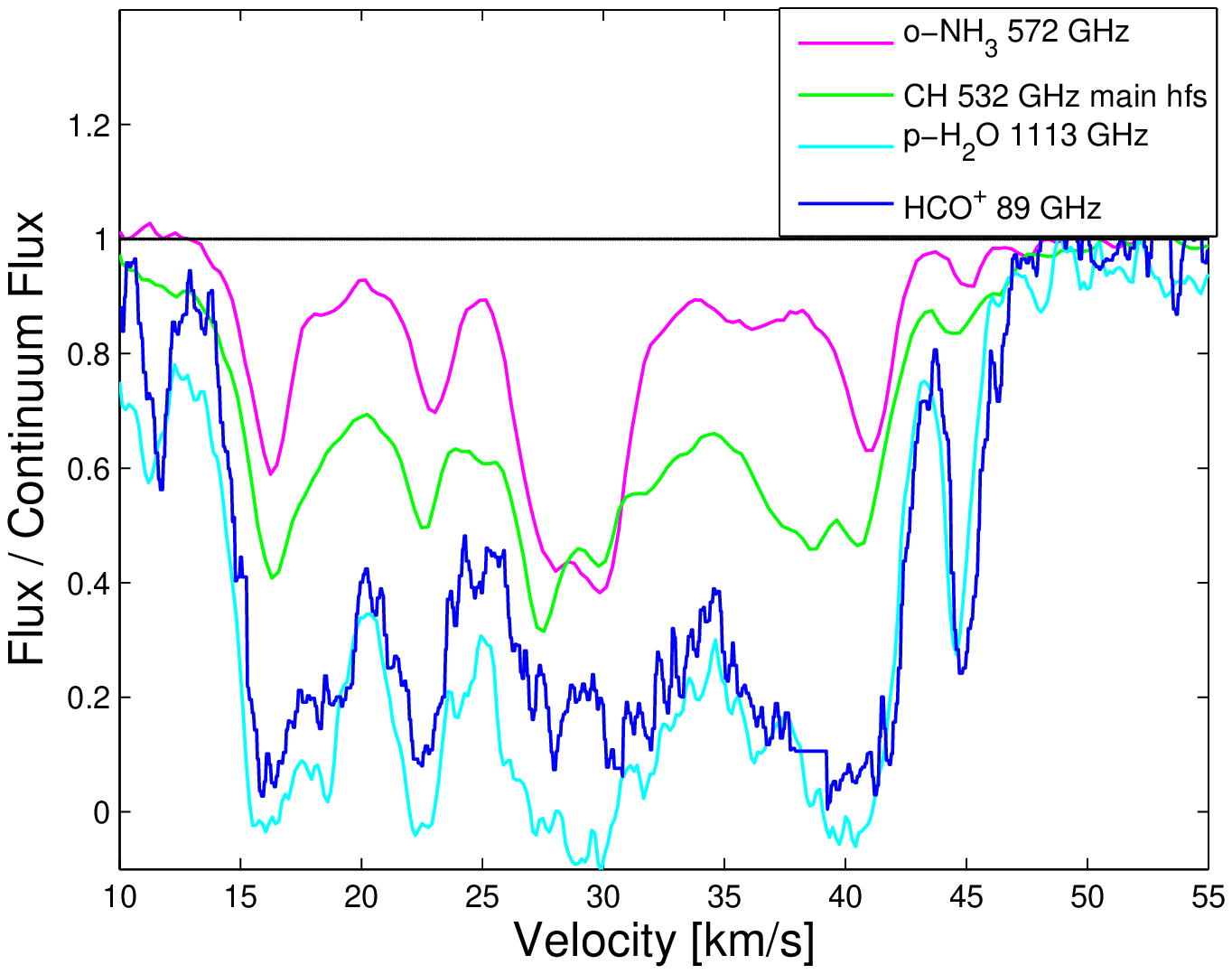}}
\caption{\emph{Comparison of absorption lines towards W49N and G10.6$-$0.4.} The intensities have been normalised to single sideband 
continuum.}
\label{fig: comparison of W31C-W49N 572 vs CH, water, h2o+}
\end{figure*}

\clearpage

\section{Comparison plots of different methods}

\begin{figure*}[\!ht]
\centering
\subfigure[]{
\includegraphics[width=.45\textwidth]{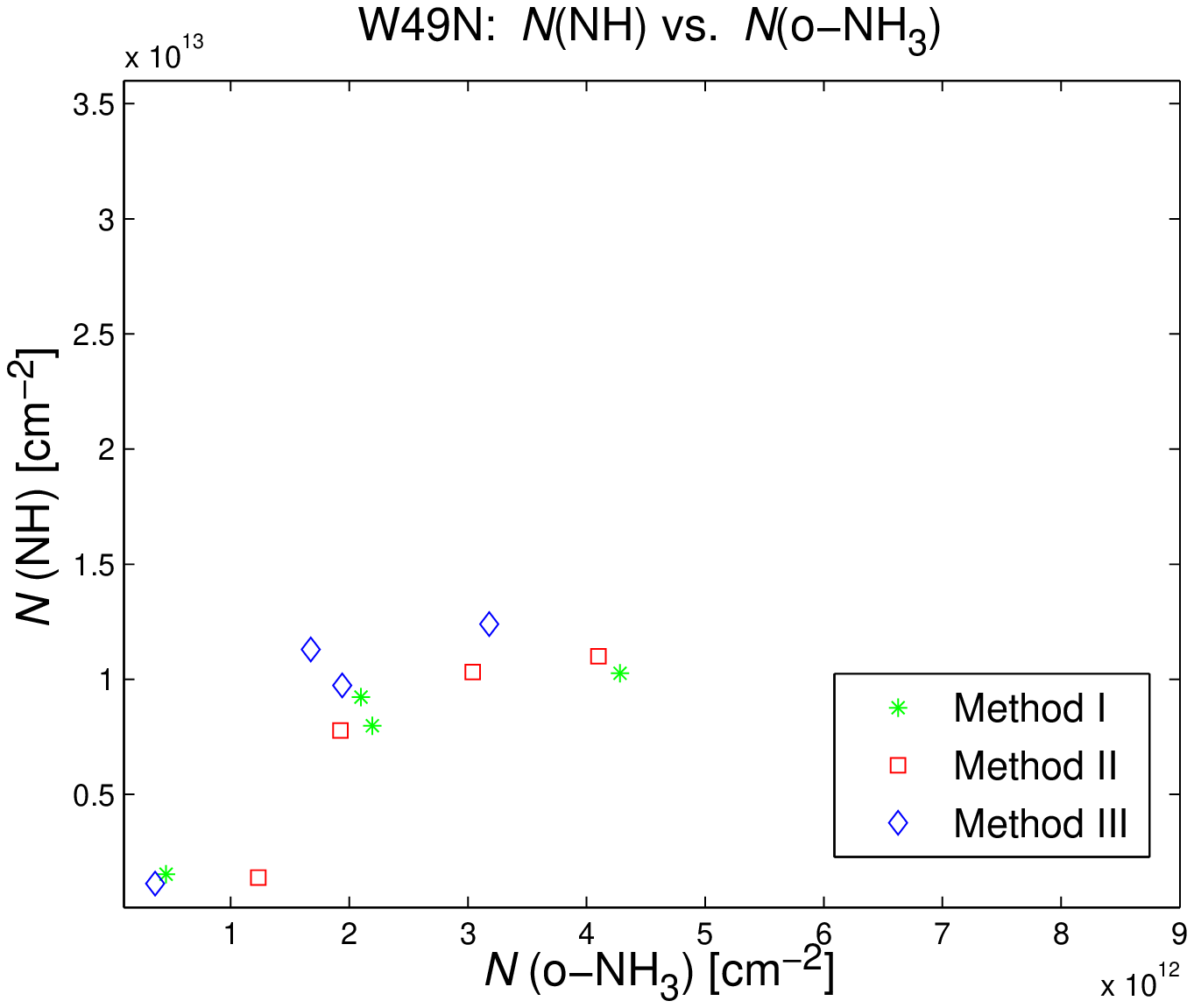}}    
\hspace{.3in}
\subfigure[]{
\includegraphics[width=.45\textwidth]{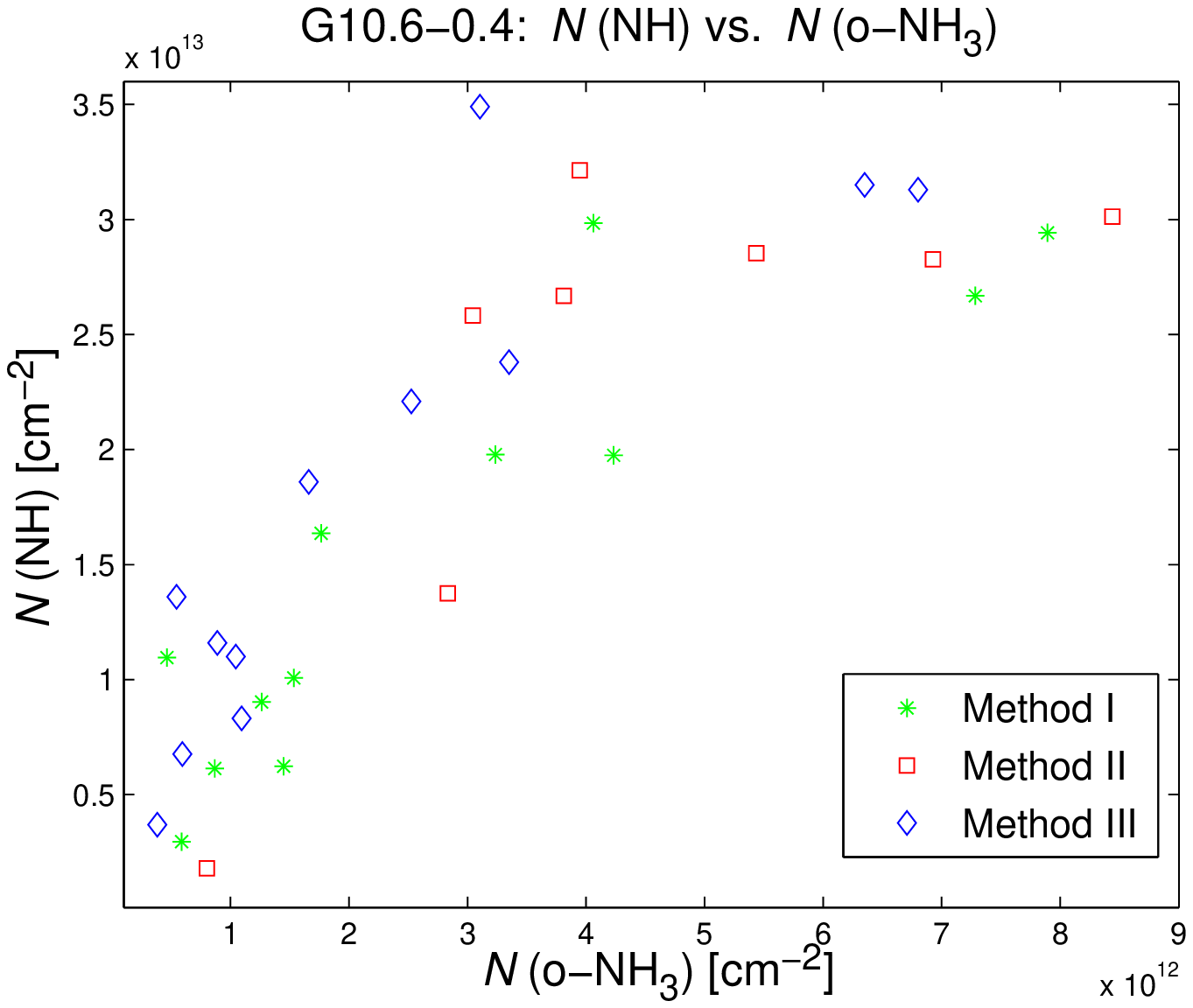}}  
\vspace{.3in}
\subfigure[]{ 
\includegraphics[width=.45\textwidth]{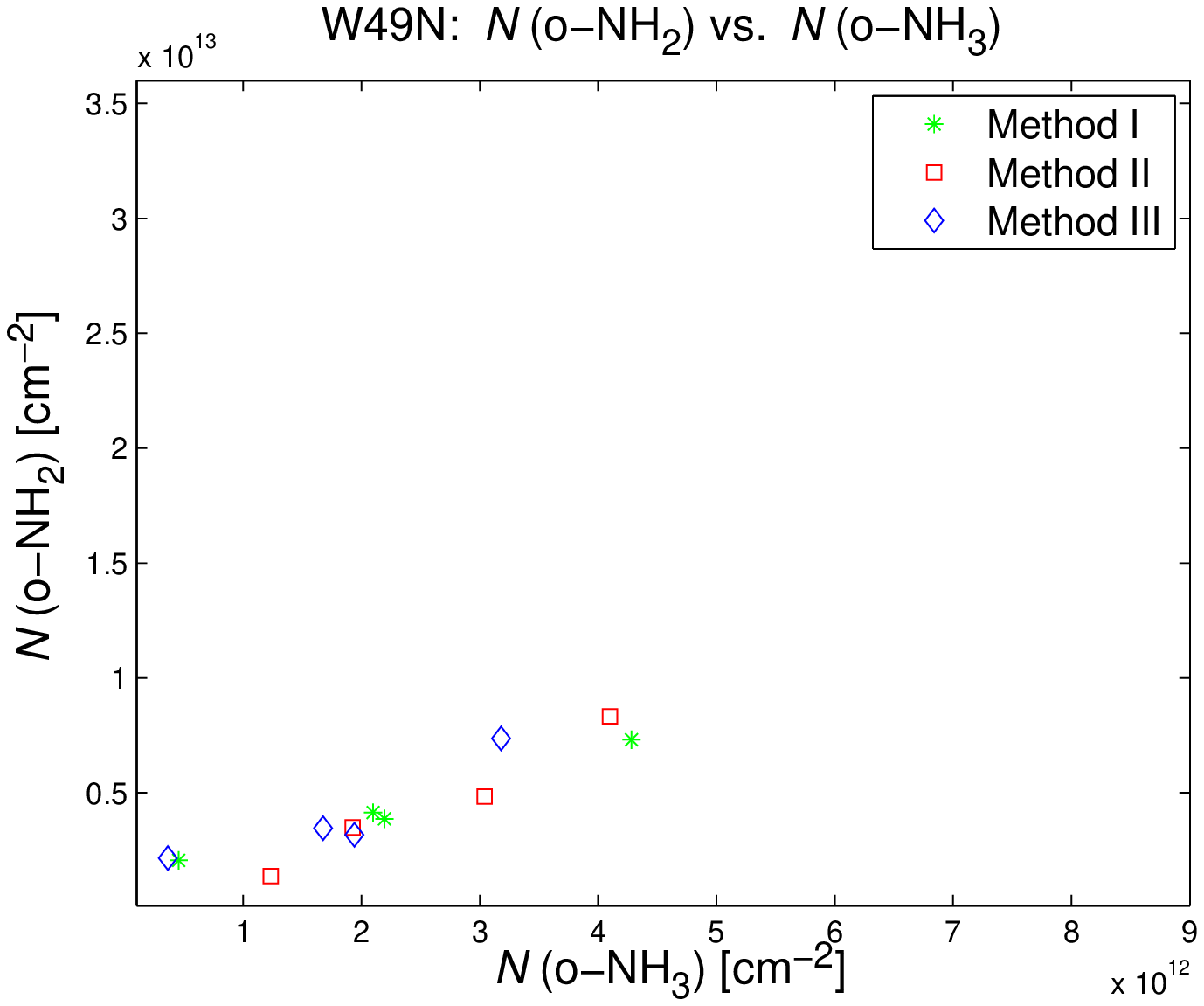}}     
\hspace{.3in}
\subfigure[]{
\includegraphics[width=.45\textwidth]{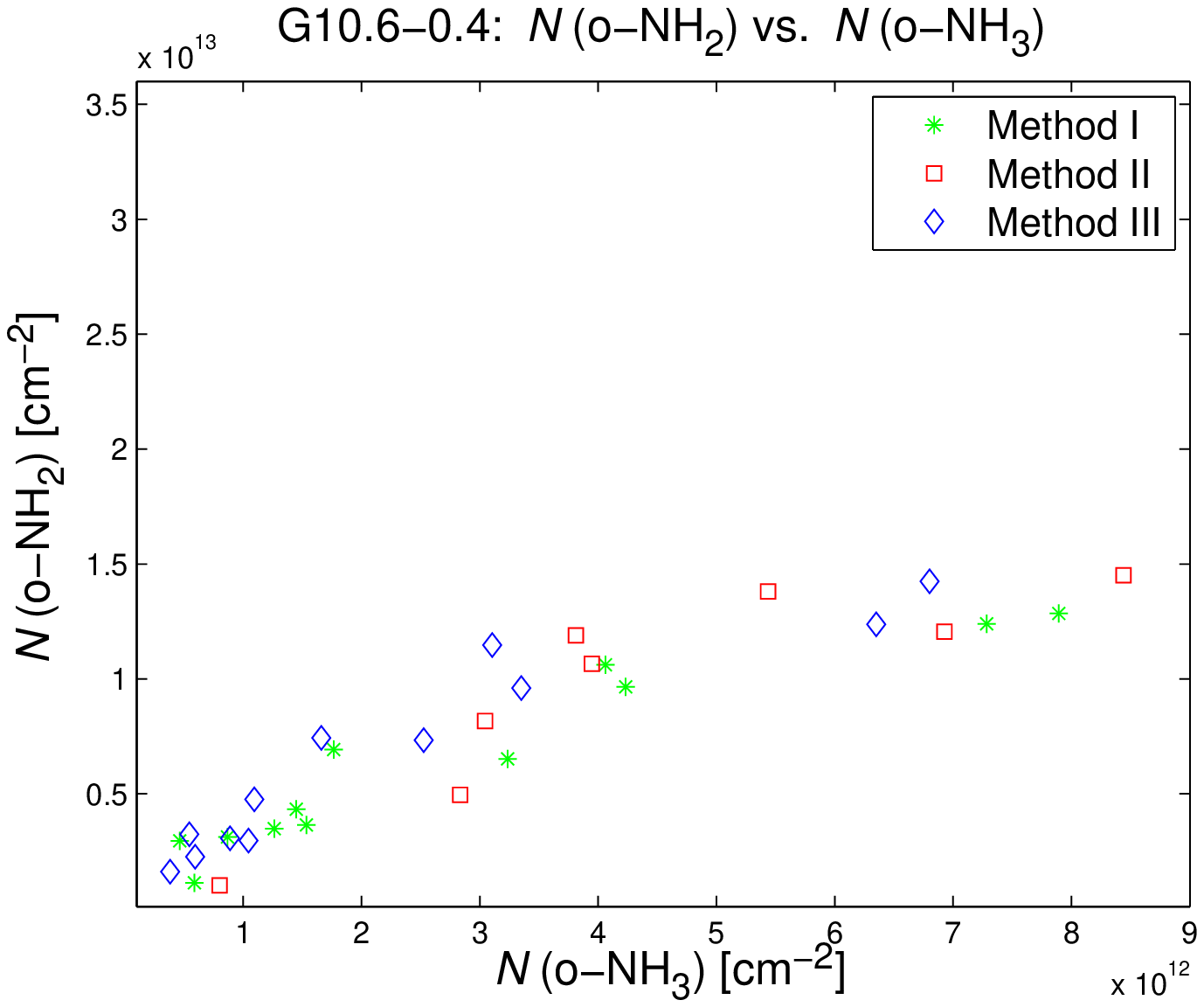}}   
\vspace{.3in}
\subfigure[]{ 
\includegraphics[width=.45\textwidth]{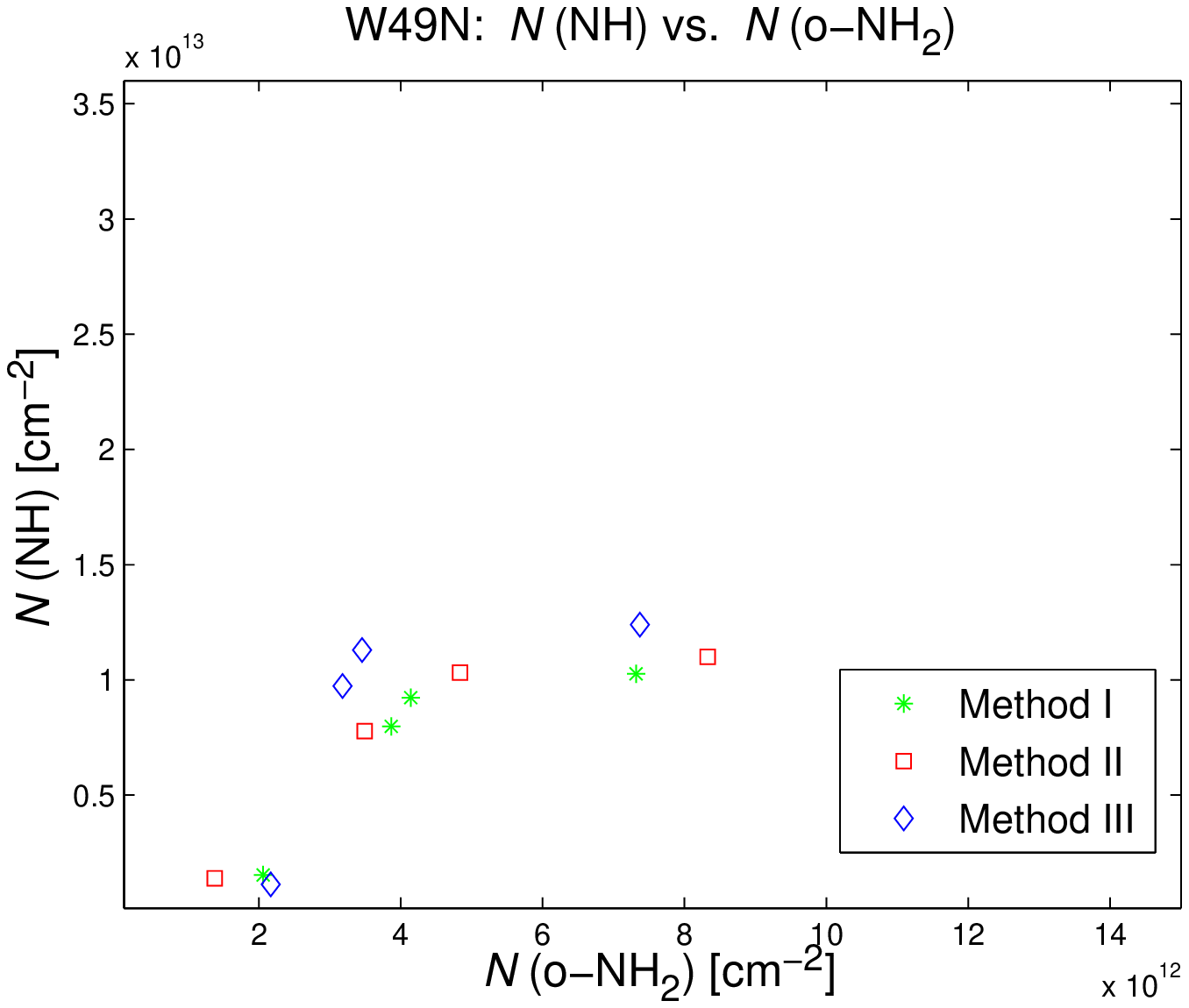}} 
\hspace{.3in}
\subfigure[]{
\includegraphics[width=.45\textwidth]{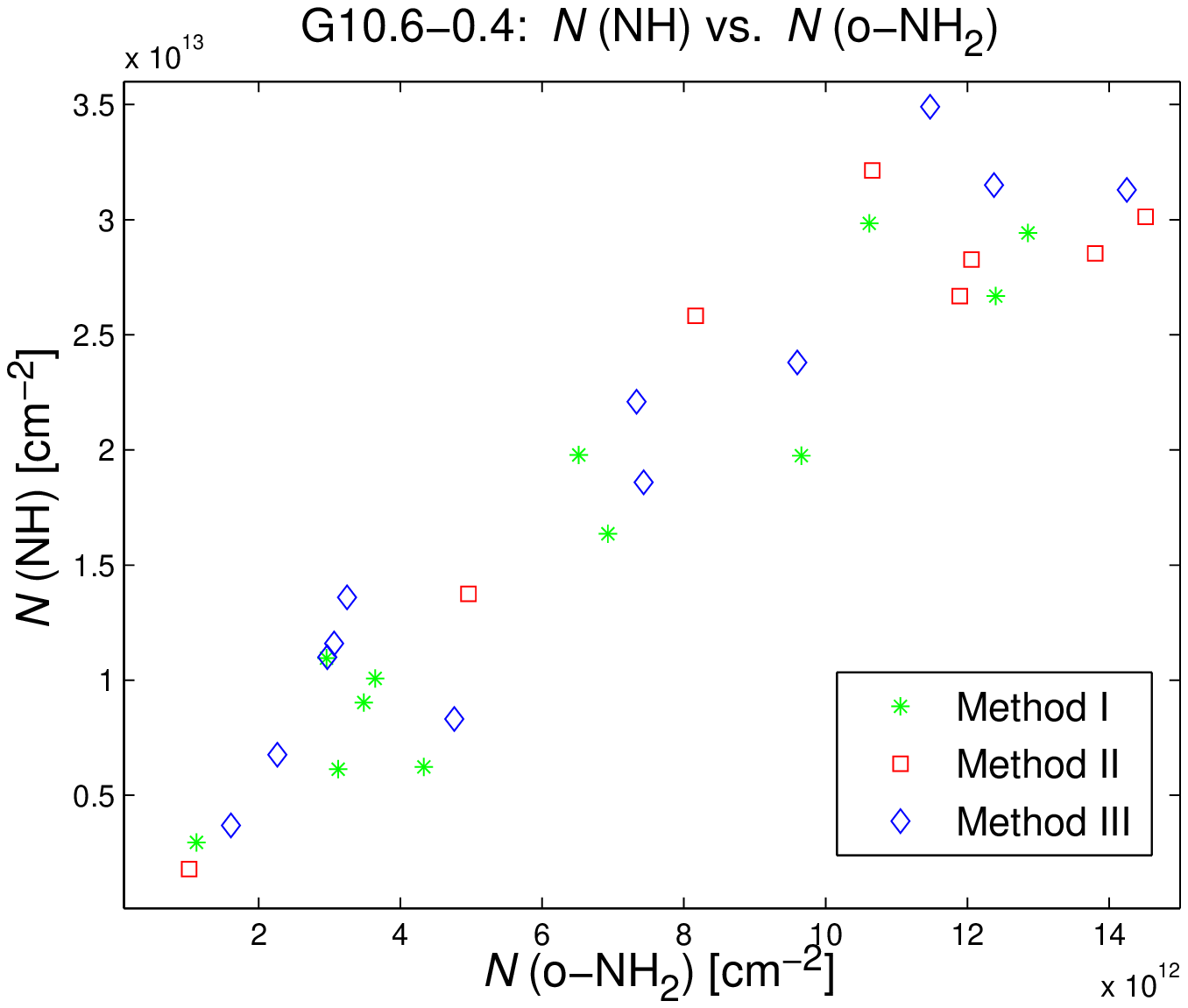}}  
\caption{\emph{Column density comparison plots.} The results from all three methods are plotted. The scatter in the results can
be considered as an estimate of the errors.}
\label{Fig: column density plots 1}
\end{figure*}

\begin{figure*}[\!ht]
\centering
\subfigure[]{
\includegraphics[width=.45\textwidth]{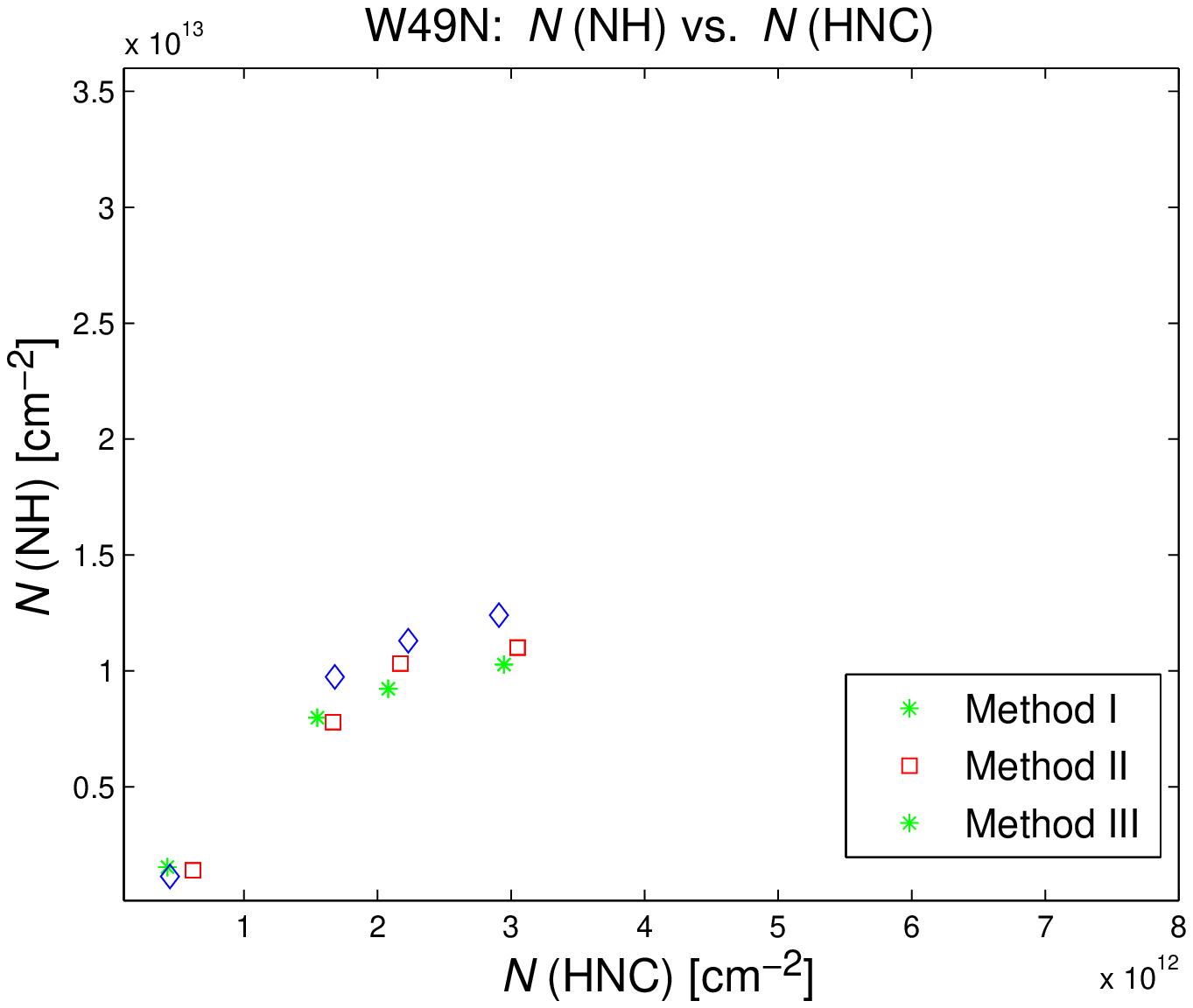}}   
\hspace{.3in}
\subfigure[]{
\includegraphics[width=.45\textwidth]{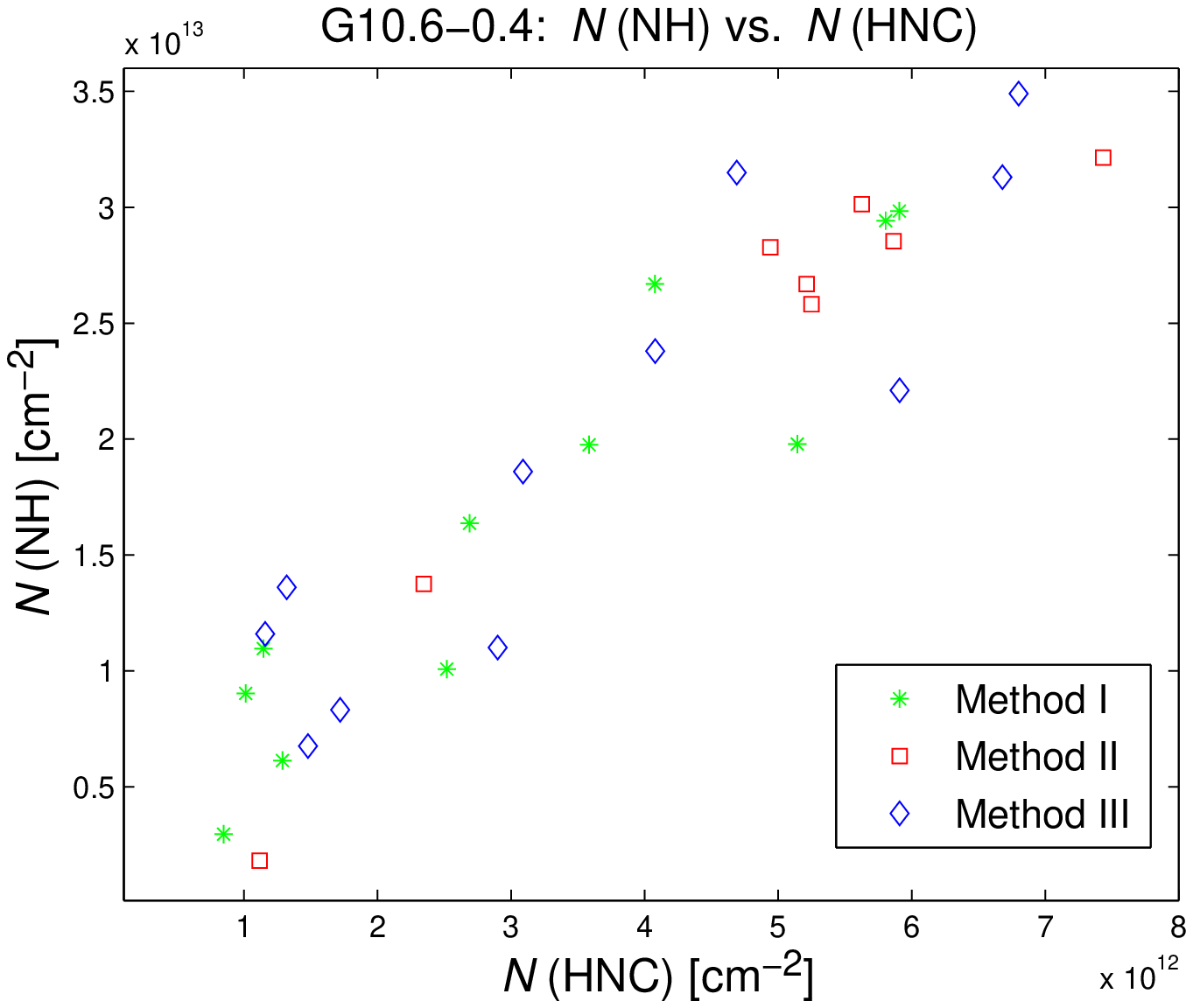}}  
\vspace{.3in}
\subfigure[]{ 
\includegraphics[width=.45\textwidth]{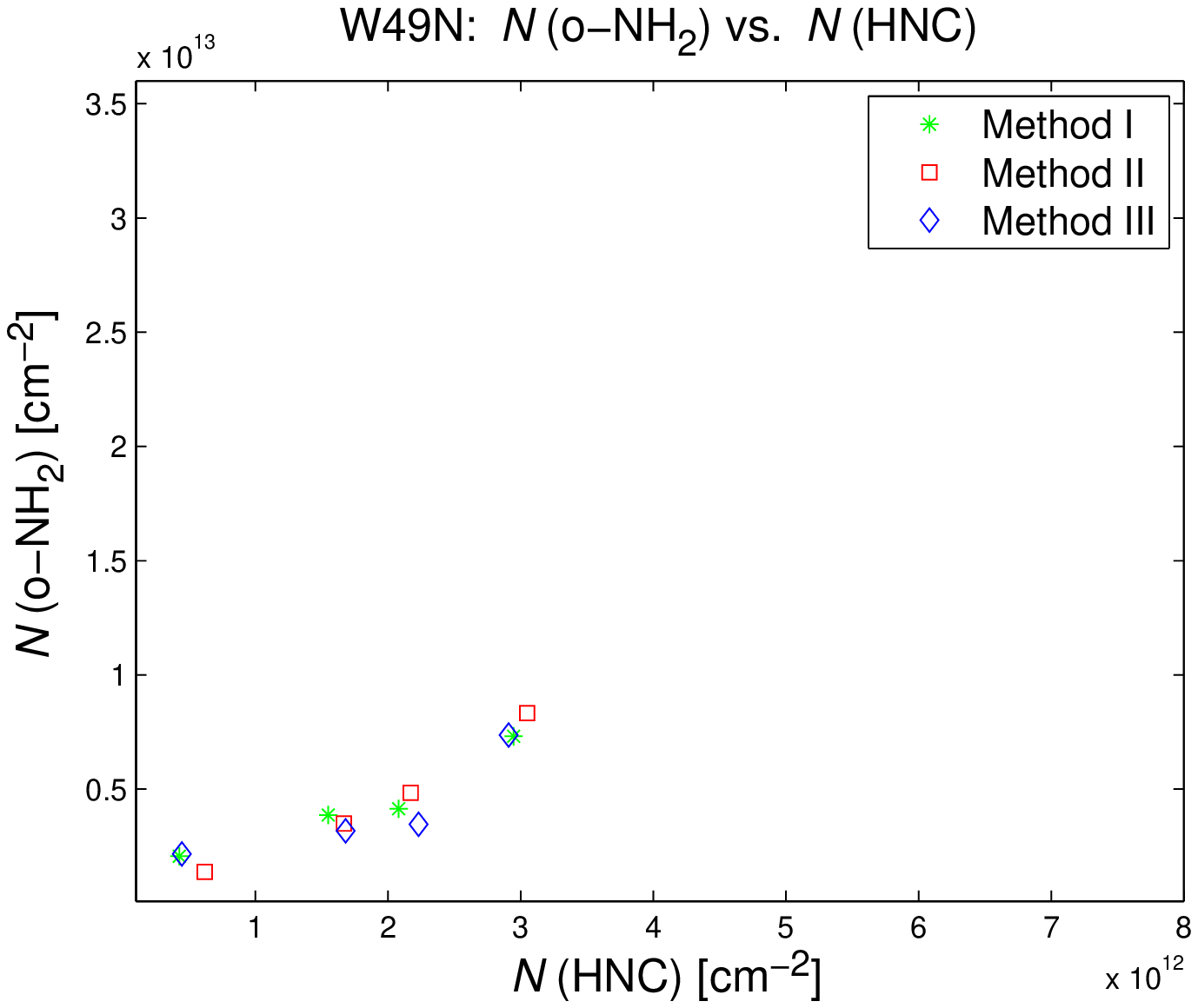}}   
\hspace{.3in}
\subfigure[]{
\includegraphics[width=.45\textwidth]{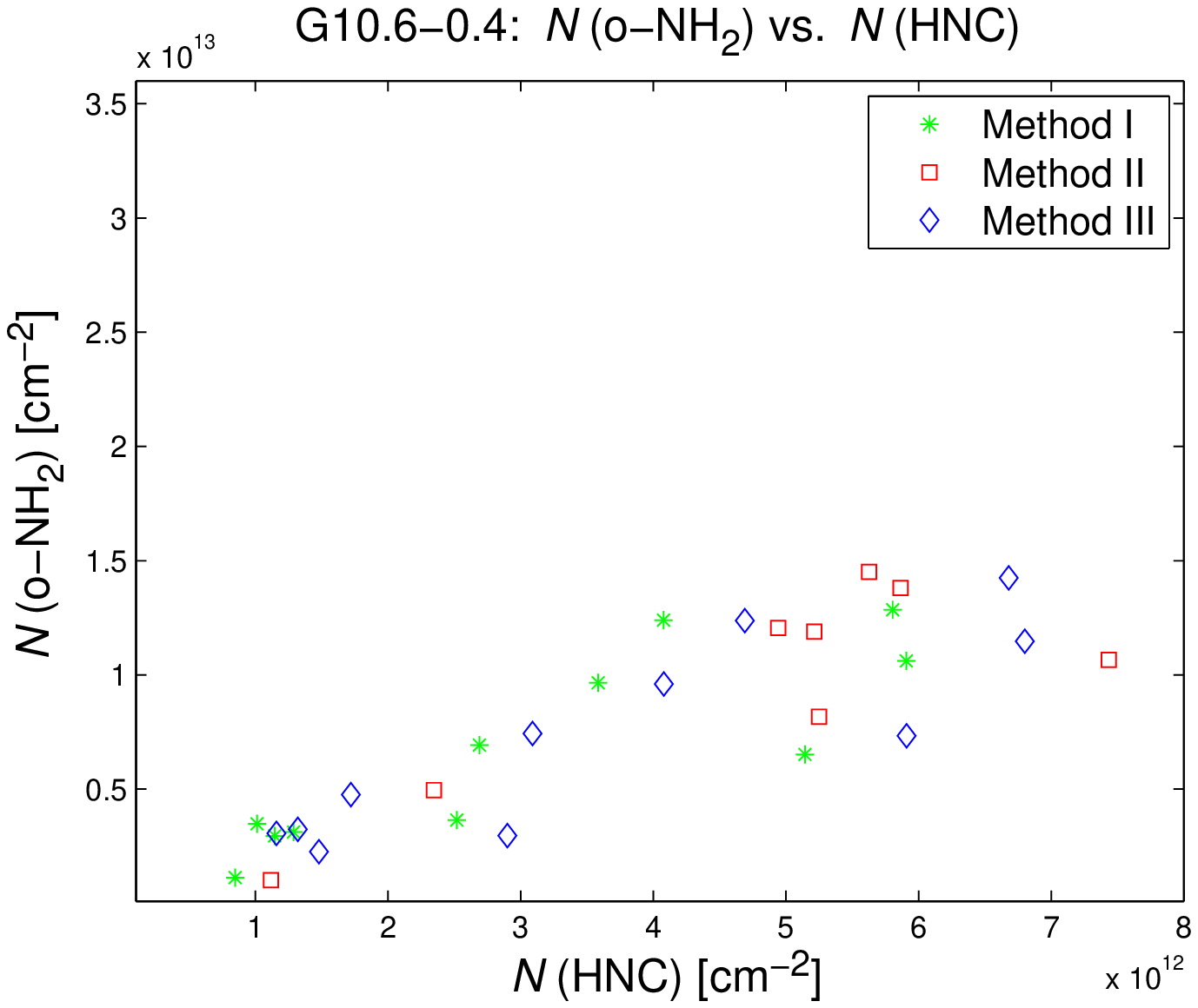}}   
\vspace{.3in}
\subfigure[]{ 
\includegraphics[width=.45\textwidth]{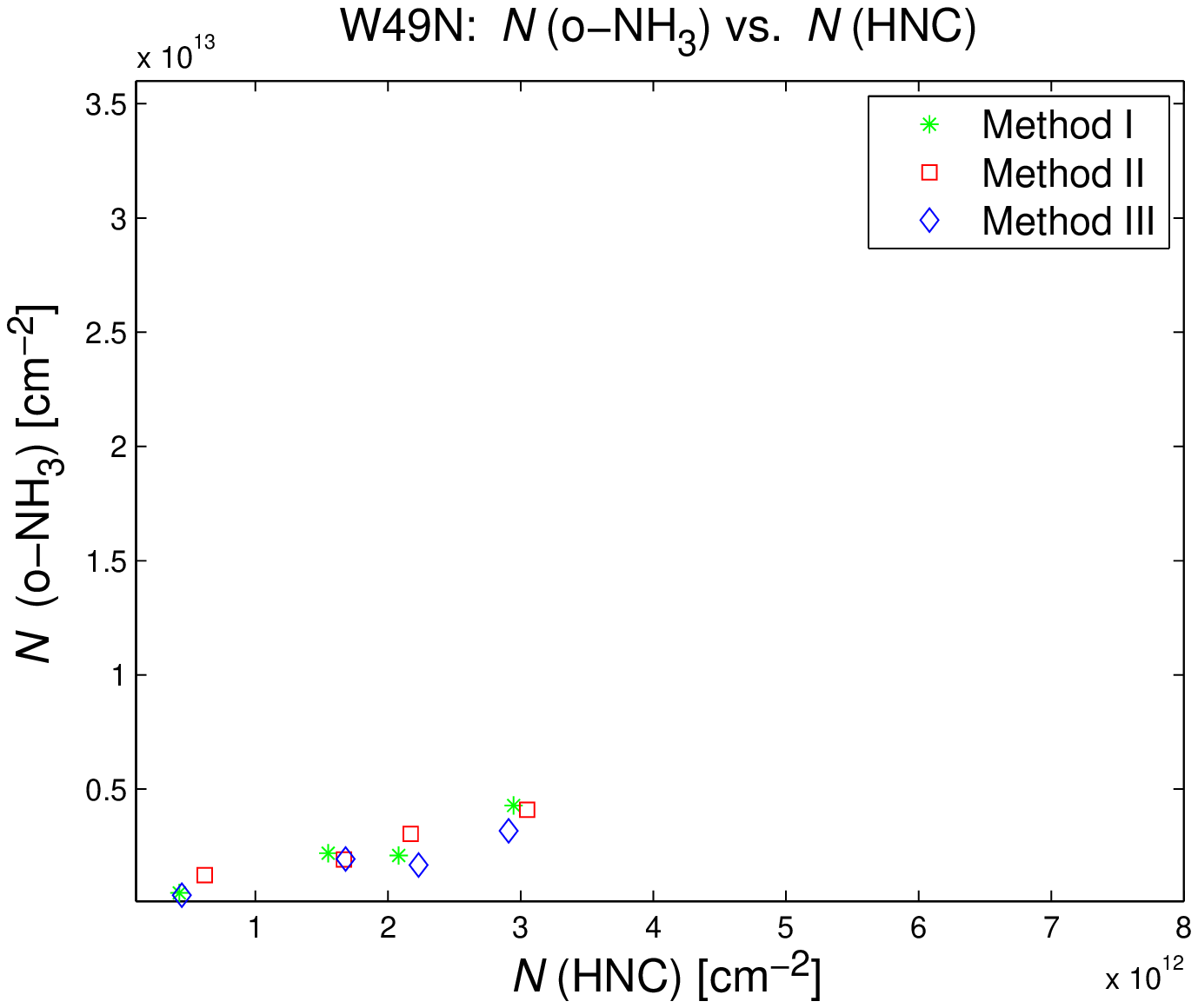}}    
\hspace{.3in}
\subfigure[]{
\includegraphics[width=.45\textwidth]{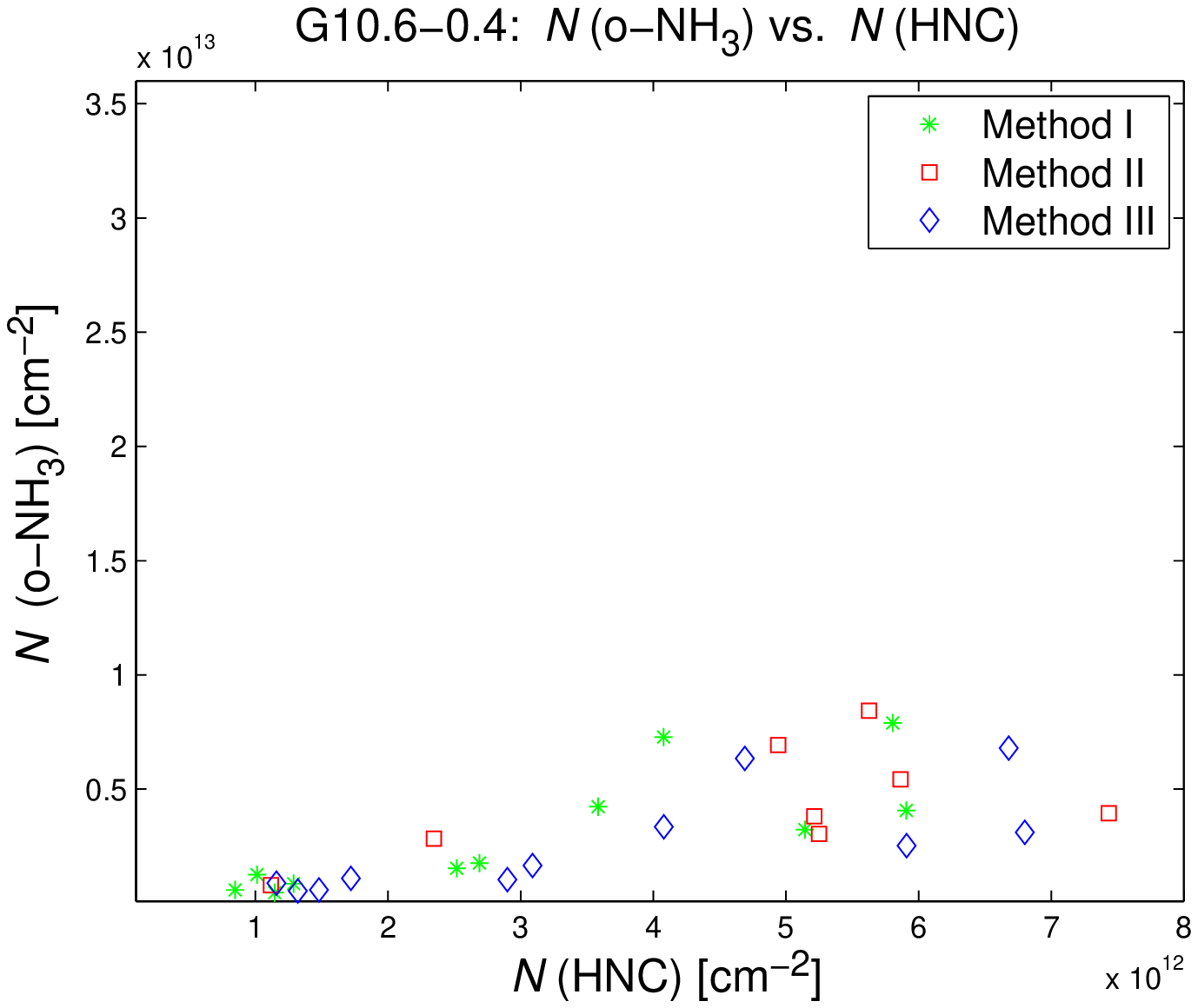}}   
\caption{\emph{Cont: Column density comparison plots~4}. Notation as in Fig.~\ref{Fig: column density plots 1}.}
\label{Fig: column density plots 4}
\end{figure*}

In Figs.~\ref{Fig: column density plots 1}\,--\,\ref{Fig: column density plots 4} we plot the resulting column densities 
from our three different methods.

Note that since Method~I and Method~III use Gaussians with   
smaller line widths than the velocity bins
of Method~II towards G10.6$-$0.4, the number of velocity components is not the same in this source. 
The +33~km~s$^{-1}$~component towards  W49N  is not used in Method~II, which on the other hand models 
a velocity bin that is difficult to fit using Gaussians in Method~I and III \mbox{(50\,--\,57~km~s$^{-1}$)}.  
The comparison of the nitrogen hydrides vs. CH in  W49N, in two of the four velocity bins, 
shows the only cases in  which the results do not agree reasonably well between the methods. 
The resulting CH column densities also show a larger spread between the methods than the other species. 
The reason is not fully understood, but is probably  caused by the different approaches to the CH modelling.
Method~II uses the  deconvolved CH spectra as a template for the nitrogen hydrides and is thereby
trying to fit the broader CH absorption to the more narrow features of the nitrogen hydrides with moderate
success.  
The other two methods use an opposite approach: they use the output of 
the fitting of the nitrogen hydrides as an input to CH to fit Gaussians only in the same parts in velocity space, 
and are thereby trying to fit narrow features to the broader CH absorption.
The fit   is good in some parts of the velocity space, and very bad in the velocity space in which CH has 
absorption but the nitrogen hydrides do not, which is expected.

\clearpage

\begin{figure*}[\!ht]
\centering
\subfigure[]{
\includegraphics[width=.45\textwidth]{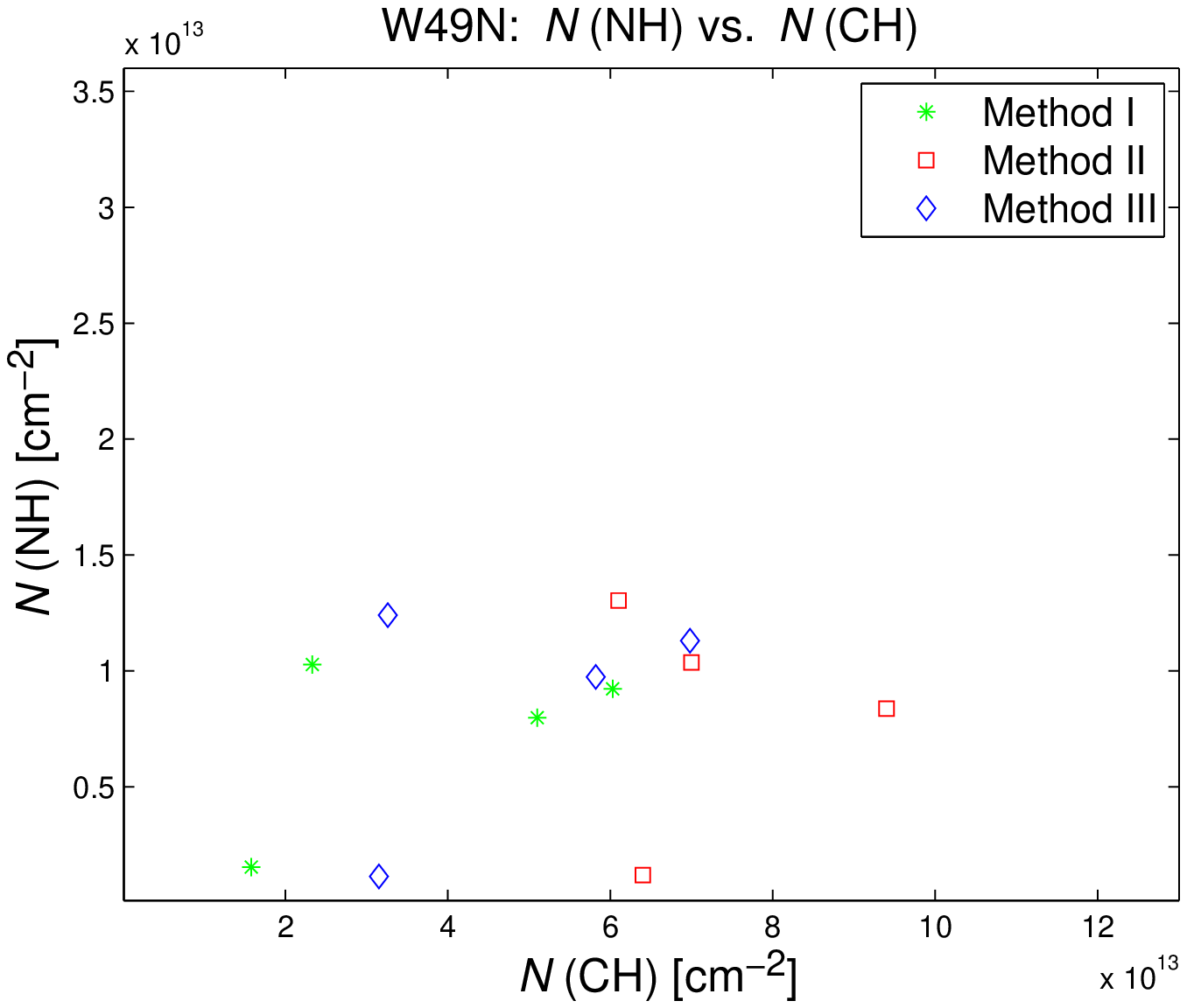}}    
\hspace{.3in}
\subfigure[]{
\includegraphics[width=.45\textwidth]{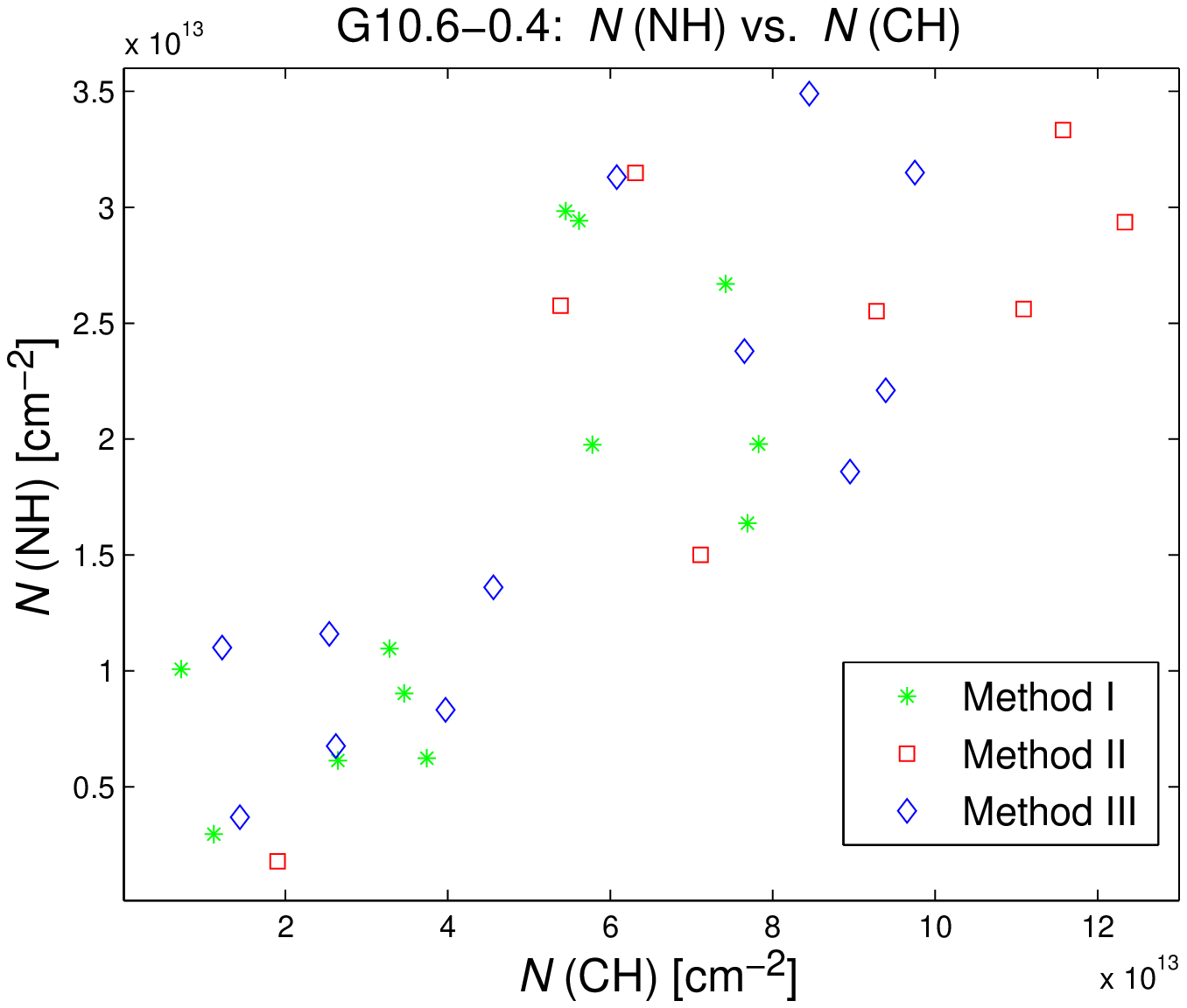}}   
\vspace{.3in}
\subfigure[]{
\includegraphics[width=.45\textwidth]{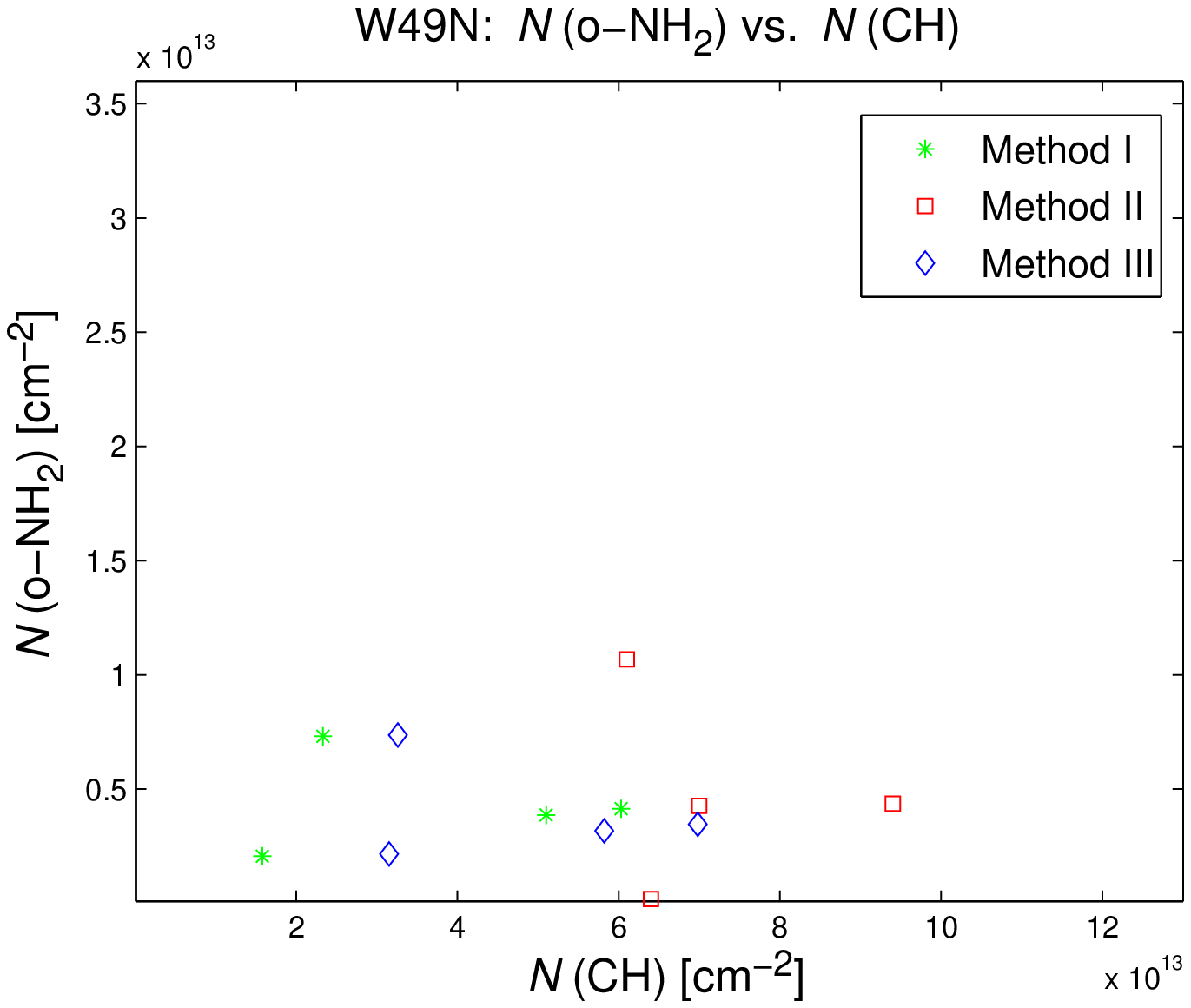}}    
\hspace{.3in}
\subfigure[]{
\includegraphics[width=.45\textwidth]{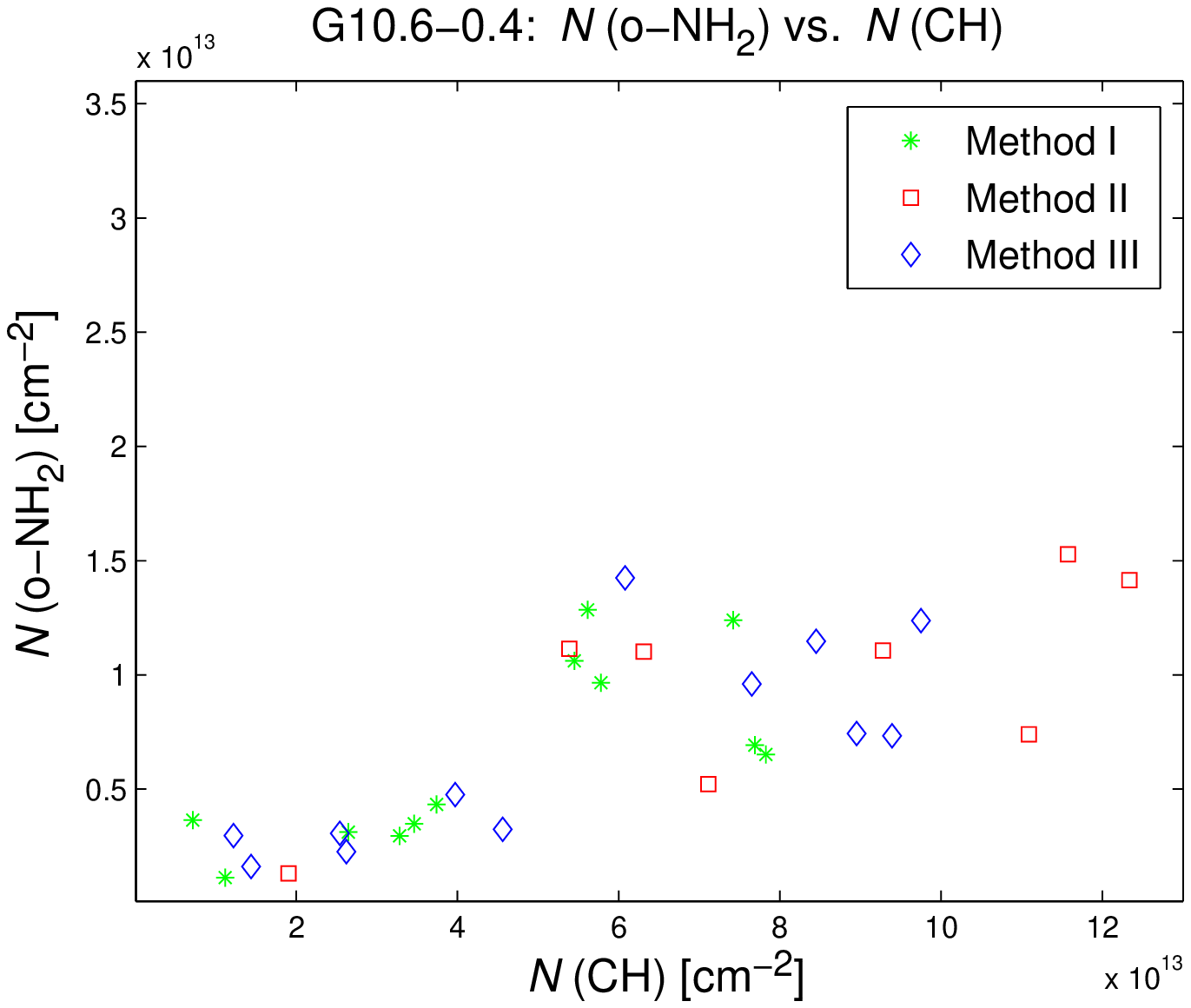}}   
\vspace{.3in}
\subfigure[]{ 
\includegraphics[width=.45\textwidth]{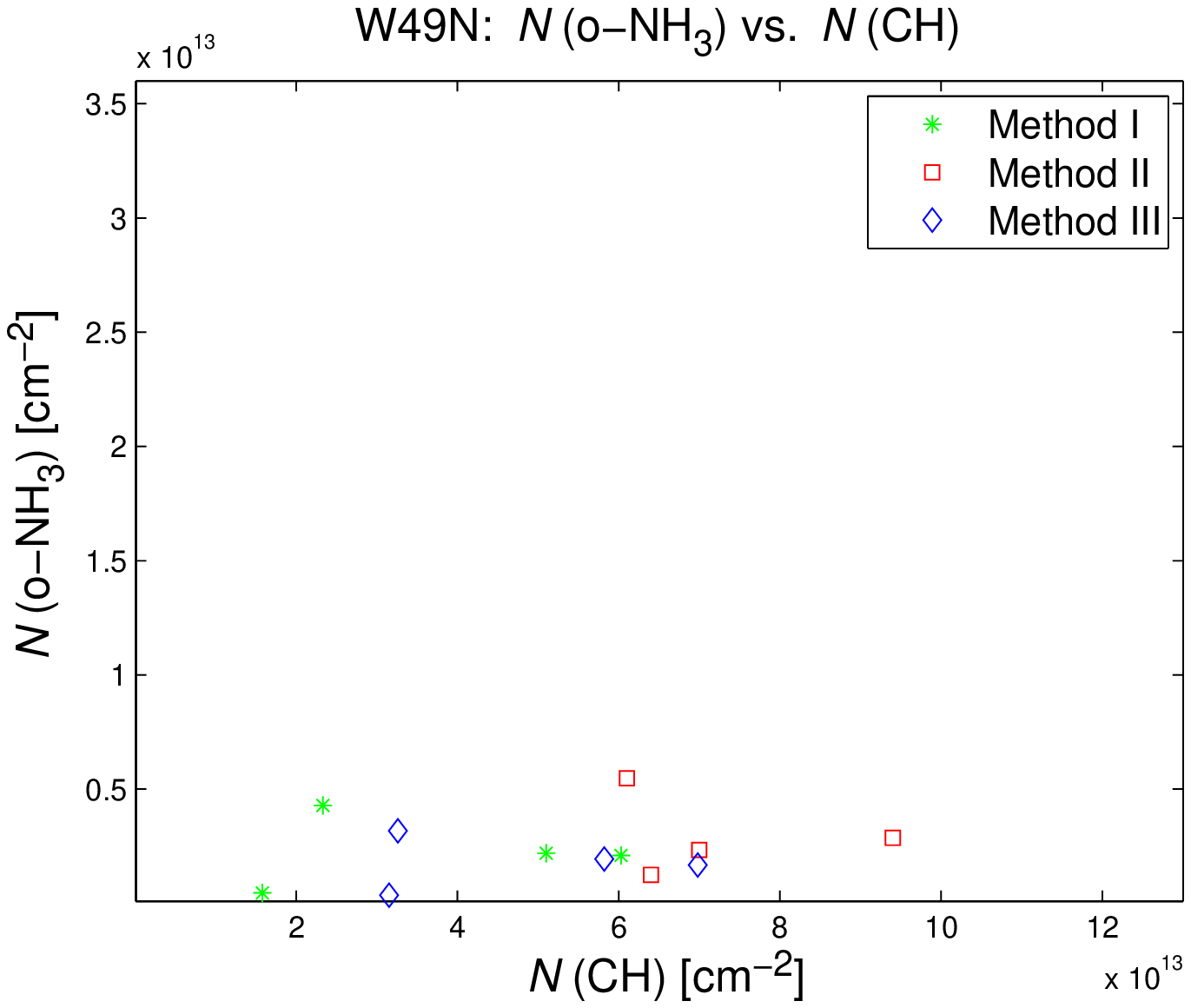}}   
\hspace{.3in}
\subfigure[]{
\includegraphics[width=.45\textwidth]{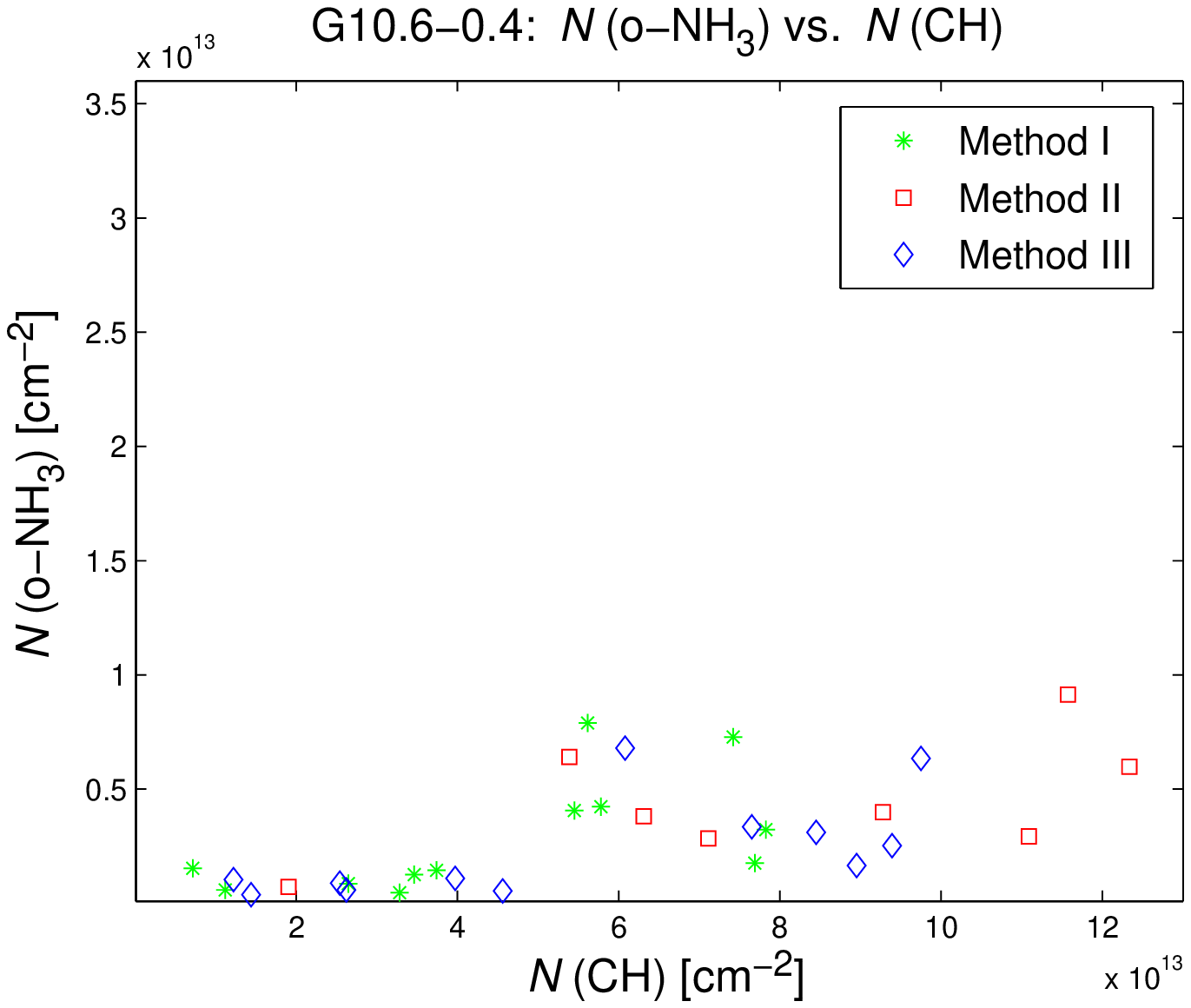}}   
\caption{\emph{Cont: Column density comparison plots~2}. Notation as in Fig.~\ref{Fig: column density plots 1}.}
\label{Fig: column density plots 2}
\end{figure*}

\clearpage

\begin{figure*}[\!ht]
\centering
\subfigure[]{
\includegraphics[width=.45\textwidth]{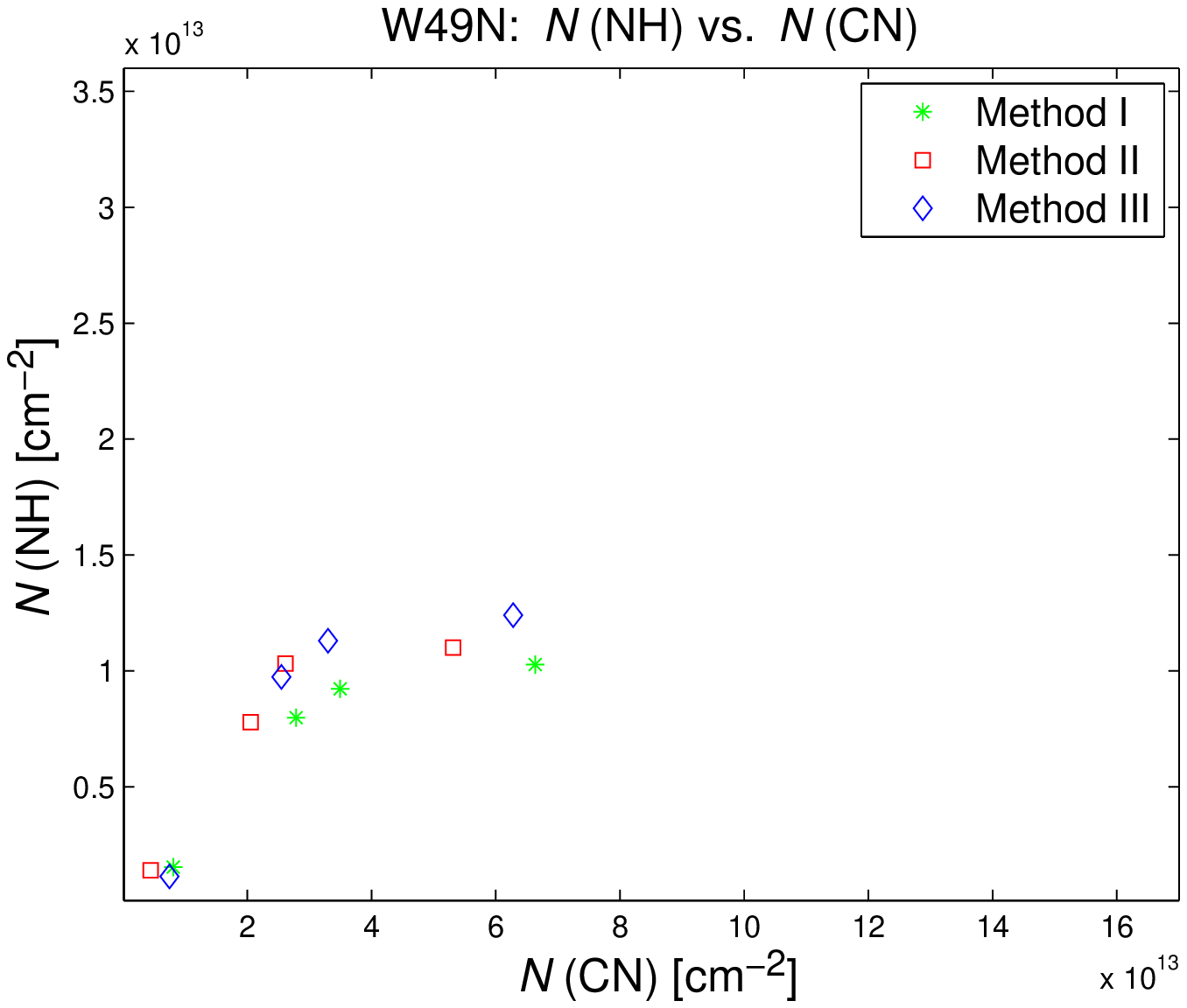}}    
\hspace{.3in}
\subfigure[]{
\includegraphics[width=.45\textwidth]{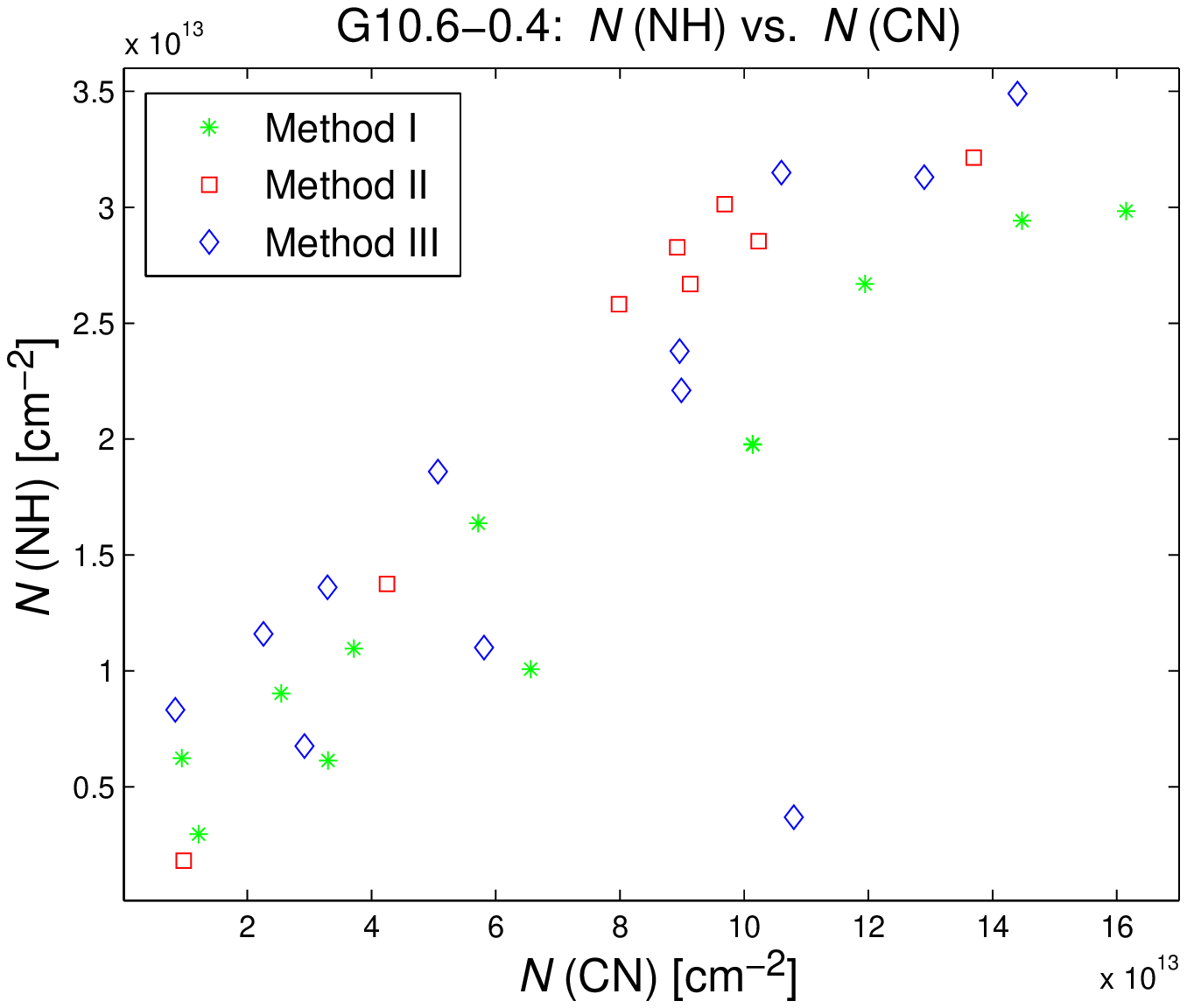}}   
\vspace{.3in}
\subfigure[]{
\includegraphics[width=.45\textwidth]{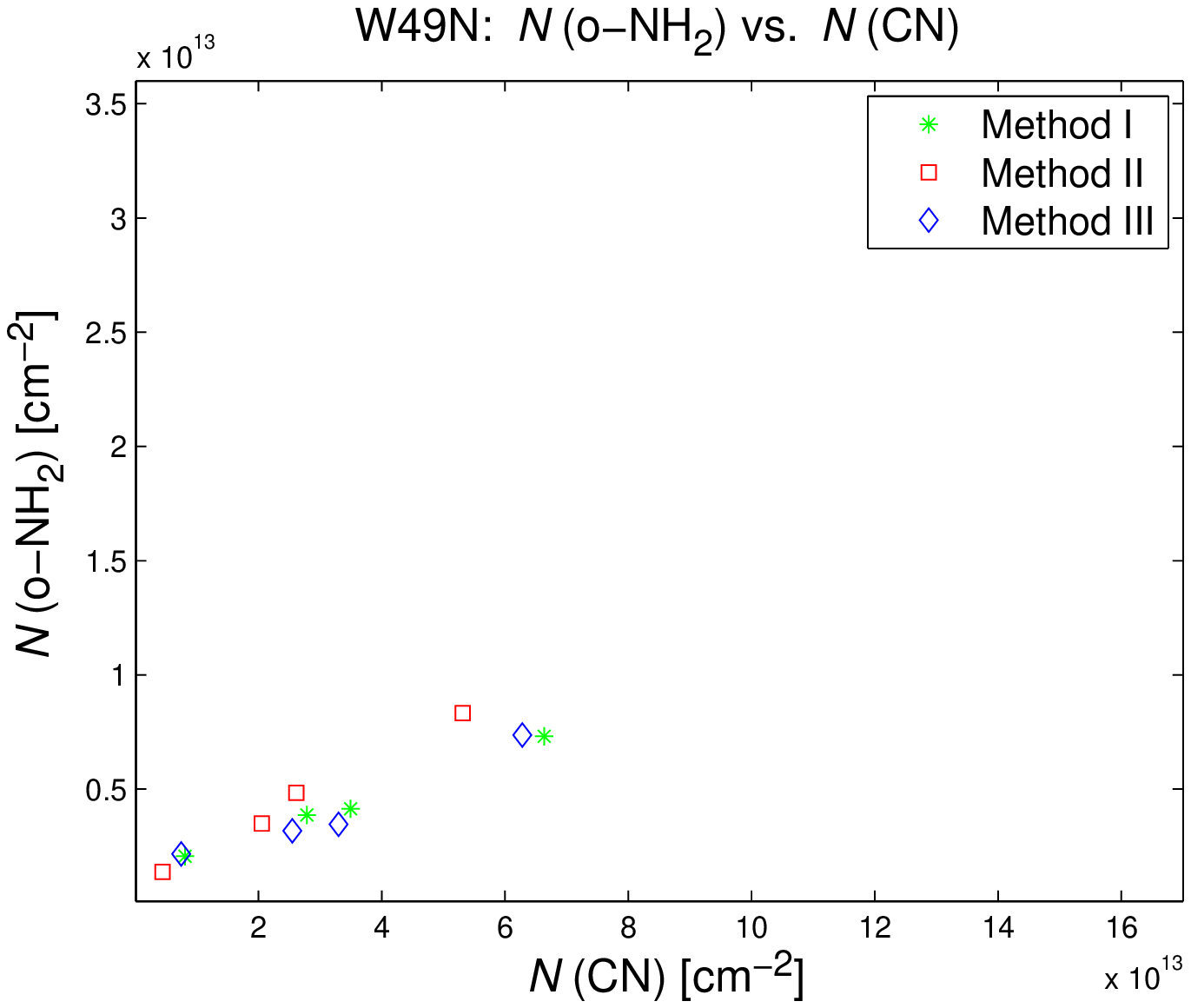}}    
\hspace{.3in}
\subfigure[]{
\includegraphics[width=.45\textwidth]{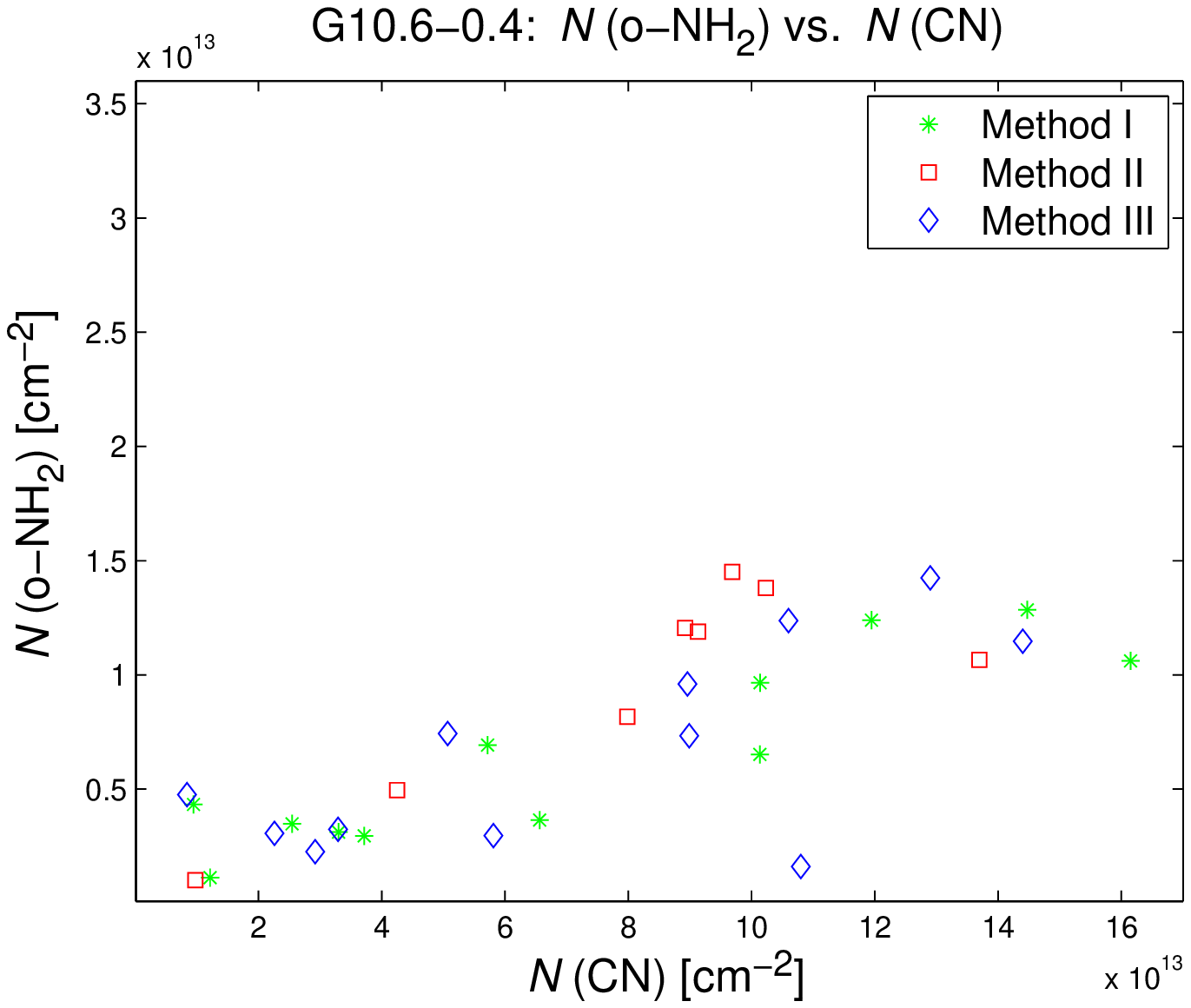}}   
\vspace{.3in}
\subfigure[]{ 
\includegraphics[width=.45\textwidth]{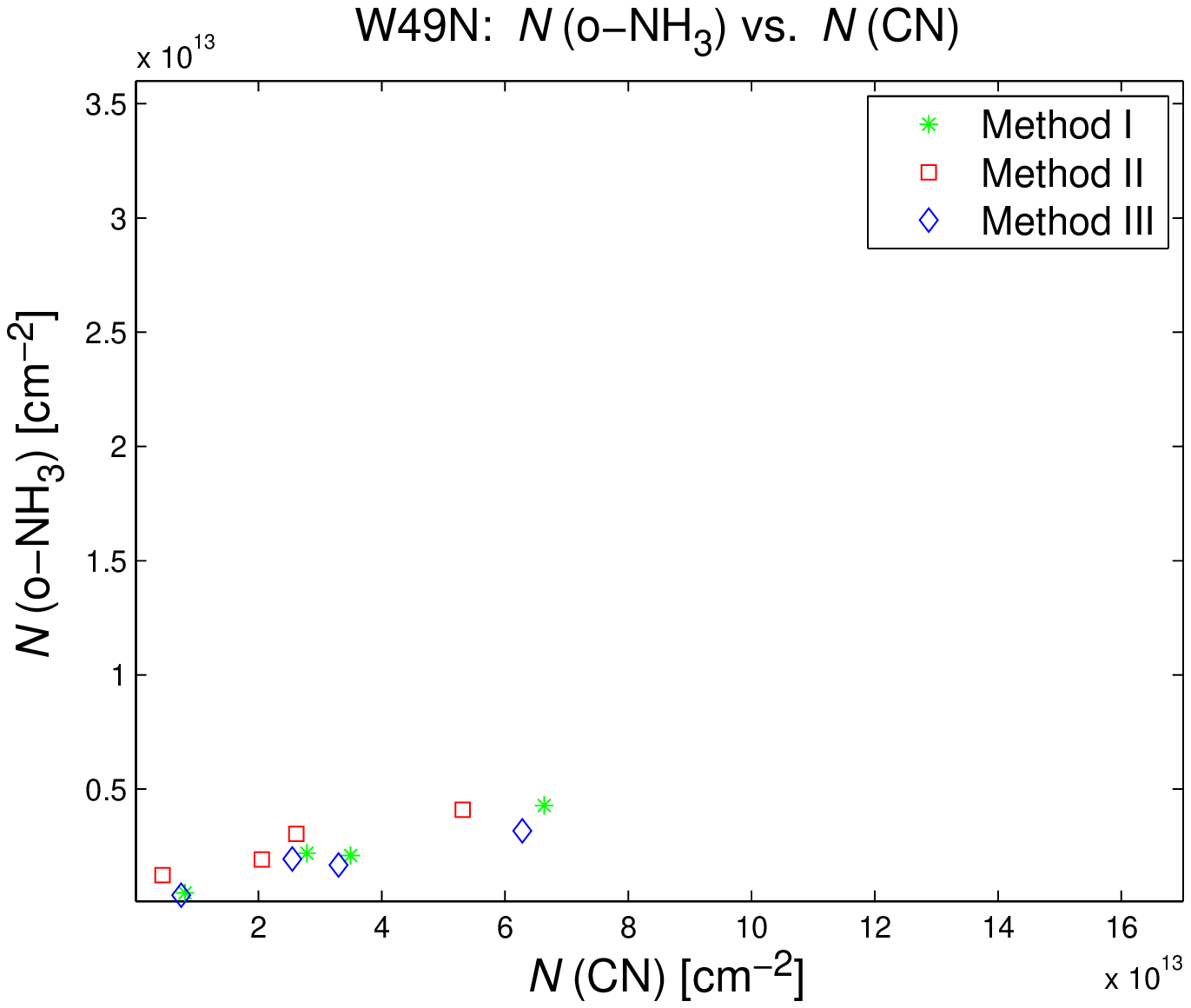}}    
\hspace{.3in}
\subfigure[]{
\includegraphics[width=.45\textwidth]{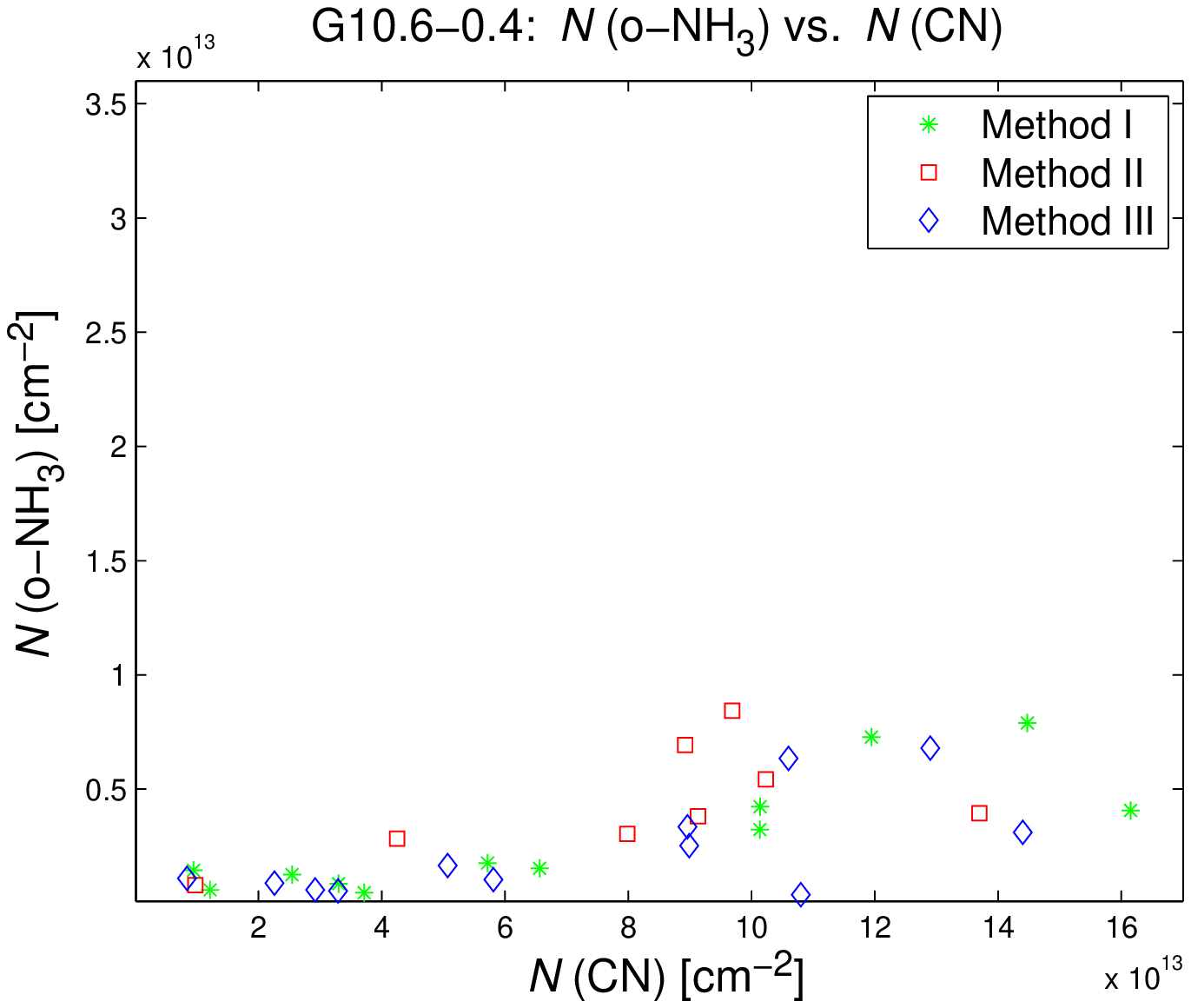}}    
\caption{\emph{Cont: Column density comparison plots~3}. Notation as in Fig.~\ref{Fig: column density plots 1}.}
\label{Fig: column density plots 3}
\end{figure*}

\clearpage

\end{document}